\documentclass[11pt,letterpaper]{article}
\usepackage{amssymb}

\usepackage{amsmath}
\usepackage{fullpage}
\usepackage{float}
\usepackage[margin=.75in]{geometry}
\usepackage[hang]{caption}
\usepackage{color}
\usepackage{graphicx}
\usepackage{psfrag}
\usepackage[english]{babel}
\usepackage{wasysym}
\usepackage{textcomp}
\usepackage{bm}
\usepackage{enumitem}
\usepackage{adjustbox}
\usepackage{amsthm,array}
\usepackage{tabularx}
\usepackage{booktabs}
\usepackage{rotating}
\usepackage{bigstrut}
\usepackage{multirow}
\usepackage{mathrsfs}
\usepackage{subfigure}
\RequirePackage[round,authoryear,longnamesfirst]{natbib}
\usepackage{hyperref}
\usepackage[svgnames]{xcolor}
\hypersetup{
	colorlinks=true,
	linkcolor=DarkRed,
	citecolor=NavyBlue}
\usepackage[toc,title,page]{appendix}

\usepackage[ruled]{algorithm2e}

\newtheorem{theorem}{Theorem}[section]

\newtheorem{assumption}{Assumption}

\newtheorem{lemma}{Lemma}[section]

\theoremstyle{definition}
\newtheorem{example}{Example}[section]
\newtheorem{rem}{Remark}[section]

\usepackage[english]{babel}
\theoremstyle{remark}

\newcommand{\diag}{\text{diag}}

\newcommand{\convP}{\stackrel{p}{\longrightarrow}}
\newcommand{\convD}{\xrightarrow{\enskip d \enskip}}

\newcommand{\N}{\mathcal{N}}
\newcommand{\eps}{\varepsilon}

\renewcommand{\eps}{\varepsilon}

\usepackage{xr}

\newlength{\bibitemsep}
\newlength{\bibparskip}
\let\oldthebibliography\thebibliography
\renewcommand\thebibliography[1]{%
	\oldthebibliography{#1}%
	\setlength{\parskip}{\bibitemsep}%
	\setlength{\itemsep}{\bibparskip}%
}

\makeatletter
\renewcommand\footnotesize{%
	\@setfontsize\footnotesize\@ixpt{11}%
	\abovedisplayskip 8\p@ \@plus2\p@ \@minus4\p@
	\abovedisplayshortskip \z@ \@plus\p@
	\belowdisplayshortskip 4\p@ \@plus2\p@ \@minus2\p@
	\def\@listi{\leftmargin\leftmargini
		\topsep 4\p@ \@plus2\p@ \@minus2\p@
		\parsep 2\p@ \@plus\p@ \@minus\p@
		\itemsep \parsep}%
	\belowdisplayskip \abovedisplayskip
}
\makeatother
\allowdisplaybreaks
\usepackage[onehalfspacing]{setspace}

\begin{document}
	\title{\huge{Wild Bootstrap Inference for Instrumental Variables Regressions with Weak and Few Clusters}\thanks{We are grateful to Aureo de Paula, Firmin Doko Tchatoka, Kirill Evdokimov, S\'ilvia Gon\c calves,  Christian Hansen, Jungbin Hwang, Qingfeng Liu, Morten Ørregaard Nielsen, Ryo Okui, Kyungchul (Kevin) Song, Naoya Sueishi, Yoshimasa Uematsu, and 
			participants at the 2021 Annual Conference of the International Association for Applied Econometrics, the 2021 Asian Meeting of the Econometric Society, the 2021 China Meeting of the Econometric Society, the 2021 Australasian Meeting of the Econometric Society, the 2021 Econometric Society European Meeting, the 37th Canadian Econometric Study Group Meetings, the 16th International Symposium on Econometric Theory and Applications, UCONN Econometrics Seminar, NUS Econometrics Seminar, the 2023 North American Winter Meeting of the Econometric Society for their valuable comments. Special thanks to James MacKinnon for very insightful discussions. Wang acknowledges the financial support from Singapore Ministry of Education Tier 1 grants RG53/20 and RG104/21. 
			Zhang acknowledges the financial support from Singapore Ministry of Education Tier 2 grant under grant MOE2018-T2-2-169 and the Lee Kong Chian fellowship. Any possible errors are our own.}}
	
	\author{
		Wenjie Wang\footnote{Division of Economics, School of Social Sciences, Nanyang Technological University.
			HSS-04-65, 14 Nanyang Drive, Singapore 637332. 
			E-mail address: wang.wj@ntu.edu.sg.}  
		\ and Yichong Zhang\footnote{School of Economics, Singapore Management University. E-mail address: yczhang@smu.edu.sg.}
	}
	
	\date{\today}
	
	\maketitle

	\begin{abstract}
		
		We study the wild bootstrap inference for instrumental variable regressions under an alternative asymptotic framework that the number of independent clusters is fixed, the size of each cluster diverges to infinity, and the within cluster dependence is sufficiently weak. We first show that the wild bootstrap Wald test controls size asymptotically up to a small error as long as the parameters of endogenous variables are strongly identified in at least one of the clusters. Second, we establish the conditions for the bootstrap tests to have power against local alternatives. We further develop a wild bootstrap Anderson-Rubin test for the full-vector inference and show that it controls size asymptotically even under weak identification in all clusters. We illustrate their good performance using simulations and provide an empirical application to a well-known dataset about US local labor markets.
		\\

		  \bigskip
		
	\end{abstract}
	
	
	\setlength{\baselineskip}{18pt}
	\setlength{\abovedisplayskip}{10pt}
	\belowdisplayskip\abovedisplayskip
	\setlength{\abovedisplayshortskip }{5pt}
	\abovedisplayshortskip \belowdisplayshortskip%
	\setlength{\abovedisplayskip}{8pt} \belowdisplayskip\abovedisplayskip%
	\setlength{\abovedisplayshortskip }{4pt} %
	\linespread{1.6}
	\large

	\section{Introduction}
	The instrument variable (IV) regression is one of the five most commonly used causal inference methods identified by \cite{Angrist-Pischke(2008)}, and it is often applied with clustered data. For example, \cite{Young(2021)} analyzes 1,359 IV regressions in 31 papers published by the American Economic Association (AEA), out of which 24 papers account for clustering of observations. 
	Three issues arise when running IV regressions with clustered data. 
	First, the strength of IVs may be heterogeneous across clusters with  a few clusters providing the main identification power. 
	Indeed, \cite{Young(2021)} finds that in the average paper of his AEA samples, with the removal of just one cluster or observation, the first-stage $F$ can decrease by 28\%,
	and 38\% of reported 0.05 significant two-stage least squares (TSLS) results can be rendered insignificant at that level.
	Second, the number of clusters is small in many IV applications.
	For instance, \cite{Acemoglu2011} cluster the standard errors at the country/polity level, resulting in 12-19 clusters, \cite{Glitz2020} cluster at the sectoral level with 16 sectors, and \cite{Rogall2021} clusters at the province (district) level with 11 provinces (30 districts), respectively. When the number of clusters is small, conventional cluster-robust inference procedures may be unreliable. Third, it is also possible that IVs are weak in all clusters, in which case researchers need to use weak-identification-robust inference methods \citep{Andrews-Stock-Sun(2019)}.

	Motivated by these issues, in this paper we study the inference for IV regressions with a small, and thus, fixed number of clusters and weak within-cluster dependence. We denote clusters in which the parameters of the endogenous variables are strongly identified as strong IV clusters.
	First, we show that a wild bootstrap Wald test, with or without the cluster-robust variance estimator (CRVE), controls size asymptotically up to a small error, as long as there exists at least one strong IV cluster. Second, we show that the wild bootstrap tests have power against local alternatives at 10\% and 5\% significance levels when there are at least five and six strong IV clusters, respectively. 
	Third, in the common case of testing a single restriction, 
	we show the bootstrap Wald test with CRVE is more powerful than that without CRVE for distant local alternatives. Fourth, we develop the full-vector inference based on a wild bootstrap \citet[AR]{Anderson-Rubin(1949)} test, which controls size asymptotically up to a small error regardless of instrument strength. Fifth, in the Online Supplement we show that for the common case with a single endogenous variable and a single IV, a wild bootstrap test based on the unstudentized Wald statistic (i.e., the one without CRVE) 
	is asymptotically equivalent to a certain wild bootstrap AR test under both null and alternative, implying that in such a case it is fully robust to weak IV. Sixth, we show in the Online Supplement that bootstrapping weak-IV-robust tests other than the AR test (e.g., the Lagrange multiplier test or the conditional quasi-likelihood ratio test) controls asymptotic size when there is at least one strong IV cluster.

	

	Our inference procedure is empirically relevant. First, we notice that besides the aforementioned examples, the numbers of clusters may also be rather small in studies that estimate the region-wise effects of certain intervention if the partition of clusters is at the state level. 
	We illustrate the usefulness of our methods in Section \ref{sec: emp} by applying them to the well-known dataset of \cite{ADH2013} in the estimation of the effects of Chinese imports on local labor markets in three US Census Bureau-designated regions (South, Midwest, and West) with 11-16 clusters at the state level.  Second, our bootstrap inference is flexible with respect to IV strength: 
	the bootstrap Wald test allows for cluster-level heterogeneity in the first stage, 
	while its AR counterpart is fully robust to weak IVs. Figure 1 reports the estimated first-stage coefficients for each cluster (state) in \citeauthor{ADH2013}'s (\citeyear{ADH2013}) dataset, which suggests that there exists substantial variation in the IV strength among states. Specifically, the first-stage coefficients of some states are quite large compared with the rest in the region, 
	while some other states have coefficients that are rather close to zero, and thus, potentially subject to weak identification. Some states even have opposite signs for their first-stage coefficients. 
	However, there is no existing proven-valid inference method for IV regressions with few clusters, where IVs may be weak in some clusters. 
	Therefore, we believe our bootstrap methods enrich practitioners' toolbox by providing a reliable inference in this context.
	Third, different from the analytical inference based on the widely used heteroskedasticity and autocorrelation consistent (HAC) estimators, 
	our approach is agnostic about the within-cluster (weak) dependence structure and thus avoids the use of tuning parameters to estimate the covariance matrix for dependent data. 
	
	\begin{figure}
		\makebox[\textwidth]{\includegraphics[width=\paperwidth,height=0.16\textheight, trim={0 0 0 1cm}]{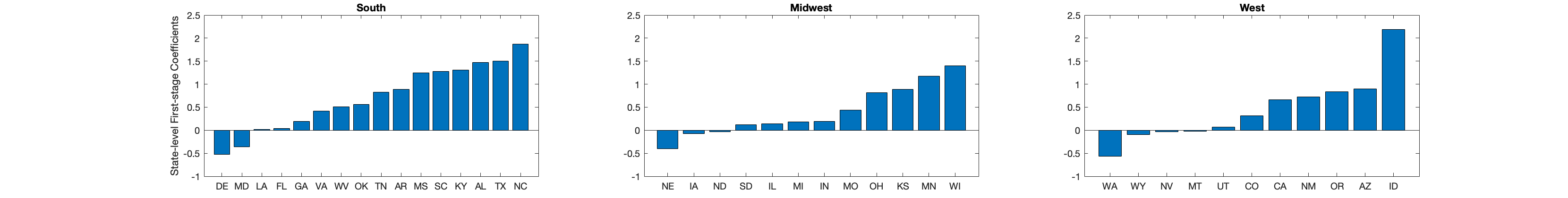}}
		\vspace*{-10mm}
		\caption{State-level First-stage Coefficients in \cite{ADH2013}}
		\label{fig:first_stage}
		{\footnotesize{Note: This figure reports the estimated state-level first-stage coefficients for South (16 states), Midwest (12 states), and West (11 states), respectively, in \cite{ADH2013}. The coefficients are obtained by running first-stage least squares regression for each state separately, controlling for the same exogenous variables as those specified in \cite{ADH2013}.  
		}}   
	\end{figure}


	
	The contributions in the present paper relate to several strands of literature. 
	First, it is related to the literature on the cluster-robust inference.\footnote{See \cite{Cameron(2008)}, \cite{Conley2011inference}, \cite{Imbens-Kolesar(2016)}, \cite{abadie2022should}, \cite{Hagemann(2017), Hagemann(2019), Hagemann2019placebo, Hagemann2020inference}, \cite{Mackinnon-Webb(2017)}, \cite{Djogbenou-Mackinnon-Nielsen(2019)}, \cite{Mackinnon-Nielsen-Webb(2019)}, 
		\cite{Ferman2019inference}, 
		\cite{Hansen-Lee(2019)},
		\cite{Menzel2021bootstrap}, 
		\cite{Mackinnon2021},
		among others, 
		and \cite{mackinnon2022cluster}
		for a recent survey.} 
	\cite{Djogbenou-Mackinnon-Nielsen(2019)}, \cite{Mackinnon-Nielsen-Webb(2019)}, and \cite{Menzel2021bootstrap} show the bootstrap validity under the asymptotic framework in which the number of clusters diverges to infinity.\footnote{We refer interested readers to \citet[Sections 4.1 and 4.2]{mackinnon2022cluster} for detailed discussions on this asymptotic framework and the alternative asymptotic framework that treats the number of clusters as fixed.} 
	\citet[IM]{Ibragimov-Muller(2010), Ibragimov-Muller(2016)},
	\citet[BCH]{Bester-Conley-Hansen(2011)},
	\citet[CRS]{Canay-Romano-Shaikh(2017)}, \cite{Hagemann(2019), Hagemann2019placebo, Hagemann2020inference}, \cite{Canay-Santos-Shaikh(2020)}, and \cite{Hwang(2020)} consider an alternative asymptotic framework in which the number of clusters is treated as fixed, while the number of observations in each cluster is relatively large and the within cluster dependence is sufficiently weak.
	However, the inference methods proposed by BCH and \cite{Hwang(2020)} require an (asymptotically) equal cluster-level sample size,\footnote{See \citet[Assumptions 3 and 4]{Bester-Conley-Hansen(2011)} and \citet[Assumptions 4 and 5]{Hwang(2020)} for details.} 
	while those proposed by IM and CRS require strong identification in all clusters. In contrast, our bootstrap Wald tests are more flexible as it does not require an equal cluster size and only needs one strong IV cluster for size control and five to six for local power, thus allowing for substantial heterogeneity in IV strength.  
	To our knowledge, there is no alternative method in the literature that remains valid in such a context.
	For full-vector inference, we further provide the bootstrap AR tests, which are fully robust to weak identification. 
	
	Second, \cite{Canay-Santos-Shaikh(2020)} prove the validity of wild bootstrap with a few large clusters by innovatively connecting it with a randomization test with sign changes. Our results rely heavily on this technical breakthrough, but also complement  \cite{Canay-Santos-Shaikh(2020)} in the following aspects. 
	First, \cite{Canay-Santos-Shaikh(2020)} focus on the linear regression with exogenous regressors and then extend their analysis to 
	a score bootstrap for the GMM estimator. 
	Instead, we focus on extending the wild restricted efficient (WRE) cluster bootstrap, which is popular among empirical researchers, for general $k$-class IV estimators.
	Therefore, our procedure cannot be formulated as a score bootstrap in the GMM setting. 
	Second, we study the local power for the Wald test both with and without CRVE. 
	The power analysis of the Wald test with CRVE is new to the literature and 
	technically involved because with a fixed number of clusters, the CRVE has a random limit. We rely on the Sherman–Morrison–Woodbury formula for the matrix inverse to analyze the behaviors of the test statistic
	and its wild bootstrap counterparts under local alternatives. Based on such a detailed asymptotic analysis, we establish the local power of the bootstrap Wald test studentized by CRVE,  
	and show it is higher than that for the Wald test without CRVE studentization for distant local alternatives in the empirically prevalent case of testing a single restriction. 
	Indeed, as the linear regression is a special case of the linear IV regression studied in this paper, we believe the same result holds for the local power of the Wald tests in \cite{Canay-Santos-Shaikh(2020)}. Third, different from its unstudentized counterpart, the local power of the Wald test with CRVE is established  without the assumption that the cluster-level first-stage coefficients have the same sign, which may not hold in some empirical studies (e.g., Figure \ref{fig:first_stage}).

	Third, our paper is related to the literature of WRE bootstrap for IV regressions.\footnote{See, for example,
		\cite{Davidson-Mackinnon(2008), Davidson-Mackinnon(2010)}, \cite{finlay2014, Finlay-Magnusson(2019)}, \cite{Roodman-Nielsen-MacKinnon-Webb(2019)}, and \cite{Mackinnon2021}.} 
	In particular, \cite{Davidson-Mackinnon(2008), Davidson-Mackinnon(2010)} first proposed the restricted efficient (RE) and WRE
	bootstrap procedures for IV regression with homoskedastic and heteroskedastic errors, respectively. \cite{finlay2014, Finlay-Magnusson(2019)} extended it 
	to clustered data, and \cite{Roodman-Nielsen-MacKinnon-Webb(2019)} incorporated it into the Stata package ``boottest". 
	Our key contribution to this strand of literature is to theoretically prove the validity of the (modified) WRE cluster bootstrap under the alternative asymptotic framework in which IVs may be weak for some clusters and the number of clusters is treated as fixed. In particular, we carefully re-design the first stage of the WRE procedure to allow for cluster-level heterogeneity of IV strength. 
	Our theoretical analysis also covers the general $k$-class IV estimators, including TSLS, the limited information maximum likelihood (LIML) estimator or the modified LIML estimator proposed by \citet[hereafter FULL estimator]{Fuller(1977)}. 
	In addition, we theoretically prove the power advantages of the bootstrap Wald test with CRVE compared with the one without. This result is new and justifies the proposal in the literature of using the bootstrap Wald test studentized by CRVE from the power perspective.

	Furthermore, our paper is related to the literature on weak identification, 
	in which various normal approximation-based inference approaches are available for nonhomoskedastic cases, among them \cite{Stock-Wright(2000)},
	\cite{Kleibergen(2005)}, \cite{Andrews-Cheng(2012)},
	\cite{Andrews(2016)}, \cite{Andrews-Mikusheva(2016)}, \cite{Andrews(2018)},
	\cite{Moreira-Moreira(2019)}, and \cite{Andrews-Guggenberger(2019)}. 
	As \citet[p.750]{Andrews-Stock-Sun(2019)} remark,
	an important question concerns the quality of the normal approximations with nonhomoskedastic observations. 
	On the other hand, when implemented appropriately, alternative approaches such as 
	bootstrap may substantially improve the inference for IV regressions.\footnote{See, for example, 
		\cite{Davidson-Mackinnon(2008), Davidson-Mackinnon(2010)}, \cite{Moreira-Porter-Suarez(2009)}, \cite{Wang-Kaffo(2016)},
		\cite{Kaffo-Wang(2017)},
		\cite{Finlay-Magnusson(2019)}, and \cite{Young(2021)}, among others.
		In addition, we notice that \cite{tuvaandorj2021robust} develops permutation versions of weak-IV-robust tests
		with (non-clustered) heteroskedastic errors.}
	We contribute to this literature by establishing formally that wild bootstrap-based AR tests control asymptotic size under both weak identification and a small number of clusters. However, contrary to the Wald tests, we find in simulations that the bootstrap AR test without CRVE has better power properties than the one studentized by CRVE. We formally establish the local power of the bootstrap AR test without CRVE. Furthermore, we show that the bootstrap validity for other weak-IV-robust tests requires at least one strong IV clusters (Sections \ref{sec: other-weak}-\ref{sec: proof-other-weak} in the Online Supplement), which provides theoretical support to the simulation results in \cite{finlay2014, Finlay-Magnusson(2019)} documenting that with clustered observations, bootstrapping the AR tests results in better finite sample size control than bootstrapping alternative robust tests.

	Finally, we note that although empirical applications often involve settings with substantial first-stage heterogeneity such as those shown in Figure \ref{fig:first_stage}, related econometric literature remains rather sparse. 
	\cite{Abadie-Gu-Shen(2019)} exploit such heterogeneity to improve the asymptotic mean squared error of IV estimators with independent and homoskedastic observations. Instead, we focus on developing inference methods that are robust to the first-stage heterogeneity for data with a small number of clusters, while allowing for (weak) within-cluster dependence and heteroskedasticity.

	The remainder of this paper is organized as follows. 
	Section \ref{sec: setup} presents the setup and the bootstrap procedures. 
	Section \ref{sec: assumption} presents main assumptions, while Sections \ref{subsec: boot-wald}-\ref{subsec: boot-robust} provide asymptotic results.  
	Simulations in Section \ref{sec: simu} suggest that our procedures have outstanding finite sample size control and, in line with our theoretical analysis, the bootstrap Wald test with CRVE has power advantages compared with the other bootstrap tests. The empirical application is presented in Section \ref{sec: emp}. We conclude in Section \ref{sec: conclu}.

	\section{Setup, Estimation, and Inference}\label{sec: setup}
	\subsection{Setup}\label{subsec: setup}
	Throughout the paper, we consider the setup of a linear IV regression with clustered data, 
	\begin{eqnarray}\label{eq: IV-model}
	y_{i,j} = X^{\top}_{i,j} \beta + W^{\top}_{i,j} \gamma + \eps_{i,j},
	\end{eqnarray}
	where the clusters are indexed by $j \in [J] = \{1, ..., J\}$ and units in the $j$-th cluster are indexed by $i \in I_{n,j} 
	= \{1, ..., n_j \}$. In \eqref{eq: IV-model}, we denote  
	$y_{i,j} \in \textbf{R}$, $X_{i,j} \in \textbf{R}^{d_x}$, and $W_{i,j} \in \textbf{R}^{d_w}$ as an outcome of interest, endogenous regressors, and exogenous regressors, respectively.
	Furthermore, we let $Z_{i,j} \in \textbf{R}^{d_z}$ 
	denote the IVs for $X_{i,j}$.
	$\beta \in \textbf{R}^{d_x}$ and $\gamma \in \textbf{R}^{d_w}$ are unknown structural parameters. 
	
	We allow the parameter of interest $\beta$ to shift with respect to (w.r.t.) the sample size, which incorporates the analyses of size and local power in a concise manner: $\beta_n = \beta_0 + \mu_{\beta}/r_n$, where $\mu_{\beta} \in \textbf{R}^{d_x}$ is the local parameter and $r_n$ is the convergence rate of the score defined later. We let $\lambda^{\top}_{\beta} \beta_0 = \lambda_0$, $\lambda_{\beta} \in \textbf{R}^{d_x \times d_r}$, $\lambda_0 \in \textbf{R}^{d_r}$, and $d_r$ denotes the number of restrictions under the null hypothesis. 
	Define $\mu = \lambda_{\beta}^{\top} \mu_{\beta}$.
	Then, the null and local alternative hypotheses studied in this paper can be written as 
	\begin{align}\label{eq: hypothesis}
	\mathcal{H}_0: \mu = 0  \quad v.s. \quad \mathcal{H}_{1,n}:  \mu \neq 0.
	\end{align}
	
	\subsection{K-Class IV Estimators}
	Throughout the paper, we consider the $k$-class IV estimators of the form: 
	\begin{align}\label{eq: k-class}
	\left(\hat{\beta}^{\top}, \hat{\gamma}^{\top}\right)^{\top} = \left( \vec{X}^{\top} \vec{X} - \hat{\kappa} \vec{X}^{\top} M_{\vec{Z}}\vec{X} \right)^{-1} \left(\vec{X}^{\top} Y - \hat{\kappa} \vec{X}^{\top} M_{\vec{Z}}Y\right), 
	\end{align}
	where $\vec{Z} = [Z:W]$, $\vec{X} = [X:W]$,
	$Y$, $X$, $Z$, and $W$ 
	are $n \times 1$, $n \times d_x$, $n \times d_z$, and $n \times d_w$-dimensional vectors and matrices formed 
	by $y_{i,j}$, $X^{\top}_{i,j}$, $Z^{\top}_{i,j}$, and $W^{\top}_{i,j}$, respectively. We also denote 
	$P_A = A(A^{\top} A)^{-1}A^{\top}$ and $M_A = I_n - P_A$, where $A$ is an $n$-dimensional matrix and $I_n$ is an $n$-dimensional identity matrix. 
	Specifically, we focus on four $k$-class estimators: (1) the TSLS estimator, 
	where $\hat{\kappa} = \hat{\kappa}_{tsls} = 1$, (2) 
	the LIML estimator, where 
	\begin{align*}
	\hat{\kappa} = \hat{\kappa}_{liml} = 
	\min_{r} r^{\top} \vec{Y}^{\top} M_W \vec{Y} r / (r^{\top} \vec{Y}^{\top} M_{\vec{Z}}\vec{Y}r), 
	\;\;  \vec{Y} = [Y:X], \;\; \text{and} \;\; r=(1,-\beta^{\top} )^{\top}, 
	\end{align*}
	(3) the FULL estimator, where 
	$\hat{\kappa} = \hat{\kappa}_{full} = \hat{\kappa}_{liml} - C/(n-d_z-d_w)$ with some constant $C$,
	and (4) the bias-adjusted TSLS (BA) estimator proposed by \cite{Nagar(1959)} and \cite{Rothenberg(1984)}, where
	$\hat{\kappa} = \hat{\kappa}_{ba} = n/(n-d_z+2)$. 
	Theoretically, we show that these four estimators are asymptotically equivalent.\footnote{More specifically, we show in Lemma S.B.1 in the Online Supplement that under our asymptotic framework, for $L \in \{\text{liml},\text{full},\text{ba}\}$,  
		$\hat{\beta}_{L} = \hat{\beta}_{tsls} + o_P(r_n^{-1})$ and $\hat{\beta}_{tsls} - \beta_n = O_P(r_n^{-1})$, where $r_n = O(n^{1/2})$.} 
	However, in simulations, we find that FULL has the best finite sample performance in the overidentified case.

	\subsection{Wild Bootstrap Inference}
	\subsubsection{Inference Procedure for Wald Statistics}
	\label{sec:wald_procedure}
	For the rest of the paper, we define $\widetilde{Z}_{i,j}$ as the residual of regressing  $Z_{i,j}$ on $W_{i,j}$ using the entire sample, and that for any random vectors $U_{i,j}$ and $V_{i,j}$, $\widehat{Q}_{UV,j} = \frac{1}{n_j}\sum_{i \in I_{n,j}}U_{i,j}V_{i,j}^\top$ and $\widehat{Q}_{UV} = \frac{1}{n}\sum_{j \in [J]}\sum_{i \in I_{n,j}}U_{i,j}V_{i,j}^\top$. 
	
	For inference, we construct Wald statistics based on the $k$-class estimator $\hat{\beta}_L$  defined in (\ref{eq: k-class}) with $\hat k = \hat k_L$ for $L \in \{\text{tsls},\text{liml},\text{full},\text{ba}\}$. When the $d_r \times d_r$ weighting matrix $\hat{A}_{r}$  is asymptotically deterministic in the sense of Assumption \ref{assumption: 4} below (such as $\hat{A}_r = I_{d_r}$, the $d_r \times d_r$ identity matrix),  we denote $T_n$ as the Wald statistic without CRVE and define it as 
	\begin{eqnarray}\label{eq: Wald-unstudentized}
	T_{n} = ||\lambda_{\beta}^{\top} \hat{\beta}_L - \lambda_0||_{\hat{A}_r},
	\end{eqnarray}
	where $||u||_A = \sqrt{u^{\top}Au}$ for a generic vector $u$ and a weighting matrix $A$. 
	
	Further define $\hat{A}_{r,CR}$ as the inverse of the CRVE:
	\begin{align}\label{eq: Wald-hat-A}
	\hat{A}_{r,CR} & = \left( \lambda^{\top}_{\beta} \widehat{V} \lambda_{\beta} \right)^{-1} ,\quad  \widehat{V} 
	=
	\widehat{Q}^{-1}
	\widehat{Q}^{\top}_{\widetilde{Z}X} \widehat{Q}^{-1}_{\widetilde{Z}\widetilde{Z}}
	\widehat{\Omega}_{CR}
	\widehat{Q}^{-1}_{\widetilde{Z}\widetilde{Z}}\widehat{Q}_{\widetilde{Z}X} 
	\widehat{Q}^{-1}, \quad \widehat{Q} = \widehat{Q}^{\top}_{\widetilde{Z}X} \widehat{Q}^{-1}_{\widetilde{Z}\widetilde{Z}}\widehat{Q}_{\widetilde{Z}X},  
	\end{align}
	$\widehat{\Omega}_{CR} = n^{-1} \sum_{j \in [J]} \sum_{i \in I_{n,j}} \sum_{k \in I_{n,j} } \widetilde{Z}_{i,j} \widetilde{Z}_{k,j}^{\top} \hat{\eps}_{i,j} \hat{\eps}_{k,j}$,
	and $\hat{\eps}_{i,j}$ is the unrestricted residual defined in (\ref{eq: unrest_residual}). 
	The corresponding Wald statistic studentized by CRVE is denoted as 
	\begin{align}\label{eq: Wald-studentized}
	T_{CR,n} = ||\lambda_\beta^\top \hat{\beta} - \lambda_0||_{\hat{A}_{r,CR}}. 
	\end{align}
	In the case of testing the value of a single endogenous variable $\beta = \beta_0$, the test statistics reduce to 
	$T_n = |\hat{\beta}_L - \beta_0|$, and  
	$T_{CR,n} = |(\hat{\beta}_L - \beta_0)/\widehat{V}^{1/2}|$,
	respectively, where $\widehat{V}$ denotes the usual cluster-robust variance estimator.

	We reject the null hypothesis in (\ref{eq: hypothesis}) at $\alpha$ significance level if the test statistics are greater than their corresponding critical values ($\hat c_n (1-\alpha)$ and $\hat c_{CR,n}(1-\alpha)$ defined below for $T_n$ and $T_{CR,n}$, respectively). We compute the critical values by a wild bootstrap procedure described below.  
	
	\begin{enumerate}
		\item[\textbf{Step 1}:] 
		For $L \in \{\text{tsls},\text{liml},\text{full},\text{ba}\}$, compute the null-restricted residual
		$$\hat{\eps}^r_{i,j} = y_{i,j} - X_{i,j}^{\top} \hat{\beta}^r_L - W_{i,j}^{\top} \hat{\gamma}^r_L,$$ where $\hat{\beta}^r_L$ and $\hat{\gamma}^r_L$ are null-restricted $k$-class IV estimators of $\beta$ and $\gamma$ from $(y_{i,j}, X^{\top}_{i,j}, W^{\top}_{i,j}, Z^{\top}_{i,j})^{\top}$,\footnote{The null-restricted $k$-class estimator is defined as 
			\begin{align*}
			& \hat{\beta}^r_L = \hat{\beta}_L - \left(X^{\top}P_{\widetilde{Z}}X -\hat{\mu}_L X^{\top}M_{\vec{Z}}X\right)^{-1} 
			\lambda_{\beta}
			\left(\lambda_{\beta}^{\top}(X^{\top}P_{\widetilde{Z}}X - \hat{\mu}_L X^{\top}M_{\vec{Z}}X)^{-1}
			\lambda_{\beta}\right)^{-1} (\lambda_{\beta}^{\top}\hat{\beta}_L - \lambda_0), \\
			& \hat{\gamma}^r_L = (W^\top W)^{-1}W^\top(Y-X\hat{\beta}_L^r), \;\; \text{where} \;\; \hat{\mu}_L = \hat{\kappa}_L-1 \;\; \text{and} \;\; \tilde{Z} = M_W Z, 
			\end{align*}
			for $L \in \{\text{tsls},\text{liml},\text{full},\text{ba}\}$; 
			e.g., see Appendix B of \cite{Roodman-Nielsen-MacKinnon-Webb(2019)} for a general formula.}
		and the unrestricted residual is defined as 
		\begin{align}\label{eq: unrest_residual}
		\hat{\eps}_{i,j} = y_{i,j} - X_{i,j}^{\top} \hat{\beta}_L - W_{i,j}^{\top} \hat{\gamma}_L, 
		\end{align}
		where $\hat{\beta}_L$ and $\hat{\gamma}_L$ are defined in (\ref{eq: k-class}) with $\hat{\kappa}_L$. 

		
		\item[\textbf{Step 2}:] 
		Construct $\overline{Z}_{i,j}$ as
		\begin{align*}
		\overline{Z}_{i,j} = \left( \widetilde{Z}^{\top}_{i,j} 1\{j=1\}, ......, \widetilde{Z}^{\top}_{i,j} 1\{j = J\} \right)^{\top}, 
		\end{align*}
		where $\widetilde{Z}_{i,j}$ is the residual of regressing $Z_{i,j}$ on $W_{i,j}$ using the entire
		sample.
		
		
		\item[\textbf{Step 3}:] Compute the first-stage residual
		\begin{align}\label{eq: boot-RE-first-stage}
		\tilde{v}_{i,j} = X_{i,j} - \widetilde{\Pi}_{\overline{Z}}^\top \overline{Z}_{i,j} -  \widetilde{\Pi}_{w}^\top W_{i,j}, 
		\end{align}
		where $\widetilde{\Pi}_{\overline{Z}}$ and $\widetilde{\Pi}_{w}$ are the OLS coefficients of $\overline{Z}_{i,j}$ and $W_{i,j}$ from regressing $X_{i,j}$ on $(\overline{Z}^{\top}_{i,j}, W^{\top}_{i,j}, \hat{\eps}_{i,j})^{\top}$ using the entire sample.

		\item[\textbf{Step 4}:] Let $\textbf{G} = \{-1, 1\}^J$ and for any $g = (g_1, ..., g_J) \in \textbf{G}$ generate 
		\begin{eqnarray}\label{eq: boot-algo-unstud-2}
		X_{i,j}^*(g) = \widetilde{\Pi}_{\overline{Z}}^\top\overline{Z}_{i,j} +  \widetilde{\Pi}_{w}^\top W_{i,j} + g_j \tilde{v}_{i,j}, \quad y_{i,j}^*(g) = X_{i,j}^{*\top}(g) \hat{\beta}^r_L + W_{i,j}^{\top} \hat{\gamma}^r_L + g_j \hat{\eps}^r_{i,j}. 
		\end{eqnarray}
		
		For each $g=(g_1, ..., g_J) \in \textbf{G}$, compute $\hat{\beta}^*_{L,g}$ and $\hat{\gamma}^*_{L,g}$, 
		the analogues of the estimators $\hat{\beta}_L$ and $\hat{\gamma}_L$
		using $\left( y_{i,j}^{*}(g), X_{i,j}^{*\top}(g) \right)^{\top}$ in place of $\left( y_{i,j}, X_{i,j}^{\top} \right)^{\top}$ and the same $(Z^{\top}_{i,j}, W^{\top}_{i,j})^{\top}$.  
		Compute the bootstrap analogue of the Wald statistics:\footnote{Let $X^*(g)$ and $Y^*(g)$ be the $n \times d_x$ matrix constructed using $X_{i,j}^*(g)$ and the $n \times 1$ vector constructed using $Y_{i,j}^*(g)$, respectively. For $L \in \{\text{tsls},\text{liml},\text{full},\text{ba}\}$ and $g \in \textbf{G}$,
			\begin{align*}
			\hat{\beta}^*_{L,g}= \left(X^{*\top}(g) P_{\widetilde{Z}} X^*(g) - \hat{\mu}^*_{L,g} X^{*\top}(g) M_{\vec{Z}} X^*(g) \right)^{-1}\left(X^{*\top}(g) P_{\widetilde{Z}} Y^*(g) - \hat{\mu}^*_{L,g} X^{*\top}(g) M_{\vec{Z}} Y^*(g)\right), \;\; \text{where} \;\; \hat{\mu}^*_{L,g} = \hat{\kappa}^*_{L,g}-1.
			\end{align*}} 
		\begin{align}\label{eq: boot-algo-unstud-3}
		T^*_{n}(g) = || \lambda_{\beta}^{\top} \hat{\beta}^*_{L,g} - \lambda_0 ||_{\hat{A}_{r}}, \quad T^*_{CR,n}(g) = ||\lambda_{\beta}^{\top} \hat{\beta}^*_{L,g} - \lambda_0||_{\hat{A}_{r,CR,g}^*}.
		\end{align} 
		$\hat{A}^*_{r,CR,g}$ is the bootstrap counterpart of $\hat A_{r,CR}$ defined as 
		\begin{align*}
		\hat{A}^*_{r,CR,g} = \left( \lambda^{\top}_{\beta} \widehat{V}^*_g \lambda_{\beta} \right)^{-1}, \quad
		\widehat{V}^*_g     
		=  \widehat{Q}_g^{*-1} 
		\widehat{Q}^{*\top}_{\widetilde{Z}X}(g) \widehat{Q}^{-1}_{\widetilde{Z}\widetilde{Z}}
		\widehat{\Omega}^*_{CR,g}
		\widehat{Q}^{-1}_{\widetilde{Z}\widetilde{Z}}\widehat{Q}^*_{\widetilde{Z}X}(g) 
		\widehat{Q}_g^{*-1},
		\end{align*}
		where for $L \in \{\text{tsls},\text{liml},\text{full},\text{ba}\}$, 
		\begin{align}
		& \widehat{\Omega}^*_{CR,g} = \frac{1}{n} \sum_{j \in [J]} \sum_{i \in I_{n,j}} \sum_{k \in I_{n,j} } \widetilde{Z}_{i,j} \widetilde{Z}_{k,j}^{\top} \hat{\eps}^*_{i,j}(g) \hat{\eps}^*_{k,j}(g), \quad  \widehat{Q}^*_{\widetilde{Z}X}(g) = \frac{1}{n} \sum_{j \in [J]} \sum_{i \in I_{n,j}}  \widetilde{Z}_{i,j} X_{i,j}^{*\top}(g),
		\notag \\
		&   \hat{\eps}^*_{i,j}(g) = y^*_{i,j}(g) - X_{i,j}^{*\top}(g) \hat{\beta}^*_{L,g} - W_{i,j}^{\top} \hat{\gamma}^*_{L,g}, \quad \text{and} \quad \widehat{Q}^*_g = \widehat{Q}_{\widetilde{Z}X}^{*\top}(g) \widehat{Q}_{\widetilde{Z}\widetilde{Z}}^{-1} \widehat{Q}^*_{\widetilde{Z}X}(g). \label{eq:Qstar}
		\end{align}
		
		
		\item[\textbf{Step 5}:]
		\hypertarget{Step 5}{We} compute the $1-\alpha$ quantiles of $\{ T^*_{n}(g): g \in \textbf{G} \}$ and $\{ T^*_{CR,n}(g): g \in \textbf{G} \}$: 
		\begin{align*}
		& \hat{c}_{n}(1-\alpha) = \inf 
		\left\{ x \in \textbf{R}: \frac{1}{|\textbf{G}|} \sum_{g \in \textbf{G}}1 \{ T^*_{n}(g) \leq x \} \geq 1-\alpha
		\right\},\\
		& \hat{c}_{CR,n}(1-\alpha) = \inf 
		\left\{ x \in \textbf{R}: \frac{1}{|\textbf{G}|} \sum_{g \in \textbf{G}}1 \{ T^*_{CR,n}(g) \leq x \} \geq 1-\alpha
		\right\},
		\end{align*}
		where $1\{E\}$ equals one whenever the event $E$ is true and equals zero otherwise and $|\textbf{G}| = 2^J$. 
		The bootstrap test for $\mathcal{H}_0$ rejects whenever $T_{CR,n}$ exceeds the critical value $\hat{c}_{CR,n}(1-\alpha)$ and $T_{n}$ exceeds $\hat{c}_{n}(1-\alpha)$ for Wald statistics with and without CRVE, respectively. 
	\end{enumerate}
	
	
	%
	
	
	
	Several remarks are in order. 
	
	\noindent
	\begin{rem}
		Step 1 imposes null when computing the residuals in the structural equation (\ref{eq: IV-model}), which is advocated by \cite{Cameron(2008)},
		\cite{Davidson-Mackinnon(2010)}, \cite{mackinnon2022cluster}, and \cite{Canay-Santos-Shaikh(2020)}, among others. 
		The estimators $\widetilde{\Pi}_{\overline{Z}}$ and $\widetilde{\Pi}_w$ in Step 3 are similar to the efficient reduced-form estimators in the WRE bootstrap procedures 
		which have superior finite sample performance for IV regressions, even when the instruments are rather weak. 
		In this paper, we focus on extending the WRE cluster bootstrap procedure\footnote{See, e.g., \citet[Section 4]{Mackinnon2021} for a detailed discussion of the WRE cluster bootstrap procedure and an efficient computation algorithm.} because (1) we find the resulting bootstrap also has excellent finite sample performance for IV regressions with a small number of clusters, and (2) we want to be consistent with the suggestions in the literature.
	\end{rem}
	
	\noindent
	\begin{rem}
		To adapt to the new framework, we carefully modify the original WRE cluster bootstrap procedure and use the fully interacted IVs $\overline{Z}_{i,j}$ in Step 3, which effectively estimate the cluster-level first-stage coefficient for $\widetilde Z_{i,j}$.  Such a modification is necessary because we allow for the IV strength, and thus, the first stage coefficient of $\widetilde Z_{i,j}$ to be different across clusters.\footnote{\cite{Finlay-Magnusson(2009), Finlay-Magnusson(2019)}, \cite{Roodman-Nielsen-MacKinnon-Webb(2019)}, and \cite{Mackinnon2021}
			considered the IV model with clustered data in which the first-stage coefficient is homogeneous across clusters.
		} We then need to preserve such a heterogeneity in the bootstrap sample by estimating the first-stage coefficient within each cluster. If we further assume the identification strength is homogeneous across clusters, then we can just use $\widetilde Z_{i,j}$ to generate $X_{i,j}^*(g)$ in Step 3. We emphasize that in the whole procedure, the fully interacted IVs $\overline{Z}_{i,j}$ are only needed to construct $X_{i,j}^*(g)$, 
		and we still use the \textit{uninteracted} IVs $\widetilde Z_{i,j}$ when computing $(\hat{\beta}_L^{\top}, \hat{\gamma}_L^{\top})^{\top}$ in (\ref{eq: k-class}), where $L \in \{\text{tsls},\text{liml},\text{full},\text{ba}\}$, and their null-restricted and bootstrap counterparts (i.e., $(\hat{\beta}^{r\top}_L, \hat{\gamma}_L^{r\top})^{\top}$ in Step 1 and $(\hat{\beta}^{*\top}_{L,g}, \hat{\gamma}^{*\top}_{L,g})^{\top}$ in Step 4). 
		
	\end{rem}

	
	\begin{rem}
		Our second modification of the original WRE cluster bootstrap procedure is that when regressing $X_{i,j}$ on $(\overline{Z}^{\top}_{i,j}, W^{\top}_{i,j}, \hat{\eps}_{i,j})^{\top}$ in Step 3, we use the unrestricted residuals $\hat{\eps}_{i,j}$ instead of the null-restricted residuals $\hat{\eps}_{i,j}^r$. This modification is required to establish the asymptotic power results (Theorems \ref{theo: boot-t-power} and \ref{theo: boot-stud-t-power}).    
	\end{rem}

	\subsubsection{Inference Procedure for Weak-instrument-robust Statistics}
	\label{sec:AR_procedure}
	In this section, we describe a full-vector wild bootstrap procedure for Anderson-Rubin (AR) type weak-IV-robust statistics with or without CRVE.  Recall that $\beta_n = \beta_0 + \mu_{\beta}/r_n$.
	Under the null, we have $\mu_\beta  = 0$, or equivalently, $\beta_n =\beta_0$. First, define the AR statistic without CRVE as
	\begin{eqnarray}\label{eq: AR-unstud-definition}
	AR_{n} = \big\Vert  \widehat{f} \big\Vert_{\hat{A}_z}, \quad
	\widehat{f}  =  n^{-1} \sum_{j \in [J]} \sum_{i \in I_{n,j}} f_{i,j},
	\end{eqnarray}
	where 
	$\hat{A}_z$ is a $d_z \times d_z$ weighting matrix with an asymptotically deterministic limit as specified in Assumption \ref{assumption: AR_further_assumptions} below,
	$f_{i,j} = \widetilde{Z}_{i,j}\bar{\eps}^r_{i,j}$, 
	$\bar{\eps}^r_{i,j} = y_{i,j} - X^{\top}_{i,j} \beta_0 - W_{i,j}^{\top} \bar{\gamma}^r $, 
	and $\bar{\gamma}^r$ is the null-restricted ordinary least squares (OLS) estimator of $\gamma$:  
	$$\bar{\gamma}^r = \left( \sum_{j \in [J]} \sum_{i \in I_{n,j}} W_{i,j} W_{i,j}^{\top} \right)^{-1}\sum_{j \in [J]} \sum_{i \in I_{n,j}} W_{i,j} 
	(y_{i,j} - X^{\top}_{i,j} \beta_0).$$
	Second, we also define the AR statistic with the (null-imposed) CRVE as
	\begin{eqnarray*}
		AR_{CR,n} = \big\Vert \widehat{f} \big\Vert_{\hat{A}_{CR}}, 
		\quad 
		\hat{A}_{CR} =  \left( n^{-1} \sum_{j \in [J]} \sum_{i \in I_{n,j}} \sum_{k \in I_{n,j} } f_{i,j} f_{k,j}^{\top} \right)^{-1}. 
	\end{eqnarray*}
	
	Our wild bootstrap procedure for the AR statistics 
	is defined as follows.
	\begin{enumerate}
		
		\item[\textbf{Step 1}:] Compute the null-restricted residual $\bar{\eps}^r_{i,j} = y_{i,j} - X_{i,j}^{\top} \beta_0 - W_{i,j}^{\top} \bar{\gamma}^r.$
		
		
		\item[\textbf{Step 2}:] Let $\textbf{G} = \{-1, 1\}^J$ and, for any $g = (g_1, ..., g_J) \in \textbf{G}$, define 
		\begin{eqnarray*}
			\widehat{f}^*_g = n^{-1} \sum_{j \in [J]} \sum_{i \in I_{n,j}} f^*_{i,j}(g_j),\quad
			\text{and} \quad f^*_{i,j}(g_j) = \widetilde{Z}_{i,j}\eps^*_{i,j}(g_j),
		\end{eqnarray*}
		where 
		$\eps^*_{i,j}(g_j) = g_j \bar{\eps}^r_{i,j}$. Compute the bootstrap statistics:
		\begin{align*}
		AR^*_{n}(g) = \big\Vert \widehat{f}^*_g \big\Vert_{\hat{A}_{z}} \quad \text{and} \quad
		AR^*_{CR,n}(g) = \big\Vert \widehat{f}^*_g \big\Vert_{\hat{A}_{CR}}. \quad
		\end{align*}

		\item[\textbf{Step 3}:] 
		Let $\hat{c}_{AR,n}(1-\alpha)$ and $\hat{c}_{AR,CR,n}(1-\alpha)$
		denote the $(1-\alpha)$-th quantile of $\{AR_{n}^*(g)\}_{g \in \textbf{G}}$ and $\{AR^*_{CR,n}(g)\}_{g \in \textbf{G}}$, respectively.
	\end{enumerate}
	
	We note that unlike the $T_{CR,n}$-based Wald test in Section \ref{sec:wald_procedure}, there is no need to bootstrap the CRVE for the $AR_{CR,n}$ test even though $\hat{A}_{CR}$ also admits a random limit. This is because $\hat{A}_{CR}$, unlike $\hat{A}_{r,CR}$ defined in \eqref{eq: Wald-hat-A}, is invariant to the sign changes. 
	Two remarks are in order. 
	\begin{rem}
		If $2^J$ is too large, the researcher can replace $\textbf{G} = \{-1,1\}^J$ by  $\hat{\textbf{G}}$ where $\hat{\textbf{G}} = \{g_1,\cdots,g_B\}$, $g_1 = \iota_J$, $\iota_J$ is a $J$-dimension vector of ones, and $g_2,\cdots,g_B$ are i.i.d. Uniform $\textbf{G}$, which is equivalent to generating each element of $g$ by a Rademacher random variable. This is known as the stochastic approximation and recommended by \cite{Canay-Romano-Shaikh(2017)} and \cite{Canay-Santos-Shaikh(2020)} for the inference under the asymptotic framework with a fixed number of clusters. All the theoretical results in this paper remains true if $\textbf{G}$ is replaced by $\hat{\textbf{G}}$ by letting $B \rightarrow \infty$. See \citet[Section 15.2.1]{LR06} for more discussion on the validity of the stochastic approximation.      
	\end{rem}
	
	\begin{rem}
		In the following, we study the size and power properties of the proposed wild bootstrap procedure applied to the $k$-class estimators. There are three main takeaways for applied researchers from our theoretical investigation. First, for inference based on Wald statistics, we recommend the bootstrap Wald test studentized by CRVE because of its superior power properties. This method is denoted as W-B-S, and we provide its pseudo code in Section \ref{sec:code} in the Online Supplement. Second, for the over-identified case, we recommend using the bootstrap Wald tests with Fuller's modified LIML estimator over TSLS because the former has a smaller finite sample bias. Third, for the full-vector inference, if researchers are concerned that all clusters are weak, and thus, would like to implement a weak-identification-robust procedure, then we recommend the bootstrap AR test without CRVE studentization because of its power advantage over alternative asymptotic and bootstrap AR tests that are studentized by CRVE.  
	\end{rem}

	\section{Main Assumptions and Several Examples}\label{sec: assumption}
	In this section, we introduce the assumptions that will be used in our analysis of the wild bootstrap tests under a small number of clusters in Sections \ref{subsec: boot-wald}-\ref{subsec: boot-robust}.  Define $\widehat{Q} = \widehat{Q}_{\widetilde{Z}X}^{\top} \widehat{Q}_{\widetilde{Z}\widetilde{Z}}^{-1}
	\widehat{Q}_{\widetilde{Z}X}$
	and $Q$ as the probability limits of $\widehat{Q}$. 
	
	\begin{assumption}\label{assumption: 1}
		\noindent (i) For each $j \in [J]$, 
		\begin{align}
		\widehat Q_{\widetilde ZW,j} = o_P(1)
		\label{eq:QZWj}
		\end{align}
		and 
		\begin{align}
		\frac{r_n}{n} \sum_{i \in I_{n,j}} W_{i,j}\eps_{i,j} = O_P(1),
		\label{eq:Wej}
		\end{align}

		where the convergence rate $r_n$ satisfies $r_n \rightarrow \infty$ and $r_n = O(\sqrt{n})$.
		
		\noindent (ii) There exists a collection of independent random variables $\{ \mathcal{Z}_j : j \in [J] \}$, 
		where $\mathcal{Z}_j \sim N(0, \Sigma_j)$ with positive definite $\Sigma_j$ for all $j \in [J]$ such that 
		\begin{align} \label{eq:mathcalZj}
		\left\{ 
		\frac{r_n}{n} \sum_{i \in I_{n,j}} \widetilde{Z}_{i,j}\eps_{i,j}
		: j \in [J] \right\} \xrightarrow{\enskip d \enskip}  
		\left\{ 
		\mathcal{Z}_{j} 
		: j \in [J] \right\}.
		\end{align}

		\noindent (iii) For each $j \in [J]$, $n_j/n \rightarrow \xi_j >0$. 
		
		\noindent (iv) $\frac{1}{n}\sum_{j \in [J]} \sum_{i \in I_{n,j}}W_{i,j}W_{i,j}^\top$ is invertible. 
	\end{assumption}
	
	\begin{rem}
		We have \eqref{eq:QZWj} if 
		\begin{eqnarray}\label{eq: ZW_proj}
		\frac{1}{n_j} \sum_{i \in I_{n,j}} \Big\Vert W_{i,j}^{\top} \left( \widehat{\Gamma}_n - \widehat{\Gamma}_{n,j} \right) \Big\Vert^2  
		\convP  0,
		\end{eqnarray}
		where $\widehat{\Gamma}_n$ and $\widehat{\Gamma}_{n,j}$ are the $d_w \times d_z$ matrices that satisfy the following orthogonality conditions: 
		$\sum_{j \in [J]} \sum_{i \in I_{n,j}}W_{i,j}(Z_{i,j} - \widehat{\Gamma}_n^\top W_{i,j})^\top =0$, and $\sum_{i \in I_{n,j}}W_{i,j}(Z_{i,j} - \widehat{\Gamma}_{n,j}^\top W_{i,j})^\top =0.$
		\citet[Assumption 2(iv) in Section A]{Canay-Santos-Shaikh(2020)} impose the same condition as (\ref{eq: ZW_proj}) and point out that one sufficient but not necessary condition for (\ref{eq: ZW_proj}) is that the distributions of $(Z^{\top}_{i,j}, W^{\top}_{i,j})_{i \in I_{n,j}}$ are the same across clusters. In our empirical application in Section \ref{sec: emp}, we compute $\widehat{Q}_{\widetilde{Z}W,j}$ for each cluster and find that they are all close to zero. We refer to Tables \ref{tab:check_assumption_South}--\ref{tab:check_assumption_West} in the Online Supplement for more details.{\footnote{It is also possible to rigorously test whether $Q_{\widetilde ZW,j} $ is close to zero for $j \in [J]$ if we can consistently estimate the covariance matrix of $\widehat Q_{\widetilde ZW,j}$ under a more detailed within-cluster dependence structure.}} 
		Assumption \ref{assumption: 1}(iii) gives the restriction on cluster sizes, and Assumption \ref{assumption: 1}(iv) ensures $\widetilde Z_{i,j}$ is uniquely defined.

	\end{rem}
	\begin{rem}
		We have \eqref{eq:Wej} and \eqref{eq:mathcalZj} hold if the within cluster dependence is sufficiently weak for some type of CLT to hold. We provide three examples of data structure below that satisfy our requirements. We focus on \eqref{eq:mathcalZj} and let $U_{i,j} = \widetilde Z_{i,j}\eps_{i,j}$, but similar results apply to (\ref{eq:Wej}) by letting $U_{i,j} = W_{i,j}\eps_{i,j}$. 
		
	\end{rem} 
	
	\begin{example} [Serial Dependence] \label{ex:panel}
		We use $i$ and $j$ to index time period and cluster, respectively, so that observations have serial dependence over time and are asymptotically independent across clusters. Such settings were considered in BCH (Lemma 1 and Section 4.1), IM (Section 3.1), and CRS (Section S.1) for time series data and IM (Section 3.2) for panel data,\footnote{Specifically, for time series data, they propose to divide the full sample into $J$ (approximately) equal sized consecutive blocks (clusters). 
			For panel data, assuming independence across individuals, then one may treat the observations for each individual as a cluster.} among others. In this setup, we can verify \eqref{eq:mathcalZj} under different levels of serial dependence.
		\begin{enumerate}
			\item ($L_q$-Mixingale) Let $U_{i,j}^{(k)}$ denote the $k$-th element of $U_{i,j}$. Suppose there exists a filtration $\mathcal{F}_{i,j}$ that satisfies the following conditions: for some $q\geq 3$ and any $l\geq 0$ and $j \in [J]$
			\begin{align*}
			\left\Vert       \mathbb E (U_{i,j}^{(k)} \mid \mathcal{F}_{i-l,j}) \right\Vert_q \leq c_{n_j,i}\psi_l, \quad \left\Vert   U_{i,j}^{(k)}-    \mathbb E (U_{i,j}^{(k)} \mid \mathcal{F}_{i+l,j}) \right\Vert_q \leq c_{n_j,i}\psi_{l+1},
			\end{align*}
			and $\max_{j \in [J]}\max_{i \in I_{n,j}}c_{n_j,i} = o(n^{1/2})$.     Then, \citet[Theorem 4]{LL20} implies \eqref{eq:mathcalZj} holds with $r_n = \sqrt{n}$. In fact, they show that the partial sum process of $\{U_{i,j}\}_{i \in I_{n,j}}$ can be approximated by a martingale, and thus, is called a mixingale. It forms a very general class of models, including martingale differences, linear processes and various types of mixing and near-epoch dependence processes as special cases. 
			\item (Long Memory) Suppose $U_{i,j} = \sum_{l=0}^{\infty}\Psi_l a_{i-l,j}$ where the innovations $a_{i,j} = (a_{i,j}^{(1)},\cdots,a_{i,j}^{(K)})^\top$ are $K$-dimensional martingale difference with respect to an filtration $\mathcal{F}_{i,j}$ such that 
			\begin{align*}
			\max_{i,j,k}\mathbb E (|a_{i,j}^{(k)}|^{2+d}\mid \mathcal{F}_{i-1,j}) < \infty,~a.s. \quad \text{and} \quad \mathbb E (a_{i,j}a_{i,j}^\top \mid \mathcal{F}_{i-1,j}) = \Sigma_a,~ a.s.   
			\end{align*}
			The $K \times K$ matrix coefficient $\Psi_l$ can be approximated by 
			\begin{align*}
			\Psi_l \sim \frac{l^{d-1}}{\Gamma(d)} \Pi, \quad \text{as} \quad l \rightarrow \infty,
			\end{align*}
			where $\Gamma(\cdot)$ is the gamma function, $\Pi$ is a non-singular $K \times K$ matrix of constants that are independent of $l$, and $d \in (0,0.5)$ is the memory parameter. Then, \citet[Theorem 1]{C02} implies (\ref{eq:mathcalZj}) holds with $r_n = n^{1/2-d}$. 
		\end{enumerate}
		
	\end{example}
	
	\begin{example}[Spatial Dependence] \label{ex:spatial} This example is proposed by \cite{Bester-Conley-Hansen(2011)}. Suppose we have $n$ individuals indexed by $l$. For $l$-th individual, its location is denoted as $s_l$, which is an $m$-dimensional integer. The distance between individual $l_1$ and $l_2$ is measured by the maximum coordinatewise metric $\text{dist}(l_1,l_2)  = ||s_{l_1}-s_{l_2}||_\infty$. Both $\widetilde Z$ and $\eps$ are indexed by the location so that $(\widetilde Z_l,\eps_l) = (\widetilde Z_{s_l},\eps_{s_l})$ for $l \in [n]$. The clusters $I_{n,j}$ for $j \in [J]$ are defined as disjoint regions ($\Lambda_1,\cdots,\Lambda_J$). Let $\mathcal{F}_{\Lambda}$ be the $\sigma$-field generated by a given random field $(\widetilde Z_s,\eps_s)$, $s \in \Lambda$ with $\Lambda$ compact and let $|\Lambda|$ be the number of $s \in \Lambda$. Let $\Upsilon_{\Lambda_1,\Lambda_2}$ denote the minimum distance from an element of $\Lambda_1$ to an element of $\Lambda_2$ where the  distance is measured by the maximum coordinatewise metric. The mixing coefficient is then 
		\begin{align*}
		\alpha_{k_1,k_2}(l) & = \sup\{ \mathbb P(A \cap B) - \mathbb P(A)\mathbb P( B) \}, \\
		& s.t. \quad A \in \mathcal{F}_{\Lambda_1},~ B \in \mathcal{F}_{\Lambda_2},~ |\Lambda_1| \leq k_1,~ \quad |\Lambda_2| \leq k_2,~ \Upsilon(\Lambda_1,\Lambda_2) \geq l. 
		\end{align*}
		\cite{Bester-Conley-Hansen(2011)} assume the mixing coefficients satisfy (1) $\sum_{l = 1}^\infty l^{m-1} \alpha_{1,1}(l)^{\delta_1/(2+\delta_1)}<\infty$, (2) $\sum_{l = 1}^\infty l^{m-1} \alpha_{k_1,k_2}(l)<\infty$ for $k_1+k_2\leq 4$, and (3) $\alpha_{1,\infty}(l) = O(l^{-m-\delta_2})$ for some $\delta_1>0$ and $\delta_2>0$. Under this assumption and other regularity conditions in their Assumptions 1 and 2, \citet[Lemma 1]{Bester-Conley-Hansen(2011)} verifies \eqref{eq:mathcalZj} with a finite number of clusters ($J$ fixed) and $r_n = \sqrt{n}$. 
	\end{example}
	
	
	\begin{example}[Network Dependence]\label{ex:network} Suppose we observe $n$ units indexed by $\ell \in [n]$ and an adjacency matrix $\mathcal{A} = \{A_{\ell,\ell'}\}$ where $A_{\ell,\ell'} = 1$ means units $\ell$ and $\ell'$ are linked and $A_{\ell,\ell'} = 0$ means otherwise. We consider a linear-in-means social interaction model studied by \cite{BDF09}. Specifically, we have 
		\begin{align} \label{eq:linear-in-mean}
		y_\ell = \alpha + \beta \frac{\sum_{\ell': A_{\ell,\ell'}=1}y_{\ell'}}{n_\ell} + \gamma B_\ell + \delta \frac{\sum_{\ell': A_{\ell,\ell'}=1}B_{\ell'}}{n_\ell} + \eps_\ell,
		\end{align}
		where $n_\ell = |\ell': A_{\ell,\ell'}=1|$ denote node $\ell$th's number of friends and $B_\ell$ represents node $\ell$th's background characteristics, and we assume $\mathbb E(\eps_\ell|B_\ell) = 0$. In this setup, we have $X_\ell = \frac{\sum_{\ell': A_{\ell,\ell'}=1}y_{\ell'}}{n_\ell}$ and $W_\ell = (1, B_\ell, \frac{\sum_{\ell': A_{\ell,\ell'}=1}B_{\ell'}}{n_\ell})^\top$. Following the literature, we assume the adjacency matrix is independent of $\{B_\ell,\eps_\ell\}_{\ell \in [n]}$. Then, \cite{BDF09} show that one can use $Z = (\tilde A^2 B, \tilde A^3 B)$  as IVs where $\tilde A$ is the $n \times n$ normalized adjacency matrix with a typical entry $\tilde A_{\ell,\ell'} = A_{\ell,\ell'}/n_{\ell}$ and $B$ is a $n \times 1$ vector of $\{B_\ell\}_{\ell \in [n]}$.  The clusters $I_{n,j}$, $j \in [J]$ form a partition of $[n]$. For a subset of indexes $S \subset [n]$, define the conductance of $S$ as $\phi_{\mathcal{A}}(S) = \frac{|\partial_{\mathcal{A}}(S)|}{vol_A(S)}$, where $|\partial_{\mathcal{A}}(S)| = \sum_{\ell \in S}\sum_{\ell' \in [n]/S}{\mathcal{A}}_{\ell,\ell‘}$ is the number of links involving a unit in $S$ and a unit not in $S$ and $vol_{\mathcal{A}}(S) = \sum_{\ell \in S}\sum_{\ell' \in [n]}\mathcal{A}_{\ell,\ell‘}$ is the sum of degrees $\sum_{\ell' \in [n]}{\mathcal{A}}_{\ell,\ell'}$ of units $\ell \in S$. Then, \cite{L22} shows \eqref{eq:mathcalZj} holds with a finite number of clusters ($J$ fixed) and $r_n = \sqrt{n}$ when $\max_{j \in [J]}\phi_\mathcal{A}(I_{n,j}) (\frac{1}{n}\sum_{\ell \in [n]}\sum_{\ell' \in [n]}\mathcal{A}_{\ell,\ell'})\rightarrow 0$ as $n \rightarrow \infty$ and the observations exhibit weak network dependence in the sense of \citet[Assumption 5]{L22}. If the clusters $\{I_{n,j}\}_{j \in [J]}$ are latent, \cite{L22} further shows it is possible to recover the clusters by spectral clustering, a method that clusters the leading $J$ eigenvectors of network graph Laplacian by the k-means algorithm. We provide more details in Section \ref{sec: simu}.   
	\end{example}
	
	\begin{rem}
		As mentioned in the Introduction, our asymptotic framework follows previous studies such as \cite{Ibragimov-Muller(2010)}, \cite{Bester-Conley-Hansen(2011)}, and \cite{Canay-Romano-Shaikh(2017)}, which proposed valid inference methods under the setting that treats the number of clusters as fixed.  
		However, we find in simulations (e.g., see Section \ref{sec: simu}) that for IV regressions, these methods may not control size under data generating processes with heterogeneous cluster sizes and IV strengths, weak IV clusters, or a small number of clusters. 
		The motivation of our study is to propose wild bootstrap-based inference methods that perform well even in such cases and provide the corresponding regularity conditions.
		
		We emphasize that the main restriction of the asymptotic framework that treats the number of clusters as fixed is it requires the within-cluster dependence to be sufficiently weak for some type of CLT to hold within each cluster (as illustrated in Examples \ref{ex:panel}-\ref{ex:network}).
		For example, as pointed out by \citet[Sections 3.1, 3.2, and 4.2]{mackinnon2022cluster}, this requirement rules out the case that $\eps_{i,j}$ follows a factor structure, i.e.,  
		\begin{align}\label{eq: factor-structure}
		\eps_{i,j} = \lambda_{i,j} \eta_{j} + u_{i,j},
		\end{align}
		where $u_{i,j}$ denotes the idiosyncratic error, $\eta_j$ denotes the cluster-wide shock, and the effect of $\eta_j$ on $\eps_{i,j}$ is given by the factor loading $\lambda_{i,j}$. 
		By contrast, such a dependence structure can be handled under the asymptotic framework that lets the number of clusters $J$ to diverge to infinity.  
		Under this framework, a CLT is applied to (appropriately normalized) $\sum_{j \in [J]}\sum_{i \in I_{n,j}}\widetilde{Z}_{i,j}\eps_{i,j}$ 
		and it is thus possible to allow for arbitrary within-cluster dependence under suitable conditions on the cluster heterogeneity; e.g., see \cite{Djogbenou-Mackinnon-Nielsen(2019)} and \cite{Hansen-Lee(2019)} for the regularity conditions.   
		In particular, the factor structure in (\ref{eq: factor-structure}) is allowed as long as $n^{-1}\sup_{j \in [J]}n_j \rightarrow 0$. 
		We also refer interested readers to \citet[Sections 4.1 and 4.2]{mackinnon2022cluster} for an excellent discussion on the two alternative asymptotic frameworks.
	\end{rem}
	
	\begin{rem}
		If there exist cluster fixed effects in the IV model, we can transform our data by projecting out the fixed effects so that our $(y,X,W,Z)$ are expressed as deviations from cluster means and assume that our conditions hold for the model involving the transformed data. 
		Such a practice is also recommended by \citet[Section 2.2]{Djogbenou-Mackinnon-Nielsen(2019)} and \citet[Section 3.2]{mackinnon2022cluster}.
		We refer to Example \ref{ex:1} below for more details. 
		Panel data is a special case of Example \ref{ex:1} in which the clusters are individual units. 
	\end{rem}

	\begin{assumption}\label{assumption: 2}
		\noindent (i) The quantities $\widehat{Q}_{\widetilde{Z}X,j}$, $\widehat{Q}_{\widetilde{Z}\widetilde{Z},j}$, $\widehat{Q}_{\widetilde{Z}X}$, and $\widehat{Q}_{\widetilde{Z}\widetilde{Z}}$ 
		converge in probability to deterministic matrices, which are denoted as $Q_{\widetilde{Z}X,j}$, $Q_{\widetilde{Z}\widetilde{Z},j}$, $Q_{\widetilde{Z}X}$, and $Q_{\widetilde{Z}\widetilde{Z}}$, respectively. 
		
		\noindent (ii) The matrices $Q_{\widetilde{Z}\widetilde{Z},j}$ is invertible for $j \in [J]$.
		
		\noindent (iii) For all $j \in [J]$, $\widehat{Q}_{XX,j} = O_P(1)$,
		$\widehat{Q}_{X\eps,j} = O_P(1)$, and $\widehat{Q}_{\dot{\eps}\dot{\eps},j} \geq c>0$ for constant $c$ with probability approaching one, 
		where $\dot{\eps}_{i,j}$ is the residual from the cluster-level projection of $\eps_{i,j}$ on $W_{i,j}$.
	\end{assumption}
	
	\begin{rem}
		Assumption \ref{assumption: 2} is required for our analysis of the Wald test, but not for the AR test. Assumption \ref{assumption: 2}(i) holds if the dependence of units within clusters is weak enough to render some type of LLN to hold. Assumption \ref{assumption: 2}(ii) is standard in the literature and holds regardless of the IV strength. 
	\end{rem}

	We conclude this section with more concrete examples. Specifically, Assumptions \ref{assumption: 1} and \ref{assumption: 2} hold in Examples \ref{ex:1} and \ref{ex:2}, and we explain why they may not hold in Examples \ref{ex:3} and \ref{ex:4}. For Example \ref{ex:6}, Assumption \ref{assumption: 1} holds while Assumption \ref{assumption: 2} does not. Consequently, the wild bootstrap AR tests are still applicable while the Wald tests are not. 
	
	\begin{example}[Cluster Fixed Effects]\label{ex:1}
		Suppose 
		\begin{align}
		\mathcal Y_{i,j} = \mathcal X_{i,j}^\top \beta + \mathcal W_{i,j}^\top \theta + \eta_{j} + \mathcal E_{i,j},
		\label{eq:ex1}
		\end{align}
		where $\mathcal Y_{i,j}$ is the outcome variable, $\mathcal X_{i,j}$ contains the endogenous variables, $\eta_{j}$ denotes cluster fixed effects, $\mathcal W_{i,j}$ contains the individual-level exogenous variables, $\mathcal E_{i,j}$ is the individual-level idiosyncratic error, and $\mathcal Z_{i,j}$ denotes the IVs. As recommended in \citet[Section 2.2]{Djogbenou-Mackinnon-Nielsen(2019)} and \citet[Section 3.2]{mackinnon2022cluster},
		we can transform the data by projecting out the fixed effects so that our $(y,X,W,Z)$ are expressed as deviations from cluster means.
		Specifically, let $y_{i,j} = \mathcal Y_{i,j} - \frac{1}{n_j} \sum_{i \in I_{n,j}}\mathcal Y_{i,j}$, $X_{i,j} = \mathcal X_{i,j} - \frac{1}{n_j} \sum_{i \in I_{n,j}}\mathcal X_{i,j}$, $W_{i,j} = \mathcal W_{i,j} - \frac{1}{n_j} \sum_{i \in I_{n,j}}\mathcal W_{i,j}$, $e_{i,j} = \mathcal E_{i,j} - \frac{1}{n_j} \sum_{i \in I_{n,j}}\mathcal E_{i,j}$, and $Z_{i,j} = \mathcal Z_{i,j} - \frac{1}{n_j} \sum_{i \in I_{n,j}}\mathcal Z_{i,j}$ be the cluster-level demeaned versions. 
		Then, we can rewrite \eqref{eq:ex1} as \eqref{eq: IV-model} and assume our conditions hold for the model involving the transformed data. 

	\end{example}

	\begin{example}[Heterogeneous IV Strength across Clusters]\label{ex:2}
		We allow for cluster-level heterogeneity with regard to IV strength. Consider the following first stage regression:
		\begin{align*}
		X_{i,j} =     Z_{i,j}^\top \Pi_{z,j,n} + W_{i,j}^\top \Pi_{w,j,n} + V_{i,j},
		\end{align*}
		where $\Pi_{z,j,n}$ and $\Pi_{w,j,n}$ are the coefficients of $Z_{i,j}$ and $W_{i,j}$, respectively, via the cluster-level population projection of $X_{i,j}$ on $Z_{i,j}$ and $W_{i,j}$, for each $j \in [J]$.\footnote{We note $\Pi_{z,j,n}$ and $\Pi_{w,j,n}$ depend on the sample size because the underlying distribution is indexed by $n$.}  Then, our model \eqref{eq: IV-model} allows for both $\Pi_{z,j,n}$ and $\Pi_{w,j,n}$ to vary across clusters. 
		In particular, we allow for some of $\Pi_{z,j,n}$ to decay to or be zero for the bootstrap Wald tests and all $\Pi_{z,j,n}$ to decay to or be zero for the bootstrap AR tests. We will come back to this point in Sections \ref{subsec: boot-wald}--\ref{subsec: boot-robust} with more details. 
	\end{example}
	
	\begin{example}[Heterogeneous Slope for the Endogenous Variable]\label{ex:3}
		Similar to \citet[Example 2]{Canay-Santos-Shaikh(2020)}, we cannot allow for $\beta$ to be heterogeneous across clusters. As a stylized example, let 
		\begin{align}
		y_{i,j} = X_{i,j}^\top \beta_j + W_{i,j}^\top \gamma_j +e_{i,j}, 
		\label{eq:ex2}
		\end{align}
		where $W_{i,j}$ is just the cluster dummies. For $\beta$ equal to a suitable weighted average of $\beta_j$'s, we may rewrite \eqref{eq:ex2} as \eqref{eq: IV-model} with $\eps_{i,j} = X_{i,j}^\top(\beta_j - \beta) + e_{i,j}$, which implies  
		\begin{align*}
		\frac{r_n}{n}  \sum_{i \in I_{n,j}} \widetilde{Z}_{i,j}\eps_{i,j} =  \frac{r_n}{n}  \sum_{i \in I_{n,j}} (Z_{i,j} - \bar Z_j) (X_{i,j}^\top(\beta_j - \beta) + e_{i,j} ).
		\end{align*}
		We then see that Assumption \ref{assumption: 1}(ii) is violated unless $\beta_j = \beta$. 
	\end{example}
	
	\begin{example}[Difference-in-Difference]\label{ex:4} 
		We do not allow for the regression with cluster-level IVs, which usually occurs in difference-in-difference analysis with endogenous treatments. 
		In this setting, the treatment status and group assignment are interpreted as our $X_{i,j}$ and $Z_{i,j}$, respectively, and they are different due to imperfect compliance. 
		In addition, when the assignment is at the cluster level (i.e., some clusters are assigned to the treatment group while the others are assigned to the control group), the value of $Z_{i,j}$
		is invariant within each cluster. If there exists cluster fixed effects which needs to be partialled out as in Example \ref{ex:1}, we have $\tilde Z_{i,j} = 0$, which means our Assumption \ref{assumption: 1}(ii) is violated because $\Sigma_j$ is zero, and thus, degenerate.
		Following the suggestion by \citet[Section S.2 of the supplemental appendix for their Example 3]{Canay-Santos-Shaikh(2020)}, it may be possible to avoid such an issue by clustering more coarsely (e.g., pairing one treated cluster with one control cluster). 
	\end{example}

	\begin{example}[Cluster-level Endogenous Variables]\label{ex:6}
		If $X_{i,j}$ is a cluster-level variable (say, $X_j$), 
		then the resulting within-cluster limiting Jacobian matrix $Q_{\widetilde{Z}X,j}$ may be random and potentially correlated with the within-cluster score component $\mathcal{Z}_j$ 
		as $X_{j}$ is endogenous, which violates Assumption \ref{assumption: 2}(i). We notice that similar issues can arise for the approaches by \cite{Bester-Conley-Hansen(2011)}, \cite{Hwang(2020)}, IM, and CRS. On the other hand, our wild bootstrap AR tests ($AR_n$ and $AR_{CR,n}$) only requires Assumption \ref{assumption: 1} but not Assumption \ref{assumption: 2}, and thus, remain valid in this case.
	\end{example}

	\section{Asymptotic Results for the Wald Tests without CRVE}
	\label{subsec: boot-wald}
	For the Wald test, we further assume the following assumption. 
	\begin{assumption}\label{assumption: 3}
		\noindent (i) $Q_{\widetilde{Z}X}$ is of full column rank. 
		
		\noindent (ii) One of the following two conditions holds: 
		(1) $d_x = 1$,
		or (2) there exists a scalar $a_j$ for each $j \in [J]$ such that  $Q_{\widetilde{Z}X,j} = a_j Q_{\widetilde{Z}X}$. 
		
	\end{assumption}

	Several remarks are in order.
	
	\begin{rem}
		Assumption \ref{assumption: 3}(i) requires (overall) strong identification for $\beta_n$. 
		Assumption \ref{assumption: 3}(ii)(1) states that if there is only \textit{one endogenous variable}, no further restrictions are required. A single endogenous variable is the leading case in empirical applications involving IV regressions. 
		In fact, 101 out of 230 specifications in \citeauthor{Andrews-Stock-Sun(2019)}'s (\citeyear{Andrews-Stock-Sun(2019)}) sample and 1,087 out of 1,359 in \citeauthor{Young(2021)}'s (\citeyear{Young(2021)}) sample has one endogenous regressor and one IV. \cite{lee2021} find that among 123 papers published in \textit{AER} between 2013 and 2019 that include IV regressions, 61 employ single-IV regressions.\footnote{They point out that the single-IV case ``includes applications such as randomized trials with imperfect compliance (estimation of LATE, \cite{Imbens_Angrist_1994}), fuzzy regression discontinuity designs
			(see discussion in \cite{Lee_Lemieux_2010}), and fuzzy regression kink designs (see discussion in
			\cite{Card_et_al_2015})."} 
		\cite{Angrist2021one} also point out that ``most studies using IV (including \cite{Angrist(1990)} and \cite{Angrist-Krueger(1991)}) report just-identified IV estimates computed with a single instrument."\footnote{In our empirical application, we revisit the influential study by \cite{ADH2013}, which also has only one IV.} 
		Assumption \ref{assumption: 3}(ii)(1) further allows for the case of single endogenous regressor and multiple IVs.     
	\end{rem}
	
	\begin{rem}
		Assumption \ref{assumption: 3}(ii)(2) is only needed if we have \textit{multiple endogenous variables}. The condition is similar to that in \citet[Assumption 2(iii)]{Canay-Santos-Shaikh(2020)}, which restricts the type of heterogeneity of the within-cluster Jacobian matrices. 
		However, it is still weaker than restrictions assumed in the literature for cluster-robust Wald tests with few clusters. For example, 
		\cite{Bester-Conley-Hansen(2011)} and \cite{Hwang(2020)} provide asymptotic approximations that are based on $t$ and $F$ distributions for the Wald statistics with CRVE.  
		Their conditions require the within-cluster Jacobian matrices to have the same limit for all clusters (i.e., Assumption \ref{assumption: 3}(ii)(2) to hold with $a_j=1$ for all $j \in [J]$).\footnote{For example, see \citet[Assumptions 3 and 4]{Bester-Conley-Hansen(2011)} and \citet[Assumptions 4 and 5]{Hwang(2020)} for details.}
		They also impose that the cluster sizes are approximately equal for all clusters and the cluster-level scores in Assumption \ref{assumption: 1}(ii) have the same normal limiting distribution for all clusters, which are not necessary for the wild bootstrap. Finally, Assumption \ref{assumption: 3} will not be needed for the bootstrap AR tests in Section \ref{subsec: boot-robust}, which require neither strong identification nor homogeneity Jacobian matrices.

		To further clarify our setting, we can relate the Jacobian matrices with the first-stage projection coefficient. 
		Specifically, recall $\Pi_{z,j,n}$ is the coefficient of $Z_{i,j}$ via the cluster-level population projection of $X_{i,j}$ on $Z_{i,j}$ and $W_{i,j}$ as defined in Example \ref{ex:2}.
		Then, we have $\lim_{n \rightarrow \infty}\Pi_{z,j,n} = \Pi_{z,j} := Q_{\widetilde{Z}\widetilde{Z},j}^{-1} Q_{\widetilde{Z}X,j}$ under our framework. Also define $\Pi_z = Q_{\widetilde{Z}\widetilde{Z}}^{-1} Q_{\widetilde{Z}X}$.
		Assumption \ref{assumption: 3}(i) ensures that overall we have strong identification as $Q_{\widetilde{Z}X}$ (and $\Pi_z$) is of full column rank. 
		Furthermore, we call the clusters in which $Q_{\widetilde{Z}X,j}$ (and $\Pi_{z,j}$) are of full column rank the strong IV clusters, i.e., $\beta_n$ is strongly identified in these clusters. 
		On the other hand, strong identification for $\beta_n$ is not ensured in the rest of the clusters. We can draw three observations from Assumption \ref{assumption: 3}. First, given the number of clusters is fixed, only one strong IV cluster is needed for Assumption \ref{assumption: 3}(i) to hold. Second, Assumption \ref{assumption: 3}(ii)(2) implies that when $a_j \neq 0$, $Q_{\widetilde{Z}X,j}$ (and $\Pi_{z,j}$) is of full column rank, so that the $j$-th cluster is a strong IV cluster. Third, Assumption \ref{assumption: 3}(ii)(2) excludes the case that $Q_{\widetilde{Z}X,j}$ is of a reduced rank but is not a zero matrix when we have multiple endogenous variables. 
		It is possible to select out the clusters with Jacobian matrices of reduced rank 
		\citep{RR00,KP06,CF19}. To formally extend these rank tests to the case with a fixed number of large clusters would merit a separate paper and is thus left  as a topic for future research.
		
	\end{rem}
	
	\begin{assumption}\label{assumption: 4}
		Suppose  $|| \hat{A}_r - A_r ||_{op} = o_P(1)$, where 
		$A_r$ is a $d_r \times d_r$ symmetric deterministic weighting matrix such that $0 < c \leq \lambda_{min}(A_r) \leq \lambda_{max}(A_r) \leq C < \infty$ for some constants $c$ and $C$, $\lambda_{\min}(A)$ and  $\lambda_{\max}(A)$ are the minimum and maximum eigenvalues of the generic matrix $A$, and 
		$||A||_{op}$ denotes the operator norm of the matrix $A$. 
	\end{assumption}
	
	Assumption \ref{assumption: 4} requires that the weighting matrix $\hat{A}_r$ in (\ref{eq: Wald-unstudentized})
	has a deterministic limit.  It rules out the case that $\hat{A}_r$ equals the inverse of CRVE, which has a random limit under a small number of clusters. We will discuss the bootstrap Wald test with CRVE in Section \ref{subsec: boot-wald-stud}. 
	
	
	
	
	\begin{theorem}\label{theo: boot-t}
		Suppose that Assumptions \ref{assumption: 1}-\ref{assumption: 4} hold. Then under $\mathcal{H}_0$, for all four estimation methods (namely, TSLS, LIML, FULL, and BA), 
		\begin{eqnarray*}
			\alpha - \frac{1}{2^{J-1}} \leq \liminf_{n \rightarrow \infty} \mathbb{P} \{ T_{n} > \hat{c}_{n}(1-\alpha) \}
			\leq \limsup_{n \rightarrow \infty} \mathbb{P} \{ T_{n} > \hat{c}_{n}(1-\alpha) \} \leq 
			\alpha + \frac{1}{2^{J-1}}.
		\end{eqnarray*}
	\end{theorem}
	
	Two remarks are in order. 
	
	\begin{rem}
		
		First, Theorem \ref{theo: boot-t} states that as long as there exists at least one strong IV cluster, the $T_n$-based wild bootstrap test has limiting null rejection probability no greater than $\alpha+1/2^{J-1}$ and no smaller than $\alpha - 1/2^{J-1}$. For the test to have power against local alternatives, we later show that at least 5 or 6 strong IV clusters are needed. The error $1/2^{J-1}$ can be viewed as the upper bound for the asymptotic size distortion, which vanishes exponentially with the total number of clusters rather than the number of strong IV clusters. Intuitively, although the weak IV clusters do not contribute to the identification of $\beta_n$, 
		the scores of such clusters still contribute to the limiting distributions of the IV estimators, which in turn determines the total number of sign changes in the bootstrap Wald statistics. We note that  $1/2^{J-1}$ equals $1.56\%$ and $0.2\%$ when $J=7$ and $10$, respectively. If such an distortion is still of concern, researchers can replace $\alpha$ in our context by $\alpha - 1/2^{J-1}$ to ensure null rejection rate.
	\end{rem}

	\begin{rem}
		It is well known that estimators such as LIML and FULL have reduced finite sample bias relative to TSLS in the over-identified case, especially when the IVs are not strong. Since the validity of the randomization with sign changes requires a distributional symmetry around zero, the LIML and FULL-based bootstrap Wald tests may therefore achieve better finite sample size control than that based on TSLS. This is also confirmed by our simulation experiments.\footnote{To theoretically document the asymptotic bias due to the dimensionality of IVs, one needs to consider an alternative framework in which the number of clusters is fixed but the number of IVs tends to infinity, following the literature on many/many weak instruments \citep{Bekker(1994), Chao-Swanson(2005), MS22}. We leave this direction of investigation for future research.}

	\end{rem}

	We next examine the power of the 
	wild bootstrap test against local alternatives.
	\begin{theorem}\label{theo: boot-t-power}
		Suppose that Assumptions \ref{assumption: 1}-\ref{assumption: 4} hold. Further suppose that there exists a subset $\mathcal{J}_s$ of $[J]$ such that $a_j>0$ for each $j \in \mathcal{J}_s$, $a_j = 0$ for $j \in [J] \backslash \mathcal{J}_s$,	and $\lceil |\textbf{G}|(1-\alpha) \rceil \leq |\textbf{G}|-2^{J-J_s+1}$, where $|\textbf{G}| = 2^J$, $J_s = |\mathcal{J}_s|$, and $a_j$ is defined in Assumption \ref{assumption: 3}.\footnote{When $d_x= 1$, we define $a_j = Q^{-1}Q_{\widetilde{Z}X}^\top Q_{\widetilde{Z}\widetilde{Z}}^{-1} Q_{\widetilde{Z}X,j}$.}
		Then under $\mathcal{H}_{1,n}$ in (\ref{eq: hypothesis}), for all four estimation methods (namely, TSLS, LIML, FULL, and BA), 
		\begin{eqnarray*}
			\lim_{|| \mu ||_2 \rightarrow \infty} \liminf_{n \rightarrow \infty} \mathbb{P} \{ T_{n} > \hat{c}_{n}(1-\alpha) \} =1. 
		\end{eqnarray*}
	\end{theorem}
	
	Several remarks are in order. 
	
	\begin{rem}
		
		To establish the power of the $T_n$-based wild bootstrap test against $r_n$-local alternatives, 
		we need 
		homogeneity of the signs of Jacobians for the strong IV clusters (i.e., $a_j>0$ for each $j \in \mathcal J_s$).
		For example, in the case with a single IV, it requires $\Pi_{z,j}$ to have the same sign across all the strong IV clusters. 
		We notice that this condition is not needed for the bootstrap Wald test studentized by CRVE described in Section \ref{subsec: boot-wald-stud}.

	\end{rem}
	
	\begin{rem}
		
		Our test compares the test statistic $T_n$ with the critical value $\hat c_n(1-\alpha)$, where $T_n$ is asymptotically equivalent to $T^*_n (\iota_J)$, i.e., the bootstrap test statistic with $g$ equal to a $J \times 1$ vector of ones, and the critical value $\hat c_n(1-\alpha)$ is just the $\lceil |\textbf{G}|(1-\alpha) \rceil$th order statistic of $\{T^*_n (g)\}_{ g \in \textbf{G}}$. 
		We show that, when $|| \mu ||_2 \rightarrow \infty$ and the signs of $g_j$ for all strong IV clusters are the same (only weak IV clusters have different signs),
		$T^*_n (g)$ is equivalent to $T^*_n (\iota_J)$, and thus, $T_n$, even under the alternative. 
		Intuitively, as $|| \mu ||_2 \rightarrow \infty$, the asymptotic behaviour of $T_n^*(g)$ can be affected by the sign changes only through the strong IV clusters (the effect of sign changes from the weak IV clusters becomes negligible). 
		
		Let us denote the set of such $g$'s that only flip the sign of weak IV clusters as $\textbf{G}_w = \{g: g_j = g_{j'}, \forall j, j' \in \mathcal J_s\}$. Given there are $J_s$ strong IV clusters, the cardinality of $\textbf{G}_w$ is $2^{J-J_s+1}$. 
		To establish the power against $\mathcal{H}_{1,n}$ in Theorem \ref{theo: boot-t-power}, we request that our bootstrap critical value $\hat c_n(1-\alpha)$ does not take values of $T^*_n(g)$ for $g \in \textbf{G}_w$ because otherwise the test statistic $T_n$ and the critical value are asymptotically equivalent even under the alternative. 
		This implies the inequality that 
		\begin{align*}
		\lceil |\textbf{G}|(1-\alpha) \rceil \leq |\textbf{G}|-2^{J-J_s+1}.
		\end{align*}
		Therefore, we need a sufficient number of strong IV clusters to establish the power result. 
		For instance, the condition $\lceil |\textbf{G}|(1-\alpha) \rceil \leq |\textbf{G}|-2^{J-J_s+1}$ requires that 
		$J_s \geq 5$ and $J_s \geq 6$ for $\alpha=10\%$ and $5\%$, respectively. Theorem \ref{theo: boot-t-power} suggests that although the size of the wild bootstrap test is well controlled even with only one strong IV cluster, its power depends on the number of strong IV clusters.
		
	\end{rem}
	
	\begin{rem}
		The wild bootstrap test has resemblance to the group-based $t$-test in IM and the randomization test with sign changes in CRS. IM and CRS approaches separately estimate the parameters using the samples in each cluster (say, $\hat{\beta}_{1}, ..., \hat{\beta}_J$), and therefore requires $\beta_n$ to be strongly identified in all clusters. In contrast, our method allows for some but not all clusters to have weak IVs in the sense of \cite{Staiger-Stock(1997)}, where $\Pi_{z,j,n}$ has the same order of magnitude as $n^{-1/2}$.\footnote{The cluster-level IV estimators of such weak clusters would become inconsistent and have highly nonstandard limiting distributions.} Also, if there exist both strong and ``semi-strong" IV clusters, in which the (unknown) convergence rates of IV estimators can vary among clusters \citep{Andrews-Cheng(2012)}, then the estimators with the slowest convergence rate\footnote{The clusters corresponding to these estimators will become effective (dominant) clusters among the $J$ clusters.} will dominate in the test statistics that are based on the cluster-level estimators. In such cases, IM and CRS can become invalid while our bootstrap test remains valid.  
		On the other hand, if $\beta_n$ is strongly identified in all clusters and the cluster-level IV estimators have minimal finite sample bias, IM and CRS have an advantage over the wild bootstrap when there are multiple endogenous variables as they do not require Assumption \ref{assumption: 3}(ii). 
		Therefore, the two types of approaches could be considered as complements, and practitioners may choose between them according to the characteristics of their data and models.
	\end{rem}

	\section{Asymptotic Results for the Wald Tests with CRVE}\label{subsec: boot-wald-stud}
	
	Now we consider a wild bootstrap test for the Wald statistic studentized by  CRVE. 

	\begin{theorem}\label{theo: boot-stud-t}
		Suppose that Assumptions \ref{assumption: 1}-\ref{assumption: 3} hold, and $J> d_r$. 
		Then under $\mathcal{H}_0$, for all four estimation methods (namely, TSLS, LIML, FULL, and BA),
		\begin{align*}
		\alpha - \frac{1}{2^{J-1}} & \leq \liminf_{n \rightarrow \infty} \mathbb{P} \{ T_{CR,n} > \hat{c}_{CR,n}(1-\alpha) \} \\
		& \leq \limsup_{n \rightarrow \infty} \mathbb{P} \{ T_{CR,n} > \hat{c}_{CR,n}(1-\alpha) \} \leq \alpha + \frac{1}{2^{J-1}}.
		\end{align*} 
	\end{theorem}
	
	We require $J>d_r$ because otherwise CRVE and its bootstrap counterpart are not invertible. Theorem \ref{theo: boot-stud-t} states that with at least one strong IV cluster, the $T_{CR,n}$-based wild bootstrap test controls size asymptotically up to a small error. Next, we turn to the local power.

	\begin{theorem}\label{theo: boot-stud-t-power}
		(i) Suppose that Assumptions \ref{assumption: 1}-\ref{assumption: 3} hold, $J-1> d_r$, and that there exists a subset $\mathcal{J}_s$ of $[J]$ such that $\min_{j \in \mathcal J_s} |a_j| > 0$,
		$a_j=0$ for each $j \in [J] \backslash \mathcal{J}_s$, 
		and $\lceil |\textbf{G}|(1-\alpha) \rceil \leq |\textbf{G}|-2^{J-J_s+1}$, where $|\textbf{G}| = 2^J$, $J_s = |\mathcal{J}_s|$, and $a_j$ is defined in Assumption \ref{assumption: 3}.\footnote{When $d_x= 1$, we define $a_j = Q^{-1}Q_{\widetilde{Z}X}^\top Q_{\widetilde{Z}\widetilde{Z}}^{-1} Q_{\widetilde{Z}X,j}$.}
		Then under $\mathcal{H}_{1,n}$ in (\ref{eq: hypothesis}), for all four estimation methods (namely, TSLS, LIML, FULL, and BA),
		\begin{eqnarray*}
			\lim_{|| \mu ||_2 \rightarrow \infty} \liminf_{n \rightarrow \infty} \mathbb{P} \{ T_{CR,n} > \hat{c}_{CR,n}(1-\alpha) \} =1.
		\end{eqnarray*}
		(ii) Further suppose that $d_r=1$. Then under $\mathcal{H}_{1,n}$ in (\ref{eq: hypothesis}), 
		for any $\delta>0$, there exists a constant $c_{\mu}>0$ such that when $|\mu| >c_\mu$, 
		\begin{align*}
		\liminf_{n \rightarrow \infty}\mathbb{P}(\phi^{cr}_n \geq \phi_n) \geq 1-\delta,
		\end{align*}
		where $\phi^{cr}_n = 1 \{ T_{CR,n} > \hat{c}_{CR,n}(1-\alpha)\}$ and $\phi_n = 1 \{ T_{n} > \hat{c}_{n}(1-\alpha) \}$.
	\end{theorem}

	\begin{rem}
		Different from Theorem \ref{theo: boot-t-power}, the power result in Theorem \ref{theo: boot-stud-t-power} does not require the homogeneity condition on the sign of first-stage coefficients for the strong IV clusters (i.e., it only requires $\min_{j \in \mathcal J_s} |a_j|>0$). 
		Therefore, the bootstrap Wald test studentized by CRVE has an advantage over its unstudentized counterpart, given that the homogeneous sign condition may not hold in some empirical studies (e.g., Figure \ref{fig:first_stage}).
		To study the local power without the sign condition, we rely on arguments very different from those in \cite{Canay-Santos-Shaikh(2020)}.
		
	\end{rem}
	
	\begin{rem}   
		We further establish in Theorem \ref{theo: boot-stud-t-power}(ii) that in the case of a $t$-test (i.e., when the null hypothesis involves one restriction), the rejection of the $T_{CR,n}$-based bootstrap test dominates that based on $T_n$ with a large probability when the local parameter $\mu$ is sufficiently different from zero. To see this, we note that, when there is just one restriction, $
		1\{T_n > \hat{c}_{n}(1-\alpha)\} = 1\{T_{CR,n} >  \tilde{c}_{CR,n}(1-\alpha)\},$ where $\tilde{c}_{CR,n}(1-\alpha)$ denotes the $(1-\alpha)$ quantile of $\left\{|(\lambda_{\beta}^{\top}\hat \beta_{L,g}^* - \lambda_0)|/\sqrt{\lambda_{\beta}^{\top}\widehat{V}\lambda_{\beta}}: g \in \textbf{G}\right\}.$ Then, Theorem \ref{theo: boot-stud-t-power}(ii) follows because $\tilde{c}_{CR,n}(1-\alpha) > \hat{c}_{CR,n}(1-\alpha)$ with large probability as $||\mu||_2$ becomes sufficiently large.
	\end{rem}

	\section{Asymptotic Results for Weak-instrument-robust Tests}\label{subsec: boot-robust}
	
	The size control of the bootstrap Wald tests with or without CRVE relies on Assumption \ref{assumption: 3}(i), which rules out overall weak identification in which all clusters are weak. In the case that the parameter of interest may be weakly identified in all clusters, we can consider the bootstrap AR tests for the full vector of $\beta_n$ as defined in Section \ref{sec:AR_procedure}.

	\begin{assumption}\label{assumption: AR_further_assumptions}
		$|| \hat{A}_z - A_z ||_{op} = o_P(1)$, where 
		$A_z$ is a $d_z \times d_z$ symmetric deterministic weighting matrix such that $0 < c \leq \lambda_{min}(A_z) \leq \lambda_{max}(A_z) \leq C < \infty$ for some constants $c$ and $C$.
	\end{assumption}
	
	
	Theorem \ref{theo: boot-AR} below shows that, in the general case with multiple IVs, the limiting null rejection probability of the $AR_{n}$-based bootstrap test does not exceed the nominal level $\alpha$, and that of the $AR_{CR,n}$ test does not exceed $\alpha$ by more than $1/2^{J-1}$ when $J>d_z$, irrespective of IV strength.
	
	
	
	
	\begin{theorem}\label{theo: boot-AR}
		Suppose Assumption \ref{assumption: 1} holds and $\beta_n=\beta_0$.
		For $AR_n$, further suppose Assumption \ref{assumption: AR_further_assumptions} holds. 
		For $AR_{CR,n}$, further suppose $J> d_z$.
		Then, 
		\begin{align*}
		\alpha - \frac{1}{2^{J-1}} & \leq \liminf_{n \rightarrow \infty} \mathbb{P} \{ AR_{n} > \hat{c}_{AR,n}(1-\alpha) \} \notag \\
		& \leq \limsup_{n \rightarrow \infty} \mathbb{P} \{ AR_{n} > \hat{c}_{AR,n}(1-\alpha) \} \leq \alpha, \; \text{and}
		\notag \\
		\alpha - \frac{1}{2^{J-1}} & \leq  \liminf_{n \rightarrow \infty} \mathbb{P} \{ AR_{CR,n} > \hat{c}_{AR,CR,n}(1-\alpha) \} \\
		& \leq  \limsup_{n \rightarrow \infty} \mathbb{P} \{ AR_{CR,n} > \hat{c}_{AR,CR,n}(1-\alpha) \} \leq \alpha + \frac{1}{2^{J-1}}.   
		\end{align*}
	\end{theorem}
	
	\noindent

	\begin{rem}
		
		First, for the bootstrap AR test studentized by the CRVE, we require the number of IVs to be smaller than the number of clusters because otherwise, the CRVE is not invertible.  
		Second, the behavior of wild bootstrap for other weak-IV-robust statistics proposed in the literature is more complicated as they depend on an adjusted sample Jacobian matrix
		(e.g., see \cite{Kleibergen(2005)}, \cite{Andrews(2016)}, \cite{Andrews-Mikusheva(2016)}, and \cite{Andrews-Guggenberger(2019)}, among others). We study the asymptotic properties of wild bootstrap for these statistics in Section  \ref{sec: other-weak} in the Online Supplement and show that their validity would require at least one strong IV cluster. 
		This provides theoretical support to the
		findings in the literature that the bootstrap AR test has better finite sample size properties than other bootstrap weak-IV-robust tests for clustered data 
		(e.g., see \cite{finlay2014, Finlay-Magnusson(2019)}). 
		Third, for the weak-IV-robust subvector inference, one may use a projection approach
		\citep{Dufour-Taamouti(2005)}
		after implementing the wild bootstrap AR tests for $\beta_n$, but the result may be conservative.\footnote{
			Alternative subvector inference methods (e.g., see Section 5.3 in \cite{Andrews-Stock-Sun(2019)} and the references therein) provide a power improvement over the projection approach with a large number of observations/clusters.
			However, it is unclear whether they can be applied to the current setting.
			It is unknown whether the asymptotic critical values given by these approaches will still be valid with a small number of clusters. 
			Also, \cite{Wang-Doko(2018)} show that bootstrap tests based on the subvector statistics therein may not be robust to weak IVs 
			even under conditional homoskedasticity.}
	\end{rem}
	

	Next, to study the power of the $AR_n$-based bootstrap test against the local alternative, 
	we let $\lambda_{\beta} = I_{d_x}$ for $\mathcal{H}_{1,n}$ in (\ref{eq: hypothesis}) 
	so that $\mu = \mu_{\beta}$ and impose the following condition.  
	
	\begin{assumption}
		\label{ass:boot-AR-power}
		(i) $Q_{\widetilde{Z}X} \neq 0$. 
		(ii) There exists a scalar $a_j$ for each $j \in [J]$ such that $Q_{\widetilde{Z}X,j} = a_jQ_{\widetilde{Z}X}$.
	\end{assumption}
	
	If $d_x = d_z = 1$, Assumption \ref{ass:boot-AR-power}(i) implies strong identification. 
	When $d_z>1$ or $d_x >1$, Assumption \ref{ass:boot-AR-power}(i) only rules out the case that $Q_{\widetilde{Z}X}$ is a zero matrix, 
	while allowing it to be nonzero but not of full column rank. 
	As noted below, this means the AR test can still have local power in some direction even without strong identification. 
	Assumption \ref{ass:boot-AR-power}(ii) is similar to Assumption \ref{assumption: 3}(ii). In particular, it holds automatically if Assumption \ref{ass:boot-AR-power}(i) holds and
	$d_x= d_z=1$ (i.e., single endogenous regressor and single IV).
	
	\begin{theorem}\label{theo: boot-AR-power}
		Suppose Assumptions \ref{assumption: 1}, \ref{assumption: AR_further_assumptions}, and \ref{ass:boot-AR-power} hold.  
		Further suppose that there exists a subset $\mathcal{J}_s$ of $[J]$ such that $\min_{j \in \mathcal J_s} a_j > 0$,
		$a_j=0$ for each $j \in [J] \backslash \mathcal{J}_s$, 
		and $\lceil |\textbf{G}|(1-\alpha) \rceil \leq |\textbf{G}|-2^{J-J_s+1}$, where $J_s = |\mathcal{J}_s|$ and $a_j$ is defined in Assumption \ref{ass:boot-AR-power}.  
		Then, under $\mathcal{H}_{1,n}$ with $\lambda_{\beta} = I_{d_x}$, 
		\begin{eqnarray*}
			\lim_{|| Q_{\widetilde{Z}X}\mu ||_2 \rightarrow \infty} \liminf_{n \rightarrow \infty} \mathbb{P} \{ AR_{n} > \hat{c}_{AR,n}(1-\alpha) \} =1.
		\end{eqnarray*}
	\end{theorem}
	
	\begin{rem}
		Notice that the bootstrap AR test not studentized by CRVE has power against local alternatives as long as $||Q_{\widetilde{Z}X}\mu||_2 \rightarrow \infty$, which may hold even when $\beta_n$ is not strongly identified and $Q_{\widetilde{Z}X}$ is not of full column rank. 
	\end{rem}
	
	\begin{rem}
		When $d_z = 1$, we have
		$1\{AR_n > \hat{c}_{AR,n}(1-\alpha) \} = 1\{AR_{CR,n} > \hat{c}_{AR,CR,n}(1-\alpha) \}$,
		which implies the $AR_n$ and $AR_{CR,n}$-based bootstrap tests have the same power against local alternatives. However, such a power equivalence does not hold when $d_z>1$. 
		Unlike the Wald test with CRVE in Section \ref{subsec: boot-wald-stud}, 
		we cannot establish the power of the AR statistics with CRVE for the general case due to the fact that the CRVE for the Wald test is computed using the estimated residual $\hat{\eps}_{i,j}$ whereas the CRVE for the AR test is computed by imposing the null (see Section \ref{sec:AR_procedure}). Consequently, unlike the Wald test, the AR test with CRVE is not theoretically guaranteed to have better power than the AR test without CRVE. 
		Indeed, we observe in Section \ref{sec: simu} that the $AR_{CR,n}$-based bootstrap test has inferior finite sample power properties compared with its $AR_n$-based counterpart.
		
	\end{rem}
	
	\begin{rem}
		
		In Section \ref{sec: other-weak} in the Online Supplement, we further show that for the common case with a single endogenous variable and a single IV, a wild bootstrap test based on the unstudentized Wald statistic is asymptotically equivalent to a certain wild bootstrap AR test under both null and alternative. 
		Second, we show that bootstrapping the Lagrange multiplier test or conditional quasi-likelihood ratio test controls asymptotic size when there is at least one strong IV cluster.
	\end{rem}

	\section{Monte Carlo Simulations}\label{sec: simu}
	In this section, we investigate the finite sample performance of the wild bootstrap tests and alternative inference methods. We consider four different data generating processes (DGPs) with four dependence structures, namely cluster fixed effect (DGP 1), network dependence (DGP 2), serial dependence (DGP 3), and spatial dependence (DGP 4).  
	For all four DGPs, the number of bootstrap replications equals 399. For DGPs 1, 3, and 4, the number of Monte Carlo replications is 50,000. The second DGP requires implementing spectral clustering for each Monte Carlo replication, which is computationally heavy. For this reason, we set the number of replications as 10,000. The nominal level $\alpha$ is set at 10\%. In addition, following the recommendation in \citet[Section 2.2]{Djogbenou-Mackinnon-Nielsen(2019)} and \citet[Section 3.2]{mackinnon2022cluster}, we conduct cluster-level demean for all DGPs. For the conciseness of the paper, we only report the results for DGPs 1 and 2 in the paper and relegate the results for DGPs 3 and 4 to Section \ref{sec: further-simu} in the Online Supplement. The overall patterns observed from DGPs 3 and 4 are very similar to those from DGPs 1 and 2.

	\subsection{Simulation Designs}\label{subsec:simu_design}
	\textbf{DGP 1.} The first DGP is similar to that in Section IV of \cite{Canay-Santos-Shaikh(2020)} and extend theirs to the IV model. The data are generated as
	\begin{eqnarray*}
		X_{i,j} = \gamma + Z^{\top}_{i,j} \Pi_{z,j} + \sigma(Z_{i,j}) \left( a_{v,j} + v_{i,j} \right), \quad
		y_{i,j} = \gamma + X_{i,j} \beta + \sigma(Z_{i,j}) \left( a_{\eps,j} + \eps_{i,j} \right),  
	\end{eqnarray*}
	for $i=1, ..., n$ and $j=1, ..., J$. 
	We set the number of clusters $J \in \{6,9,12,20\}$. 
	The total number of observations $n$ is equal to 500.
	To allow for unbalanced clusters, we follow \cite{Djogbenou-Mackinnon-Nielsen(2019)} and \cite{Mackinnon2021} and set the cluster sizes as
	\begin{align*}
	n_j = \left[ n \frac{\exp( r \cdot j /J)}{\sum_{j \in [J]} \exp(r \cdot j /J) }  \right], \;\; \text{for} \;\; j = 1, ..., J-1,  
	\end{align*}
	and $n_J = n - \sum_{j \in [j]} n_j$. 
	We let $r=4$ to generate substantial heterogeneity in cluster sizes.\footnote{$r=4$ corresponds to the highest value of the heterogeneity parameter considered in the simulations of \cite{Djogbenou-Mackinnon-Nielsen(2019)} and \cite{mackinnon2022cluster}.} 
	For example, when $J=6$, the cluster sizes are $8,17,33,65,127,$ and $250$, respectively.\footnote{When $J=20$, the cluster sizes are $2,2,3,3,4,5,6,8,10,12,15,18,22,27,33,41,50,61,75$, and $103$, respectively. We use this setting to investigate the robustness of the inference methods 
		under a scenario that is quite different from the fixed-$J$ asymptotic framework. In addition, when $J=20$ and $d_z=3$, we combine the two clusters with $n_j=2$ and combine the two clusters with $n_j=3$, 
		so that IM and CRS can also be implemented.}  
	In addition, $(\eps_{i,j}, v_{i,j})$, $(a_{\eps,j}, a_{v,j})$, $Z_{i,j}$, and $\sigma(Z_{i,j})$ are specified as follows:
	\begin{align*}
	& (\eps_{i,j}, u_{i,j})^{\top} \sim N(0, I_2), \quad
	v_{i,j} = \rho_{\eps v} \eps_{i,j} + (1-\rho_{\eps v}^2)^{1/2} u_{i,j}, \quad 
	(a_{\eps,j}, a_{u,j})^{\top} \sim N(0, I_2), \quad \notag \\
	& a_{v,j} = \rho_{\eps v} a_{\eps, j} + (1-\rho_{\eps v}^2)^{1/2} a_{u,j}, \quad
	Z_{i,j} \sim N(0, I_{d_z}), \quad \text{and} \quad \sigma(Z_{i,j}) = \left(\sum_{k=1}^{d_z} Z_{i,j,k}\right)^2/d_z, 
	\end{align*}
	where 
	$I_{d_z}$ is a $d_z \times d_z$ identity matrix
	and $Z_{i,j,k}$ denotes the $k$-th element of $Z_{i,j}$.\footnote{As mentioned above, we project out the cluster fixed effects. Specifically, we can define $\widetilde{Z}_{i,j} = Z_{i,j} - \frac{1}{n_j}\sum_{i \in I_{n,j}}Z_{i,j}$.  
		Note that $\frac{1}{\sqrt{n_j}}\sum_{i \in I_{n,j}}\widetilde{Z}_{i,j}\sigma(Z_{i,j})(a_{\eps,j}+\eps_{i,j})$ is asymptotically normal conditional on $a_{\eps,j}$, as $n_j$ diverges to infinity. This is because $\left(\frac{1}{\sqrt{n_j}}\sum_{i \in I_{n,j}}Z_{i,j}\left(\sum_{k=1}^{d_z}Z_{i,j,k}^2\right), \left[\frac{1}{\sqrt{n_j}}\sum_{i \in I_{n,j}}Z_{i,j}\right] \left[\frac{1}{n_j} \sum_{i \in I_{n,j}} \left(\sum_{k=1}^{d_z}Z_{i,j,k}^2\right)\right],  \frac{1}{\sqrt{n_j}}\sum_{i \in I_{n,j}}Z_{i,j}\left(\sum_{k=1}^{d_z}Z_{i,j,k}^2\right)\eps_{i,j}\right)$ are jointly asymptotically normal given $(a_{\eps,j})_{j \in [J]}$. Then, we have
		\begin{align*}
		\frac{1}{\sqrt{n_j}}\sum_{i \in I_{n,j}}\widetilde{Z}_{i,j}\sigma(Z_{i,j})(a_{\eps,j}+\eps_{i,j}) & = \frac{a_{\eps,j}}{\sqrt{n_j}}\sum_{i \in I_{n,j}}Z_{i,j}\left(\sum_{k=1}^{d_z}Z_{i,j,k}^2\right) + \frac{1}{\sqrt{n_j}}\sum_{i \in I_{n,j}}Z_{i,j}\left(\sum_{k=1}^{d_z}Z_{i,j,k}^2\right)\eps_{i,j} \\
		& - \left[\frac{1}{\sqrt{n_j}}\sum_{i \in I_{n,j}}Z_{i,j}\right] \left[\frac{a_{\eps,j}}{n_j}\sum_{i \in I_{n,j}}  \left(\sum_{k=1}^{d_z}Z_{i,j,k}^2\right)\right]+ o_P(1),     
		\end{align*}
		which is asymptotically normal given $(a_{\eps,j})_{j \in [J]}$ as the cluster size diverges to infinity, so that Assumption \ref{assumption: 1}(ii) is satisfied.}
	The number of IVs is set to be $d_z \in \{1,3\}$,
	and  $\rho_{\eps v} \in \{0.3, 0.5, 0.7\}$ corresponds to the degree of endogeneity. 
	In addition, we introduce cluster heterogeneity to the $d_z \times 1$ first-stage coefficients $\Pi_{z,j}$ by letting $\Pi_{z,j} = (\Pi/2, ..., \Pi/2)^{\top}$ for $1\leq j \leq J/3$, $\Pi_{z,j} = (\Pi, ..., \Pi)^{\top}$ for $J/3 < j \leq 2J/3$, 
	and $\Pi_{z,j} = (2\Pi, ..., 2\Pi)^{\top}$ for $2J/3 < j \leq J$, 
	where $\Pi \in \{0.25, 0.375, 0.5\}$.  
	Such a DGP satisfies our assumptions and allows for heterogeneity in both cluster sizes and IV strengths. 
	The values of $\beta$ and $\gamma$ are both set to $1$.
	In Section \ref{sec: further-simu} in the Online Supplement, we further report the simulation results for the case where $\Pi$ remains the same across clusters, so that the cluster-level heterogeneity in identification strength originates solely from the heterogeneity in cluster size. We find that the overall patterns remain very similar to those reported in Section \ref{subsec: simu-results} below. 
	

	\noindent \textbf{DGP 2.} We consider the linear-in-mean social interaction model detailed in Example \ref{ex:network}. Specifically, following \cite{L22}, we generate the network $\mathcal A$ with $n=500$ nodes as $\mathcal A_{i,j} = 1\{||\eta_{i} - \eta_{j}||_2 \geq (7/(\pi n))^{1/2}\}$, where $\eta_i \stackrel{i.i.d.}{\sim} \text{Uniform}[0,1]^2$. Following \cite{BDF09}, we generate data according to  \eqref{eq:linear-in-mean} in which $\eps_i \stackrel{i.i.d.}{\sim} \N(0,0.1)$, 
	\begin{align*}
	B_i = \begin{cases}
	0 & \text{ with probability $1/2$} \\
	\exp( - \log(2) + (\log(4))^{1/2}\eps_i') & \text{ with probability $1/2$} 
	\end{cases},
	\end{align*}
	where $\eps_i' \stackrel{i.i.d.}{\sim}\N(0,1)$ and $(\eps_i,\eps_i',\eta_i)$ are independent. The observables $(y_i,X_i,W_i,Z_i)$ are as defined in Example \ref{ex:3} with $d_z=2$.  
	According the calibration study by \cite{BDF09}, we set the parameters $(\alpha,\beta,\delta) = (0.7683,0.4666,0.1507)$. Because the IV strength scales with $\gamma \beta + \delta$, we study the performance of our method under various identification strength by setting this value to be $(0.11, 0.1498, 0.1896)$,
	which correspond to values 
	$(-0.0872,   -0.0019,    0.0834)$
	for $\gamma$. For comparison, we note that \cite{BDF09} set $\gamma = 0.0834$. We follow the classification procedure proposed by \cite{L22} to obtain the clusters. 
	\begin{enumerate}
		\item Input: a positive integer $L$ and network $\mathcal A$. 
		\item Compute all separated components (no links between two components) of the network denoted as  $\{\mathcal V_{h}\}_{h \in \{0\} \cup [H]}$, where $\{\mathcal V_{h}\}_{h \in \{0\} \cup [H]}$ is a partition of $[n]$ (n vertexes) and they are sorted in ascending order according to their sizes. We keep all the components whose sizes are greater than 5. Denote the number of components left as $L'+1$ for some $L' \geq 0$.  
		\item Suppose the biggest component $\mathcal V_0$ has size $\tilde n_0$. By permuting labels, we suppose  $\mathcal V_0 = [\tilde n_0]$ and denote its  adjacency matrix as $\mathcal A_0$. Then, we compute the graph Laplacian as 
		\begin{align*}
		\mathcal L_0 = I_{\tilde n_0} - D_0^{-1/2} \mathcal A_0 D_0^{-1/2},    
		\end{align*}
		where   $D_0 = \diag(\sum_{i \in [\tilde n_0]} A_{0,1,i},\cdots,\sum_{i \in [\tilde n_0]} A_{0,\tilde n_0,i})$ is an $\tilde n_0 \times \tilde n_0$ diagonal matrix of degrees.  
		\item Obtain the top $L$ eigenvector matrix of $\mathcal L_0$ corresponding to its $L$ largest eigenvalues and denote it as $ V = [V_1^\top,\cdots,V_{\tilde n_0}^\top]^\top,$
		where $V_{i} \in \Re^{L}$ for $i \in [\tilde n_0]$. 
		\item Apply K-means algorithm to $V$ and divide $\mathcal V_0 = [\tilde n_0]$ into $L$ groups, denoted as $\mathcal V_{0,1},\cdots,\mathcal V_{0,L}$. 
		\item Output: we obtain $J = L+L'$ clusters $(\mathcal V_{0,1},\cdots,\mathcal V_{0,L},\mathcal V_{1},\cdots,\mathcal V_{L'})$. 
	\end{enumerate}
	We set $n=500$ and let $L$ be $[5,10,20,30]$. The number of clusters ($J$) depends on $L'$ which varies across simulation replications. We report the average $J$ below.

	
	\subsection{Inference Methods}\label{sec: inf-method}

	We investigate the finite sample performance of the following inference methods. 
	\begin{enumerate}
		\item IM: This inference method is proposed by \cite{Ibragimov-Muller(2010)}, which compares their group-based $t$-test statistic with the critical value of a $t$-distribution with $J-1$ degrees of freedom. The test statistic is constructed by separately running the IV regression using the samples in each cluster.  
		\item CRS: The approximate randomization test proposed by \cite{Canay-Romano-Shaikh(2017)}, which compares IM's test statistic with the critical value of the sign changes-based randomization distribution of the statistic. 
		\item ASY: Compare $T_{CR,n}$, the Wald statistic studentized by CRVE as described in Section \ref{sec:wald_procedure}, with a standard normal critical value (as $r=1$ here). 
		This is the standard Wald test.  
		\item BCH: This inference method is proposed by \cite{Bester-Conley-Hansen(2011)}, which compares $T_{CR,n}$ with $\sqrt{\frac{J}{J-1}}$ times 
		the critical value of a $t$-distribution with $J-1$ degrees of freedom.
		\item WRE: Compare the studentized Wald statistic $T_{CR,n}$ with the critical value generated by the WRE cluster (WREC) bootstrap procedure, as described in \cite{Davidson-Mackinnon(2010)}, \cite{Finlay-Magnusson(2019)}, \cite{Roodman-Nielsen-MacKinnon-Webb(2019)}, and \cite{Mackinnon2021}. 
		\item W-B: Compare the unstudentized Wald statistic $T_n$, which sets the weighting matrix $\hat A_r = 1$ in (\ref{eq: Wald-unstudentized}), with the wild bootstrap critical value $\hat c_n$ as described in Section \ref{sec:wald_procedure}. Therefore, this bootstrap procedure does not involve the CRVE-based studentization. 
		\item W-B-S: Compare the studentized Wald statistic $T_{CR,n}$ with the wild bootstrap critical value $\hat c_{CR,n}$ as described in Section \ref{sec:wald_procedure}.  Its pseudo code is provided in Section \ref{sec:code} in the Online Supplement. 
		\item AR-ASY: Compare $AR_{CR,n}$ with $\chi^2(d_z)$ critical value, 
		where $AR_{CR,n}$ is the AR statistic studentized by CRVE and $d_z$ is the number of instruments.  
		\item AR-B: Compare the unstudentized AR statistic $AR_n$, which sets the weighting matrix $\hat A_z = I_{d_z}$ in (\ref{eq: AR-unstud-definition}), with the wild bootstrap critical value $\hat c_{AR,n}$ as described in Section \ref{sec:AR_procedure}. 
		\item AR-B-S: Compare $AR_{CR,n}$ with the wild bootstrap critical value $\hat c_{AR,CR,n}$ as described in Section \ref{sec:AR_procedure}.       
	\end{enumerate}
	Except for the methods AR-ASY, AR-B, and AR-B-S, all other methods require the estimation of parameter $\beta$. All four k-class estimators mentioned in the paper can be used. Here, we report the results regarding the TSLS estimator for simulations with one IV and report those regarding the FULL estimator for simulations with multiple IVs, as it is known that FULL has reduced finite sample bias than TSLS in the over-identified case.\footnote{The performance of LIML is very similar to FULL in our simulations. We therefore omit the LIML results but they are available upon request.}

	
	\subsection{Simulation Results}\label{subsec: simu-results}
	
	\noindent \textbf{DGP 1.} Tables \ref{tab:dgp1_size_1}-\ref{tab:dgp1_size_2} report the null empirical rejection frequencies of the ten inference methods described above with $J=6$ or $12$. For succinctness, we report the results for $J=9$ or $20$ in Section \ref{subsec: further-simu-DGP1} of the Online Supplement.  
	Also as mentioned above, the results of the Wald tests are based on TSLS when $d_z=1$ and based on FULL 
	when $d_z=3$, respectively. Following the recommendation in the literature, we set the tuning parameter of FULL to be 1.\footnote{In this case FULL is best unbiased to a second order among $k$-class estimators under normal errors \citep{Rothenberg(1984)}.} Several observations are in order. 
	\begin{enumerate}
		\item In general, size distortions increase when the first-stage coefficient $\Pi$ become small, the degree of endogeneity $\rho_{\eps v}$ becomes high, or the IV model becomes over-identified ($d_z=3$).
		
		\item  IM and CRS tests can have considerable over-rejections when the number of clusters is relatively large. For example, when $J=12$, the maximum null empirical rejection frequencies for IM are 0.185 in Table \ref{tab:dgp1_size_1} and 0.209 in Table \ref{tab:dgp1_size_2}, while the maximum rejection frequencies for CRS are 0.137 and 0.242, respectively. 
		Similarly, when $J=20$, 
		the maximum null rejection frequencies for IM are 0.209
		in Table \ref{tab:dgp1_size_1_9_20}
		and 0.286 in Table \ref{tab:dgp1_size_2_9_20}, while those for CRS 
		are 0.142 and 0.302, respectively, in Section \ref{subsec: further-simu-DGP1} of the Online Supplement.  
		
		\item ASY and BCH tests typically have large size distortions when the number of clusters is relatively small. For example, when $J=6$ and $\Pi = 0.25$, for $\rho_{\eps v}=0.3, 0.5,$ and $0.7$, the null empirical rejection frequencies of ASY are 0.211, 0.219, and 0.228, respectively, in Table \ref{tab:dgp1_size_1}
		and 0.254, 0.258, and 0.265, respectively, in Table \ref{tab:dgp1_size_2}. BCH improves upon ASY, but still has corresponding rejection frequencies equal to 0.147, 0.154, and 0.168, respectively, in Table \ref{tab:dgp1_size_1}, and 0.180, 0.185, and 0.193, respectively, in Table \ref{tab:dgp1_size_2}. 
		
		\item Among the Wald-based inference methods (IM, CRS, ASY, BCH, WRE, W-B, and W-B-S), only the three wild bootstrap tests (WRE, W-B, and W-B-S) have good size control across different settings of $J$, $\Pi$, $\rho_{\eps v}$, and $d_z$.
		
		\item All the three AR-based inference methods control the size across different settings, but AR-ASY, which is based on the chi-squared critical values, can be rather conservative in the over-identified case. In particular, in Table \ref{tab:dgp1_size_2}, it does not reject the null when $J=6$ and has very low rejection frequencies when $J=9$ in Table \ref{tab:dgp1_size_2_9_20} in the Online Supplement.\footnote{We notice that the null rejection probabilities of the $AR_{CR,n}$-based asymptotic test decrease toward zero when $d_z$ approaches $J$. If $d_z$ is equal to $J$, the value of $AR_{CR,n}$ will be exactly equal to $d_z$ (or $J$), and thus has no variation (for $\bar{f} = \left(\bar{f}_{1}, ..., \bar{f}_J \right)^{\top}$ 
			and $\bar{f}_{j} = n^{-1} \sum_{i \in I_{n,j}} f_{i,j},$ 
			$
			AR_{CR,n} = 
			\iota_J^{\top} \bar{f} \left( \bar{f}^{\top} \bar{f} \right)^{-1}
			\bar{f}^{\top} \iota_J = \iota_J^{\top} \iota_J = d_z
			$
			as long as $\bar{f}$ is invertible, where $\iota_J$ denotes a $J$-dimensional vector of ones). By contrast, the $AR_{n}$-based bootstrap test works well even when $d_z$ is larger than $J$.}    
		
	\end{enumerate}
	
\begin{figure}[h] 
		\vspace{-2cm}
		\makebox[\textwidth]{\includegraphics[width=1\paperwidth,height=0.43\textheight]{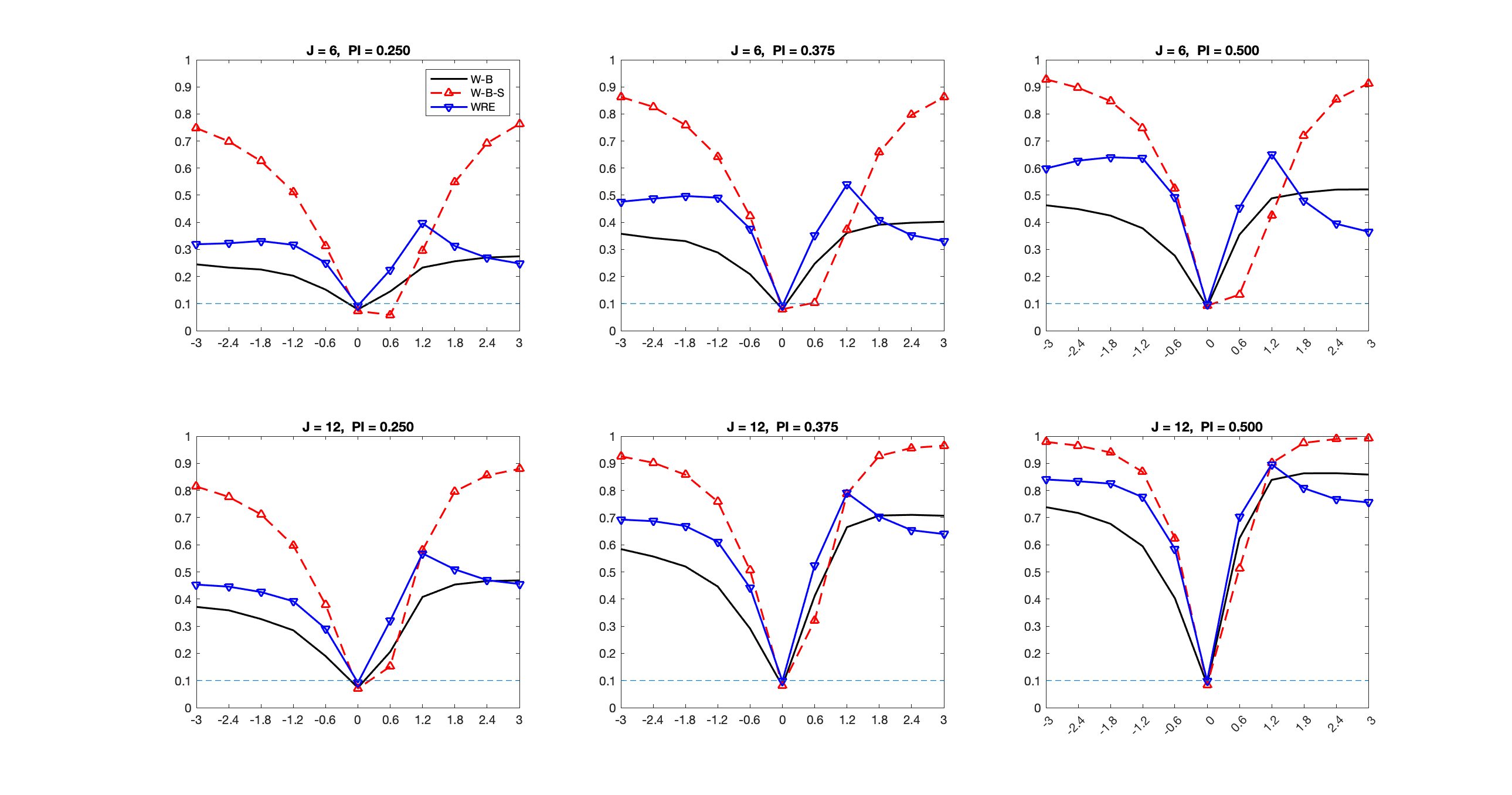}}
		\vspace*{-18mm}
		\caption{Power Comparison for Wald Tests under DGP 1 with $d_z=1$ and $J=6$ or $12$}
		\label{fig:power-dgp1-wald-K1}
		{\footnotesize{Note: W-B: dark solid line; W-B-S: red dashed line with upward-pointing triangle; WRE: blue solid line with downward-pointing triangle.}}
		
		\makebox[\textwidth]{\includegraphics[width=1\paperwidth,height=0.53\textheight]{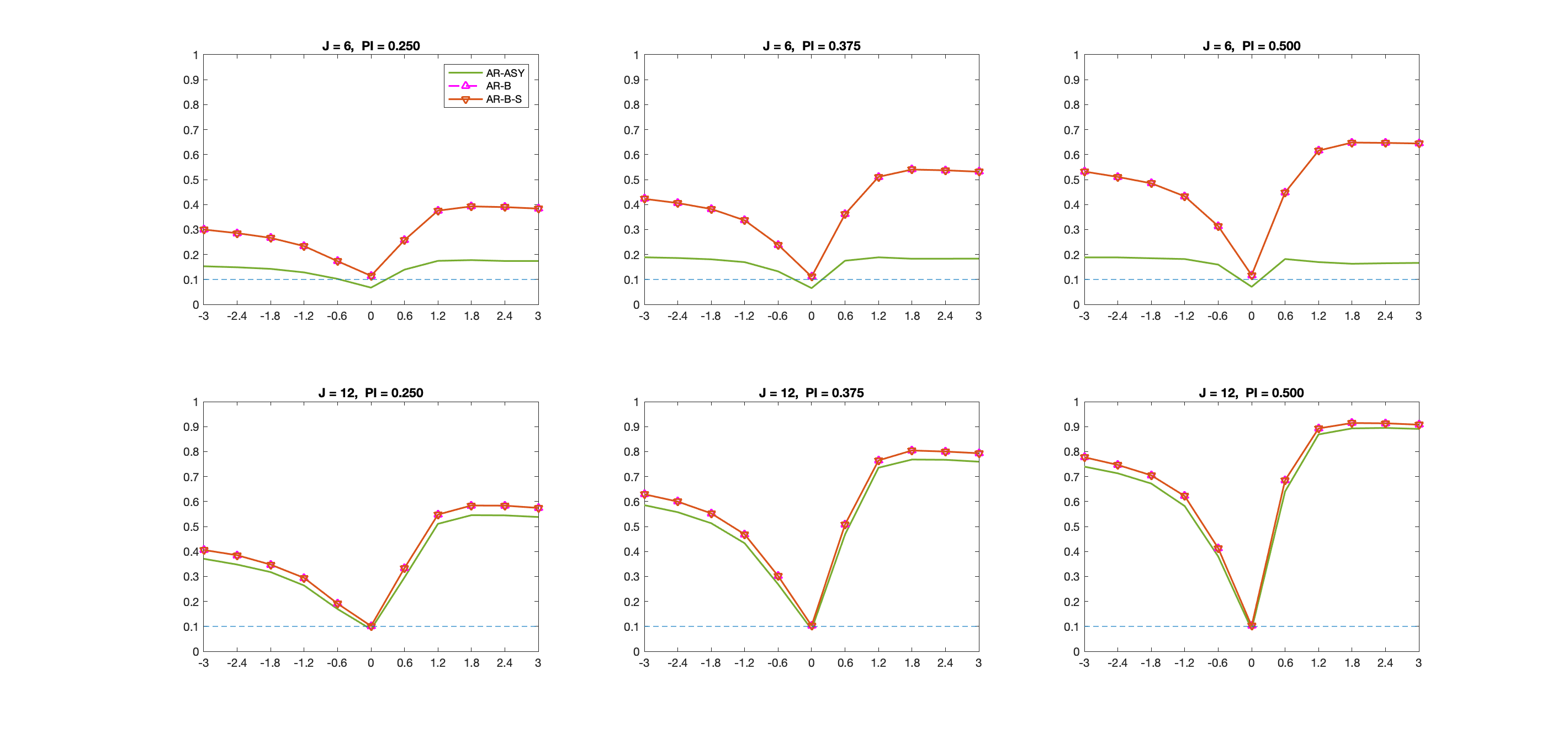}}
		\vspace*{-18mm}
		\caption{Power Comparison for AR Tests under DGP 1 with $d_z=1$ and $J=6$ or $12$}
		\label{fig:power-dgp1-ar-K1}
		{\footnotesize{Note: AR-ASY: dark green solid line; AR-B: magenta dashed line with upward-pointing triangle; AR-B-S: gold solid line with downward-pointing triangle.}}
	\end{figure}

\begin{figure}[H] 
	\vspace{-2cm}
	\makebox[\textwidth]{\includegraphics[width=1\paperwidth,height=0.5\textheight]{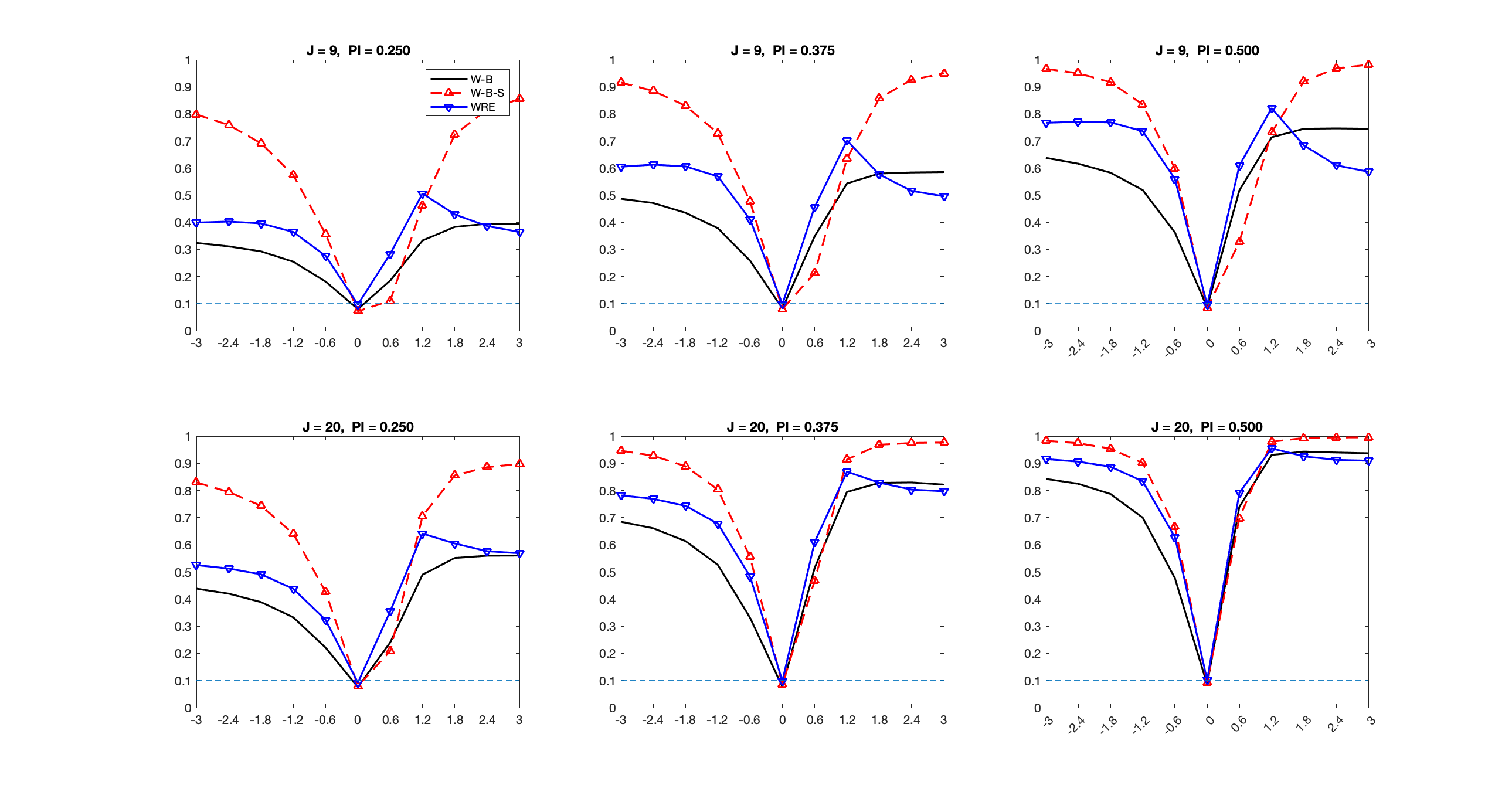}}
	\vspace*{-12mm}
	\caption{Power Comparison for Wald Tests under DGP 1 with $d_z=1$ and $J=9$ or $20$}
	\label{fig:power-dgp1-wald-K1-J920}
	{\footnotesize{Note: W-B: dark solid line; W-B-S: red dashed line with upward-pointing triangle; WRE: blue solid line with downward-pointing triangle.}}
	
	\makebox[\textwidth]{\includegraphics[width=1\paperwidth,height=0.5\textheight]{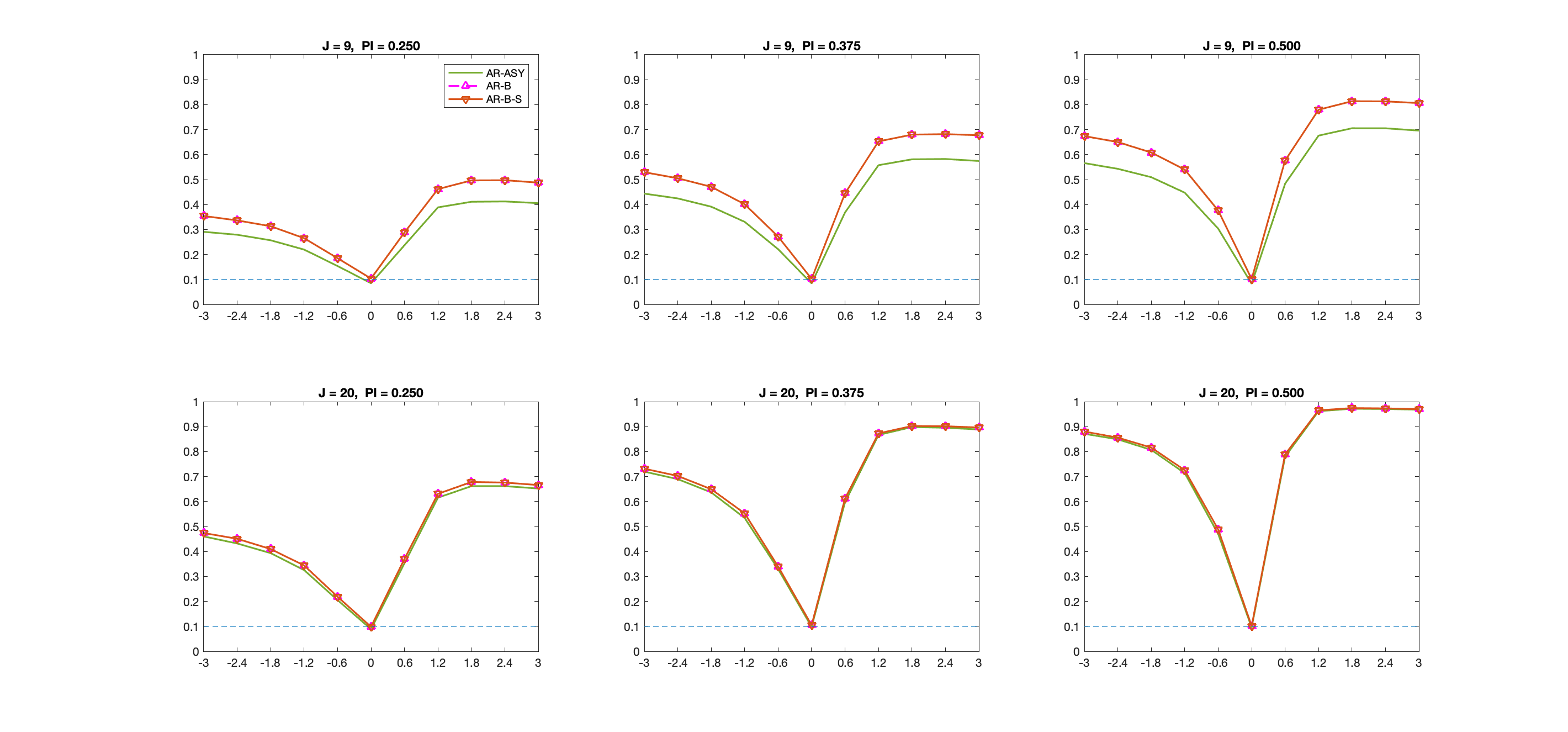}}
	\vspace*{-12mm}
	\caption{Power Comparison for AR Tests under DGP 1 with $d_z=1$ and $J=9$ or $20$}
	\label{fig:power-dgp1-ar-K1-J920}
	{\footnotesize{Note: AR-ASY: dark green solid line; AR-B: magenta dashed line with upward-pointing triangle; AR-B-S: gold solid line with downward-pointing triangle. }}
\end{figure}

\begin{figure}[H] 
	\vspace{-2cm}
	\makebox[\textwidth]{\includegraphics[width=1\paperwidth,height=0.5\textheight]{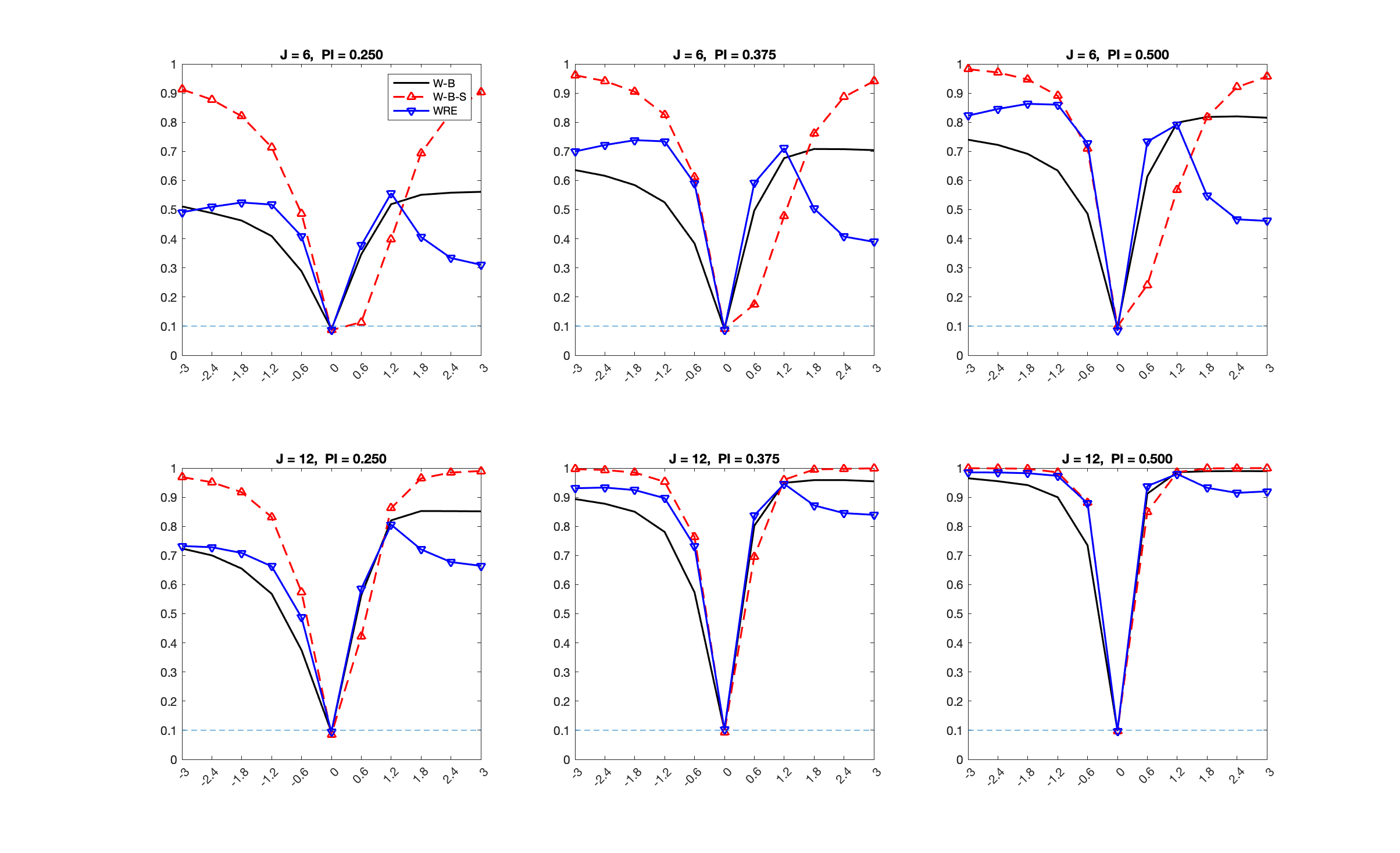}}
	\vspace*{-12mm}
	\caption{Power Comparison for Wald Tests under DGP 1 with $d_z=3$ and $J=6$ or $12$}
	\label{fig:power-dgp1-wald-K3-J612}
	{\footnotesize{Note: W-B: dark solid line; W-B-S: red dashed line with upward-pointing triangle; WRE: blue solid line with downward-pointing triangle.}}
	
	\makebox[\textwidth]{\includegraphics[width=1\paperwidth,height=0.5\textheight]{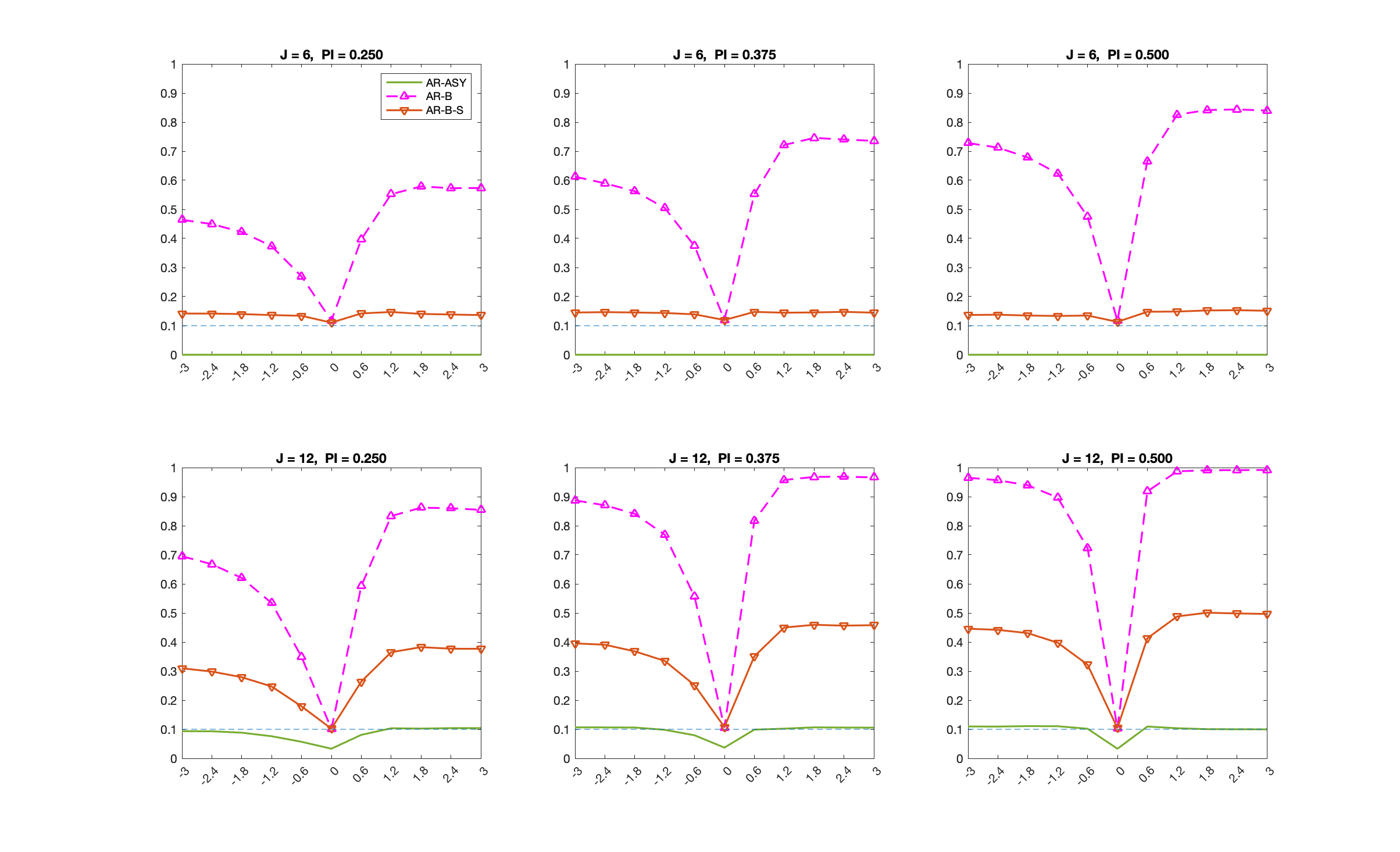}}
	\vspace*{-12mm}
	\caption{Power Comparison for AR Tests under DGP 1 with $d_z=3$ and $J=6$ or $12$}
	\label{fig:power-dgp1-ar-K3-J612}
	{\footnotesize{Note: AR-ASY: dark green solid line; AR-B: magenta dashed line with upward-pointing triangle; AR-B-S: gold solid line with downward-pointing triangle. }}
\end{figure}

\begin{figure}[H] 
	\vspace{-2cm}
	\makebox[\textwidth]{\includegraphics[width=1\paperwidth,height=0.5\textheight]{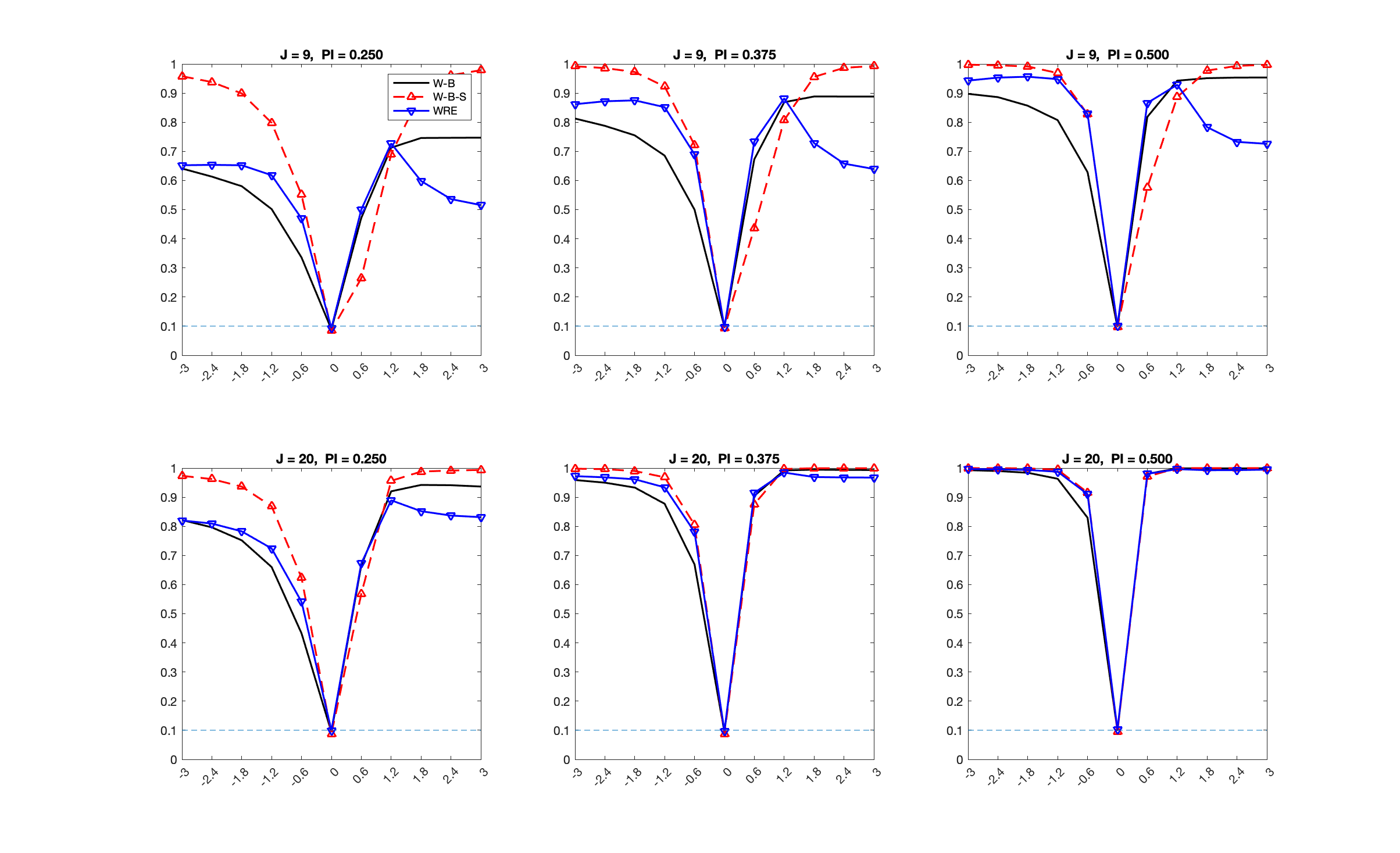}}
	\vspace*{-12mm}
	\caption{Power Comparison for Wald Tests under DGP 1 with $d_z=3$ and $J=9$ or $20$}
	\label{fig:power-dgp1-wald-K3-J920}
	{\footnotesize{Note: W-B: dark solid line; W-B-S: red dashed line with upward-pointing triangle; WRE: blue solid line with downward-pointing triangle.}}
	
	\makebox[\textwidth]{\includegraphics[width=1\paperwidth,height=0.5\textheight]{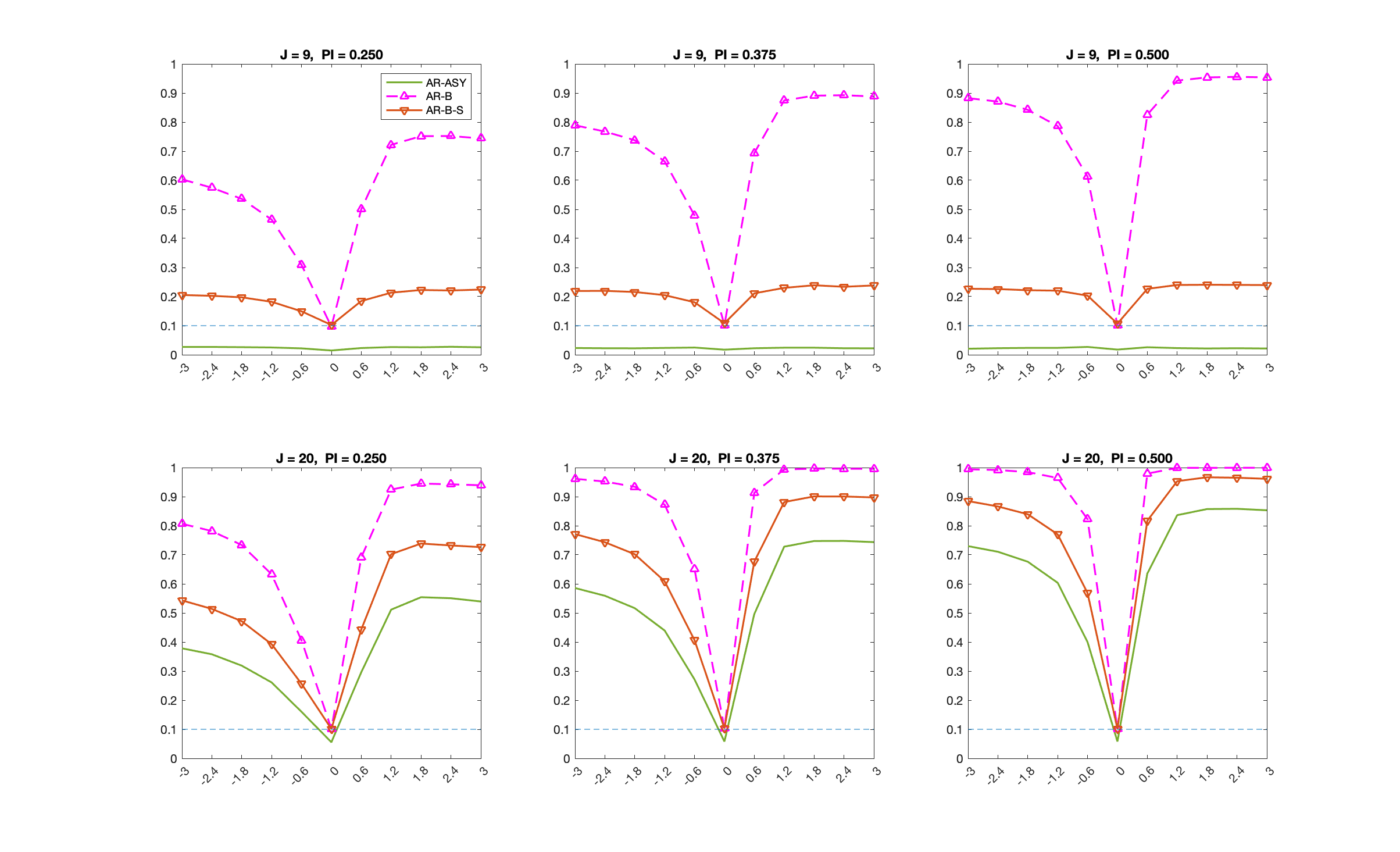}}
	\vspace*{-12mm}
	\caption{Power Comparison for AR Tests under DGP 1 with $d_z=3$ and $J=9$ or $20$}
	\label{fig:power-dgp1-ar-K3-J920}
	{\footnotesize{Note: AR-ASY: dark green solid line; AR-B: magenta dashed line with upward-pointing triangle; AR-B-S: gold solid line with downward-pointing triangle. }}
\end{figure}

	Furthermore, for power comparisons we focus on the inference methods that control size throughout the above simulations (namely, WRE, W-B, W-B-S, AR-ASY, AR-B, and AR-B-S).
	The true value of $\beta$ is set equal to 1 and we vary the value of $\beta_0-\beta$ from $-3$ to $3$. 
	$\rho_{\eps v}$ is set equal to $0.5$.
	Figures \ref{fig:power-dgp1-wald-K1} and \ref{fig:power-dgp1-ar-K1} show the power curves of the bootstrap Wald tests and the AR tests, respectively, with $d_z=1$ and $J=6$ or $12$. 
	Figures \ref{fig:power-dgp1-wald-K1-J920} and \ref{fig:power-dgp1-ar-K1-J920} in the Online Supplement show the power curves with $d_z=1$ and $J=9$ or $20$. 
	In addition, Figures \ref{fig:power-dgp1-wald-K3-J612}--\ref{fig:power-dgp1-ar-K3-J920} report those for $d_z=3$. We highlight several observations below. 
	
	\begin{enumerate}
		\item In general, all the tests become more powerful when the IV strength becomes stronger and/or the number of clusters becomes larger. 
		
		\item According to Figures \ref{fig:power-dgp1-wald-K1}, \ref{fig:power-dgp1-wald-K1-J920},
		\ref{fig:power-dgp1-wald-K3-J612}, and \ref{fig:power-dgp1-wald-K3-J920}, when the alternative is sufficiently distant, W-B-S (the bootstrap Wald test studentized by CRVE) is more powerful than W-B, especially when the identification is relatively weak and/or the number of clusters is small. This is in line with our theoretical result in Theorem \ref{theo: boot-stud-t-power}(ii) of Section \ref{subsec: boot-wald-stud}.   
		
		\item In Figures \ref{fig:power-dgp1-wald-K1}, \ref{fig:power-dgp1-wald-K1-J920}, 
		\ref{fig:power-dgp1-wald-K3-J612}, and \ref{fig:power-dgp1-wald-K3-J920}, the power curves of WRE are between those of W-B-S and W-B in most cases and highest for certain alternatives, but they may decrease when the alternative becomes more distant, especially in the cases with relatively weak IVs and few clusters. 
		
		\item Figures \ref{fig:power-dgp1-ar-K1}, \ref{fig:power-dgp1-ar-K1-J920}, \ref{fig:power-dgp1-ar-K3-J612}, and \ref{fig:power-dgp1-ar-K3-J920} show that among the AR tests, AR-B (the wild bootstrap AR tests without CRVE studentization) has the highest power, followed by AR-B-S.\footnote{In the case with one IV ($d_z=1$), AR-B and AR-B-S are equivalent so that their power curves coincide with each other.} AR-ASY has the lowest power among the three, which is in line with the under-rejections observed in the size results. 
		
		\item Comparing Figures \ref{fig:power-dgp1-wald-K1} and \ref{fig:power-dgp1-ar-K1}, we find that overall W-B-S has the best power properties among these tests. Similar observation can be made by comparing the corresponding figures of bootstrap Wald and AR tests in the Online Supplement. 
	\end{enumerate}
	
	\noindent \textbf{DGP 2.}
	Table \ref{tab:dgp4_size} reports the size for the network data.
	Note the identification strength increases with $\gamma$. 
	Similar to DGP 1, we find that ASY and BCH does not control size when the number of clusters is small, while IM and CRS  have large size distortions when the number of clusters is large so that within each cluster, the number of observations is not sufficiently large to render asymptotic normality. 
	Instead, the wild bootstrap-based inference methods (WRE, W-B, W-B-S, AR-B, and AR-B-S) have good size control regardless of the number of clusters. 
	In Section \ref{subsec: further-simu-DGP2}, we further report the power results for DGP 2. 
	Consistent with DGP 1 and other simulation designs, we find that our W-B-S has the best power, especially against distant alternatives. 
	
	\begin{table}[h]
		\adjustbox{max width=\textwidth}{%
			\centering
			\begin{tabular}{ll|lll|lll|lll}
				&        &         & $\rho_{\eps v} = 0.3$  &        &         & $\rho_{\eps v} = 0.5$  &        &         & $\rho_{\eps v} = 0.7$  &        \\
				&        & $\Pi=0.25$ & $\Pi=0.375$ & $\Pi=0.5$ & $\Pi=0.25$ & $\Pi=0.375$ & $\Pi=0.5$ & $\Pi=0.25$ & $\Pi=0.375$ & $\Pi=0.5$ \\
				\hline
				$J=6$  & IM        & 0.063 & 0.057 & 0.055 & 0.083 & 0.078 & 0.071 & 0.139 & 0.114 & 0.093 \\
				& CRS       & 0.117 & 0.112 & 0.103 & 0.136 & 0.135 & 0.131 & 0.133 & 0.123 & 0.129 \\
				& ASY       & 0.211 & 0.236 & 0.259 & 0.219 & 0.240 & 0.257 & 0.228 & 0.241 & 0.255 \\
				& BCH       & 0.147 & 0.165 & 0.183 & 0.154 & 0.173 & 0.186 & 0.168 & 0.175 & 0.189 \\
				& WRE       & 0.093 & 0.092 & 0.086 & 0.090 & 0.095 & 0.093 & 0.097 & 0.097 & 0.097 \\
				& W-B & 0.078 & 0.086 & 0.091 & 0.072 & 0.086 & 0.093 & 0.074 & 0.079 & 0.089 \\
				& W-B-S   & 0.058 & 0.063 & 0.066 & 0.071 & 0.083 & 0.093 & 0.112 & 0.106 & 0.105 \\
				& AR-ASY    & 0.068 & 0.067 & 0.069 & 0.068 & 0.072 & 0.073 & 0.066 & 0.068 & 0.071 \\
				& AR-B & 0.117 & 0.115 & 0.115 & 0.110 & 0.123 & 0.122 & 0.114 & 0.113 & 0.118 \\
				& AR-B-S   & 0.117 & 0.115 & 0.115 & 0.110 & 0.123 & 0.122 & 0.114 & 0.113 & 0.118 \\
				\hline
				$J=12$ & IM        & 0.079 & 0.070 & 0.068 & 0.119 & 0.107 & 0.095 & 0.185 & 0.149 & 0.143 \\
				& CRS       & 0.126 & 0.122 & 0.117 & 0.137 & 0.130 & 0.129 & 0.134 & 0.132 & 0.127 \\
				& ASY       & 0.115 & 0.157 & 0.167 & 0.142 & 0.156 & 0.175 & 0.162 & 0.163 & 0.173 \\
				& BCH       & 0.091 & 0.127 & 0.137 & 0.119 & 0.128 & 0.143 & 0.139 & 0.138 & 0.143 \\
				& WRE       & 0.092 & 0.103 & 0.099 & 0.095 & 0.098 & 0.099 & 0.098 & 0.093 & 0.099 \\
				& W-B & 0.070 & 0.089 & 0.091 & 0.074 & 0.080 & 0.089 & 0.080 & 0.073 & 0.085 \\
				& W-B-S   & 0.064 & 0.069 & 0.074 & 0.080 & 0.084 & 0.086 & 0.110 & 0.098 & 0.102 \\
				& AR-ASY    & 0.086 & 0.095 & 0.088 & 0.086 & 0.091 & 0.093 & 0.089 & 0.089 & 0.090 \\
				& AR-B & 0.096 & 0.107 & 0.103 & 0.099 & 0.104 & 0.103 & 0.099 & 0.102 & 0.104 \\
				& AR-B-S   & 0.096 & 0.107 & 0.103 & 0.099 & 0.104 & 0.103 & 0.099 & 0.102 & 0.104 \\
		\end{tabular}}
		\caption{Size Comparison for DGP 1 with $d_z=1$ and $J=6$ or $12$}
		\label{tab:dgp1_size_1}
		{\footnotesize{Note: IM, CRS, ASY, BCH, WRE, W-B, W-B-S, AR-ASY, AR-B, and AR-B-S denote the ten inference methods described in Section \ref{sec: inf-method}. $J$ denotes the number of clusters. The nominal level is 10\%. }}
	\end{table}
	
	\begin{table}[h]
		\adjustbox{max width=\textwidth}{%
			\centering
			\begin{tabular}{ll|lll|lll|lll}
				&        &         & $\rho_{\eps v} = 0.3$  &        &         & $\rho_{\eps v} = 0.5$  &        &         & $\rho_{\eps v} = 0.7$  &        \\
				&        & $\Pi=0.25$ & $\Pi=0.375$ & $\Pi=0.5$ & $\Pi=0.25$ & $\Pi=0.375$ & $\Pi=0.5$ & $\Pi=0.25$ & $\Pi=0.375$ & $\Pi=0.5$ \\
				\hline
				$J=6$ & IM     & 0.067 & 0.061 & 0.059 & 0.086 & 0.065 & 0.063 & 0.131 & 0.093 & 0.074 \\
				& CRS    & 0.111 & 0.110 & 0.111 & 0.136 & 0.117 & 0.119 & 0.190 & 0.157 & 0.135 \\
				& ASY    & 0.254 & 0.268 & 0.274 & 0.258 & 0.268 & 0.280 & 0.265 & 0.276 & 0.263 \\
				& BCH    & 0.180 & 0.191 & 0.196 & 0.185 & 0.191 & 0.201 & 0.193 & 0.202 & 0.189 \\
				& WRE    & 0.092 & 0.088 & 0.082 & 0.093 & 0.088 & 0.084 & 0.100 & 0.089 & 0.083 \\
				& W-B    & 0.084 & 0.085 & 0.094 & 0.092 & 0.088 & 0.096 & 0.096 & 0.090 & 0.096 \\
				& W-B-S  & 0.065 & 0.071 & 0.084 & 0.090 & 0.093 & 0.097 & 0.110 & 0.105 & 0.098 \\
				& AR-ASY & 0.000 & 0.000 & 0.000 & 0.000 & 0.000 & 0.000 & 0.000 & 0.000 & 0.000 \\
				& AR-B   & 0.111 & 0.116 & 0.113 & 0.118 & 0.108 & 0.117 & 0.119 & 0.114 & 0.116 \\
				& AR-B-S & 0.115 & 0.115 & 0.116 & 0.116 & 0.114 & 0.111 & 0.118 & 0.113 & 0.116 \\
				\hline
				$J=12$ & IM     & 0.089 & 0.081 & 0.073 & 0.120 & 0.104 & 0.093 & 0.209 & 0.162 & 0.131 \\
				& CRS    & 0.112 & 0.107 & 0.103 & 0.149 & 0.132 & 0.125 & 0.242 & 0.196 & 0.167 \\
				& ASY    & 0.173 & 0.182 & 0.188 & 0.183 & 0.184 & 0.187 & 0.184 & 0.177 & 0.186 \\
				& BCH    & 0.139 & 0.148 & 0.152 & 0.153 & 0.151 & 0.157 & 0.156 & 0.148 & 0.155 \\
				& WRE    & 0.095 & 0.099 & 0.101 & 0.102 & 0.099 & 0.101 & 0.097 & 0.101 & 0.099 \\
				& W-B    & 0.097 & 0.097 & 0.100 & 0.099 & 0.094 & 0.096 & 0.094 & 0.095 & 0.096 \\
				& W-B-S  & 0.073 & 0.082 & 0.092 & 0.090 & 0.091 & 0.094 & 0.107 & 0.101 & 0.098 \\
				& AR-ASY & 0.033 & 0.037 & 0.034 & 0.036 & 0.032 & 0.034 & 0.033 & 0.034 & 0.034 \\
				& AR-B   & 0.105 & 0.102 & 0.102 & 0.103 & 0.102 & 0.103 & 0.099 & 0.104 & 0.097 \\
				& AR-B-S & 0.101 & 0.105 & 0.098 & 0.106 & 0.100 & 0.103 & 0.099 & 0.103 & 0.100 \\
		\end{tabular}}
		\caption{Size Comparison for DGP 1 with $d_z=3$ and $J=6$ or $12$}
		\label{tab:dgp1_size_2}
		{\footnotesize{Note: IM, CRS, ASY, BCH, WRE, W-B, W-B-S, AR-ASY, AR-B, and AR-B-S denote the ten inference methods described in Section \ref{sec: inf-method}. $J$ denotes the number of clusters. The nominal level is 10\%. }}
	\end{table}
	
	\begin{table}[h]
		\begin{center}
			\begin{tabular}{c|rrrrrrrrrr}
				& \multicolumn{1}{l}{IM} & \multicolumn{1}{l}{CRS} & \multicolumn{1}{l}{ASY} & \multicolumn{1}{l}{BCH} & \multicolumn{1}{l}{WRE} & \multicolumn{1}{l}{W-B} & \multicolumn{1}{l}{W-B-S} & \multicolumn{1}{l}{AR-B} & \multicolumn{1}{l}{AR-B-S} & \multicolumn{1}{l}{AR-ASY} \\ \hline
				$L$        & \multicolumn{10}{c}{$\gamma =  -0.0872$  } \\ \hline
				5   &  0.245 & 0.216 & 0.248  & 0.169 & 0.117 & 0.097 & 0.080  & 0.115 & 0.100 & 0.021   \\
				10  &  0.183 & 0.187 &   0.192&  0.158& 0.118 & 0.105 & 0.109  & 0.108 & 0.106 & 0.072   \\
				20  &  0.190 & 0.196 &  0.162 & 0.147& 0.112 & 0.101 & 0.116  & 0.104 & 0.105 & 0.084  \\
				30  &  0.700 & 0.703 & 0.153 & 0.142 & 0.112 & 0.103 & 0.122  & 0.105 & 0.103 & 0.087 \\ \hline
				$L$      & \multicolumn{10}{c}{$\gamma =  -0.0019$ } \\ \hline
				5   & 0.236 & 0.215 & 0.257 & 0.176 & 0.113 & 0.090 & 0.083  & 0.118 & 0.103 & 0.020   \\
				10  &  0.169 & 0.174 & 0.189 & 0.155 & 0.112 & 0.098 & 0.100  & 0.105 & 0.112 & 0.073   \\
				20  &  0.196 & 0.204 & 0.160 & 0.146 & 0.105 & 0.097 & 0.104  & 0.102 & 0.102 & 0.083   \\
				30  & 0.689 & 0.693  & 0.153 & 0.142 & 0.106 & 0.096 & 0.106  & 0.101 & 0.099 & 0.082  \\ \hline
				$L$      & \multicolumn{10}{c}{$\gamma = 0.0834$} \\ \hline 
				5   &  0.216 & 0.201 & 0.262 & 0.180  & 0.111 & 0.092 & 0.088 & 0.117 & 0.105 & 0.021   \\
				10  &  0.164 & 0.166 & 0.197 & 0.162  & 0.108 & 0.096 & 0.098  & 0.105 & 0.109 & 0.076  \\
				20  &  0.200 & 0.208 & 0.164 & 0.145 & 0.103 & 0.097 & 0.101  & 0.104 & 0.097 & 0.077   \\
				30  & 0.698 & 0.703  & 0.153 &  0.142 & 0.108 & 0.100 & 0.106  & 0.108 & 0.104 & 0.089   \\
			\end{tabular}%
			\caption{Size Comparison for DGP 2}
			\label{tab:dgp4_size}%
		\end{center}
		\vspace*{-4mm}
		{\footnotesize{Note: IM, CRS, ASY, BCH, WRE, W-B, W-B-S, AR-ASY, AR-B, and AR-B-S denote the ten inference methods described in Section \ref{sec: inf-method}. $L$ denotes the target number of clusters in the algorithm of spectral clustering defined in DGP 2. The nominal level is 10\%. }}
	\end{table}%

	\section{Empirical Application}\label{sec: emp}
	In an influential study, \cite{ADH2013} analyze the effect of rising Chinese import competition on US local labor markets between 1990 and 2007, when the share of total US spending on Chinese goods increased substantially from 0.6\%  to 4.6\%. The dataset of \cite{ADH2013} includes 722 commuting zones (CZs) that cover the entire mainland US.
	In this section, we further analyze the region-wise effects of such import exposure by applying IV regression with the proposed wild bootstrap procedures to three Census Bureau-designated regions: South, Midwest and West, with 16, 12, and 11 states, respectively, in each region.\footnote{The Northeast region is not included in the study because of the relatively small number of states ($9$) and small number of CZs in each state (e.g., Connecticut and Rhode Island have only 2 CZs).}
	
	Following \cite{ADH2013}, we consider the linear IV model
	\begin{align}
	y_{i,j} = X_{i,j}\beta + W_{i,j}^{\top}\gamma + \eps_{i,j},
	\end{align}
	where the outcome variable $y_{i,j}$ denote the decadal change in average individual log weekly wage in a given CZ and the endogenous variable $X_{i,j}$ is the change in Chinese import exposure per worker in a CZ, where imports are apportioned to the CZ according to its share of national industry employment. The Bartik-type instrument $Z_{i,j}$ 
	(e.g., see \cite{Goldsmith(2020)})
	proposed by \cite{ADH2013}
	is Chinese import growth in other high-income countries, where imports are apportioned to the CZs in the same way as that for $X_{i.j}$.\footnote{See Sections I.B and III.A in \cite{ADH2013} for a detailed definition of these variables.} 
	As can be seen from Figure \ref{fig:first_stage}, the predictive power of the IV varies considerably among states. 
	The exogenous variables $W_{i,j}$ include the characteristic variables of CZs and decade specified in \cite{ADH2013} as well as state dummies.
	Our regressions are based on the CZ samples in each region, and the samples are clustered at the state level, following \cite{ADH2013}.
	Besides the results for the full sample, we also report those for female and male samples separately. 
	Tables \ref{tab:check_assumption_South}--\ref{tab:check_assumption_West} in the Online Supplement show the values of $\widehat{Q}_{\tilde{Z}W,j}$ for each cluster in the three regions, and all of them are close to zero. (\ref{eq:QZWj}) in Assumption \ref{assumption: 1}(i) is thus plausible in this application. 
	
	The main result of the IV regression for the three regions is given in Table \ref{tab:app-IV},
	with the number of observations ($n$) and clusters ($J$) for each region. 
	As we have only one endogenous variable and one IV, the TSLS estimator is used throughout this section. 
	We further report the conventional 90\% confidence intervals based on CRVE with the standard normal critical value (ASY) and the 90\% bootstrap confidence sets (CSs) by inverting the corresponding $T_n$ (W-B), $T_{CR,n}$ (W-B-S), and $AR_n$ (AR-B) bootstrap tests with a 10\% nominal level.\footnote{AR-B and AR-B-S are numerically equivalent with one instrument.}
	The computation of the bootstrap CSs is conducted over the parameter space $[-10,10]$ with a step size of 0.01, and the number of bootstrap draws is set at 2,000 for each step. We highlight the main findings below. 
	
	\begin{enumerate}
		\item Table \ref{tab:app-IV} reports that the TSLS estimates of the average effect of Chinese imports on wages equal $-0.97$ and $-1.05$ for the South and West regions, respectively, while equals $-0.025$ for the Midwest region.\footnote{That is, a \$$1,000$ per worker increase in a CZ's exposure to Chinese imports is estimated to reduce average weekly earnings by $0.97$, $1.05$, and $0.025$ log points, respectively, for the three regions (the corresponding TSLS estimate in \cite{ADH2013} for the entire mainland US is $-0.76$).} In addition, we find that the values of the effective first-stage F statistics \citep{Olea-Pflueger(2013)}  are 85.26, 8.69, and 63.45 for South, Midwest, and West, respectively, suggesting the identification is relatively weak for Midwest compared with the other two regions.\footnote{We used CRVE as the variance estimator of the effective first-stage F statistic. We also note that the asymptotic results and critical values for the effective F statistic in \cite{Olea-Pflueger(2013)} are based on the consistency of the variance estimator. In the case of CRVE, this would require the number of clusters to diverge to infinity. Therefore, when the number of available clusters is small, the critical values provided by \cite{Olea-Pflueger(2013)} might not have satisfactory performance. Still, we can learn from the values of the effective F statistics that the identification for Midwest may be relatively weak.}
		
		\item All the three types of bootstrap CSs (W-B, W-B-S, and AR-B) are longer than the conventional asymptotic CSs (ASY) for all cases in Table \ref{tab:app-IV}. 
		For example, the studentized Wald bootstrap CSs (W-B-S) are $22.3\%$, $26.2\%$, and $26.4\%$ longer than the ASY CSs for South, Midwest, and West, respectively, in the all-sample case. Similar differences in the CS length can also be observed for female and male samples, respectively. This is in line with the simulation results in Section \ref{sec: simu}, suggesting that conventional CSs can be too short in such cases and result in empirically relevant under-coverage.\footnote{We also computed CSs with BCH's critical values, which are based on $t$ distributions instead of the standard normal distribution, and find that the three types of bootstrap CSs are longer than the BCH CSs in all cases.}

		\item Only the effect in the South region is significantly different from zero at the 10\% level under all three bootstrap CSs, while the effects in the other two regions are not. 
		This is in line with the result of effective F statistics, which suggests that the identification is strongest for South.
		
		\item We note that the effect on West is only significant under the studentized bootstrap Wald CS (W-B-S, $[-1.22, -0.47]$). In addition, W-B-S CSs have the shortest length among the three types of bootstrap CSs in most cases in Table \ref{tab:app-IV}, which is in line with our power results in Sections \ref{subsec: boot-wald-stud} and \ref{sec: simu}. 
		
		\item Comparing the results for female and male samples, we find that across all the regions, the effects are more substantial for the male samples. Furthermore, the effects for both female and male samples in the South are significantly different from zero.
	\end{enumerate}
	
	\begin{table}[h]
		\begin{center}
			\adjustbox{max width=\textwidth}{%
				\begin{tabular}{ccccccccc}
					\hline
					\hline
					Gender & Region  & $n$   & $J$  & Estimate & ASY CS  & W-B CS          & W-B-S CS            & AR-B CS              \\
					\hline
					All & South   & 578 & 16 & -0.97  & [-1.45, -0.49]  & {[}-1.71, -0.58{]} & {[}-1.61, -0.43{]} & {[}-1.70, -0.58{]} \\
					& Midwest & 504 & 12 & -0.025 & [-0.56, 0.51] & {[}-0.69, 0.83{]}  & {[}-0.60, 0.75{]}  & {[}-0.69, 0.83{]}  \\
					& West    & 276 & 11 & -1.05  & [-1.34, -0.75] & {[}-1.50, 0.25{]}  & {[}-1.22, -0.47{]} & {[}-1.50, 0.24{]}  \\
					\hline
					Female & South   & 578 & 16 & -0.81 & [-1.25, -0.37]  & {[}-1.48, -0.41{]} & {[}-1.40, -0.26{]} & {[}-1.48, -0.41{]} \\
					& Midwest & 504 & 12 & 0.024 & [-0.45, 0.49]  & {[}-0.64, 0.74{]}  & {[}-0.56, 0.67{]}  & {[}-0.64, 0.74{]}  \\
					& West    & 276 & 11 & -0.61 & [-1.01, -0.21]  & {[}-1.46, 0.74{]}  & {[}-0.91, 0.30{]}  & {[}-1.46, 0.75{]}  \\
					\hline
					Male   & South   & 578 & 16 & -1.08 & [-1.61, -0.54] & {[}-1.90, -0.65{]} & {[}-1.78, -0.53{]} & {[}-1.91, -0.65{]} \\
					& Midwest & 504 & 12 & -0.17 & [-0.81, 0.46]  & {[}-1.02, 0.80{]}  & {[}-0.86, 0.77{]}  & {[}-1.01, 0.82{]}  \\
					& West    & 276 & 11 & -1.26 & [-1.68, -0.84]  & {[}-1.99, 0.68{]}  & {[}-1.58, 0.42{]}  & {[}-2.00, 0.69{]} \\
					\hline
					\hline
				\end{tabular}
			}
		\end{center}
		\vspace*{-8mm}
		\caption{IV regressions of \cite{ADH2013} with all, female, and male samples for three US regions}
		\label{tab:app-IV}%
	\end{table}
	
	

	\section{Conclusion}\label{sec: conclu}
	In many empirical applications of IV regressions with clustered data, the IVs may be weak for some clusters and the number of clusters may be small. 
	In this paper, we provide valid inference methods in this context.
	For the Wald tests with and without CRVE, 
	we extend the WRE cluster bootstrap procedure to allow for the setting of few clusters and cluster-level heterogeneity in IV strength. 
	For the full-vector inference, we develop wild bootstrap AR tests that control size asymptotically irrespective of IV strength, and also show that the bootstrap validity of other weak-IV-robust tests requires at least one strong IV cluster.  
	Furthermore, we establish the power properties of the Wald and AR tests.
	Finally, recent studies by \cite{hansen2022jackknife} and \cite{mackinnon2022leverage} suggest that asymptotic or bootstrap inference based on jackknife variance estimators can be more reliable than that based on the conventional CRVE for OLS models. Therefore, it may be interesting to consider extending jackknife inference to the current setting. We leave this line of investigation for future research.

\singlespacing

\section{Pseudo-Code for the W-B-S Procedure}\label{sec:code}
This section provides the pseudo-code for our recommended wild bootstrap inference procedure W-B-S. We emphasize that when constructing the bootstrapped sample $X^*_{i,j}(g)$
and $y^*_{i,j}(g)$
in Step 4.2 of Algorithm 1 below, we use the interacted instruments $\overline{Z}_{i,j}$ defined in Step 2 of the algorithm. 
However, we still use the original instruments $Z_{i,j}$ to estimate the estimator and construct the test statistic $T_{CR,n}$. 

\medskip \IncMargin{-1em}
\begin{algorithm}[H]
	\caption{Procedure for W-B-S}
	\label{algo:1}
	\SetKwInOut{Input}{input}\SetKwInOut{Output}{output}
	\Input{Data $\{y_{i,j},X_{i,j},W_{i,j},Z_{i,j}\}_{i \in [n_j], j \in [J]}$, the null hypothesis $\lambda_\beta^\top \beta_0 =\lambda_0$.}
	\Output{The decision of rejecting the null hypothesis.}
	\begin{enumerate}
		\item[\textbf{Step 1.}] Compute the studentized test statistic $T_{CR,n}$ in \eqref{eq: Wald-studentized} of the main text.
		\item[\textbf{Step 2.}]             Construct the null-restricted and unrestricted residuals $\hat{\eps}^r_{i,j}$ and $\hat{\eps}_{i,j}$, respectively, following the formulas in Section 2.3.1.  Construct $\overline{Z}_{i,j}$ as
		\begin{align*}
		\overline{Z}_{i,j} = \left( \widetilde{Z}^{\top}_{i,j} 1\{j=1\}, \cdots, \widetilde{Z}^{\top}_{i,j} 1\{j = J\} \right)^{\top}, 
		\end{align*}
		where $\widetilde{Z}_{i,j}$ is the residual of regressing $Z_{i,j}$ on $W_{i,j}$ using the entire
		sample.
		
		\item[\textbf{Step 3.}]  Compute the first-stage residual $\tilde{v}_{i,j}$ in \eqref{eq: boot-RE-first-stage} of the main text.
		
		\item[\textbf{Step 4.}] Bootstrap the test statistic. \textbf{for} $b \leftarrow 1$ to $B$ \textbf{do}
		\begin{enumerate}
			\item[\textbf{Step 4.1.}] Generate $g_b \in \{-1,1\}^J$ so that each element of $g_b$ is a Rademacher random variable and all elements of $g_b$ are independent\;
			\item[\textbf{Step 4.2.}] Use $\overline{Z}_{i,j}$,  $\tilde{v}_{i,j}$, $\hat{\eps}^r_{i,j}$, and $g_b$ to construct $X_{i,j}^*(g)$ and $y_{i,j}^*(g)$ following \eqref{eq: boot-algo-unstud-2} of the main text\;
			\item[\textbf{Step 4.3.}] Use the bootstrapped sample  $\{y_{i,j}^*(g_b),X_{i,j}^*(g_b),W_{i,j},Z_{i,j}\}_{i \in [n_j], j \in [J]}$ to construct the bootstrapped test statistic $T^*_{CR,n}(g_b)$ following \eqref{eq: boot-algo-unstud-3} of the main text. 
		\end{enumerate}
		\textbf{end}
		\item[\textbf{Step 5.}]  Compute the critical value 
		\begin{align*}
		\hat{c}_{CR,n}(1-\alpha) = \inf \left\{ x \in \textbf{R}: \frac{1}{B} \sum_{b=1}^B 1 \{ T^*_{CR,n}(g_b) \leq x \} \geq 1-\alpha
		\right\}.
		\end{align*}
		\item[\textbf{Step 6.}] Reject the null hypothesis if $T_{CR,n} >   \hat{c}_{CR,n}(1-\alpha) $.
	\end{enumerate}
\end{algorithm} \DecMargin{-1em}

\section{$\widehat{Q}_{\tilde{Z}W,j}$ in the Empirical Application}\label{subsec: cluster-level}

Tables \ref{tab:check_assumption_South}--\ref{tab:check_assumption_West} report the values of $\widehat{Q}_{\tilde{Z}W,j}$ for each cluster in the South, Midwest, and West regions, and all of them are close to zero. 
For example, in Table \ref{tab:check_assumption_South}, the column $``W_1"$ reports the values of $\widehat{Q}_{\tilde{Z}W_1,j}$, where $j=1, ..., 16$ for the South region.  
Therefore, (\ref{eq:QZWj}) in Assumption \ref{assumption: 1}(i) is plausible in this application.

\begin{table}[h]
	\begin{center}
		\begin{tabular}{cccccccc}
			\hline
			\hline
			States & $W_1$ & $W_2$ & $W_3$ & $W_4$ & $W_5$ & $W_6$ & $W_7$  \\
			\hline
			1& 0.0027  & 0.0042  & 0.0009  & 0.0035  & 0.0012  & 0.0001    & 0.0001  \\
			2& -0.0008  & 0.0000    & -0.0002 & 0.0003  & 0.0005   & 0.0000  & 0.0001    \\
			3& 0.0019  & -0.0076 & -0.0017 & -0.0026  & -0.0018  & -0.0001   & -0.0005 \\
			4& 0.0039  & 0.0035  & 0.0001  & 0.0099  & 0.0046  & 0.0002  & -0.0005 \\
			5& -0.0007 & -0.0061 & -0.0017 & -0.0066 & -0.0033 & -0.0001   & -0.0001 \\
			6& -0.0025 & -0.0028 & -0.0001 & -0.0063 & -0.0039 & 0.0001    & 0.0001    \\
			7& 0.0004  & 0.0006   & -0.0001 & 0.0006 & 0.0005  & 0.0001    & -0.0001 \\
			8& 0.0068   & -0.0614 & -0.0384 & -0.0421 & -0.0180 & -0.0015 & -0.0025 \\
			9& 0.0017  & -0.0009 & -0.0001   & 0.0003  & 0.0003   & -0.0001  & 0.0001  \\
			10& -0.0047 & 0.0061  & 0.0026  & 0.0008  & -0.0002 & -0.0001  & 0.0003  \\
			11& 0.0000  & 0.0041  & 0.0006   & 0.0044  & 0.0023  & 0.0001    & 0.0000    \\
			12& -0.0022 & 0.0002  & 0.0003  & -0.0018 & -0.0009 & -0.0001   & 0.0002  \\
			13& 0.0017  & 0.0022   & 0.0002  & 0.0024  & 0.0014  & -0.0002 & 0.0003  \\
			14& 0.0002  & 0.0042  & 0.0023  & 0.0039  & 0.0020  & 0.0001  & 0.0000    \\
			15& -0.0015 & -0.0086  & -0.0012 & -0.0090 & -0.0039  & -0.0001   & -0.0001 \\
			16& 0.0015  & 0.0027   & 0.0001    & 0.0042  & 0.0021  & -0.0001   & 0.0001   \\
			\hline
			\hline
		\end{tabular}
		\caption{$\widehat{Q}_{\tilde{Z}W,j}$ for South in \cite{ADH2013}}
		\label{tab:check_assumption_South}
	\end{center}
\end{table}

\begin{table}[h]
	\begin{center}
		\begin{tabular}{cccccccc}
			\hline
			\hline
			States & $W_1$ & $W_2$ & $W_3$ & $W_4$ & $W_5$ & $W_6$ & $W_7$  \\
			\hline
			1& 0.0041  & 0.0070  & 0.0038  & 0.0075  & 0.0048   & 0.0001  & 0.0001 \\
			2&-0.0015 & 0.0022  & 0.0005  & -0.0016 & -0.0004 & -0.0001 & 0.0002 \\
			3&0.0005  & 0.0017  & -0.0001 & 0.0022  & 0.0010    & 0.0001    & 0.0001   \\
			4&-0.0012 & -0.0024 & -0.0007 & -0.0027 & -0.0014  & 0.0000        & -0.0001 \\
			5&-0.0056 & -0.0170 & -0.0025 & -0.0187 & -0.0096 & -0.0001  & -0.0003 \\
			6& 0.0041   & 0.0090  & 0.0010  & 0.0114  & 0.0051  & 0.0001  & 0.0001   \\
			7& 0.0020  & 0.0001    & -0.0004 & 0.0022  & 0.0012  & -0.0001   & -0.0001  \\
			8&-0.0003  & -0.0002 & -0.0002 & -0.0001   & 0.0002  & 0.0001    & -0.0001  \\
			9&-0.0003 & -0.0021 & -0.0001 & -0.0024 & -0.0008 & 0.0000        & -0.0001  \\
			10&-0.0015 & 0.0002  & -0.0003 & -0.0003 & -0.0011 & -0.0001   & 0.0003 \\
			11&0.0001  & 0.0007  & 0.0001    & 0.0011  & 0.0005  & 0.0000        & -0.0001  \\
			12&-0.0015 & -0.0002 & -0.0002 & -0.0012 & -0.0007 & -0.0001   & 0.0001  \\
			\hline
			\hline
		\end{tabular}
		\caption{$\widehat{Q}_{\tilde{Z}W,j}$ for Midwest in \cite{ADH2013}}
		\label{tab:check_assumption_Midwest}
	\end{center}
\end{table}

\begin{table}[h]
	\begin{center}
		\begin{tabular}{cccccccc}
			\hline
			\hline
			States & $W_1$ & $W_2$ & $W_3$ & $W_4$ & $W_5$ & $W_6$ & $W_7$  \\
			\hline
			1&0.0152  & 0.0535  & 0.0154  & 0.0632  & 0.0341  & 0.0005  & 0.0005  \\
			2&-0.0013 & -0.0093 & 0.0016  & -0.0151 & -0.0092 & 0.0002   & 0.0001  \\
			3&0.0028  & 0.0061  & -0.0001 & 0.0061  & 0.0033  & -0.0001 & 0.0001  \\
			4&0.0015  & 0.0058  & 0.0006  & 0.0066  & 0.0031  & 0.0001    & 0.0001    \\
			5&0.0002  & 0.0016  & 0.0001  & 0.0021  & 0.0011   & -0.0001   & -0.0001   \\
			6&-0.0001   & 0.0041  & -0.0022 & 0.0081  & 0.0030  & 0.0001    & -0.0003 \\
			7&0.0016  & 0.0081  & 0.0013   & 0.0096  & 0.0051  & 0.0000        & 0.0001    \\
			8&0.0014  & 0.0048  & 0.0008  & 0.0056  & 0.0027  & -0.0001   & 0.0001    \\
			9&-0.0014  & 0.0126  & 0.0015  & 0.0120  & 0.0060    & 0.0001    & 0.0001  \\
			10&-0.0124 & -0.0510 & -0.0122 & -0.0522 & -0.0247 & -0.0004  & -0.0006 \\
			11& 0.0001    & 0.0004  & 0.0001    & 0.0006  & 0.0003  & 0.0001  & -0.0001  \\
			\hline
			\hline
		\end{tabular}
		\caption{$\widehat{Q}_{\tilde{Z}W,j}$ for West in \cite{ADH2013}}
		\label{tab:check_assumption_West}
	\end{center}
\end{table}

\section{Equivalence Among $k$-Class Estimators}
\label{sec:equiv}
We define $\hat{\beta}_L$ as the $k$-class estimator with $\hat{\kappa}_L$ for $L \in \{\text{tsls},\text{liml},\text{full},\text{ba}\}$. 
Their null-restricted and bootstrap counterparts are denoted as $\hat{\beta}_L^r$ and $\hat{\beta}_{L,g}^*$, respectively, and $\hat{\gamma}_L$, $\hat{\gamma}_L^r$, and $\hat{\gamma}_{L,g}^*$ are similarly defined. In the following, we show that $\hat{\beta}_{tsls}$, $\hat{\beta}_{liml}$, $\hat{\beta}_{full}$, $\hat{\beta}_{ba}$ are asymptotically equivalent, and so be their null-restricted and bootstrap counterparts.

\begin{lemma}
	Suppose Assumptions \ref{assumption: 1}, \ref{assumption: 2}, and \ref{assumption: 3}(i) hold.  Then, for $L \in \{\text{liml},\text{full},\text{ba}\}$,      we have
	\begin{align*}
	& \hat{\beta}_{L} = \hat{\beta}_{tsls} + o_P(r_n^{-1}), \quad  \hat{\beta}_{tsls} - \beta_n = O_P(r_n^{-1}),   \\
	& \hat{\beta}_{L}^r = \hat{\beta}_{tsls}^r + o_P(r_n^{-1}),  \quad \text{and} \quad \hat{\beta}_{tsls}^r - \beta_n = O_P(r_n^{-1}).
	\end{align*}
	\label{lem:equiv0}
\end{lemma}
\textbf{Proof.} 
First, $L \in \{\text{liml},\text{full},\text{ba}\}$, we have $ \hat{\mu}_L = \hat{\kappa}_L-1$ and 
\begin{align*}
\left( \hat{\beta}^{\top}_L, \hat{\gamma}^{\top}_L \right)^{\top} = & \left( \vec{X}^{\top} P_{\vec{Z}} \vec{X} - \hat{\mu}_L \vec{X}^{\top} M_{\vec{Z}}\vec{X} \right)^{-1} \left(\vec{X}^{\top} P_{\vec{Z}}Y - \hat{\mu}_L \vec{X}^{\top} M_{\vec{Z}}Y \right) \notag \\
= & \left( \Upsilon^{\top}\vec{X} \right)^{-1} \Upsilon^{\top}Y, 
\;\; \text{where} \;\; \Upsilon = 
\left[ P_{\vec{Z}} X - \hat{\mu}_L M_{\Vec{Z}}X : W \right]. 
\end{align*}
Then, by applying the Frisch–Waugh–Lovell Theorem we obtain that 
\begin{align*}
\hat{\beta}_L = \left(X^{\top} P_{\widetilde{Z}} X - \hat{\mu}_L X^{\top} M_{\vec{Z}} X \right)^{-1}\left(X^{\top} P_{\widetilde{Z}} Y - \hat{\mu}_L X^{\top} M_{\vec{Z}} Y\right), \; \text{where} \; \tilde{Z} = M_W Z. 
\end{align*}

By construction, $\hat{\beta}_{tsls}$ corresponds to $\hat{\mu}_{tsls} = 0$. For the LIML estimator, we have 
\begin{align*}
\hat{\mu}_{liml}=\min_{r} r^{\top} \vec{Y}^{\top} M_W Z(Z^{\top} M_W Z)^{-1} Z^{\top} M_W \vec{Y} r / (r^{\top} \vec{Y}^{\top} M_{\vec{Z}}\vec{Y}r) \quad \text{and} \quad r=(1,-\beta^{\top} )^{\top},    
\end{align*}
which implies that 
\begin{align}
n \hat{\mu}_{liml} \leq \left(\frac{1}{\sqrt{n}} \eps^{\top} M_W Z \right) \left(\frac{1}{n} Z^{\top} M_W Z\right)^{-1} \left(\frac{1}{\sqrt{n}} Z^{\top} M_W \eps\right)/
\left(\frac{1}{n}\eps^{\top}M_{\vec{Z}}\eps\right). 
\label{eq:muliml0}
\end{align}
We note that  
\begin{align}
& \frac{1}{\sqrt{n}}Z^{\top}M_W\eps 
= \frac{1}{\sqrt{n}}\sum_{j \in [J]}\sum_{i \in I_{n,j}} \widetilde{Z}_{i,j}\eps_{i,j} = O_P(\sqrt{n}/r_n) \quad \text{and} \notag \\
& \left( \frac{1}{n} Z^{\top} M_W Z \right)^{-1} = \left( \frac{1}{n} \sum_{j \in [J]} \sum_{i \in I_{n,j}} \widetilde{Z}_{i,j}\widetilde{Z}_{i,j}^{\top} \right)^{-1} = O_P(1). 
\label{eq:muliml1}
\end{align}
In addition, let $\hat{\eps}_{i,j}$ be the residual from the full sample projection of $\eps_{i,j}$ on $W_{i,j}$. Then, we have
\begin{align}
\frac{1}{n} \eps^{\top} M_{\vec{Z}} \eps & = \frac{1}{n} \eps^{\top}\eps 
- \frac{1}{n} \eps^{\top} \vec{Z} (\vec{Z}^{\top}\vec{Z})^{-1} \vec{Z}^{\top}\eps  \notag \\
& = \widehat{Q}_{\eps\eps} - \widehat{Q}_{\eps\widetilde{Z}} \widehat{Q}_{\widetilde{Z}\widetilde{Z}}^{-1} \widehat{Q}_{\eps\widetilde{Z}}^\top  - \widehat{Q}_{\eps W} \widehat{Q}_{WW}^{-1} \widehat{Q}_{\eps W}^\top \notag \\
& = \widehat{Q}_{\eps\eps}  - \widehat{Q}_{\eps W} \widehat{Q}_{WW}^{-1} \widehat{Q}_{\eps W}^\top + o_P(1) \notag \\
& = \widehat{Q}_{\tilde{\eps}\tilde{\eps}} + o_P(1)  \notag \\
& = \sum_{j \in [J]} \xi_j \widehat{Q}_{\tilde{\eps}\tilde{\eps},j} + o_P(1) \notag \\
& \geq \sum_{j \in [J]} \xi_j \widehat{Q}_{\dot{\eps}\dot{\eps},j} + o_P(1) \notag \\
& = \widehat{Q}_{\dot{\eps}\dot{\eps}} + o_P(1) \geq c~w.p.a.1,
\label{eq:muliml2}
\end{align}
where $c$ is a positive constant, the first inequality is by the definition that $\dot{\eps}_{i,j}$ is the residual from the cluster-level projection of $\eps_{i,j}$ on $W_{i,j}$,  and the second inequality is by Assumption \ref{assumption: 2}(iii). Combining \eqref{eq:muliml0}--\eqref{eq:muliml2}, we have $\hat{\mu}_{liml} = O_P(r_n^{-2})$. In addition, we have $\frac{1}{n} X^{\top} M_{\vec{Z}} X = O_P(1)$, $\frac{1}{n} X^{\top} M_{\vec{Z}} Y = O_P(1)$, and 
\begin{align}
\left(\frac{1}{n}X^{\top} P_{\widetilde{Z}} X\right)^{-1} = \left(\widehat{Q}_{X\widetilde{Z}} \widehat{Q}_{\widetilde{Z}\widetilde{Z}}^{-1}\widehat{Q}_{\widetilde{Z}X}    \right)^{-1} = O_P(1),
\end{align}
where the last equality holds by Assumptions \ref{assumption: 2}(ii) and \ref{assumption: 3}(i). This means 
\begin{align*}
\hat{\beta}_{liml} =     \hat{\beta}_{tsls} + o_P(r_n^{-1}).
\end{align*}
In addition, we note $\hat{\mu}_{full} = \hat{\mu}_{liml} - \frac{C}{n-d_z-d_w} = O_P(r_n^{-2})$ and $\hat{\mu}_{ba} = O(n^{-1})$, respectively. Therefore, we have the same results for  $(\hat{\beta}_{full},\hat{\beta}_{ba})$. 

For the second statement in the lemma, we have
\begin{align*}
(\hat{\beta}_{tsls} - \beta_n)  & 
= \left(X^{\top} P_{\widetilde{Z}} X \right)^{-1} X^{\top} P_{\widetilde{Z}} \eps = \left(\widehat{Q}_{X\widetilde{Z}} \widehat{Q}_{\widetilde{Z}\widetilde{Z}}^{-1}\widehat{Q}_{\widetilde{Z}X}    \right)^{-1} \widehat{Q}_{X\widetilde{Z}} \widehat{Q}_{\widetilde{Z}\widetilde{Z}}^{-1}\widehat{Q}_{\widetilde{Z}\eps}  = O_P(r_n^{-1}),
\end{align*}
where the last equality holds because $\widehat{Q}_{\widetilde{Z}\eps} = \frac{1}{n}\sum_{j \in [J]} \sum_{i \in I_{n,j}}\widetilde{Z}_{i,j}\eps_{i,j} = O_P(r_n^{-1})$. 

Next, we turn to the third statement in the lemma. We note that, for $L \in \{\text{tsls},\text{liml},\text{full},\text{ba}\}$ and
$\lambda_{\beta}^{\top}\hat{\beta}^r_L = \lambda_0$,  
\begin{align*}
& \hat{\beta}^r_L = \hat{\beta}_L - \left(X^{\top}(P_{\widetilde{Z}} - \hat{\mu}_L M_{\Vec{Z}})X\right)^{-1} \lambda_{\beta}
\left(\lambda_{\beta}^{\top}(X^{\top}(P_{\widetilde{Z}} - \hat{\mu}_L M_{\vec{Z}})X)^{-1}
\lambda_{\beta}\right)^{-1} (\lambda_{\beta}^{\top}\hat{\beta}_L - \lambda_0) = O_P(1).
\end{align*}
As $\hat{\mu}_{L} = O_P(r_n^{-2})$ and $\hat{\beta}_L = \hat{\beta}_{tsls}+o_P(r_n^{-1})$ for $L \in \{\text{liml},\text{full},\text{ba}\}$, we have
\begin{align*}
\hat{\beta}^r_L = \hat{\beta}_{tsls} - \left(X^{\top}P_{\widetilde{Z}}X\right)^{-1} \lambda_{\beta} \left(\lambda_{\beta}^{\top} (X^{\top}P_{\widetilde{Z}}X)^{-1} \lambda_{\beta}\right)^{-1} 
(\lambda_{\beta}^{\top} \hat{\beta}_{tsls} - \lambda_0) + o_P(r_n^{-1}) = \hat{\beta}^r_{tsls} + o_P(r_n^{-1}).  
\end{align*}

For the last statement in the lemma, we note that 
\begin{align*}
\hat{\beta}^r_{tsls} - \beta_n & =   (\hat{\beta}_{tsls} - \beta_n)   - \left(X^{\top}P_{\widetilde{Z}}X\right)^{-1} \lambda_{\beta} \left(\lambda_{\beta}^{\top} (X^{\top}P_{\widetilde{Z}}X)^{-1} \lambda_{\beta}\right)^{-1} 
\lambda_{\beta}^{\top} (\hat{\beta}_{tsls} - \beta_n) \\
& - \left(X^{\top}P_{\widetilde{Z}}X\right)^{-1} \lambda_{\beta} \left(\lambda_{\beta}^{\top} (X^{\top}P_{\widetilde{Z}}X)^{-1} \lambda_{\beta}\right)^{-1} 
(\lambda_{\beta}^{\top}\beta_n - \lambda_0) = O_P(r_n^{-1}),
\end{align*}
where the last equality holds because $\lambda_{\beta}^{\top}\beta_n - \lambda_0= \mu r_n^{-1}$ by construction and $\hat{\beta}_{tsls} - \beta_n = O_P(r_n^{-1})$. $\blacksquare$

\begin{lemma}
	Suppose Assumptions \ref{assumption: 1}, \ref{assumption: 2}, and \ref{assumption: 3}(i) hold and $\widehat{Q}_{\widetilde{Z}W,j}(\hat{\gamma}_L^r-\gamma) = o_P(r_n^{-1})$. Then, for $L \in \{\text{liml},\text{full},\text{ba}\}$ and $g \in \textbf{G}$, we have
	\begin{align}
	\hat{\beta}_{L,g}^* = \hat{\beta}_{tsls,g}^* + o_P(r_n^{-1}) \quad \text{and} \quad \hat{\beta}_{tsls,g}^*- \beta_n = O_P(r_n^{-1}). \label{eq:equiv_b}
	\end{align}
	\label{lem:equivalence}
\end{lemma}
\textbf{Proof.} By the same argument in the proof of Lemma \ref{lem:equiv0}, for $L \in \{\text{tsls},\text{liml},\text{full},\text{ba}\}$ and $g \in \textbf{G}$, we have
\begin{align}
\hat{\beta}^*_{L,g}= \left(X^{*\top}(g) P_{\widetilde{Z}} X^*(g) - \hat{\mu}^*_{L,g} X^{*\top}(g) M_{\vec{Z}} X^*(g) \right)^{-1}\left(X^{*\top}(g) P_{\widetilde{Z}} Y^*(g) - \hat{\mu}^*_{L,g} X^{*\top}(g) M_{\vec{Z}} Y^*(g)\right).
\label{eq:equiv_b1}    
\end{align}
such that $\hat{\mu}_{tsls,g}^* = 0$, $\hat{\mu}_{full,g}^* = \hat{\mu}_{liml,g}^* - \frac{C}{n-d_z-d_x}$, $\hat{\mu}_{ba,g}^* = \hat{\mu}_{ba}$, and  
$$\hat{\mu}^*_{liml,g}=\min_{r} r^{\top} \vec{Y}^{*\top}(g) M_W Z(Z^{\top} M_W Z)^{-1} Z^{\top} M_W \vec{Y}^*(g) r / (r^{\top} \vec{Y}^{*\top}(g) M_{\vec{Z}}\vec{Y}^*(g)r),$$ 
where $\vec{Y}^{*}(g) = [Y^*(g) : X^*(g)]$, $Y^*(g)$ is an $n \times 1$ vector formed by $Y_{i,j}^*(g)$, and $r=(1,-\beta^{\top} )^{\top}$. 

Following the same argument previously, we have 
\begin{align}\label{eq:muhatliml*}
n \hat{\mu}_{liml,g}^* \leq \left(\frac{1}{\sqrt{n}} \eps_g^{*r\top} M_W Z \right) \left(\frac{1}{n} Z^{\top} M_W Z\right)^{-1} \left(\frac{1}{\sqrt{n}} Z^{\top} M_W \eps_g^{*r}\right)/
\left(\frac{1}{n}\eps_g^{*r\top}M_{\vec{Z}}\eps_g^{*r}\right), 
\end{align}
where $\eps_g^{*r}$ is an $n\times 1$ vector formed by $g_j\hat{\eps}_{i,j}^r$. We first note that 
\begin{align}
\frac{1}{n} \eps_g^{*r\top} M_W Z & = \frac{1}{n}\sum_{j \in [J]} \sum_{i \in I_{n,j}} g_j \widetilde{Z}_{i,j} \hat{\eps}^r_{i,j} \notag \\
& = \sum_{j \in [J]} g_j\left[\frac{1}{n} \sum_{i \in I_{n,j}} \widetilde{Z}_{i,j} \eps_{i,j} 
+ \frac{1}{n} \sum_{i \in I_{n,j}} \widetilde{Z}_{i,j}X_{i,j}^{\top}(\beta_n - \hat{\beta}^r_L) 
+ \frac{1}{n} \sum_{i \in I_{n,j}} \widetilde{Z}_{i,j}W_{i,j}^{\top}(\gamma - \hat{\gamma}^r_L) \right] \notag \\
& = O_P(r_n^{-1}), \label{eq:eps_g*r}
\end{align}
where the last line is by Assumptions \ref{assumption: 1}(ii), Lemma \ref{lem:equiv0}, and the assumption that $\widehat{Q}_{\widetilde{Z}W,j}(\hat{\gamma}_L^r-\gamma) = o_P(r_n^{-1})$. Second, we have
\begin{align}
\frac{1}{n}\eps_g^{*r\top}M_{\vec{Z}}\eps_g^{*r} & = \frac{1}{n} \eps_g^{*r\top} \eps_g^{*r} 
- \frac{1}{n} \eps_g^{*r\top} \vec{Z} (\vec{Z}^{\top}\vec{Z})^{-1} \vec{Z}^{\top} \eps_g^{*r} \notag \\
& =     \frac{1}{n} \sum_{j \in [J]}\sum_{i \in I_{n,j}}(g_j\hat{\eps}_{i,j}^{r})^2  - \left[\frac{1}{n} \sum_{j \in [J]}\sum_{i \in I_{n,j}}(g_j\hat{\eps}_{i,j}^{r} \widetilde{Z}_{i,j}^\top)\right] \widehat{Q}_{\widetilde{Z}\widetilde{Z}}^{-1}\left[\frac{1}{n} \sum_{j \in [J]}\sum_{i \in I_{n,j}}(g_j\hat{\eps}_{i,j}^{r} \widetilde{Z}_{i,j})\right] \notag \\
& - \left[\frac{1}{n} \sum_{j \in [J]}\sum_{i \in I_{n,j}}(g_j\hat{\eps}_{i,j}^{r} W_{i,j}^\top)\right] \widehat{Q}_{WW}^{-1}\left[\frac{1}{n} \sum_{j \in [J]}\sum_{i \in I_{n,j}}(g_j\hat{\eps}_{i,j}^{r} W_{i,j})\right] \notag \\
& =        \frac{1}{n} \sum_{j \in [J]}\sum_{i \in I_{n,j}}(\hat{\eps}_{i,j}^{r})^2   - \left[\frac{1}{n} \sum_{j \in [J]}\sum_{i \in I_{n,j}}(g_j\hat{\eps}_{i,j}^{r} W_{i,j}^\top)\right] \widehat{Q}_{WW}^{-1}\left[\frac{1}{n} \sum_{j \in [J]}\sum_{i \in I_{n,j}}(g_j\hat{\eps}_{i,j}^{r} W_{i,j})\right] + o_P(1), \label{eq:eps_g*r2}
\end{align}
where the last equality is by $\frac{1}{n_j} \sum_{i \in I_{n,j}}\hat{\eps}^{r}_{i,j} \widetilde{Z}_{i,j} = O_P(r_n^{-1})$, as in (\ref{eq:eps_g*r}). We further note that 
\begin{align*}
\hat{\eps}^{r}_{i,j} = \eps_{i,j} - X_{i,j}^\top(\hat{\beta}_L^r - \beta_n) - W_{i,j}^\top (\hat{\gamma}_L^r - \gamma),     
\end{align*}
where $\hat{\beta}_L^r - \beta_n = O_P(r_n^{-1})$ and 
\begin{align*}
\hat{\gamma}_L^r - \gamma = \widehat{Q}_{WW}^{-1} \left[ \frac{1}{n}\sum_{j \in [J]} \sum_{i \in I_{n,j}}W_{i,j}\left(Y_{i,j} - X_{i,j}^\top \hat{\beta}_L^r\right)  \right] - \gamma = \widehat{Q}_{WW}^{-1} \widehat{Q}_{W\eps} + O_P(r_n^{-1}). 
\end{align*}
Therefore, we have
\begin{align*}
\frac{1}{n} \sum_{j \in [J]}\sum_{i \in I_{n,j}}(\hat{\eps}_{i,j}^{r})^2 = \frac{1}{n} \sum_{j \in [J]}\sum_{i \in I_{n,j}}\mathscr{E}_{i,j}^2 + o_P(1),
\end{align*}
where $\mathscr{E}_{i,j} = \eps_{i,j} - W_{i,j}^\top\widehat{Q}_{WW}^{-1} \widehat{Q}_{W\eps}.$ Similarly, we can show that 
\begin{align*}
& \left[\frac{1}{n} \sum_{j \in [J]}\sum_{i \in I_{n,j}}(g_j\hat{\eps}_{i,j}^{r} W_{i,j}^\top)\right] \widehat{Q}_{WW}^{-1}\left[\frac{1}{n} \sum_{j \in [J]}\sum_{i \in I_{n,j}}(g_j\hat{\eps}_{i,j}^{r} W_{i,j})\right] \\
& = \left[\frac{1}{n} \sum_{j \in [J]}\sum_{i \in I_{n,j}}(g_j\mathscr{E}_{i,j} W_{i,j}^\top)\right] \widehat{Q}_{WW}^{-1}\left[\frac{1}{n} \sum_{j \in [J]}\sum_{i \in I_{n,j}}(g_j\mathscr{E}_{i,j} W_{i,j})\right]  + o_P(1),  
\end{align*}
which, combined with \eqref{eq:eps_g*r2}, implies 
\begin{align*}
\frac{1}{n}\eps_g^{*r\top}M_{\vec{Z}}\eps_g^{*r} & = \frac{1}{n} \sum_{j \in [J]}\sum_{i \in I_{n,j}}\mathscr{E}_{i,j}^2 - \left[\frac{1}{n} \sum_{j \in [J]}\sum_{i \in I_{n,j}}(g_j\mathscr{E}_{i,j} W_{i,j}^\top)\right] \widehat{Q}_{WW}^{-1}\left[\frac{1}{n} \sum_{j \in [J]}\sum_{i \in I_{n,j}}(g_j\mathscr{E}_{i,j} W_{i,j})\right]  + o_P(1) \notag \\
& = \frac{1}{n} \sum_{j \in [J]}\sum_{i \in I_{n,j}}(\mathscr{E}_{i,j} - g_j W_{i,j}^\top \hat{\theta}_g )^2 + o_P(1),
\end{align*}
where $\hat{\theta}_g = \widehat{Q}_{WW}^{-1}\left[\frac{1}{n} \sum_{j \in [J]}\sum_{i \in I_{n,j}}(g_j\mathscr{E}_{i,j} W_{i,j})\right]$ and the second equality holds because 
\begin{align*}
\widehat{Q}_{WW} = \frac{1}{n} \sum_{j \in [J]} \sum_{i \in I_{n,j}}W_{i,j}W_{i,j}^\top = \frac{1}{n} \sum_{j \in [J]} \sum_{i \in I_{n,j}}(g_jW_{i,j})(g_jW_{i,j})^\top.  
\end{align*}
Recall that $\dot{\eps}_{i,j}$ is the residual from the cluster-level projection of $\eps_{i,j}$ on $W_{i,j}$, which means there exists a vector $\hat{\theta}_j $ such that $\eps_{i,j} = \dot{\eps}_{i,j} + W_{i,j}^\top \hat{\theta}_j$ and $\sum_{i \in I_{n,j}}\dot{\eps}_{i,j}W_{i,j} = 0$. Then, we have
\begin{align*}
\frac{1}{n} \sum_{j \in [J]}\sum_{i \in I_{n,j}}(\mathscr{E}_{i,j} - g_j W_{i,j}^\top \hat{\theta}_g )^2 = \frac{1}{n} \sum_{j \in [J]} \left[\sum_{i \in I_{n,j}} \dot{\eps}_{i,j}^2 + \sum_{i \in I_{n,j}}(W_{i,j}^\top \hat{\theta}_j - g_j W_{i,j}^\top \hat{\theta}_g )^2\right] \geq \frac{1}{n} \sum_{j \in [J]} \sum_{i \in I_{n,j}}\dot{\eps}_{i,j}^2 \geq c~w.p.a.1,
\end{align*}
where the last inequality is by Assumption \ref{assumption: 2}(iii). This implies 
\begin{align}
\frac{1}{n}\eps_g^{*r\top}M_{\vec{Z}}\eps_g^{*r} & \geq c-o_P(1).
\label{eq:eps_g*r3}
\end{align}
Combining \eqref{eq:muhatliml*}, \eqref{eq:eps_g*r}, and \eqref{eq:eps_g*r3}, we obtain that $\hat{\mu}_{liml,g}^* = O_P(r_n^{-2})$, and thus, $\hat{\mu}^*_{full,g} = O_P(r_n^{-2})$. It is also obvious that $\hat{\mu}^*_{ba,g} = O_P(n^{-1})$. Given that $\hat{\mu}^*_{L,g} = O_P(r_n^{-2} + n^{-1})$, 
to establish $\hat{\beta}_{L,g}^* = \hat{\beta}_{tsls,g}^* + o_P(r_n^{-1})$, 
it suffices to show $\frac{1}{n}X^{*\top}(g) M_{\vec{Z}} X^*(g) = O_P(1)$, $\frac{1}{n}X^{*\top}(g) M_{\vec{Z}} Y^*(g) = O_P(1)$, and $(\frac{1}{n}X^{*\top}(g) P_{\widetilde{Z}} X^*(g))^{-1} = O_P(1)$. 

For $\frac{1}{n}X^{*\top}(g) M_{\vec{Z}} Y^*(g) = \frac{1}{n}X^{*\top}(g) Y^*(g) - \frac{1}{n}X^{*\top}(g) P_{\vec{Z}} Y^*(g)$, we first show 
\begin{align}\label{eq: Q_hat_X*Y*}
\frac{1}{n}X^{*\top}(g) Y^*(g) &= \frac{1}{n}\sum_{j \in [J]}\sum_{i \in I_{n,j}}
X^*_{i,j}(g)\left( X_{i,j}^{*\top}(g) \hat{\beta}^r_L + W_{i,j}\hat{\gamma}^r_L + g_j \hat{\eps}^r_{i,j} \right) \notag \\
& = \widehat{Q}^*_{XX}(g) \hat{\beta}^r_L + \widehat{Q}^*_{XW}(g) \hat{\gamma}^r_L + \widehat{Q}^*_{X\eps}(g) = O_P(1),
\end{align}
where $\widehat{Q}^*_{XX}(g) = \frac{1}{n}\sum_{j \in [J]}\sum_{i \in I_{n,j}}X^*_{i,j}(g)X^{*\top}_{i,j}(g)$, 
$\widehat{Q}^*_{XW}(g) = \frac{1}{n}\sum_{j \in [J]}\sum_{i \in I_{n,j}}X^*_{i,j}(g)W^{\top}_{i,j}$, and 
$\widehat{Q}^*_{X\eps}(g) = \frac{1}{n}\sum_{j \in [J]}\sum_{i \in I_{n,j}}g_j X^*_{i,j}(g) \hat{\eps}^r_{i,j}$. 
Notice that 
\begin{align}\label{eq: Q_hat_X*X*}
\widehat{Q}^*_{XX}(g) = \sum_{j \in [J]} \frac{n_j}{n} \frac{1}{n_j}\sum_{i \in I_{n,j}}
\left( X_{i,j} + (g_j-1)\tilde{v}_{i,j} \right)
\left( X_{i,j} + (g_j-1)\tilde{v}_{i,j} \right)^{\top} = O_P(1)
\end{align}
because $\widehat{Q}_{XX,j}=O_P(1)$, $\widehat{Q}_{X\tilde{v},j}=O_P(1)$, 
and $\widehat{Q}_{\tilde{v}\tilde{v},j}=O_P(1)$. To see these three relations, we note that 
\begin{align*}
\widehat{Q}_{X\tilde{v},j} = \widehat{Q}_{XX,j} - \widehat{Q}_{X\overline{Z},j}\widetilde{\Pi}_{\overline{Z}} - \widehat{Q}_{XW,j} \widetilde{\Pi}_{w}=O_P(1), 
\end{align*}
by $\widehat{Q}_{XX,j}=O_P(1)$, $\widehat{Q}_{X\overline{Z},j} = O_P(1)$, 
$\widehat{Q}_{XW,j}=O_P(1)$, $\tilde{\Pi}_{\overline{Z}}=O_P(1)$, and $\widetilde{\Pi}_w = O_P(1)$, 
where $\widehat{Q}_{X\overline{Z},j} = \frac{1}{n_j}\sum_{i \in I_{n,j}}X_{i,j}\overline{Z}_{i,j}^{\top}$, and 
$\widehat{Q}_{XW,j} = \frac{1}{n_j}\sum_{i \in I_{n,j}}X_{i,j}W_{i,j}^{\top}$.
Similar arguments hold for $\widehat{Q}_{\tilde{v}\tilde{v},j}$. 
We can also show that  
\begin{align}\label{eq: Q_hat_X*W}
\widehat{Q}^*_{XW}(g) = \sum_{j \in [J]}\frac{n_j}{n}\frac{1}{n_j}\sum_{i \in I_{n,j}} \left(X_{i,j} + (g_j-1)\tilde{v}_{i,j} \right) W_{i,j}^{\top} = O_P(1), 
\end{align}
by $\widehat{Q}_{XW,j}=O_P(1)$ and $\widehat{Q}_{\tilde{v}W,j}=O_P(1)$, 
where $\widehat{Q}_{XW,j} = \frac{1}{n_j}\sum_{i \in I_{n,j}}X_{i,j}W_{i,j}^{\top}$ 
and $\widehat{Q}_{\tilde{v}W,j} = \frac{1}{n_j}\sum_{i \in I_{n,j}}\tilde{v}_{i,j}W_{i,j}^{\top}$. 
In addition, we have 
\begin{align}\label{eq: Q_hat_X*eps}
\widehat{Q}^*_{X\eps}(g) = \sum_{j \in [J]}\frac{n_j}{n}\frac{1}{n_j}\sum_{i \in I_{n,j}}g_j\left(X_{i,j} + (g_j-1)\tilde{v}_{i,j}\right)\hat{\eps}^r_{i,j} = O_P(1), 
\end{align}
by $\widehat{Q}_{X\hat{\eps},j} = O_P(1)$ and $\widehat{Q}_{\tilde{v}\hat{\eps},j}=O_P(1)$ under similar arguments as those in (\ref{eq:eps_g*r}), 
where 
\begin{align*}
\widehat{Q}_{X\hat{\eps},j} = \frac{1}{n_j}\sum_{i \in I_{n,j}}X_{i,j}\hat{\eps}_{i,j}^r \quad \text{and} \quad \widehat{Q}_{\tilde{v}\hat{\eps},j} = \frac{1}{n_j}\sum_{i \in I_{n,j}}\tilde{v}_{i,j}\hat{\eps}_{i,j}^r.    
\end{align*}
Combining (\ref{eq: Q_hat_X*X*}), (\ref{eq: Q_hat_X*W}), (\ref{eq: Q_hat_X*eps}), $\hat{\beta}^r_L=O_P(1)$, 
and $\hat{\gamma}^r_L= \widehat{Q}_{WW}^{-1} \widehat{Q}_{w\eps}+o_P(1)=O_P(1)$, we obtain (\ref{eq: Q_hat_X*Y*}). Next, by the fact that $\frac{1}{n}\sum_{j \in [J]} \sum_{i \in I_{n,j}}\widetilde{Z}_{i,j}W_{i,j}^{\top} = 0$, we have
\begin{align*}
\frac{1}{n}X^{*\top}(g)P_{\vec{Z}}Y^*(g) 
= \begin{pmatrix}
\widehat{Q}_{\widetilde{Z}X}^{*\top}(g) & \widehat{Q}_{XW}^{*}(g)
\end{pmatrix} 
\begin{pmatrix}
\widehat{Q}_{\widetilde{Z}\widetilde{Z}}^{-1} & 0 \\
0 &  \widehat{Q}_{WW}^{-1}
\end{pmatrix}
\begin{pmatrix}
\widehat{Q}_{\widetilde{Z}Y}^*(g) \\
\widehat{Q}_{WY}^*(g)
\end{pmatrix},
\end{align*}
where 
\begin{align*}
&\widehat{Q}_{\widetilde{Z}X}^*(g) = \frac{1}{n}\sum_{j \in [J]} \sum_{i \in I_{n,j}}\widetilde{Z}_{i,j}X_{i,j}^{*\top}(g),  \quad \widehat{Q}_{\widetilde{Z}Y}^*(g) = \frac{1}{n}\sum_{j \in [J]} \sum_{i \in I_{n,j}}\widetilde{Z}_{i,j}Y_{i,j}^{*}(g), \\
& \text{and} \quad \widehat{Q}_{WY}^*(g) = \frac{1}{n}\sum_{j \in [J]} \sum_{i \in I_{n,j}}W_{i,j}Y_{i,j}^{*}(g).
\end{align*}
Following the same lines of reasoning, we can show that, for all $g \in \textbf{G}$, $\widehat{Q}_{\widetilde{Z}X}^*(g) = O_P(1)$, $\widehat{Q}_{\widetilde{Z}Y}^*(g)=O_P(1)$, and $\widehat{Q}_{WY}^*(g)=O_P(1)$. In addition, by Assumptions \ref{assumption: 1}(iv) and \ref{assumption: 2}(iii), $\widehat{Q}_{\widetilde{Z}\widetilde{Z}}^{-1} = O_P(1)$ and $\widehat{Q}_{WW}^{-1} = O_P(1)$, which further implies that $\frac{1}{n}X^{*\top}(g) M_{\vec{Z}} Y^*(g) = O_P(1)$.

By a similar argument, we can show $\frac{1}{n}X^{*\top}(g) M_{\vec{Z}} X^*(g) = O_P(1)$. Last, we have
\begin{align*}
\frac{1}{n}X^{*\top}(g) P_{\widetilde{Z}} X^*(g) = \widehat{Q}_{\widetilde{Z}X}^{*\top}(g) \widehat{Q}_{\widetilde{Z}\widetilde{Z}}^{-1}\widehat{Q}_{\widetilde{Z}X}^{*}(g)
= & \widehat{Q}_{\widetilde{Z}X}^{\top}\widehat{Q}_{\widetilde{Z}\widetilde{Z}}^{-1}\widehat{Q}_{\widetilde{Z}X} + o_P(1) \convP Q_{\widetilde{Z}X}^{\top}Q_{\widetilde{Z}\widetilde{Z}}^{-1}Q_{\widetilde{Z}X},
\end{align*}
where we use the fact that $\widehat{Q}_{\widetilde{Z}X}^{*}(g) = \widehat{Q}_{\widetilde{Z}X} + o_P(1)$, which is established in Step 1 in the proof of Theorem \ref{theo: boot-t}. In addition, $Q_{\widetilde{Z}X}^{\top}Q_{\widetilde{Z}\widetilde{Z}}^{-1}Q_{\widetilde{Z}X}$ is invertible by Assumptions \ref{assumption: 2}(ii) and \ref{assumption: 3}(i). This implies $(\frac{1}{n}X^{*\top}(g) P_{\widetilde{Z}} X^*(g))^{-1} = O_P(1)$, which further implies 
$\hat{\beta}_{L,g}^* = \hat{\beta}_{tsls,g}^* + o_P(r_n^{-1})$. 

For the second result, we note that 
\begin{align*}
\hat{\beta}_{tsls,g}^* - \beta_n & =    \left[ \widehat{Q}_{\widetilde{Z}X}^{*\top}(g) \widehat{Q}_{\widetilde{Z}\widetilde{Z}}^{-1}\widehat{Q}_{\widetilde{Z}X}^{*}(g)\right]^{-1}\widehat{Q}_{\widetilde{Z}X}^{*\top}(g) \widehat{Q}_{\widetilde{Z}\widetilde{Z}}^{-1} \left[\frac{1}{n}\sum_{j \in [J]}\sum_{i \in I_{n,j}}\widetilde{Z}_{i,j}y_{i,j}^*(g)\right] - \beta_n \\
& = \hat{\beta}_{tsls}^r - \beta_n + \left[ \widehat{Q}_{\widetilde{Z}X}^{*\top}(g) \widehat{Q}_{\widetilde{Z}\widetilde{Z}}^{-1}\widehat{Q}_{\widetilde{Z}X}^{*}(g)\right]^{-1}\widehat{Q}_{\widetilde{Z}X}^{*\top}(g) \widehat{Q}_{\widetilde{Z}\widetilde{Z}}^{-1} \left[\frac{1}{n}\sum_{j \in [J]}\sum_{i \in I_{n,j}}g_j\widetilde{Z}_{i,j}\hat{\eps}_{i,j}^r\right],
\end{align*}
where the second equality holds because $\sum_{j \in [J]}\sum_{i \in I_{n,j}}g_j\widetilde{Z}_{i,j}W_{i,j}^\top = 0.$ In addition, note that 
\begin{align*}
\frac{1}{n_j}\sum_{i \in I_{n,j}}\widetilde{Z}_{i,j}\hat{\eps}_{i,j}^r = \frac{1}{n_j}\sum_{i \in I_{n,j}}\widetilde{Z}_{i,j}\eps_{i,j} - \widehat{Q}_{\widetilde{Z}X,j}(\hat{\beta}_{tsls}^r - \beta_n) -\widehat{Q}_{\widetilde{Z}W,j}(\hat{\gamma}_{tsls}^r - \gamma) = O_P(r_n^{-1}).  
\end{align*}
Combining this with the fact that $\hat{\beta}_{tsls}^r - \beta_n = O_P(r_n^{-1})$ as shown in Lemma \ref{lem:equiv0}, we have $\hat{\beta}_{tsls,g}^* - \beta_n = O_P(r_n^{-1})$. $\blacksquare$

\begin{lemma}
	Suppose Assumptions \ref{assumption: 1} and \ref{assumption: 2} hold. Then, we have, for $j \in [J]$, 
	$$\widehat{Q}_{\widetilde{Z}W,j}(\bar{\gamma}^r - \gamma) = o_P(r_n^{-1}).$$
	If, in addition, Assumption \ref{assumption: 3}(i) holds, then we have,  For $j \in [J]$, $L \in \{\text{tsls},\text{liml},\text{full},\text{ba}\}$, and $g \in \textbf{G}$, 
	\begin{align*}
	\widehat{Q}_{\widetilde{Z}W,j} (\hat{\gamma}_L - \gamma) = o_P(r_n^{-1}), \quad \widehat{Q}_{\widetilde{Z}W,j} (\hat{\gamma}^r_L - \gamma) = o_P(r_n^{-1}), \quad \text{and} \quad \widehat{Q}_{\widetilde{Z}W,j} (\hat{\gamma}^*_{L,g} - \gamma) = o_P(r_n^{-1}).
	\end{align*}
	\label{lem:equiv2}
\end{lemma}
\textbf{Proof.} 
Note that we assume $\widehat{Q}_{\widetilde{Z}W,j}=o_P(1)$ and $\frac{1}{n}\sum_{j \in [J]} \sum_{i \in I_{n,j}}W_{i,j}\eps_{i,j} = O_P(r_n^{-1})$. The first statement holds because 
\begin{align}
\bar{\gamma}^r - \gamma & = \widehat{Q}_{WW}^{-1} \left[\frac{1}{n}\sum_{j \in [J]}\sum_{i \in I_{n,j}}W_{i,j}(Y_{i,j} - X_{i,j}^\top \beta_0) \right] - \gamma   \notag \\ 
& = \widehat{Q}_{WW}^{-1} \left[\frac{1}{n}\sum_{j \in [J]}\sum_{i \in I_{n,j}}W_{i,j}\eps_{i,j}\right] - \widehat{Q}_{WW}^{-1}\widehat{Q}_{WX} (\beta_n- \beta_0) = O_P(r_n^{-1}),
\label{eq:gamma_bar}
\end{align}

Next, if Assumption \ref{assumption: 3}(i) also holds, then 
\begin{align}
\hat{\gamma}_L - \gamma & = \widehat{Q}_{WW}^{-1} \left[\frac{1}{n}\sum_{j \in [J]}\sum_{i \in I_{n,j}}W_{i,j}(Y_{i,j} - X_{i,j}^\top\hat{\beta}_L) \right] - \gamma   \notag \\ 
& = \widehat{Q}_{WW}^{-1} \left[\frac{1}{n}\sum_{j \in [J]}\sum_{i \in I_{n,j}}W_{i,j}\eps_{i,j}\right] - \widehat{Q}_{WW}^{-1}\widehat{Q}_{WX} (\hat{\beta}_L- \beta_n) = O_P(r_n^{-1}),
\label{eq:gamma_L}
\end{align}
where the last equality holds by Lemma \ref{lem:equiv0}. This implies $\widehat{Q}_{\widetilde{Z}W,j}(\hat{\gamma}_L-\gamma) = o_P(r_n^{-1})$. In the same manner, we can show $\widehat{Q}_{\widetilde{Z}W,j}(\hat{\gamma}_L^r-\gamma) = o_P(r_n^{-1})$. Last, given Assumptions \ref{assumption: 1}, \ref{assumption: 2}, \ref{assumption: 3}(i), and the fact that $\widehat{Q}_{\widetilde{Z}W,j}(\hat{\gamma}_L^r-\gamma) = o_P(r_n^{-1})$, Lemma \ref{lem:equivalence} shows $\hat{\beta}_{L,g}^* - \beta_n = (\hat{\beta}_{L,g}^* - \hat{\beta}_{tsls,g}^*) + (\hat{\beta}_{tsls,g}^* - \beta_n) = O_P(r_n^{-1})$. Then, following the same argument in \eqref{eq:gamma_L}, we can show $\hat{\gamma}_{L,g}^* - \gamma = O_P(r_n^{-1})$, which leads to the desired result. 
$\blacksquare$

\section{Proof of Theorem \ref{theo: boot-t}}
\label{sec:pf_boot-t}
When $d_x =1$, we define $a_j = Q^{-1}Q_{\widetilde{Z}X}^\top Q_{\widetilde{Z}\widetilde{Z}}^{-1} Q_{\widetilde{Z}X,j}$. For $d_x>2$, $a_j$ is defined in Assumption \ref{assumption: 3}. Let $\mathbb{S} \equiv   
\textbf{R}^{d_z \times d_x} \times \textbf{R}^{d_z \times d_z} \times \otimes_{j \in [J]} \textbf{R}^{d_z} \times \textbf{R}^{d_r \times d_r}$ and write an element $s \in \mathbb{S}$ 
by $s = \left(s_1, s_2, \{s_{3,j}: j \in [J]\}, s_4 \right)$ where $s_{3,j} \in \textbf{R}^{d_z}$ for any $j \in [J]$. 
Define the function $T$: $\mathbb{S} \rightarrow \textbf{R}$ to be given by 
\begin{align}\label{eq: T_W_u_s}
T(s) = \bigg\Vert \lambda_{\beta}^{\top} \left(s_1^{\top} s_2^{-1} s_1\right)^{-1} s_1^{\top} s_2^{-1} \left(\sum_{j \in [J]} s_{3,j} \right)\bigg\Vert_{s_4}
\end{align}
for any $s \in \mathbb{S}$ such that $s_2$ and $s_1^{\top}s_2^{-1}s_1$ are invertible and let $T(s)=0$ otherwise. 
We also identify any $(g_1, ..., g_q) = g \in \textbf{G} = \{-1, 1\}^J$ with an action on $s \in \mathbb{S}$ 
given by $gs = \left( s_1, s_2, \{ g_j s_{3,j} : j \in [J] \}, s_4 \right)$. 
For any $s \in \mathbb{S}$ and $\textbf{G'} \subseteq \textbf{G}$, denote the ordered values of 
$\{ T(gs) : g \in \textbf{G'} \}$ by
$T^{(1)}(s | \textbf{G'}) \leq \ldots \leq T^{(|\textbf{G'}|)} (s | \textbf{G'})$.
In addition, for any $\textbf{G'} \subseteq \textbf{G}$, denote the ordered values of 
$\{ T^*_n(g) : g \in \textbf{G'} \}$ by
$T^{*(1)}_n(\textbf{G'}) \leq \ldots \leq T_n^{*(|\textbf{G'}|)}(\textbf{G'})$.

Given this notation we can define the statistics $S_n, \widehat{S}_n \in \mathbb{S}$ as 
\begin{align*}
S_n & =  
\left( \widehat{Q}_{\widetilde{Z}X}, \widehat{Q}_{\widetilde{Z}\widetilde{Z}}, \left\{ \frac{r_n}{n} \sum_{i \in I_{n,j}} \widetilde{Z}_{i,j} \eps_{i,j} : j \in [J] \right\}, \hat{A}_r \right), \quad \widehat{S}_n =
\left( \widehat{Q}_{\widetilde{Z}X}, \widehat{Q}_{\widetilde{Z}\widetilde{Z}}, \left\{ \frac{r_n}{n} \sum_{i \in I_{n,j}} \widetilde{Z}_{i,j} \hat{\eps}^r_{i,j}: j \in [J] \right\}, \hat{A}_r \right). 
\end{align*}

Let $E_n$ denote the event 
$
E_n = I\left\{\text{$\widehat{Q}_{\widetilde{Z}X}$ is of full rank value and $\widehat{Q}_{\widetilde{Z}\widetilde{Z}}$ is invertible}\right\}, 
$
and Assumptions \ref{assumption: 2}-\ref{assumption: 3} imply that 
$\liminf_{n \rightarrow \infty} \mathbb{P} \{ E_n=1 \} =1.$
Also let $T^*_n(g)=0$ if $E_n=0$.

We first give the proof for the Wald statistic based on TSLS. 
Note that whenever $E_n=1$ and $\mathcal{H}_0$ is true, 
the Frisch-Waugh-Lovell theorem implies that  
\begin{align}\label{eq: W_u_S}
r_nT_{n} &    = \bigg\Vert r_n \left(\lambda_{\beta}^{\top} \hat{\beta}_{tsls} - \lambda_0 \right) \bigg\Vert_{\hat{A}_r} 
=\bigg\Vert r_n \lambda_{\beta}^{\top} \left(\hat{\beta}_{tsls} - \beta_n \right) \bigg\Vert_{\hat{A}_r} \notag \\
&    = \bigg\Vert \lambda_{\beta}^{\top} \widehat{Q}^{-1}   \widehat{Q}^{\top}_{\widetilde{Z}X} \widehat{Q}_{\widetilde{Z}\widetilde{Z}}^{-1} \sum_{j \in [J]} \frac{r_n}{n}
\sum_{i \in I_{n,j}} \widetilde{Z}_{i,j} \eps_{i,j} \bigg\Vert_{\hat{A}_r} 
=  T(\iota_J S_n),
\end{align}
where $\widehat{Q} = \widehat{Q}^{\top}_{\widetilde{Z}X} \widehat{Q}_{\widetilde{Z}\widetilde{Z}}^{-1} \widehat{Q}_{\widetilde{Z}X}$ and $\iota_J \in \textbf{G}$ is a $J \times 1$ vector of ones.

In the following, we divide the proof into three steps. In the first step, we show 
\begin{eqnarray}\label{eq: T_n*_T_g_S_n_hat}
T_n^*(g) = T(g \widehat{S}_n) + o_P(1) \;\; \text{for any $g \in \textbf{G}$}. 
\end{eqnarray}
In the second step, we show 
\begin{align}\label{eq: T_g_S_n_hat_T_g_S_n}
T(g \widehat{S}_n) = T(g S_n) + o_P(1) \;\; \text{for any $g \in \textbf{G}$}. 
\end{align}
In the last step, we prove the desired result. 

\textbf{Step 1.}
By the continuous mapping theorem, it suffices to show 
$\widehat{Q}^*_{\widetilde{Z}X}(g) = \widehat{Q}_{\widetilde{Z}X} + o_P(1)$. Note that  
\begin{eqnarray*}
	&& \widehat{Q}^*_{\widetilde{Z}X}(g) = \frac{1}{n} \sum_{j \in [J]} \sum_{i \in I_{n,j}} \widetilde{Z}_{i,j}X_{i,j}^{*\top}(g) =
	\frac{1}{n}\sum_{j \in [J]}\sum_{i \in I_{n,j}}\widetilde{Z}_{i,j}\left(X_{i,j} + (g_j-1)\tilde{v}_{i,j}\right)^{\top}.
\end{eqnarray*}
Therefore, 
it suffices to show $\frac{1}{n_j} \sum_{i \in I_{n,j}}  \widetilde{Z}_{i,j} \tilde{v}_{i,j} = o_P(1)$ for all $j \in [J]$. Recall $\overline{Z}_{i,j}$ is just $\widetilde{Z}_{i,j}$ interacted with all the cluster dummies and $\widetilde{\Pi}_{\overline{Z}}$ is the OLS coefficient of $\overline{Z}_{i,j}$ defined in (\ref{eq: boot-RE-first-stage}). Denote $\widetilde{\Pi}_{\overline{Z},j}$ as the $j$-th block of $\widetilde{\Pi}_{\overline{Z}}$, which corresponds to the OLS coefficient of $\widetilde{\Pi}_{\overline{Z},j}$. We have
\begin{align}\label{eq: Z_til_v_til}
\frac{1}{n_j} \sum_{i \in I_{n,j}} \widetilde{Z}_{i,j} \tilde{v}_{i,j} 
& =  \frac{1}{n_j} \sum_{i \in I_{n,j}}\left( \widetilde{Z}_{i,j}X_{i,j} -  \widetilde{Z}_{i,j}\widetilde{Z}_{i,j}^\top \widetilde{\Pi}_{\overline{Z},j}- \widetilde{Z}_{i,j}W_{i,j}^\top\widetilde{\Pi}_w  \right) 
= \widehat{Q}_{\widetilde{Z}X,j} - \widehat{Q}_{\widetilde{Z}\widetilde{Z},j} \widetilde{\Pi}_{\overline{Z},j} + o_P(1),
\end{align}
where the second equality in \eqref{eq: Z_til_v_til} holds by $\frac{1}{n_j}\sum_{i \in I_{n,j}}\widetilde{Z}_{i,j}W_{i,j}^{\top}=o_P(1)$ and $\widetilde{\Pi}_{w} = O_P(1)$. 
In particular, 
\begin{align*}
\widetilde{\Pi}_w = \left( \widehat{Q}_{\widetilde{W}\widetilde{W}} - \widehat{Q}_{\widetilde{W}\hat{\eps}} \widehat{Q}^{-1}_{\hat{\eps}\hat{\eps}} \widehat{Q}_{\hat{\eps}\widetilde{W}}\right)^{-1}
\left( \widehat{Q}_{\widetilde{W}X} - \widehat{Q}_{\widetilde{W}\hat{\eps}} \widehat{Q}^{-1}_{\hat{\eps}\hat{\eps}}\widehat{Q}_{\hat{\eps}X}\right), 
\end{align*}
where $\widehat{Q}_{\widetilde{W}\widetilde{W}} = \frac{1}{n}\sum_{j \in [J]}\sum_{i \in I_{n,j}}\widetilde{W}_{i,j}\widetilde{W}_{i,j}^{\top}$, 
$\widehat{Q}_{\widetilde{W}\hat{\eps}} = \frac{1}{n}\sum_{j \in [J]}\sum_{i \in I_{n,j}}\widetilde{W}_{i,j}\hat{\eps}_{i,j}$, 
$\widehat{Q}_{\hat{\eps}\hat{\eps}} = \frac{1}{n}\sum_{j \in [J]}\sum_{i \in I_{n,j}}\hat{\eps}_{i,j}^2$, 
$\widetilde{W}_{i,j} = W_{i,j} - \widehat{\Gamma}_{w,j}^{\top} \widetilde{Z}_{i,j}$, 
and $\widehat{\Gamma}_{w,j} = \widehat{Q}^{-1}_{\widetilde{Z}\widetilde{Z},j}\widehat{Q}_{\widetilde{Z}W,j}$.
Notice that by $\widehat{Q}_{\widetilde{Z}W,j} = o_P(1)$, $\widehat{\Gamma}_{w,j} = o_P(1)$ so that 
\begin{align}\label{eq:Q_WWtil}
\widehat{Q}_{\widetilde{W}\widetilde{W}} = \widehat{Q}_{WW} + o_P(1). 
\end{align}
Similarly, we have
\begin{align}\label{eq:Q_Wetil}
\widehat{Q}_{\widetilde{W}\hat{\eps}} = \widehat{Q}_{W\hat{\eps}} + o_P(1) = o_P(1), 
\end{align}
where the second equality follows from $\widehat{Q}_{W\hat{\eps}}=0$ by the first-order condition of the $k$-class estimators. 
Furthermore, $\widehat{Q}_{\hat{\eps}\hat{\eps}} \geq c > 0$ by Assumption \ref{assumption: 2}(iii)
and $\widehat{Q}_{\hat{\eps}\hat{\eps}} \geq \frac{1}{n}\sum_{j \in [J]}\sum_{i \in I_{n,j}}\dot{\eps}^2_{i,j}$, 
where $\dot{\eps}_{i,j}$ is the residual from the cluster-level projection of $\eps_{i,j}$ on $W_{i,j}$. 
Therefore, $\widehat{Q}_{\widetilde{W}\widetilde{W}} - \widehat{Q}_{\widetilde{W}\hat{\eps}} \widehat{Q}^{-1}_{\hat{\eps}\hat{\eps}} \widehat{Q}_{\hat{\eps}\widetilde{W}} = \widehat{Q}_{WW} + o_P(1)$ by combining (\ref{eq:Q_WWtil}) and (\ref{eq:Q_Wetil}), and further by Assumption \ref{assumption: 1}(iv),  
\begin{align}\label{eq:Q_WWtil_inv}
\left( \widehat{Q}_{\widetilde{W}\widetilde{W}} - \widehat{Q}_{\widetilde{W}\hat{\eps}} \widehat{Q}^{-1}_{\hat{\eps}\hat{\eps}} \widehat{Q}_{\hat{\eps}\widetilde{W}}\right)^{-1} = O_P(1).
\end{align}
Next, we define 
$\widehat{Q}_{\widetilde{Z}\dot{X},j} = \frac{1}{n_j}\sum_{i \in I_{n,j}}\widetilde{Z}_{i,j}\dot{X}_{i,j}^{\top}$ and recall  $\widehat{Q}_{\widetilde{Z}\widetilde{Z},j} = \frac{1}{n_j}\sum_{i \in I_{n,j}}\widetilde{Z}_{i,j}\widetilde{Z}_{i,j}^{\top}$, where $\dot{X}_{i,j} = X_{i,j} - \widetilde{\Pi}_{w}^{\top}W_{i,j} - \widetilde{\Pi}_{\hat{\eps}}^{\top}\hat{\eps}_{i,j}$, and $\widetilde{\Pi}_{w}$ and $\widetilde{\Pi}_{\hat{\eps}}$ are the OLS coefficients of $W_{i,j}$ and $\hat{\eps}_{i,j}$, respectively,  from regressing $X_{i,j}$ on $(\overline{Z}^{\top}_{i,j}, W^{\top}_{i,j}, \hat{\eps}_{i,j})^{\top}$ using the entire sample. Then, we have 
\begin{align}
\label{eq:Pitilde_{z,j}}
\widetilde{\Pi}_{\overline{Z},j} & = \widehat{Q}^{-1}_{\widetilde{Z}\widetilde{Z},j} \widehat{Q}_{\widetilde{Z}\dot{X},j}  = \widehat{Q}^{-1}_{\widetilde{Z}\widetilde{Z},j}\left[\widehat{Q}_{\widetilde{Z}X,j} - \widehat{Q}_{\widetilde{Z}W,j} \widetilde{\Pi}_w - \widehat{Q}_{\widetilde{Z}\hat{\eps},j} \widetilde{\Pi}_{\hat{\eps}}\right] = \widehat{Q}^{-1}_{\widetilde{Z}\widetilde{Z},j}\widehat{Q}_{\widetilde{Z}X,j} + o_P(1), 
\end{align}
where we use the facts that $\widehat{Q}_{\widetilde{Z}W,j} = o_P(1)$, $\widetilde{\Pi}_w = O_P(1)$, $\widetilde{\Pi}_{\hat{\eps}} = O_P(1)$, and $\widehat{Q}_{\widetilde{Z}\hat{\eps},j} = o_P(1)$. 
In particular, 
\begin{align*}
\widetilde{\Pi}_{\hat{\eps}} = \left( \widehat{Q}_{\tilde{\eps}\tilde{\eps}} - \widehat{Q}_{\tilde{\eps}W} \widehat{Q}^{-1}_{WW} \widehat{Q}_{W\tilde{\eps}}\right)^{-1}
\left( \widehat{Q}_{\tilde{\eps}X} - \widehat{Q}_{\tilde{\eps}W} \widehat{Q}^{-1}_{WW}\widehat{Q}_{WX}\right), 
\end{align*}
where $\widehat{Q}_{\tilde{\eps}\tilde{\eps}} = \frac{1}{n}\sum_{j \in [J]}\sum_{i \in I_{n,j}}\tilde{\eps}^2_{i,j}$, 
$\widehat{Q}_{\tilde{\eps}W} = \frac{1}{n}\sum_{j \in [J]}\sum_{i \in I_{n,j}}\tilde{\eps}_{i,j}W^{\top}_{i,j}$, 
$\tilde{\eps}_{i,j} = \hat{\eps}_{i,j} - \widehat{\Gamma}_{\hat{\eps},j}^{\top} \widetilde{Z}_{i,j}$, 
and $\widehat{\Gamma}_{\hat{\eps},j} = \widehat{Q}_{\widetilde{Z}\widetilde{Z},j}^{-1}\widehat{Q}_{\widetilde{Z}\hat{\eps},j}$.
Then, by using similar arguments as those for (\ref{eq:Q_WWtil}), (\ref{eq:Q_Wetil}), and (\ref{eq:Q_WWtil_inv}), we obtain 
\begin{align*}
\widehat{Q}_{\tilde{\eps}\tilde{\eps}} = \widehat{Q}_{\hat{\eps}\hat{\eps}} + o_P(1), \; \widehat{Q}_{\tilde{\eps}W} = \widehat{Q}_{\hat{\eps}W} + o_P(1) = o_P(1), \; 
\text{and} \; \left( \widehat{Q}_{\tilde{\eps}\tilde{\eps}} - \widehat{Q}_{\tilde{\eps}W} \widehat{Q}^{-1}_{WW} \widehat{Q}_{W\tilde{\eps}}\right)^{-1}=O_P(1). 
\end{align*}

To see the last equality in \eqref{eq:Pitilde_{z,j}}, we note that
\begin{align*}
\widehat{Q}_{\widetilde{Z}\hat{\eps},j}    & = \frac{1}{n_j}\sum_{i \in I_{n,j}} \widetilde{Z}_{i,j}\hat{\eps}_{i,j} \\
& =     \frac{1}{n_j}\sum_{i \in I_{n,j}} \widetilde{Z}_{i,j}\eps_{i,j} - \frac{1}{n_j}\sum_{i \in I_{n,j}} \widetilde{Z}_{i,j}X_{i,j}^\top(\hat{\beta}_{tsls} - \beta_n) -\frac{1}{n_j}\sum_{i \in I_{n,j}} \widetilde{Z}_{i,j}W_{i,j}^\top(\hat{\gamma}_{tsls} - \gamma) = o_P(1), 
\end{align*}
where the last equality holds by Assumption \ref{assumption: 1}(ii), Lemma \ref{lem:equiv2}, and Lemma \ref{lem:equiv0}. 
Plugging \eqref{eq:Pitilde_{z,j}} into \eqref{eq: Z_til_v_til}, we obtain the desired result that $\frac{1}{n_j} \sum_{i \in I_{n,j}} \widetilde{Z}_{i,j} \tilde{v}_{i,j}  = o_P(1)$.

\textbf{Step 2.}
We note that whenever $E_n=1$, for every $g \in \textbf{G}$,
\begin{align}\label{eq: difference_W_u_S_n_hat_S_n}
\left|T(gS_n) - T(g\widehat{S}_n)\right| 
&\leq 
\bigg\Vert \lambda_{\beta}^{\top} \widehat{Q}^{-1}\widehat{Q}^{\top}_{\widetilde{Z}X} \widehat{Q}_{\widetilde{Z}\widetilde{Z}}^{-1} \sum_{j \in [J]} \frac{n_j}{n} \frac{1}{n_j} \sum_{i \in I_{n,j}} g_j \widetilde{Z}_{i,j}X_{i,j}^{\top} r_n(\beta_n-\hat{\beta}^r_{tsls}) \bigg\Vert_{\hat{A}_r} \notag \\
& + \bigg\Vert \lambda_{\beta}^{\top} \widehat{Q}^{-1}\widehat{Q}^{\top}_{\widetilde{Z}X} \widehat{Q}_{\widetilde{Z}\widetilde{Z}}^{-1} \sum_{j \in [J]} \frac{n_j}{n} \frac{1}{n_j} \sum_{i \in I_{n,j}} g_j \widetilde{Z}_{i,j}W_{i,j}^{\top} r_n(\gamma-\hat{\gamma}^r_{tsls}) \bigg\Vert_{\hat{A}_r}. 
\end{align}
By Lemma \ref{lem:equiv2}, 
we have
\begin{align}\label{eq: limsup_W_u_0_gamma}
&   \limsup_{n \rightarrow \infty} \mathbb{P} \left\{  \bigg\Vert \lambda_{\beta}^{\top} \widehat{Q}^{-1}\widehat{Q}^{\top}_{\widetilde{Z}X} \widehat{Q}_{\widetilde{Z}\widetilde{Z}}^{-1} \sum_{j \in [J]} \frac{n_j}{n} \frac{1}{n_j} \sum_{i \in I_{n,j}} g_j \widetilde{Z}_{i,j}W_{i,j}^{\top} r_n(\gamma-\hat{\gamma}^r_{tsls}) \bigg\Vert_{\hat{A}_r } >\eps; E_n=1 \right\} =0. 
\end{align}


Note under both cases in Assumption \ref{assumption: 3}(ii), we have $\sum_{j \in [J]} \xi_j a_j = 1$ and 
\begin{align}\label{eq:a_j*}
\lambda_\beta^\top Q^{-1}Q_{\widetilde{Z}X}^\top Q_{\widetilde{Z}\widetilde{Z}}^{-1} Q_{\widetilde{Z}X,j}   = a_j \lambda_\beta^\top.     
\end{align}

Then, for any $\eps >0$, we have 
\begin{align}\label{eq: limsup_W_u_0_beta}
&\limsup_{n \rightarrow \infty} \mathbb{P} \left\{  \bigg\Vert \lambda_{\beta}^{\top} \widehat{Q}^{-1}\widehat{Q}^{\top}_{\widetilde{Z}X} \widehat{Q}_{\widetilde{Z}\widetilde{Z}}^{-1} \sum_{j \in [J]} \frac{n_j}{n} \frac{1}{n_j} \sum_{i \in I_{n,j}} g_j \widetilde{Z}_{i,j}X_{i,j}^{\top} r_n(\beta_n -\hat{\beta}^r_{tsls}) \bigg\Vert_{\hat{A}_r} > \eps; E_n=1 \right\} \notag \\
&    =  \limsup_{n \rightarrow \infty} \mathbb{P} \left\{ \bigg\Vert  \sum_{j \in [J]} \xi_j g_j a_j r_n(\lambda_{\beta}^{\top}\beta_n -\lambda_{\beta}^{\top}\hat{\beta}^r_{tsls})\bigg\Vert_{\hat{A}_r} > \eps; E_n=1 \right\} =0, 
\end{align}
where the last equality holds because 
$\lambda_{\beta}^{\top}\hat{\beta}^r_{tsls}=\lambda_0$ under $\mathcal{H}_0$.

Note that $T(g \widehat{S}_n) = T(g S_n)$ whenever $E_n=0$ as we have defined 
$T(s)=0$ for any $s=\left(s_1, s_2, \{s_{3,j}: j \in [J]\}, s_4 \right)$ whenever 
$s_2$ or $s_1^{\top}s_2^{-1}s_1$ is not invertible.
Therefore, results in (\ref{eq: difference_W_u_S_n_hat_S_n}), (\ref{eq: limsup_W_u_0_gamma}) and (\ref{eq: limsup_W_u_0_beta}) imply \eqref{eq: T_g_S_n_hat_T_g_S_n}. 


\textbf{Step 3.} Note that by  \hyperlink{assumption: 1}{Assumptions \ref{assumption: 1}}, \hyperlink{assumption: 2}{\ref{assumption: 2}}, \ref{assumption: 4},
and the continuous mapping theorem, we have 
\begin{align}\label{eq: W_u_S_limit}
\left( \widehat{Q}_{\widetilde{Z}X}, \widehat{Q}_{\widetilde{Z}\widetilde{Z}}, 
\left\{ \frac{r_n}{n} \sum_{i \in I_{n,j}} \widetilde{Z}_{i,j} \eps_{i,j} : j \in [J] \right\}, \hat{A}_r 
\right)
\xrightarrow{\enskip d \enskip}
\left( Q_{\widetilde{Z}X}, Q_{\widetilde{Z}\widetilde{Z}}, 
\left\{ \mathcal{Z}_j: j \in [J]\right\}, A_r 
\right) \equiv S, 
\end{align}
where $\xi_j > 0$ for all $j \in [J]$ by Assumption \ref{assumption: 1}(iii). 
Therefore, we obtain from (\ref{eq: T_n*_T_g_S_n_hat}), (\ref{eq: T_g_S_n_hat_T_g_S_n}), (\ref{eq: W_u_S_limit}), and the continuous mapping theorem that
\begin{align*}
\left(T(S_n), \left\{ T^*_n(g) : g \in \textbf{G} \right\} \right) 
\xrightarrow{\enskip d \enskip} 
\left(T(S), \{ T(gS) : g \in \textbf{G} \} \right). 
\end{align*}

For any $x \in \textbf{R}$ letting $\lceil x \rceil$ denote the smallest integer larger than $x$ and 
$k^* \equiv \lceil |\textbf{G}| (1-\alpha) \rceil$, 
we obtain from (\ref{eq: W_u_S}) that 
\begin{align}\label{eq: W_u_c_hat}
1\left\{ T_{n} > \hat{c}_{n}(1-\alpha) \right\} 
= 1\left\{ T(S_n) > T_n^{*(k^*)}( \textbf{G}) \right\}.
\end{align}
Since $\liminf_{n \rightarrow \infty} P\{ E_n =1\} =1$, we have
\begin{align*}
&   \limsup_{n \rightarrow \infty} \mathbb{P} \{ T_n > \hat{c}_n(1-\alpha))\} = \limsup_{n \rightarrow \infty} \mathbb{P} \{ T_n > \hat{c}_n(1-\alpha)); E_n=1 \}  \notag \\
& \leq \limsup_{n \rightarrow \infty} \mathbb{P} \{ T(S_n) \geq T_n^{*(k^*)} (\textbf{G}); E_n=1\}
\leq \mathbb{P} \{ T(S) \geq  T^{(k^*)} (S | \textbf{G}) \}  \leq \alpha + 2^{1-q},
\end{align*}
where the second inequality is due to the Portmanteau's theorem. 
To see the last inequality, we note that for all $g \in \textbf{G}$, $\mathbb{P}\{ T(gS) = T(-gS)\} = 1$,
$\mathbb{P}\{ T(gS) = T(\tilde{g}S)\} = 0$ for $\tilde{g} \notin \{ g, -g \}$, 
and under the null, $T(gS)$ has the same distribution across $g \in \textbf{G}$. Let $|\textbf{G}| = 2^q$. Then, we have
\begin{align*}
|\textbf{G}|\mathbb{E}1\{ T(S) \geq  T^{(k^*)} (S | \textbf{G}) \} = \mathbb{E} \sum_{g \in \textbf{G}}1\{ T(gS) \geq  T^{(k^*)} (S | \textbf{G}) \} \leq |\textbf{G}|-k^*+2,
\end{align*}
which implies 
\begin{align*}
\mathbb{E}1\{ T(S) \geq  T^{(k^*)} (S | \textbf{G}) \} \leq 1 - \frac{k^*}{|\textbf{G}|} + \frac{2}{|\textbf{G}|} \leq   \alpha + \frac{1}{2^{J-1}}.
\end{align*}

For the lower bound, first note that $k^* > |\textbf{G}| - 2$
implies that $\alpha - \frac{1}{2^{J-1}} \leq 0$, in which case the result trivially follows. 
Now assume $k^*  \leq |\textbf{G}| - 2$, then
\begin{align*}
\liminf_{n \rightarrow \infty}
\mathbb{P} \{ T_n > \hat{c}_n(1-\alpha) \}
= \liminf_{n \rightarrow \infty} \mathbb{P} \{ T(S_n) > T_n^{*(k^*)} (\textbf{G}) \} 
\geq   \mathbb{P} \{ T(S) > T^{(k^*)} (S | \textbf{G}) \} \notag \\
\geq   \mathbb{P} \{ T(S) > T^{(k^*+2)} (S | \textbf{G}) \} +  \mathbb{P} \{ T(S) = T^{(k^*+2)} (S | \textbf{G}) \}   
\geq  \alpha - \frac{1}{2^{J-1}},
\end{align*}
where the first equality follows from (\ref{eq: W_u_c_hat}), 
the first inequality follows from Portmanteau's theorem, 
the second inequality holds because 
$\mathbb{P} \{ T^{(\textbf{z}+2)} (S | \textbf{G}) > T^{(\textbf{z})}(S | \textbf{G}) \} =1$ 
for any integer $\textbf{z} \leq |\textbf{G}| -2$
by (\ref{eq: T_W_u_s}) and Assumption \ref{assumption: 1}, 
and the last inequality follows from noticing that $k^*+2 = \lceil |\textbf{G}|((1-\alpha)+2/|\textbf{G}|) \rceil
= \lceil |\textbf{G}| (1-\alpha') \rceil$ with $\alpha' = \alpha - \frac{1}{2^{J-1}}$ 
and the properties of randomization tests. 

Lemma \ref{lem:equiv0} and \ref{lem:equivalence} in the Supplement further show the other $k$-class estimators and their null-restricted and bootstrap counterparts are asymptotically equivalent to those of the TSLS estimator. Therefore, the results for LIML, FULL, and BA estimators can be derived in the same manner. $\blacksquare$

\section{Proof of Theorem \ref{theo: boot-t-power}}
\label{sec:boot-t-power}
For the power of the $T_n$-based wild bootstrap test, we focus on the TSLS estimator. As shown in the end of the proof of Theorem \ref{theo: boot-t}, the test statistics constructed using LIML, FULL, and BA estimators and their bootstrap counterparts are asymptotically equivalent to those constructed based on TSLS estimator, which leads to the desired result. Note that 
\begin{align*}
|| r_n (\lambda_{\beta}^{\top} \hat{\beta}_{tsls} - \lambda_0) ||_{\hat{A}_r} & =   || r_n \lambda_{\beta}^{\top} (\hat{\beta}_{tsls} - \beta_n) 
+ r_n \lambda_{\beta}^{\top} (\beta_n - \hat{\beta}^r_{tsls}) ||_{\hat{A}_r}  \notag \\
& = \Big\Vert \lambda_{\beta}^{\top} \widehat{Q}^{-1}  \widehat{Q}_{\widetilde{Z}X}^{\top} \widehat{Q}_{\widetilde{Z}\widetilde{Z}}^{-1} 
\sum_{j \in [J]} \sum_{i \in I_{n,j}} \frac{r_n \widetilde{Z}_{i,j} \eps_{i,j}}{n} + r_n \lambda_{\beta}^{\top} (\beta_n - \hat{\beta}^r_{tsls} ) 
\Big\Vert_{\hat{A}_r}  \notag \\
& = \Big\Vert  \lambda_{\beta}^{\top} \widehat{Q}^{-1}  \widehat{Q}_{\widetilde{Z}X}^{\top} \widehat{Q}_{\widetilde{Z}\widetilde{Z}}^{-1} \sum_{j \in [J]} \sum_{i \in I_{n,j}} \left(  \frac{r_n\widetilde{Z}_{i,j} \eps_{i,j}}{n} + \frac{\widetilde{Z}_{i,j} X_{i,j}^{\top}}{n}
r_n  (\beta_n - \hat{\beta}^r_{tsls} )  \right) \Big\Vert_{\hat{A}_r}. 
\end{align*}
Notice that Assumptions \ref{assumption: 1}, \ref{assumption: 2}, \ref{assumption: 3}(i), and Lemma \ref{lem:equiv0} imply that 
$r_n(\hat{\beta}^r_{tsls} - \beta_n)$ is bounded in probability. This implies 
\begin{align}
r_n T_n = || r_n(\lambda_{\beta}^{\top} \hat{\beta}_{tsls} - \lambda_0) ||_{\hat{A}_r} \xrightarrow{\enskip d \enskip} \Big\Vert \sum_{j \in [J]} \left[  \lambda_{\beta}^{\top} Q^{-1}  Q_{\widetilde{Z}X}^{\top} Q_{\widetilde{Z}\widetilde{Z}}^{-1}  \mathcal{Z}_j \right]+  \mu  \Big\Vert_{A_r}.
\label{eq:Tn}
\end{align}

Recall $\widehat{Q}$ and $\widehat{Q}^{*}_g$ defined in \eqref{eq: Wald-hat-A} and \eqref{eq:Qstar}, respectively. We have $\widehat{Q}^{*-1}_g  = \widehat{Q}^{-1}+o_P(1)$ and 
\begin{align}\label{eq: boot-unstud-power-2}
& || r_n \lambda_{\beta}^{\top} (\hat{\beta}^*_{tsls,g} - \hat{\beta}^r_{tsls} ) ||_{\hat{A}_r}  \notag \\
&  = \Big\Vert  \lambda_{\beta}^{\top} \widehat{Q}^{*-1}_g  \widehat{Q}_{\widetilde{Z}X}^{*\top}(g) \widehat{Q}_{\widetilde{Z}\widetilde{Z}}^{-1} \sum_{j \in [J]} \sum_{i \in I_{n,j}} g_j \left(   \frac{r_n \widetilde{Z}_{i,j} \eps_{i,j}}{n} + \frac{\widetilde{Z}_{i,j} X_{i,j}^{\top}}{n}
r_n  (\beta_n - \hat{\beta}^r_{tsls} )  
+ \frac{\widetilde{Z}_{i,j} W_{i,j}^{\top}}{n}
r_n  (\gamma - \hat{\gamma}^r_{tsls} ) \right) \Big\Vert_{\hat{A}_r} \notag \\
&  = \Big\Vert  \lambda_{\beta}^{\top} \widehat{Q}^{-1}  \widehat{Q}_{\widetilde{Z}X}^{\top} \widehat{Q}_{\widetilde{Z}\widetilde{Z}}^{-1} \sum_{j \in [J]} \sum_{i \in I_{n,j}} g_j \left(   \frac{r_n \widetilde{Z}_{i,j} \eps_{i,j}}{n} + \frac{\widetilde{Z}_{i,j} X_{i,j}^{\top}}{n}
r_n  (\beta_n - \hat{\beta}^r_{tsls} )  \right) \Big\Vert_{\hat{A}_r} + o_P(1), 
\end{align}
where the last equality follows from Lemma \ref{lem:equiv2}
and $\widehat{Q}^*_{\widetilde{Z}X}(g) = \widehat{Q}_{\widetilde{Z}X} + o_P(1)$. Furthermore, we notice that 
\begin{align}\label{eq: boot-unstud-power-3}
\hat{\beta}^r_{tsls} & =  \hat{\beta}_{tsls} - \widehat{Q}^{-1} \lambda_{\beta} \left( \lambda_{\beta}^{\top} \widehat{Q}^{-1} \lambda_{\beta} \right)^{-1} \left( \lambda_{\beta}^{\top} \hat{\beta}_{tsls} - \lambda_0 \right) \notag \\
& = \hat{\beta}_{tsls} - \widehat{Q}^{-1} \lambda_{\beta} \left\{ \left( \lambda_{\beta}^{\top} \widehat{Q}^{-1} \lambda_{\beta} \right)^{-1} \lambda_{\beta}^{\top} (\hat{\beta}_{tsls} - \beta_n) +   \left( \lambda_{\beta}^{\top} \widehat{Q}^{-1} \lambda_{\beta} \right)^{-1} (\lambda_{\beta}^{\top} \beta_n - \lambda_0) \right\}. 
\end{align}
Therefore, employing (\ref{eq: boot-unstud-power-3}) with $r_n (\lambda_{\beta}^{\top} \beta_n - \lambda_0) = \lambda_{\beta}^{\top} \mu_{\beta}$, 
we conclude that whenever $E_n=1$, 
\begin{align*}
\sum_{i \in I_{n,j}} \frac{\widetilde{Z}_{i,j} X_{i,j}^{\top}}{n} r_n (\beta_n - \hat{\beta}^r_{tsls}) & =
\sum_{i \in I_{n,j}} \frac{\widetilde{Z}_{i,j} X_{i,j}^{\top}}{n}  \left\{  \left( I_{d_x} - \widehat{Q}^{-1} \lambda_{\beta} 
\left( \lambda_{\beta}^{\top} \widehat{Q}^{-1} \lambda_{\beta} \right)^{-1} \lambda_{\beta}^{\top}\right)  r_n (\beta_n - \hat{\beta}_{tsls}) \right. \notag \\
& \left.
+ \widehat{Q}^{-1} \lambda_{\beta} \left( \lambda_{\beta}^{\top} \widehat{Q}^{-1} \lambda_{\beta} \right)^{-1} \lambda_{\beta}^{\top} \mu_{\beta} \right\}.
\end{align*}
Together with $(\ref{eq: boot-unstud-power-2})$, this implies that 
\begin{align*}
& \limsup_{n \rightarrow \infty} \mathbb{P} \begin{pmatrix}
\Big\Vert r_n \lambda_{\beta}^{\top} (\hat{\beta}^*_{tsls,g} - \hat{\beta}^r_{tsls} ) 
- \lambda_{\beta}^{\top} \widehat{Q}^{-1}  \widehat{Q}_{\widetilde{Z}X}^{\top} \widehat{Q}_{\widetilde{Z}\widetilde{Z}}^{-1} \\
\times \sum_{j \in [J]} \sum_{i \in I_{n,j}} g_j \left(  \frac{r_n \widetilde{Z}_{i,j}\eps_{i,j}}{n}  + \xi_j \widehat{Q}_{\widetilde{Z}X,j} \widehat{Q}^{-1}  \lambda_{\beta} \left( \lambda_{\beta}^{\top} \widehat{Q}^{-1} \lambda_{\beta} \right)^{-1} \lambda_{\beta}^{\top} \mu_{\beta} \right) \Big\Vert_{\hat{A}_r} > \eps; E_n=1 
\end{pmatrix} \\
&  = 
\limsup_{n \rightarrow \infty} \mathbb{P} \begin{pmatrix}
& \Big\Vert r_n \lambda_{\beta}^{\top} (\hat{\beta}^*_{tsls,g} - \hat{\beta}^r_{tsls} ) 
- \lambda_{\beta}^{\top} Q^{-1}  Q_{\widetilde{Z}X}^{\top} Q_{\widetilde{Z}\widetilde{Z}}^{-1} \\
& \times \sum_{j \in [J]} g_j \left[  \mathcal{Z}_j + \xi_j Q_{\widetilde{Z}X,j} Q^{-1}  \lambda_{\beta} \left( \lambda_{\beta}^{\top} Q^{-1} \lambda_{\beta} \right)^{-1} \mu \right] \Big\Vert_{\hat{A}_r} > \eps; E_n=1 
\end{pmatrix} \\
& =
\limsup_{n \rightarrow \infty} \mathbb{P} \begin{pmatrix}
& \Big\Vert r_n \lambda_{\beta}^{\top} (\hat{\beta}^*_{tsls,g} - \hat{\beta}^r_{tsls} ) 
-  
\sum_{j \in [J]} g_j \left[  \lambda_{\beta}^{\top} Q^{-1}  Q_{\widetilde{Z}X}^{\top} Q_{\widetilde{Z}\widetilde{Z}}^{-1}  \mathcal{Z}_j + \xi_j a_j \mu \right] \Big\Vert_{\hat{A}_r} > \eps; E_n=1 
\end{pmatrix} 
=0.
\end{align*}
This implies 
\begin{align}
r_nT_n^*(g) = \Big\Vert r_n \lambda_{\beta}^{\top} (\hat{\beta}^*_{tsls,g} - \hat{\beta}^r_{tsls} ) \Big \Vert_{\hat{A}_r}     \xrightarrow{\enskip d \enskip} \Big\Vert \sum_{j \in [J]} g_j\left[   \lambda_{\beta}^{\top} Q^{-1}  Q_{\widetilde{Z}X}^{\top} Q_{\widetilde{Z}\widetilde{Z}}^{-1}  \mathcal{Z}_j +  \xi_j a_j \mu \right] \Big\Vert_{A_r}.
\label{eq:Tnstar}
\end{align}

In addition, let $\textbf{G}_s = \textbf{G} \backslash \textbf{G}_w$, where $\textbf{G}_w = \{g \in \textbf{G}: g_{j} = g_{j'}, \forall j,j' \in \mathcal{J}_s\}$.
To establish the power result for the bootstrap test, we request that $\hat c_n(1-\alpha)$ does not take values of $T^*_n(g)$ for $g \in \textbf{G}_w$ because otherwise the test statistic and the critical value are asymptotically equivalent even under the alternative.
We note that the following condition is imposed in the theorem: $|\textbf{G}_s| = |\textbf{G}| - 2^{J-J_s+1} \geq k^*$.
Therefore, based on \eqref{eq:Tn} and \eqref{eq:Tnstar}, to establish the desired result, it suffices to show that as $||\mu||_2 \rightarrow \infty$, 
\begin{align*}
\liminf_{n \rightarrow \infty} \mathbb{P} \{ T_n > \max_{g \in \textbf{G}_s} T^*_n(g) \} \rightarrow 1,
\end{align*}
which follows under similar arguments as those employed in the proof of Theorem 3.2 in \cite{Canay-Santos-Shaikh(2020)}. $\blacksquare$

\section{Proof of Theorem \ref{theo: boot-stud-t}}
Following the same argument in the proof of Theorem \ref{theo: boot-stud-t-power}, we can show that
\begin{align*}
(\sqrt{n}T_{CR,n},\{\sqrt{n}T_{CR,n}^*(g)\}_{g \in \textbf{G}}) \xrightarrow{\enskip d \enskip}  (T_{CR,\infty}(\iota_J),\{T_{CR,\infty}(g)\}_{g \in \textbf{G}})    
\end{align*}
where $T_{CR,n}(g)$ is defined in 
the proof of Theorem \ref{theo: boot-stud-t-power} with $\mu=0$ as we are under the null. 
Then, the rest of the proof is similar to Step 3 in the proof of Theorem \ref{theo: boot-t}. 
We omit the detail for brevity. $\blacksquare$

\section{Proof of Theorem \ref{theo: boot-stud-t-power}}
\label{sec:pf_boot-stud-t-power}
For the power analysis, we focus on the TSLS estimator. The results for other $k$-class estimators can be derived in the same manner given Lemma \ref{lem:equivalence}. Recall $a_j$ defined in Assumption \ref{assumption: 3}. We further define
\begin{align}\label{eq: thm-2-2-T_CR_g}
T_{CR,\infty}(g) = \left\Vert \lambda_\beta^\top \tilde{Q} \left[\sum_{j \in [J]} g_j  \mathcal{Z}_j \right]+ c_{0,g} \mu \right\Vert_{A_{r,CR,g}},
\end{align}
where $\tilde{Q} = Q^{-1} Q_{\widetilde{Z}X}^{\top} Q_{\widetilde{Z}\widetilde{Z}}^{-1}$, $Q = Q_{\widetilde{Z}X}^\top Q_{\widetilde{Z}\widetilde{Z}}Q_{\widetilde{Z}X}$,  
$c_{0,g} =  \sum_{j \in [J]} \xi_j g_j a_j$, and 

\begin{align*}
A_{r,CR,g}  = \begin{pmatrix}
\sum_{j \in [J]} \left\{\lambda_\beta^\top \tilde{Q} \left[g_j\mathcal{Z}_j - a_j\xi_j \sum_{\tilde{j} \in [J]} g_{\tilde{j}}\mathcal{Z}_{\tilde{j}}\right]+\xi_j (g_j - c_{0,g} )a_j \mu \right\} \\
\times \left\{\lambda_\beta^\top \tilde{Q} \left[g_j\mathcal{Z}_j - a_j\xi_j \sum_{\tilde{j} \in [J]} g_{\tilde{j}}\mathcal{Z}_{\tilde{j}}\right]+\xi_j (g_j - c_{0,g} )a_j \mu \right\}^\top
\end{pmatrix}^{-1}.
\end{align*}
We order $\{T_{CR,\infty}(g)\}_{g \in \textbf{G}}$ in ascending order: 
$(T_{CR,\infty})^{(1)} \leq \cdots \leq (T_{CR,\infty})^{|\textbf{G}|}$. In the proof of Theorem \ref{theo: boot-t-power}, we have already shown that, under $\mathcal{H}_{1,n}$, 
\begin{align*}
r_n(\lambda_\beta^\top \hat{\beta}_{tsls} - \lambda_0) \xrightarrow{\enskip d \enskip} \sum_{j \in [J]} \left[ \lambda_\beta^\top \tilde{Q} \mathcal{Z}_j \right] + \mu, \quad
r_n(\lambda_\beta^\top \hat{\beta}^*_{tsls,g} - \lambda_0) \xrightarrow{\enskip d \enskip} \sum_{j \in [J]} g_j \left[ \lambda_\beta^\top \tilde{Q} \mathcal{Z}_j \right] + c_{0,g}\mu.
\end{align*}
Next, we derive the limit of $\hat{A}_{r,CR}$ and $\hat{A}^*_{r,CR,g}$. We first note that 
\begin{align*}
\frac{r_n}{n}\sum_{i \in I_{n,j}} \widetilde{Z}_{i,j}\hat{\eps}_{i,j} & = \frac{r_n}{n}\sum_{i \in I_{n,j}} \widetilde{Z}_{i,j}\left[\eps_{i,j} -X_{i,j}^\top(\hat{\beta}_{tsls} - \beta_n) - W_{i,j}^\top(\hat{\gamma}_{tsls} -\gamma)\right]\\
& = \frac{r_n}{n}\sum_{i \in I_{n,j}} \widetilde{Z}_{i,j}\eps_{i,j} - \xi_j\widehat{Q}_{\widetilde{Z}X,j} r_n(\hat{\beta}_{tsls} - \beta_n) + o_P(1) \\
& = \frac{r_n}{n}\sum_{i \in I_{n,j}} \widetilde{Z}_{i,j}\eps_{i,j} - \xi_j\widehat{Q}_{\widetilde{Z}X,j} \widehat{Q}^{-1}\widehat{Q}_{\widetilde{Z}X}^\top \widehat{Q}_{\widetilde{Z}\widetilde{Z}}^{-1}\sum_{\tilde{j} \in [J]} \frac{r_n}{n} \sum_{i \in I_{n,\tilde{j}}} \widetilde{Z}_{i,\tilde{j}}\eps_{i,\tilde{j}} + o_P(1) \\
& \xrightarrow{\enskip d \enskip} \mathcal{Z}_j -  \xi_jQ_{\widetilde{Z}X,j}Q^{-1}Q_{\widetilde{Z}X}^\top Q_{\widetilde{Z}\widetilde{Z}}^{-1} \sum_{\tilde{j} \in [J]} \mathcal{Z}_{\tilde{j}} + o_P(1),
\end{align*}
where the first equality holds by Lemma \ref{lem:equiv2}. This implies 
\begin{align*}
& \frac{r_n^2}{n}\hat{\Omega}_{CR} \convD \sum_{j \in [J]} \left(\mathcal{Z}_j -  \xi_jQ_{\widetilde{Z}X,j}\tilde{Q} \sum_{\tilde{j} \in [J]} \mathcal{Z}_{\tilde{j}}\right)\left(\mathcal{Z}_j -  \xi_jQ_{\widetilde{Z}X,j}\tilde{Q} \sum_{\tilde{j} \in [J]} \mathcal{Z}_{\tilde{j}}\right)^\top, \\
& \frac{r_n^2}{n}\hat{A}_{r,CR}     \convD A_{r,CR,\iota_J}, \quad \text{and} \quad T_{CR,n} \convD T_{CR,\infty}(\iota_J),
\end{align*}
where $\iota_J$ is a $J \times 1$ vector of ones. Similarly, we have
\begin{align}
& \frac{r_n}{n}\sum_{i \in I_{n,j}} \widetilde{Z}_{i,j}\hat{\eps}^*_{i,j}(g) \notag \\
& = \frac{r_n}{n}\sum_{i \in I_{n,j}} \widetilde{Z}_{i,j}\left[g_j\hat{\eps}_{i,j}^r - X_{i,j}^{*\top}(g)(\hat{\beta}_{tsls,g}^* - \hat{\beta}^r_{tsls}) - W_{i,j}^\top(\hat{\gamma}_{tsls,g}^* - \hat{\gamma}^r_{tsls})\right] \notag \\
& = \frac{g_jr_n}{n}\sum_{i \in I_{n,j}} \widetilde{Z}_{i,j}\hat{\eps}_{i,j}^r - \xi_j\widehat{Q}_{\widetilde{Z}X,j}^*(g) r_n(\hat{\beta}_{tsls,g}^* - \hat{\beta}^r_{tsls}) + o_P(1) \notag \\
& = \frac{g_jr_n}{n}\sum_{i \in I_{n,j}} \widetilde{Z}_{i,j}\eps_{i,j} - \xi_j g_j\widehat{Q}_{\widetilde{Z}X,j}r_n(\hat{\beta}^r_{tsls} - \beta_n)-
\xi_j\widehat{Q}_{\widetilde{Z}X,j} r_n(\hat{\beta}_{tsls,g}^* - \hat{\beta}^r_{tsls}) + o_P(1) \notag \\
& = \frac{g_jr_n}{n}\sum_{i \in I_{n,j}} \widetilde{Z}_{i,j}\eps_{i,j} - \xi_j g_j Q_{\widetilde{Z}X,j}r_n(\hat{\beta}^r_{tsls} - \beta_n)-
\xi_j Q_{\widetilde{Z}X,j} r_n(\hat{\beta}_{tsls,g}^* - \hat{\beta}^r_{tsls}) + o_P(1),
\label{eq:Zeps^star}
\end{align}
where $\widehat{Q}_{\widetilde{Z}X,j}^*(g) = \frac{1}{n_j}\sum_{i \in I_{n,j}}\widetilde{Z}_{i,j}X_{i,j}^{*\top}(g)$, the second equality is by Lemma \ref{lem:equiv2}, and the last equality holds because $\widehat{Q}_{\widetilde{Z}X,j} = Q_{\widetilde{Z}X,j} + o_P(1)$,  $\widehat{Q}_{\widetilde{Z}X,j}^*(g) = Q_{\widetilde{Z}X,j} + o_P(1)$ proved in Step 1 of the proof of Theorem \ref{theo: boot-t}, $r_n(\hat{\beta}^r_{tsls}-\beta_n) = O_P(1)$, and $r_n(\hat{\beta}^*_{tsls,g} - \beta_n) = O_P(1)$. In addition, following the same arguments that lead to \eqref{eq: boot-unstud-power-2}, we have
\begin{align}
r_n(\hat{\beta}_{tsls,g}^* - \hat{\beta}^r_{tsls}) & = \tilde{Q} \sum_{\tilde{j} \in [J]} \sum_{i \in I_{n,\tilde{j}}}  \frac{r_n g_{\tilde{j}} \widetilde{Z}_{i,\tilde{j}}\eps_{i,\tilde{j}}}{n} + \tilde{Q}\sum_{\tilde{j} \in [J]} \xi_{\tilde{j}} g_{\tilde{j}} Q_{\widetilde{Z}X,\tilde{j}}r_n(\beta_n - \hat{\beta}^r_{tsls})  + o_P(1) \notag \\
& = \tilde{Q} \sum_{\tilde{j} \in [J]} \sum_{i \in I_{n,\tilde{j}}}  \frac{r_n g_{\tilde{j}} \widetilde{Z}_{i,\tilde{j}} \eps_{i,\tilde{j}}}{n} +\sum_{\tilde{j} \in [J]} \xi_{\tilde{j}}g_{\tilde{j}}a_{\tilde{j}}r_n(\beta_n - \hat{\beta}^r_{tsls})  + o_P(1).
\label{eq:b^*-b^r}
\end{align}
Note $\widehat{Q}^{*-1}_g\widehat{Q}_{\widetilde{Z}X}^{*\top}(g)\widehat{Q}_{\widetilde{Z}\widetilde{Z}}^{-1}  \convP     \tilde{Q}$. Therefore, combining \eqref{eq:Zeps^star} and \eqref{eq:b^*-b^r}, we have
\begin{align*}
& \lambda_\beta^\top \widehat{Q}_g^{*-1}\widehat{Q}_{\widetilde{Z}X}^{*\top}(g)\widehat{Q}_{\widetilde{Z}\widetilde{Z}}^{-1} \left[\frac{r_n}{n}\sum_{i \in I_{n,j}} \widetilde{Z}_{i,j}\hat{\eps}^*_{i,j}(g)\right] \\
& = \lambda_\beta^\top \tilde{Q}\left[\frac{g_j r_n}{n}\sum_{i \in I_{n,j}} \widetilde{Z}_{i,j}\eps_{i,j} - \xi_j g_j Q_{\widetilde{Z}X,j}r_n(\hat{\beta}^r_{tsls} - \beta_n)-
\xi_j Q_{\widetilde{Z}X,j} r_n(\hat{\beta}_{tsls,g}^* - \hat{\beta}^r_{tsls}) \right] + o_P(1) \\
& = \lambda_\beta^\top \tilde{Q}\left[\frac{g_j r_n}{n}\sum_{i \in I_{n,j}} \widetilde{Z}_{i,j}\eps_{i,j}\right] - \xi_j g_j a_j \lambda_\beta^\top r_n(\hat{\beta}^r_{tsls} - \beta_n)-
\xi_j a_j \lambda_\beta^\top r_n(\hat{\beta}_{tsls,g}^* - \hat{\beta}^r_{tsls})  + o_P(1)\\
& \convD \lambda_\beta^\top \tilde{Q} g_j \mathcal{Z}_j - \xi_j a_j \lambda_\beta^\top \tilde{Q} \sum_{\tilde{j} \in [J]}  g_{\tilde{j}} \mathcal{Z}_{\tilde{j}} + \xi_j (g_j - c_{0,g})a_j \mu, 
\end{align*}
where the second equality is by the fact that $\tilde{Q}Q_{\widetilde{Z}X,j} = a_j I_{d_x}$ and the last convergence is by the fact that $\lambda_\beta^\top r_n(\hat{\beta}^r_{tsls} - \beta_n) = -\mu$. This implies 
\begin{align*}
& \frac{r_n^2}{n}\hat{A}_{r,CR,g}^* \convD A_{r,CR,g},\quad \text{and thus,} \quad  (\sqrt{n}T_{CR,n},\{\sqrt{n}T_{CR,n}^*(g)\}_{g \in \textbf{G}}) \convD (T_{CR,\infty}(\iota_J),\{T_{CR,\infty}(g)\}_{g \in \textbf{G}}).     
\end{align*}

By the Portmanteau theorem, we have 
\begin{align*}
\liminf_{n \rightarrow \infty}\mathbb{P} \{ T_{CR,n} > \hat{c}_{CR,n}(1-\alpha) \} & \geq  \mathbb{P}\left\{ T_{CR,\infty} (\iota_J) > (T_{CR,\infty})^{(k^*)} \right\}.
\end{align*}
We aim to show that, as $||\mu ||_2\rightarrow \infty$, we have
\begin{align}
\mathbb{P}\left\{ T_{CR,\infty} (\iota_J) > \max_{g \in \textbf{G}_s}T_{CR,\infty} (g)\right\} \rightarrow 1, 
\label{eq:GsCR-liniv}
\end{align}
where $\textbf{G}_s = \textbf{G} \backslash \textbf{G}_w$, and $\textbf{G}_w = \{g \in \textbf{G}: g_{j} = g_{j'}, \forall j,j' \in \mathcal{J}_s\}$.
Then, given $|\textbf{G}_s| = |\textbf{G}|-2^{J-J_s+1}$ and $k^* = \lceil |\textbf{G}|(1-\alpha) \rceil \leq |\textbf{G}|-2^{J-J_s+1}$, \eqref{eq:GsCR-liniv} implies that   as $ ||\mu||_2 \rightarrow \infty$,
\begin{align*}
\mathbb{P}\left\{ T_{CR,\infty}(\iota_J) > (T_{CR,\infty})^{(k^*)}  \right\} \geq \mathbb{P}\left\{ T_{CR,\infty}(\iota_J) > \max_{g \in \textbf{G}_s}T_{CR,\infty}(g) \right\} \rightarrow 1.
\end{align*}
Therefore, it suffices to establish \eqref{eq:GsCR-liniv}.

By (\ref{eq: thm-2-2-T_CR_g}), we see that  
$$T_{CR,\infty}(\iota_J) = \left\Vert \lambda_\beta^\top \tilde{Q} \left[\sum_{j \in [J]}  \mathcal{Z}_j \right]+ \mu \right\Vert_{A_{r,CR,\iota_J}},$$
and $A_{r,CR,\iota_J}$ is independent of $\mu$ as $c_{0, \iota_J}=1$. In addition, we have 
$\lambda_{\min}(\tilde{Q}^\top \lambda_\beta^\top A_{r,CR,\iota_J} \lambda_\beta \tilde{Q})>0$ with probability one.  
Therefore, for any $\delta>0$, we can find a sufficiently small constant $c>0$ and a sufficiently large constant $M>0$ such that with probability greater than $1-\delta$, for any $\mu$, 
\begin{align}
T_{CR,\infty}(\iota_J) \geq  c ||\mu||^2_2 - M. 
\label{eq:IV-Tiota}
\end{align}
On the other hand, for $g \in \textbf{G}_s$, 
we can write  $T_{CR,\infty}(g)$ as
\begin{align*}
T_{CR,\infty}(g)& = \biggl\{(N_{0,g} + c_{0,g} \mu )^\top \left[\sum_{j \in [J]}(N_{j,g} + c_{j,g} \mu )(N_{j,g} + c_{j,g} \mu )^\top\right]^{-1} (N_{0,g} + c_{0,g} \mu )\biggr\}, 
\end{align*}
where for $j \in [J]$,
$N_{0,g} = \lambda_\beta^\top \tilde{Q} \left[\sum_{j \in [J]} g_j \mathcal{Z}_j \right]$, 
$N_{j,g} =  \lambda_\beta^\top \tilde{Q} \left[g_j\mathcal{Z}_j - a_j \xi_j \sum_{\tilde{j} \in [J]} g_{\tilde{j}}\mathcal{Z}_{\tilde{j}} \right]$,
and $c_{j,g} =  \xi_j (g_j - c_{0,g})a_j$. 

We claim that for $g \in \textbf{G}_s$, $c_{j,g} \neq 0$ for some $j \in \mathcal J_s$. Suppose it does not hold, then it implies that $g_j = c_{0,g}$ for all $j \in \mathcal J_s$, i.e., 
for all $j \in \mathcal J_s$, $g_j$ shares the same sign, and thus, contradicts the definition of $\textbf{G}_s$. Therefore, combining the claim with the assumption that $\min_{j \in \mathcal J_s}|a_j| > 0$, we have 
$\min_{g \in \textbf{G}_s} \sum_{j \in [J]} c_{j,g}^2 >0.$

In addition, we note that  
\begin{eqnarray*}
	&& \sum_{j \in [J]} (N_{j,g} + c_{j,g} \mu)(N_{j,g} + c_{j,g} \mu )^\top  \\
	& =&  \sum_{j \in [J]} N_{j,g} N_{j,g}^\top + \sum_{j \in [J]} c_{j,g} N_{j,g} \mu^\top 
	+ \sum_{j \in [J]} c_{j,g} \mu N_{j,g}^\top + (\sum_{j \in [J]} c_{j,g}^2) \mu \mu^\top \\
	&\equiv&  M_1 + M_2 \mu^\top +  \mu M_2^\top + \overline{c}^2 \mu \mu^\top,
\end{eqnarray*}
where we denote
$M_1 = \sum_{j \in [J]} N_{j,g} N_{j,g}^\top$, 
$M_2 = \sum_{j \in [J]} c_{j,g} N_{j,g}$,  and $\overline{c}^2 = \sum_{j \in [J]} c_{j,g}^2$. 
For notation ease, we suppress the dependence of $(M_1,M_2,\overline{c})$ on $g$.  
Then, we have
\begin{align*}
M_1 + M_2 \mu^\top +  \mu M_2^\top + \overline{c}^2 \mu\mu^\top = M_1 - \frac{M_2 M_2^\top}{\overline{c}^2} + \left(\frac{M_2}{\overline{c}} + \overline{c} \mu \right)\left(\frac{M_2}{\overline{c}} + \overline{c}\mu \right)^\top.
\end{align*}
Note for any $d_r \times 1$ vector $u$, by the Cauchy–Schwarz inequality, 
\begin{align*}
u^\top \left( M_1 - \frac{M_2 M_2^\top}{\overline{c}^2}\right)   u = \sum_{j \in [J]} (u^\top N_{j,g})^2 - \frac{\left(\sum_{j \in [J]} u^\top N_{j,g} c_{j,g}\right)^2}{\sum_{j \in [J]} c_{j,g}^2} \geq 0,
\end{align*}
where the equal sign holds if and only if there exist $(u,g,c) \in \Re^{d_r} \times \textbf{G}_s \times \Re$ such that 
\begin{align*}
u^\top N_{j,g} = c c_{1,j},~j \in [J], 
\end{align*}
or equivalently, 
\begin{align*}
\left(I_J + (-a_1\xi_1,\cdots,-a_J\xi_J)^\top \iota_J^\top \right) (g_1\mathcal Z_1,\cdots,g_J \mathcal Z_j)^\top \tilde Q^\top \lambda_\beta u     = c(c_{1,g},\cdots,c_{J,g})^\top.
\end{align*}
This occurs with probability zero if $J - 1> d_r$ as $\{g_j \mathcal Z_j\}_{j \in [J]}$ are independent and non-degenerate normal vectors and the matrix $\left(I_J + (-a_1\xi_1,\cdots,-a_J\xi_J)^\top\iota_J^\top \right)$ has rank $J-1$. Therefore, the matrix $\mathbb{M} \equiv M_1 - \frac{M_2 M_2^\top}{\overline{c}^2}$ is invertible with probability one. Specifically, denote $\mathbb{M}$ as $\mathbb{M}(g)$ to highlight its dependence on $g$. 
We have $\max_{g \in \textbf{G}_s }(\lambda_{\min}(\mathbb{M}(g)))^{-1} = O_P(1)$. 
In addition, denote $\frac{M_2}{\overline{c}} + \overline{c} \mu $ as $\mathbb{V}$, which is a $d_r \times 1$ vector. Then,  we have
\begin{align*}
& \left[\sum_{j \in [J]} (N_{j,g} + c_{j,g} \mu )(N_{j,g} + c_{j,g} \mu )^\top\right]^{-1} 
= [\mathbb{M} + \mathbb{V}\mathbb{V}^\top]^{-1} 
= \mathbb{M}^{-1} - \mathbb{M}^{-1} \mathbb{V} (1+ \mathbb{V}^\top \mathbb{M}^{-1}\mathbb{V})^{-1} \mathbb{V}^\top \mathbb{M}^{-1},
\end{align*}
where the second equality is due to the Sherman–Morrison–Woodbury formula. 

Next, we note that 
\begin{align*}
N_{0,g} + c_{0,g} \mu  = N_{0,g} + c_{0,g} \left(\frac{\mathbb{V}}{\overline{c}} - \frac{M_2}{\overline{c}^2}\right)
\equiv \mathbb{M}_0 + \frac{c_{0,g}}{\overline{c}}\mathbb{V},
\end{align*}
where 
$
\mathbb{M}_0 = N_{0,g} - \frac{c_{0,g} M_2}{\overline{c}^2} = N_{0,g} - \frac{c_{0,g} (\sum_{j \in [J]} c_{j,g} N_{j,g})}{\sum_{j \in [J]} c_{j,g}^2}.$
With these notations, we have
\begin{align*}
& (N_{0,g} + c_{0,g} \mu )^\top \left[\sum_{j \in [J]} (N_{j,g} + c_{j,g} \mu )(N_{j,g} + c_{j,g} \mu )^\top\right]^{-1} 
(N_{0,g} + c_{0,g} \mu ) \\
&  = \left( \mathbb{M}_0 + \frac{c_{0,g}}{\overline{c}}\mathbb{V} \right)^\top(\mathbb{M}^{-1} - \mathbb{M}^{-1} \mathbb{V} (1+ \mathbb{V}^\top \mathbb{M}^{-1}\mathbb{V})^{-1} \mathbb{V}^\top \mathbb{M}^{-1})\left( \mathbb{M}_0 + \frac{c_{0,g}}{\overline{c}}\mathbb{V} \right) \\
& \leq 2\mathbb{M}_0^\top (\mathbb{M}^{-1} - \mathbb{M}^{-1} \mathbb{V} (1+ \mathbb{V}^\top \mathbb{M}^{-1}\mathbb{V})^{-1} \mathbb{V}^\top \mathbb{M}^{-1})\mathbb{M}_0 \\
& +  \frac{2c_{0,g}^2}{\overline{c}^2} \mathbb{V}^\top (\mathbb{M}^{-1} - \mathbb{M}^{-1} \mathbb{V} (1+ \mathbb{V}^\top \mathbb{M}^{-1}\mathbb{V})^{-1} \mathbb{V}^\top \mathbb{M}^{-1})\mathbb{V}\\
& \leq  2\mathbb{M}_0^\top \mathbb{M}^{-1} \mathbb{M}_0 + \frac{2c_{0,g}^2}{\overline{c}^2} \frac{ \mathbb{V}^\top \mathbb{M}^{-1}\mathbb{V}}{1+\mathbb{V}^\top \mathbb{M}^{-1}\mathbb{V}} 
\leq 2\mathbb{M}_0^\top \mathbb{M}^{-1} \mathbb{M}_0 + \frac{2c_{0,g}^2}{\overline{c}^2} \\
& \leq 2 (\lambda_{\min}(\mathbb{M}))^{-1}\left\Vert N_{0,g} - \frac{c_{0,g} (\sum_{j \in [J]} c_{j,q}N_{j,g})}{\sum_{j \in [J]} c_{j,g}^2}\right\Vert^2  + \frac{2c_{0,g}^2}{\sum_{j \in [J]} c_{j,g}^2} 
\equiv C(g),
\end{align*}
where the first inequality holds because $(u+v)^\top A (u+v) \leq 2 (u^\top A u + v^\top A v)$ for some $d_r\times d_r$ positive semidefinite matrix $A$ and $u,v \in \Re^{d_r}$, the second inequality holds due to the fact that $\mathbb{M}^{-1} \mathbb{V} (1+ \mathbb{V}^\top \mathbb{M}^{-1}\mathbb{V})^{-1} \mathbb{V}^\top \mathbb{M}^{-1}$ is positive semidefinite, the third inequality holds because $\mathbb{V}^\top \mathbb{M}^{-1}\mathbb{V}$ is a  nonnegative scalar, and the last holds by substituting in the expressions for $\mathbb{M}_0$ and $\overline{c}$. 

Then, we have
\begin{align}
\max_{g \in \textbf{G}_s} T_{CR,\infty}(g)  \leq   \max_{g \in \textbf{G}_s} C(g).
\label{eq:IV-Tiota*}
\end{align}
Combining \eqref{eq:IV-Tiota} and \eqref{eq:IV-Tiota*}, we have, as $||\mu||_2 \rightarrow \infty$, 
\begin{align*}
& \liminf_{n \rightarrow \infty}\mathbb{P} \{ T_{CR,n} > \hat{c}_{CR,n}(1-\alpha) \} 
\geq  \mathbb{P}\left\{ T_{CR,\infty}(\iota_J) > (T_{CR,\infty})^{(k^*)}  \right\} 
\geq \mathbb{P}\left\{ T_{CR,\infty}(\iota_J) > \max_{g \in \textbf{G}_s} T_{CR,\infty}(g)  \right\} \\
&  = 1- \mathbb{P}\left\{ T_{CR,\infty}(\iota_J) \leq \max_{g \in \textbf{G}_s} T_{CR,\infty}(g)  \right\} 
\geq 1- \mathbb{P}\left\{ c||\mu||^2_2 - M \leq \max_{g \in \textbf{G}_s} C(g) \right\}-\delta \rightarrow 1-\delta, 
\end{align*}
where the second inequality is by the fact that $k^* \leq |\textbf{G}_s|$, and thus, $(T_{CR,\infty})^{(k^*)} \leq \max_{g \in \textbf{G}_s} T_{CR,\infty}(g)$ and the last convergence holds because $\max_{g \in \textbf{G}_s} C(g) = O_P(1)$ and does not depend on $\mu$. As $\delta$ is arbitrary, we can let $\delta \rightarrow 0$ and obtain the desired result. 

For the proof of part (ii), let $\tilde{c}_{CR,n}(1-\alpha)$ denote the $(1-\alpha)$ quantile
of 
\begin{align*}
\left\{|(\lambda_{\beta}^{\top}\hat{\beta}^*_{tsls,g} - \lambda_0)|/\sqrt{\lambda_{\beta}^{\top}\widehat{V}\lambda_{\beta}}: g \in \textbf{G}\right\},
\end{align*}
i.e., the bootstrap statistic $T_n^*(g)$ studentized by the original CRVE instead of the bootstrap CRVE. 
Because $d_r = 1$, we have
\begin{align*}
1\{T_n > \hat{c}_{n}(1-\alpha)\} = 1\{T_{CR,n} >  \tilde{c}_{CR,n}(1-\alpha)\}.
\end{align*}
Therefore, we have
\begin{align*}
\liminf_{n \rightarrow \infty}\mathbb{P}(\phi^{cr}_n \geq \phi_n) & = \liminf_{n \rightarrow \infty}\mathbb{P}(1\{T_{CR,n} >  \hat {c}_{CR,n}(1-\alpha)\} \geq 1\{T_{CR,n} >  \tilde{c}_{CR,n}(1-\alpha)\}) \\
& = 1 - \limsup_{n \rightarrow \infty}\mathbb{P}(1\{T_{CR,n} >  \hat {c}_{CR,n}(1-\alpha)\} < 1\{T_{CR,n} >  \tilde{c}_{CR,n}(1-\alpha)\}) \\
& \geq 1 - \limsup_{n \rightarrow \infty}\mathbb{P}(\sqrt{n}\hat {c}_{CR,n}(1-\alpha)>  \sqrt{n}\tilde{c}_{CR,n}(1-\alpha)) \\
& \geq 1 - \limsup_{n \rightarrow \infty}\mathbb{P}\left(\left( \frac{r_n^2}{n}\lambda_{\beta}^{\top} \widehat{V} \lambda_{\beta} \right)^{1/2} \sqrt{n} \hat {c}_{CR,n}(1-\alpha)  + \max_{g \in \textbf{G}}|r_n\lambda_{\beta}^{\top}(\hat{\beta}^*_{tsls,g} - \beta_n)|> |\mu| \right),
\end{align*}
where we use the fact that 
\begin{align*}
\sqrt{n}    \tilde{c}_{CR,n}(1-\alpha) \geq \left( \frac{r_n^2}{n}\lambda_{\beta}^{\top} \widehat{V} \lambda_{\beta} \right)^{-1/2}\left[|\mu| -\max_{g \in \textbf{G}}|r_n\lambda_{\beta}^{\top}(\hat{\beta}^*_{tsls,g} - \beta_n)| \right].  
\end{align*}

Further note that $\max_{g \in \textbf{G}}|r_n\lambda_{\beta}^{\top}(\hat{\beta}^*_{tsls,g} - \beta_n)| = O_P(1)$ and does not depend on $\mu$, and
\begin{align*}
\left( \frac{r_n^2}{n}\lambda_{\beta}^{\top} \widehat{V} \lambda_{\beta} \right)^{-1}    \convD A_{r,CR,\iota_J}
\end{align*}
and $A_{r,CR,\iota_J}$ does not depend on $\mu$ either. 

Last, although $\hat{c}_{CR,n}(1-\alpha)$ depends on $\mu$, we have 
\begin{align*}
\sqrt{n}\hat{c}_{CR,n}(1-\alpha) \leq \sqrt{n}\max_{g \in \textbf{G}_s} T_{CR,n}^*(g) \convD \max_{g \in \textbf{G}_s} T_{CR,\infty}(g) \leq \max_{g \in \textbf{G}_s} C(g) = O_P(1)
\end{align*}
for some $C(g)$ that does not depend on $\mu$ as has been proved in Part (i). 
Therefore, for any $\delta>0$, there exists a constant $c_{\mu}>0$, such that when $|\mu| > c_\mu$, 
\begin{align*}
&   \limsup_{n \rightarrow \infty}\mathbb{P}\left(\left( \frac{r_n^2}{n}\lambda_{\beta}^{\top} \widehat{V} \lambda_{\beta} \right)^{1/2}  \sqrt{n}\hat {c}_{CR,n}(1-\alpha)  + \max_{g \in \textbf{G}}|r_n\lambda_{\beta}^{\top}(\hat{\beta}^*_{tsls,g} - \beta_n)|> |\mu| \right) \\
& \leq \limsup_{n \rightarrow \infty}\mathbb{P}\left(\left( \frac{r_n^2}{n}\lambda_{\beta}^{\top} \widehat{V} \lambda_{\beta} \right)^{1/2} \sqrt{n}\max_{g \in \textbf{G}_s} T_{CR,n}(g)  + C > |\mu| \right) + \delta/2  \\
& \leq \mathbb P\left(A_{r,CR,\iota_J}^{-1/2} \max_{g \in \textbf{G}_s} T_{CR,\infty}(g)  + C \geq |\mu| \right)+ \delta/2 \\
& \leq \mathbb P\left(A_{r,CR,\iota_J}^{-1/2} \max_{g \in \textbf{G}_s} C(g)  + C \geq c_\mu \right)+ \delta/2
\leq \delta,
\end{align*}
where $C$ in the first inequality is a constant such that for $n$ being sufficiently large, 
\begin{align*}
\mathbb P \left(\max_{g \in \textbf{G}}|r_n\lambda_{\beta}^{\top}(\hat{\beta}^*_{tsls,g} - \beta_n)| \geq C  \right) \leq \delta/2,
\end{align*}
the second inequality is by Portmanteau theorem, and the last inequality holds if $c_\mu$ is sufficiently large. This concludes the proof. $\blacksquare$

\section{Proof of  Theorem \ref{theo: boot-AR}}
\label{sec:pf_sec2end}
The proof for the $AR_n$-based wild bootstrap test follows similar arguments as those in  \hyperlink{theo: boot-t}{Theorem \ref{theo: boot-t}}, 
and thus we keep exposition more concise. 
Let $\mathbb{S} \equiv \otimes_{j \in [J]} \textbf{R}^{d_z} \times \textbf{R}^{d_z \times d_z}$ and write an element $s \in \mathbb{S}$ 
by $s = ( \{s_{1j} : j \in [J]\}, s_2)$ where $s_{1j} \in \textbf{R}^{d_z}$ for any $j \in [J]$. 
Define the function $T_{AR}$: $\mathbb{S} \rightarrow \textbf{R}$ to be given by 
\begin{align}\label{eq: T_AR_u_s}
T_{AR}(s) = \bigg\Vert \sum_{j \in [J]} s_{1j} \bigg\Vert_{s_2}. 
\end{align}
Given this notation we can define the statistics $S_n, \widehat{S}_n \in \mathbb{S}$ as 
\begin{align*}
S_n =   
\left\{ \frac{r_n}{n}  \sum_{i \in I_{n,j}} \widetilde{Z}_{i,j} \eps_{i,j} : j \in [J], \hat{A}_z \right\}, 
\quad
\widehat{S}_n = 
\left\{ \frac{r_n}{n} \sum_{i \in I_{n,j}} \widetilde{Z}_{i,j} \bar{\eps}^r_{i,j}: j \in [J], \hat{A}_z \right\},
\end{align*}
where $\bar{\eps}^r_{i,j} = y_{i,j}-X_{i,j}^{\top}\beta_0-W^{\top}_{i,j}\bar{\gamma}^r$.
Note that by the Frisch-Waugh-Lovell theorem, 
\begin{align}\label{eq: AR_u_n_S}
r_n AR_{n} 
&    = \bigg\Vert \sum_{j \in [J]} \frac{r_n}{n}  \sum_{i \in I_{n,j}} \widetilde{Z}_{i,j} \eps_{i,j} \bigg\Vert_{\hat{A}_z} = T_{AR}(S_n).
\end{align}

Therefore, 
letting $k^* \equiv \lceil |\textbf{G}| (1-\alpha) \rceil$, we obtain from (\ref{eq: AR_u_n_S})-(\ref{eq: AR_u_n_S_hat}) that 
\begin{align*}
1 \left\{ AR_{n} > \hat{c}_{AR,n}(1-\alpha) \right\} = 1 \left\{ T_{AR}(S_n) > T^{(k^*)}_{AR}( \widehat{S}_n | \textbf{G}) \right\}.
\end{align*}

Furthermore, note that we have
\begin{align}\label{eq: T_AR_u_S_n}
T_{AR}\left( - \iota_J \widehat{S}_n \right) &    = T_{AR}\left( \iota_J \widehat{S}_n \right) 
=  \bigg\Vert \sum_{j \in [J]}  \frac{r_n}{n}  \sum_{i \in I_{n,j}} \widetilde{Z}_{i,j}  \left(y_{i,j} - X_{i,j}^{\top} \beta_0 - W_{i,j}^{\top} \bar{\gamma}^r \right) \bigg\Vert_{\hat{A}_z} \notag \\
&    = \bigg\Vert \sum_{j \in [J]}  \frac{r_n}{n}  \sum_{i \in I_{n,j}} \widetilde{Z}_{i,j}  \left( \eps_{i,j} - W_{i,j}^{\top} (\bar{\gamma}^r - \gamma) \right) \bigg\Vert_{\hat{A}_z}
= T_{AR}\left(S_n\right),
\end{align}
where the third equality follows from $\sum_{j \in [J]} \sum_{i \in I_{n,j}} \widetilde{Z}_{i,j}W_{i,j}^{\top} = 0$. 
(\ref{eq: T_AR_u_S_n}) implies that if $k^* 
> |\textbf{G}| -2$,
then $1 \{ T_{AR}(S_n) > T_{AR}^{(k^*)} (\widehat{S}_n | \textbf{G}) \} = 0$, 
and this gives the upper bound in Theorem  \hyperlink{theo: boot-AR}{\ref{theo: boot-AR}}. 
We therefore assume that $k^* \leq |\textbf{G}| -2$, in which case 
\begin{align}\label{eq: limsup_T_AR_u_n}
\limsup_{n \rightarrow \infty} \mathbb{P} \{ T_{AR}(S_n) > T^{(k^*)}_{AR} (\widehat{S}_n | \textbf{G}) \} 
&  = \limsup_{n \rightarrow \infty} \mathbb{P} \{ T_{AR}(S_n) > T^{(k^*)}_{AR} (\widehat{S}_n | \textbf{G} \setminus \{\pm  \iota_J \})\} \notag \\
& \leq \limsup_{n \rightarrow \infty} \mathbb{P} \{ T_{AR}(S_n) \geq T^{(k^*)}_{AR} (\widehat{S}_n | \textbf{G}  \setminus \{\pm \iota_J \})\}.
\end{align}

Then, to examine the right hand side of (\ref{eq: limsup_T_AR_u_n}), 
first note that by Assumptions \ref{assumption: 1} and \ref{assumption: AR_further_assumptions},  
and the continuous mapping theorem we have 
\begin{align}\label{eq: AR_u_S_1}
\left\{ \frac{r_n}{n}  \sum_{i \in I_{n,j}} \widetilde{Z}_{i,j} \eps_{i,j} : j \in [J], \hat{A}_z \right\} 
\xrightarrow{\enskip d \enskip}
\left\{ \mathcal{Z}_j: j \in [J], A_z \right\} \equiv S,
\end{align}
where $\xi_j > 0$ for all $j \in [J]$. Furthermore, by Assumptions \ref{assumption: 1}(i), Lemma \ref{lem:equiv2}, and $\beta_n=\beta_0$, we have 
\begin{align*}
\frac{r_n}{n} \sum_{i \in I_{n,j}}\widetilde{Z}_{i,j}\bar{\eps}_{i,j}^r = \frac{r_n}{n} \sum_{i \in I_{n,j}}\widetilde{Z}_{i,j}\eps_{i,j} - \frac{1}{n}\sum_{i \in I_{n,j}}\widetilde{Z}_{i,j}W_{i,j}^\top r_n(\bar{\gamma}^r - \gamma) = \frac{r_n}{n} \sum_{i \in I_{n,j}}\widetilde{Z}_{i,j}\eps_{i,j} + o_P(1),
\end{align*}
and thus, for every $g \in \textbf{G}$,
\begin{align}\label{eq: g_T_AR_u_S_n}
T_{AR}(g \widehat{S}_n)  
= T_{AR}(g S_n) + o_P(1). 
\end{align}
We thus obtain from results in (\ref{eq: AR_u_S_1})-(\ref{eq: g_T_AR_u_S_n}) and the continuous mapping theorem that 
\begin{align*}
\left( T_{AR}(S_n), \left\{ T_{AR}(g \widehat{S}_n) : g \in \textbf{G} \right\} \right) \xrightarrow{\enskip d \enskip} 
\left(T_{AR}(S), \{ T_{AR}(gS) : g \in \textbf{G} \} \right). 
\end{align*}
Then, by the Portmanteau’s theorem and the properties of randomization tests, we have 
\begin{align*}
\limsup_{n \rightarrow \infty} \mathbb{P} \left\{ AR_{n} > \hat{c}_{AR,n}(1-\alpha) \right\} & \leq \mathbb{P} \left\{ T_{AR}(S) \geq  T_{AR}^{k^*}(\text{G} \setminus \{\pm \iota_J\}) \right\} 
= \mathbb{P} \left\{ T_{AR}(S) >  T_{AR}^{k^*}(\text{G}) \right\} \leq \alpha. 
\end{align*}
The lower bounds follow by applying similar arguments as those for \hyperlink{theo: boot-t}{Theorem \ref{theo: boot-t}}. 

For the $AR_{CR,n}$-based wild bootstrap test, define the statistics $S_{CR,n}, \widehat{S}_{CR,n} \in \mathbb{S}$ as
\begin{align*}
S_{CR,n} =   
\left\{ \frac{r_n}{n} \sum_{i \in I_{n,j}} \widetilde{Z}_{i,j} \eps_{i,j} : j \in [J], \frac{r_n^2}{n}\hat{A}_{CR} \right\}, 
\widehat{S}_{CR,n} = 
\left\{ \frac{r_n}{n}  \sum_{i \in I_{n,j}} \widetilde{Z}_{i,j} \bar{\eps}^r_{i,j}: j \in [J], \frac{r_n^2}{n}\hat{A}_{CR} \right\}.
\end{align*}
Then, we have for any action $g \in \textbf{G}$ that 
\begin{align}\label{eq: AR_u_n_S_hat}
\sqrt{n}AR^*_{n}(g) &    = \bigg\Vert \sum_{j \in [J]} \frac{r_n}{n}  \sum_{i \in I_{n,j}} g_j \widetilde{Z}_{i,j} \bar{\eps}^r_{i,j} \bigg\Vert_{\frac{r_n^2}{n} \hat{A}_{CR}} = T_{AR}(g \widehat{S}_n).
\end{align}

We set $E_n \in \textbf{R}$ to equal 
$E_n \equiv 1 \left\{ \frac{r_n^2}{n^2} \sum_{j\in J} \sum_{i \in I_{n,j}} \sum_{k \in I_{n,j}} \widetilde{Z}_{i,j} \widetilde{Z}_{k,j}^{\top} \bar{\eps}^r_{i,j} \bar{\eps}^r_{k,j} \; \text{is invertible} \right\},$
and have 
\begin{align}\label{eq: AR-P_A_n=1}
\liminf_{n \rightarrow \infty} \mathbb{P} \{ E_n = 1\} =1.
\end{align}

In addition, 
similar to the case with $AR_n$ and $AR^*_n(g)$, we have under $\beta_n = \beta_0$,
\begin{align*}
T_{AR} \left( g \widehat{S}_{CR,n} \right) = T_{AR} \left(g S_{CR,n} \right) + o_P(1) \;\; \text{for every $g \in \textbf{G}$}, 
\end{align*}
\begin{align*}
S_{CR,n}
\xrightarrow{\enskip d \enskip}
\left\{ \mathcal{Z}_j: j \in [J], A_{CR} \right\} \equiv S_{CR},
\end{align*}
where $A_{CR} = \sum_{j \in [J]} \mathcal{Z}_j \mathcal{Z}_j^{\top}$, 
and 
\begin{align}\label{eq: AR_n_AR^*_n_T_S_n_T_gS_n}
\left( T_{AR}(S_{CR,n}), \{ T_{AR}(g \widehat{S}_{CR,n}) : g \in \textbf{G} \} \right) 
\xrightarrow{\enskip d \enskip}  \left( T_{AR}(S_{CR}), \{ T_{AR} (g S_{CR}) : g \in \textbf{G} \} \right).
\end{align}

Therefore, we have
\begin{align*}\label{eq: AR_portmanteau}
\limsup_{n \rightarrow \infty} \mathbb{P} \left\{ AR_{CR,n} > \hat{c}_{AR,CR,n}(1-\alpha) \right\} 
& \leq  
\limsup_{n \rightarrow \infty} \mathbb{P} \left\{ AR_{CR,n} \geq \hat{c}_{AR,CR,n}(1-\alpha); E_n = 1 \right\} \notag \\
&  \leq 
\mathbb{P} \left\{ T_{AR}(S_{CR}) \geq 
T_{AR}^{(k^*)} (S_{CR} | \textbf{G}) \right\},
\end{align*}
which follows from (\ref{eq: AR-P_A_n=1}), (\ref{eq: AR_n_AR^*_n_T_S_n_T_gS_n}), the continuous mapping theorem 
and Portmanteau's theorem. 
The claim of the upper bound in the theorem then follows from similar arguments as those in Theorem \ref{theo: boot-t}.
$\blacksquare$

\section{Proof of Theorem \ref{theo: boot-AR-power}}
\label{sec:boot-AR-power}
Define $AR_{\infty}(g) = || \sum_{j \in [J]}g_j \mathcal{Z}_j + \sum_{j \in [J]}\xi_jg_ja_jQ_{\widetilde{Z}X}\mu ||_{A_z}$. 
In particular, notice that $AR_{\infty}(\iota_J) = || \sum_{j \in [J]} \mathcal Z_j + Q_{\widetilde{Z}X} \mu ||_{A_z}$ since $\sum_{j \in [J]} \xi_j a_j =1$.
Following same arguments in the proof of Theorem \ref{theo: boot-AR}, we can show that, under $\mathcal{H}_{1,n}$ with $\lambda_{\beta} = I_{d_x}$, 
\begin{align*}
(r_nAR_{n}, \{r_nAR_n^*(g)\}_{g \in \textbf{G}}    ) \xrightarrow{\enskip d \enskip} (AR_{\infty}(\iota_J), \{AR_{\infty}(g)\}_{g \in \textbf{G}}). 
\end{align*}

Similar to the proofs of Theorems \ref{theo: boot-t-power} and \ref{theo: boot-stud-t-power}, in order to establish Theorem \ref{theo: boot-AR-power}, it suffices to show that as $||Q_{\widetilde{Z}X}\mu||_2 \rightarrow \infty$, 
\begin{align*}
\mathbb{P} \{ AR_{\infty}(\iota_J) > \max_{g \in \textbf{G}_s} AR_{\infty}(g) \} \rightarrow 1.
\end{align*}
By Triangular inequality, we have
$AR_{\infty}(\iota_J) \geq ||Q_{\widetilde{Z}X}\mu||_{A_z} - O_P(1)$, 
and $$\max_{g \in \textbf{G}_s}AR_{\infty}(g) \leq \max_{g \in \textbf{G}_s}|\sum_{j \in [J]}g_j \xi_j a_j|  ||Q_{\widetilde{Z}X} \mu||_{A_z} + O_P(1).$$ 
In addition, $\max_{g \in \textbf{G}_s} |\sum_{j \in [J]} g_j \xi_j a_j|<1$ so that as $||Q_{\widetilde{Z}X}\mu||_2 \rightarrow \infty$, $||Q_{\widetilde{Z}X}\mu||_{A_z} \rightarrow \infty$ and 
\begin{align*}
||Q_{\widetilde{Z}X}\mu||_{A_z} - \max_{g \in \textbf{G}_s} |\sum_{j \in [J]} g_j \xi_j a_j| ||Q_{\widetilde{Z}X}\mu||_{A_z} \rightarrow \infty. 
\end{align*}
This concludes the proof. $\blacksquare$




\section{Further Results for the Weak-IV-Robust Statistics}\label{sec: other-weak}

First, we show that in the specific case with one endogenous regressor and one IV, if the wild bootstrap procedure for the $AR_n$ statistic is applied to the unstudentized Wald statistic $T_n$, then the resulting test will be asymptotically equivalent to the $AR_n$-based bootstrap test, both under the null and the alternative.
More precisely, for $T_n = | \hat{\beta} - \beta_0 |$, where $\hat{\beta}$ is the TSLS estimator, the wild bootstrap generates
\begin{equation}\label{eq: score-boot}
T^{s*}_{n}(g) = \left|\widehat{Q}_{\widetilde{Z}X}^{-1} \frac{1}{n} \sum_{j \in [J]} \sum_{i \in I_{n,j}} g_j \widetilde{Z}_{i,j} \hat{\eps}^r_{i,j}  \right|.
\end{equation}
Notice that in this case, the restricted TSLS estimator $\hat{\gamma}^r$ and the restricted OLS estimator $\bar{\gamma}^r$ are the same, which implies the residuals $\hat{\eps}^r_{i,j}$ and $\Bar{\eps}^r_{i,j}$ defined in Sections \ref{subsec: boot-wald} and \ref{subsec: boot-robust}, respectively, are the same. 
Let $\hat{c}^s_n(1-\alpha)$ denote the $(1-\alpha)$-th quantile of 
$\{ T^{s*}_n(g) \}_{g \in \textbf{G}}$. 
We show this equivalence below. 

\begin{theorem}\label{theo: AR_Wald_equivalent}
	Suppose that $d_x=d_z=1$, $\liminf_{n \rightarrow \infty}\mathbb{P}(\widehat{Q}_{\widetilde{Z}X} \neq 0) = 1$, $\hat{\beta}$ in the Wald test is computed via TSLS, and Assumption \ref{assumption: 1}(iv) holds. Then, 
	\begin{eqnarray*}
		\liminf_{n \rightarrow \infty}
		\mathbb{P} \{ \phi_n^s = \phi_n^{ar} \} =1,
	\end{eqnarray*}
	where $\phi_n^s = 1 \{ T_n > \hat{c}^s_n(1-\alpha) \}$ and $\phi_n^{ar} = 1 \{ AR_n > \hat{c}^{ar}_n(1-\alpha) \}$. 
\end{theorem}

Several remarks are in order.  
First, with one endogenous variable and one IV, the Jacobian matrix $\widehat{Q}_{\widetilde{Z}X}$ is just a scalar, which shows up in both the Wald statistic $T_n$ and the bootstrap critical value. After the cancellation of $\widehat{Q}_{\widetilde{Z}X}$, 
$T_n$ and its critical value are numerically the same as their AR counterparts, which leads to Theorem \ref{theo: AR_Wald_equivalent}. 
Second, for $\widehat{Q}_{\widetilde{Z}X}$ to be cancelled, we only need 
$\liminf_{n \rightarrow \infty}\mathbb{P} \{ \widehat{Q}_{\widetilde{Z}X} \neq 0 \} = 1$, which is very mild. It holds when $\widehat{Q}_{\widetilde ZX}$ is continuously distributed. 
Even when both $\widetilde{Z}_{i,j}$ and $X_{i,j}$ are discrete, it still holds if there exists at least one strong IV cluster, i.e., $Q_{\widetilde{Z}X} \neq 0$, where $Q_{\widetilde{Z}X}$ is the probability limit of $\widehat{Q}_{\widetilde{Z}X}$. 
When both $\widetilde{Z}_{i,j}$ and $X_{i,j}$ are discrete and $Q_{\widetilde{Z}X} = 0$, this condition still holds if some type of CLT holds such that $r_n \widehat{Q}_{\widetilde{Z}X} \convD N(c,\sigma^2)$, as $N(c,\sigma^2)$ is continuous. Third, the robustness of the $T_n$-based bootstrap test in (\ref{eq: score-boot}) does not carry over to the general case with multiple IVs as $T_n$ and its bootstrap statistic can no longer be reduced to their AR counterparts. 
Fourth, the robustness to weak IV cannot be extended to the $T_{CR,n}$-based bootstrap test, for which we have to further bootstrap the CRVE. 

Second, we discuss wild bootstrap inference with other weak-IV-robust statistics. 
To introduce the test statistics, 
we define the sample Jacobian as
\begin{align*}
\widehat{G}  
= \left( \widehat{G}_{1}, ..., \widehat{G}_{d_x} \right) \in \textbf{R}^{d_z \times d_x}, 
\quad
\widehat{G}_{l} & =  n^{-1} \sum_{j \in [J]} \sum_{i \in I_{n,j}} \widetilde{Z}_{i,j} X_{i,j,l},  \; \text{for} \; l=1, ..., d_x,
\end{align*}
and define the orthogonalized sample Jacobian as
\begin{align*}
\widehat{D} &=
\left( \widehat{D}_{1}, ..., \widehat{D}_{d_x} \right) \in \textbf{R}^{d_z \times d_x}, \; 
\quad
\widehat{D}_{l} =  \widehat{G}_{l} - \widehat{\Gamma}_{l}\widehat{\Omega}^{-1} 
\widehat{f} \in \textbf{R}^{d_z}, \;
\end{align*} 
where 
$\widehat{\Omega} = 
n^{-1}\sum_{j \in [J]}\sum_{i \in I_{n,j}}\sum_{k \in I_{n,j}}f_{i,j}f_{k,j}^{\top} $,
and $\widehat{\Gamma}_{l}  =  n^{-1} \sum_{j \in [J]} \sum_{i \in I_{n,j}} \sum_{k \in I_{n,j}}
\left( \widetilde{Z}_{i,j} X_{i,j,l} \right) f_{k,j}^\top,$ 
for $l=1, ..., d_x$.
Therefore, under the null $\beta_n = \beta_0$ 
and the framework where the number of clusters tends to infinity,
$\widehat{D}$
equals the sample Jacobian matrix $\widehat{G}$ adjusted to be asymptotically independent of $\widehat{f}$.   

Then, 
the cluster-robust version of \cite{Kleibergen(2002),Kleibergen(2005)}'s LM statistic is defined as
\begin{align*}
LM_n & = n \widehat{f}^{\top} \widehat{\Omega}^{-1/2}
P_{\widehat{\Omega}^{-1/2} \widehat{D}} \widehat{\Omega}^{-1/2} \widehat{f}.
\end{align*}
In addition, 
the conditional quasi-likelihood ratio (CQLR) statistic in \cite{Kleibergen(2005)}, \cite{Newey-Windmeijer(2009)}, 
and \cite{Guggenberger-Ramalho-Smith(2012)} are adapted from \cite{Moreira(2003)}'s 
conditional likelihood ratio (CLR) test,
and its cluster-robust version takes the form 
\begin{align*}
LR_n & = 
\frac{1}{2} \left( AR_{CR,n} - rk_n 
+ \sqrt{\left(AR_{CR,n} - rk_n\right)^2 + 4LM_n \cdot rk_n}\right),\notag 
\end{align*}
where $rk_n$ 
is a conditioning statistic and 
we let $rk_n = n \widehat{D}^\top \widehat{\Omega}^{-1} \widehat{D}$.\footnote{This choice follows \cite{Newey-Windmeijer(2009)}.
	\cite{Kleibergen(2005)} uses alternative formula for $rk_n$, and 
	\cite{Andrews-Guggenberger(2019)} introduce alternative CQLR test statistic.} 

The wild bootstrap procedure for the LM and CQLR tests is as follows. 
We compute
\begin{align*}
\widehat{D}^*_g & = 
\left( \widehat{D}^*_{1,g}, ..., \widehat{D}^*_{d_x,g} \right),
\quad
\widehat{D}^*_{l,g}    = \widehat{G}_{l} - \widehat{\Gamma}^*_{l,g} 
\widehat{\Omega}^{-1} 
\widehat{f}^*_g, \;  \notag \\
\widehat{\Gamma}^*_{l,g} &  = n^{-1} \sum_{j \in [J]} \sum_{i \in I_{n,j}} \sum_{k \in I_{n,j}}
\left(\widetilde{Z}_{i,j}X_{i,j,l} \right) f^*_{k,j}(g_j)^{\top}, \; l=1, ..., d_x,
\end{align*}
for any $g = (g_1, ..., g_q) \in \textbf{G}$, 
where the definition of $\widehat{f}_g^*$ and $f^*_{k,j}(g_j)$ is the same as that in Section \ref{subsec: boot-robust}.
Then, we compute the bootstrap analogues of the test statistics as
\begin{align*}
LM^*_n(g) &    = 
n (\widehat{f}^*_g)^\top \widehat{\Omega}^{-1/2}
P_{\widehat{\Omega}^{-1/2} \widehat{D}^*_g} \widehat{\Omega}^{-1/2} \widehat{f}^*_g, \notag \\
LR^*_n(g) & = 
\frac{1}{2} \left( AR^*_{CR,n}(g) - rk_n 
+ \sqrt{\left(AR^*_{CR,n}(g) - rk_n\right)^2 + 4LM^*_n(g) \cdot rk_n}\right).
\end{align*}
Let $\hat{c}_{LM,n}(1-\alpha)$ and $\hat{c}_{LR,n}(1-\alpha)$ 
denote the $(1-\alpha)$-th quantile of $\{LM_{n}^*(g)\}_{g \in \textbf{G}}$ and $\{LR^*_{n}(g)\}_{g \in \textbf{G}}$, respectively.
We notice that with at least one strong cluster, 
\begin{align*}
LM_n \xrightarrow{\enskip d \enskip} \bigg\Vert 
\left( 
\widetilde{D}^{\top} \left( \sum_{j \in [J]} \xi_j \mathcal{Z}_{\eps,j} \mathcal{Z}_{\eps,j}^{\top} \right)^{-1} \widetilde{D}
\right)^{-1/2} 
\widetilde{D}^{\top} \left( \sum_{j \in [J]} \xi_j \mathcal{Z}_{\eps,j} \mathcal{Z}_{\eps,j}^{\top} \right)^{-1} 
\sum_{j \in [J]} \sqrt{\xi_j} \mathcal{Z}_{\eps,j} \bigg\Vert^2, \notag
\end{align*}
where $\widetilde{D} = \left( \widetilde{D}_1, ..., \widetilde{D}_{d_x}\right)$, and for 
$l=1, ..., d_x$,
$$
\widetilde{D}_l = Q_{\widetilde{Z}X} - \left\{ \sum_{j \in [J]} \left(\xi_j Q_{\widetilde{Z}X,j,l} \right) \left( \sqrt{\xi_j} \mathcal{Z}_{\eps,j} \right) \right\}
\left\{ \sum_{j \in [J]} \xi_j \mathcal{Z}_{\eps,j} 
\mathcal{Z}_{\eps,j}^{\top} \right\}^{-1}
\sum_{j \in [J]} \sqrt{\xi_j} \mathcal{Z}_{\eps,j}.$$
Although the limiting distribution is nonstandard, 
we are able to establish the validity results by connecting the bootstrap LM test with the randomization test and by showing the asymptotic equivalence of the bootstrap LM and CQLR tests in this case. 
We conjecture that similar results can also be established for other weak-IV-robust statistics proposed in the literature.  

\begin{theorem}\label{theo: boot-LM-CLR}
	Suppose Assumptions \ref{assumption: 1}, \ref{assumption: 2}(i), and \ref{assumption: 3} hold, $\beta_n=\beta_0$, and $q > d_z$,  then 
	\begin{eqnarray*}
		\alpha - \frac{1}{2^{J-1}} \leq \liminf_{n \rightarrow \infty} \mathbb{P} \{LM_n > \hat{c}_{LM,n}(1-\alpha) \}
		\leq \limsup_{n \rightarrow \infty} \mathbb{P} \{LM_n > \hat{c}_{LM,n}(1-\alpha) \} \leq \alpha + \frac{1}{2^{J-1}}; \\
		\alpha - \frac{1}{2^{J-1}} \leq \liminf_{n \rightarrow \infty} \mathbb{P} \{ LR_n > \hat{c}_{LR,n}(1-\alpha) \}
		\leq \limsup_{n \rightarrow \infty} \mathbb{P} \{ LR_n > \hat{c}_{LR,n}(1-\alpha) \} \leq \alpha + \frac{1}{2^{J-1}}. 
	\end{eqnarray*}
\end{theorem}

\section{Proof of Theorem \ref{theo: AR_Wald_equivalent}  }
\label{sec:AR_Wald_equivalent_pf}
Notice that when $d_x = d_z = 1$, $\widehat{Q}_{\widetilde{Z}X}$ is a scalar,
and the restricted TSLS estimator $\hat{\gamma}^r$ is equivalent to the restricted OLS estimator $\bar{\gamma}^r$, which is well defined by Assumption \ref{assumption: 1}(iv), 
so that $\hat{\eps}_{i,j}^r = \bar{\eps}_{i,j}^r$. 
Therefore, 
\begin{align*}
AR_n = \left| \frac{1}{n} \sum_{j \in [J]} \sum_{i \in I_{n,j}} \widetilde{Z}_{i,j} \bar{\eps}^r_{i,j}\right| 
= \left| \frac{1}{n} \sum_{j \in [J]} \sum_{i \in I_{n,j}} \widetilde{Z}_{i,j} \hat{\eps}^r_{i,j}\right|,
\end{align*}
and whenever $\widehat{Q}_{\widetilde{Z}X} \neq 0$,
\begin{align*}
T_n & = |(\hat{\beta}_{tsls} - \beta_n) + (\beta_n - \beta_0)| \\
& = \left| \widehat{Q}_{\widetilde{Z}X}^{-1}  \frac{1}{n} \sum_{j \in [J]} \sum_{i \in I_{n,j}} \left[\widetilde{Z}_{i,j} \eps_{i,j} - \widetilde{Z}_{i,j}X_{i,j}( \hat{\beta}^r_{tsls} - \beta_n )\right]\right| \\
& = \left| \widehat{Q}_{\widetilde{Z}X}^{-1}  \frac{1}{n} \sum_{j \in [J]} \sum_{i \in I_{n,j}} \left[\widetilde{Z}_{i,j} \eps_{i,j} - \widetilde{Z}_{i,j}X_{i,j}( \hat{\beta}^r_{tsls} - \beta_n ) - \widetilde{Z}_{i,j}W_{i,j}^\top( \hat{\gamma}^r_{tsls} - \gamma )\right]\right| \\
&= \left| \widehat{Q}_{\widetilde{Z}X}^{-1}  \frac{1}{n} \sum_{j \in [J]} \sum_{i \in I_{n,j}} \widetilde{Z}_{i,j} \hat{\eps}^r_{i,j} \right| = \left|\widehat{Q}^{-1}_{\widetilde{Z}X}\right| AR_n,
\end{align*}
by $\beta_0 = \hat{\beta}^r_{tsls}$, 
$\sum_{j \in [J]}\sum_{i \in I_{n,j}}\widetilde{Z}_{i,j}W_{i,j}^{\top}=0$,
and $\hat{\eps}^r_{i,j} = \eps_{i,j} - X_{i,j}^{\top}( \hat{\beta}^r_{tsls} - \beta_n ) - W_{i,j}^{\top}(\hat{\gamma}^r_{tsls} - \gamma)$. 

In addition, for the bootstrap statistics we have
\begin{align*}
AR^*_n(g) = \left| \frac{1}{n} \sum_{j \in [J]} \sum_{i \in I_{n,j}} g_j \widetilde{Z}_{i,j} \bar{\eps}^r_{i,j}\right|
= \left| \frac{1}{n} \sum_{j \in [J]} \sum_{i \in I_{n,j}} g_j \widetilde{Z}_{i,j} \hat{\eps}^r_{i,j}\right|    
\end{align*}
and whenever $\widehat{Q}_{\tilde{Z}X} \neq 0$,
\begin{align*}
T^{s*}_n(g) = \left| \widehat{Q}_{\widetilde{Z}X}^{-1} \frac{1}{n} \sum_{j \in [J]}  \sum_{i \in I_{n,j}} g_j \widetilde{Z}_{i,j} \hat{\eps}^r_{i,j}\right| = \left|\widehat{Q}_{\widetilde{Z}X}^{-1}\right|AR^*_n(g).
\end{align*}
Therefore, $1\{ T_n > \hat{c}^s_n(1-\alpha) \}$ is equal to 
$1\{ AR_n > \hat{c}_{AR,n}(1-\alpha) \}$ whenever $\widehat{Q}_{\widetilde{Z}X} \neq 0$. We conclude that 
$\liminf_{n \rightarrow \infty} \mathbb{P} \left\{ \phi_n^s = \phi_n^{ar} \right\} =1$ because  $\liminf_{n \rightarrow \infty} \mathbb{P} \left\{ \widehat{Q}_{\widetilde{Z}X} \neq 0 \right\}=1$. 
$\blacksquare$

\section{Proof of Theorem \ref{theo: boot-LM-CLR}}\label{sec: proof-other-weak}

The proof for the bootstrap LM test follows similar arguments as those for the studentized version of the bootstrap AR test.
Let $\mathbb{S} \equiv \textbf{R}^{d_z \times d_x} \times \otimes_{j \in [J]} \textbf{R}^{d_z}$, 
and write an element $s \in \mathbb{S}$ by $s = \left( \{ s_{1,j} : j \in [J] \}, \{ s_{2,j} : j \in [J] \} \right)$.
We identify any $(g_1, ..., g_q) = g \in \textbf{G} = \{-1, 1\}^J$ with an action on $s \in \mathbb{S}$ 
given by $gs = \left( \{ s_{1,j} : j \in [J] \}, \{ g_j s_{2,j} : j \in [J] \} \right)$.
We define the function $T_{LM}: \mathbb{S} \rightarrow \textbf{R}$ to be given by 
\begin{align}
T_{LM}(s) \equiv \Bigg\Vert \left( D(s)^{\top}  \left( \sum_{j \in [J]}  s_{2,j} s^{\top}_{2,j} \right)^{-1} D(s) \right)^{-1/2}
D(s)^{\top} \left( \sum_{j \in [J]}  s_{2,j} s^{\top}_{2,j} \right)^{-1}  \sum_{j \in [J]} s_{2,j} 
\Bigg\Vert^2,
\end{align}
for any $s \in \mathbb{S}$ such that $\sum_{j \in [J]}  s_{2,j} s_{2,j}^{\top}$ and 
$D(s)^{\top}  \left( \sum_{j \in [J]}  s_{2,j} s^{\top}_{2,j} \right)^{-1} D(s)$ are invertible
and set $T_{LM}(s)=0$ whenever one of the two is not invertible, 
where 
\begin{align}
D(s) &\equiv \left(D_1(s), ..., D_{d_x}(s) \right), \notag \\
D_l(s) &\equiv 
\sum_{j \in [J]} s_{1,j,l} - \left( \sum_{j \in [J]} s_{1,j,l} s_{2,j}^{\top} \right) \left(\sum_{j \in [J]} s_{2,j} s_{2,j}^{\top} \right)^{-1}
\sum_{j \in [J]} s_{2,j},
\end{align}
for $s_{1,j} = (s_{1,j,1}, ..., s_{1,j,d_x})$ and $l=1, ..., d_x$.

Furthermore, define the statistic $S_n$ as 
\begin{align}\label{eq: LM_S_n}
S_n \equiv 
\left( \left\{ \frac{1}{n} \sum_{i \in I_{n,j}} \tilde{Z}_{i,j} X_{i,j}^{\top} : j \in [J] \right\},  
\left\{ \frac{r_n}{n} \sum_{i \in I_{n,j}} \tilde{Z}_{i,j} \epsilon_{i,j} : j \in [J] \right\} \right),
\end{align}
Note that for $l = 1, ..., d_x$ and $j \in [J]$, by Assumptions \ref{assumption: 1}(iii) and \ref{assumption: 2}(i) we have
\begin{align}
\frac{1}{n} \sum_{i \in I_{n,j}} \tilde{Z}_{i,j} X_{i,j,l} 
\convP  \xi_j Q_{\tilde{Z} X, j, l},
\end{align}
where $Q_{\tilde{Z}X,j,l}$ denotes the $l$-th column of the $d_z \times d_x$-dimensional matrix $Q_{\tilde{Z}X,j}$. 
Then, by Assumptions \ref{assumption: 1}(ii) and \ref{assumption: 1}(iii), \ref{assumption: 3}(i) and \ref{assumption: 3}(ii),   
and the continuous mapping theorem we have 
\begin{align}\label{eq: LM_S}
S_n \xrightarrow{\enskip d \enskip}
\left( \left\{  \xi_j a_j Q_{\tilde{Z}X} : j \in [J] \right\} , \left\{ \mathcal{Z}_j : j \in [J]\right\} \right) \equiv S,
\end{align}
where $\xi_j > 0$ for all $j \in [J]$. 
Also notice that for $l=1, ..., d_x$, 
\begin{align}
\widehat{D}_l
&    = \sum_{j \in [J]} \left( \frac{1}{n} \sum_{i \in I_{n,j}} \tilde{Z}_{i,j} X_{i,j,l} \right)
- \left( \sum_{j \in [J]} \left( \frac{1}{n} \sum_{i \in I_{n,j}} \tilde{Z}_{i,j} X_{i,j,l} \right) 
\left( \frac{r_n}{n} \sum_{k \in I_{k,j}} \tilde{Z}_{k,j} \bar{\epsilon}^r_{k,j}\right)^\top \right) \notag \\
& \cdot \left(\sum_{j \in [J]} \left( \frac{r_n}{n} \sum_{i \in I_{i,j}} \tilde{Z}_{i,j} \bar{\epsilon}^r_{i,j}\right)   
\left( \frac{r_n}{n} \sum_{k \in I_{k,j}} \tilde{Z}_{k,j} \bar{\epsilon}^r_{k,j}\right)^\top  \right)^{-1} 
\frac{r_n}{n} \sum_{i \in I_{i,j}} \tilde{Z}_{i,j} \bar{\epsilon}^r_{i,j}, 
\end{align}
and $\frac{r_n}{n}\sum_{i \in I_{n,j}} \tilde{Z}_{i,j}\bar{\eps}_{i,j}^r = \frac{r_n}{n}\sum_{i \in I_{n,j}} \tilde{Z}_{i,j}\eps_{i,j} + o_P(1)$ 
by $\beta_n = \beta_0$ and Lemma \ref{lem:equiv2}. 
In addition, we set $A_n \in \textbf{R}$ to equal 
\begin{align}
A_n \equiv I \left\{\text{$\widehat{D}$ is of full rank value and $\widehat{\Omega}$ is invertible} \right\},
\end{align}
and we have 
\begin{align}\label{eq: P_A_n=1}
\liminf_{n \rightarrow \infty} \mathbb{P} \{ A_n = 1\} =1,
\end{align}
which holds because $\left\{ \mathcal{Z}_j : j \in [J] \right\}$ are independent and continuously distributed 
with covariance matrices that are of full rank, 
and $Q_{\tilde{Z}X,j}$ are of full column rank for all $j \in \mathcal{J}_s$, 
by Assumptions \ref{assumption: 1}(ii), \ref{assumption: 3}(i), and \ref{assumption: 3}(ii).  

It follows that whenever $A_n=1$, 
\begin{align}\label{eq: LM_n_LM^*_n_T_S_n_T_gS_n}
\left( LM_n, \{ LM^*_n(g) : g \in \textbf{G} \} \right) 
= \left( T_{LM}(S_n), \{ T_{LM} (g S_n) : g \in \textbf{G} \} \right) + o_P(1).
\end{align}
In what follows, we denote the ordered values of $\{ T_{LM}(gs): g \in \textbf{G} \}$ by 
\begin{align}
T^{(1)}_{LM}(s | \textbf{G}) \leq ... \leq T^{|\textbf{G}|}_{LM} (s | \textbf{G}).
\end{align}

Next, we have
\begin{align}\label{eq: LM_portmanteau}
& \limsup_{n \rightarrow \infty} \mathbb{P} \left\{ LM_n > \hat{c}_{LM,n}(1-\alpha) \right\} \notag \\
&   \leq  
\limsup_{n \rightarrow \infty} \mathbb{P} \left\{ LM_n \geq \hat{c}_{LM,n}(1-\alpha); A_n = 1 \right\} \notag \\
& \leq 
\mathbb{P} \left\{ T_{LM}(S) \geq \inf \left\{ u \in \textbf{R}: \frac{1}{|\textbf{G}|} \sum_{g \in \textbf{G}} I \{ T_{LM}(gS) \leq u \} \geq 1-\alpha  \right\} \right\}, 
\end{align}
where the final inequality follows from (\ref{eq: LM_S_n}), (\ref{eq: LM_S}), (\ref{eq: P_A_n=1}), (\ref{eq: LM_n_LM^*_n_T_S_n_T_gS_n}), the continuous mapping theorem 
and Portmanteau's theorem. 
Therefore, setting $k^* \equiv \lceil |\textbf{G}|(1-\alpha)\rceil$, we can obtain from (\ref{eq: LM_portmanteau}) that 
\begin{align}\label{eq: LM-studentized-portmanteau}
& 
\limsup_{n \rightarrow \infty} \mathbb{P} \left\{ LM_n > \hat{c}_{LM,n}(1-\alpha) \right\} \notag \\
& \leq 
\mathbb{P} \left\{ T_{LM}(S) > T_{LM}^{(k^*)}(S | \textbf{G} ) \right\} + \mathbb{P} \left\{ T_{LM}(S) = T_{LM}^{(k^*)} (S | \textbf{G}) \right\} \notag \\
& \leq
\alpha +  \mathbb{P} \left\{ T_{LM}(S) = T_{LM}^{(k^*)} (S | \textbf{G}) \right\},
\end{align}
where the final inequality follows by $gS \overset{d}{=} S$ for all $g \in \textbf{G}$ and the properties of randomization tests. 
Then, 
we notice that for all $g \in \textbf{G}$, $T_{LM}(gS)=T_{LM}(-gS)$ with probability 1, and $P\{ T_{LM}(gS) = T_{LM}(\tilde{g}S) \}=0$ for $\tilde{g} \notin \{g, -g\}$. 
Therefore, 
\begin{align}\label{eq: LM_1/2^{J-1}}
\mathbb{P} \left\{ T_{LM}(S) = T_{LM}^{(k^*)} (S | \textbf{G}) \right\} = \frac{1}{2^{J-1}}.
\end{align}
The claim of the upper bound in the theorem then follows from (\ref{eq: LM-studentized-portmanteau}) 
and (\ref{eq: LM_1/2^{J-1}}).
The proof for the lower bound is similar to that for the bootstrap AR test, and thus is omitted.

To prove the result for the CQLR test, we note that
\begin{align}\label{eq: equivalence-LM-LR}
LR_n 
&    = \frac{1}{2} \left\{ AR_{CR,n} - rk_n + \sqrt{ (AR_{CR,n} - rk_n)^2 + 4 \cdot LM_n \cdot rk_n } \right\} \notag \\
&    = \frac{1}{2} \left\{ AR_{CR,n} - rk_n + \left| AR_{CR,n} - rk_n \right| \sqrt{ 1+ \frac{4 \cdot LM_n \cdot rk_n}{(AR_{CR,n} - rk_n)^2}}  \right\} \notag \\
&    = \frac{1}{2} \left\{ AR_{CR,n} - rk_n + \left| AR_{CR,n} - rk_n \right| \left( 1+ 2\cdot LM_n \frac{rk_n}{(AR_{CR,n} - rk_n)^2}
(1+o_P(1))  \right) \right\} \notag \\
&    = LM_n \frac{rk_n}{rk_n - AR_{CR,n}} (1+ o_P(1)) = LM_n + o_P(1), 
\end{align}
where the third equality follows from the mean value expansion $\sqrt{1+x} = 1+ (1/2)(x+o(1))$,
the fourth and last equalities follow from $AR_{CR,n} - rk_n <0$ w.p.a.1 since $AR_{CR,n} = O_P(1)$ while $rk_n \rightarrow \infty$ w.p.a.1 under  \hyperlink{assumption: 3}{Assumption \ref{assumption: 3}(i)}. 
Using arguments similar to those in (\ref{eq: equivalence-LM-LR}), we obtain that for each $g \in \textbf{G}$,
\begin{align}
LR^*_n(g) = LM^*_n(g) \frac{rk_n}{rk_n - AR^*_{CR,n}(g)} (1+ o_P(1)) =
LM^*_n(g) + o_P(1), 
\end{align}
by $AR^*_{CR,n}(g) - rk_n <0$ w.p.a.1 since $AR^*_{CR,n}(g) = O_P(1)$ for each $g \in \textbf{G}$. 
Then, it follows that whenever $A_n=1$, 
\begin{align}\label{eq: LR_n_LR^*_n_T_S_n_T_gS_n}
\left( LR_n, \{ LR^*_n(g) : g \in \textbf{G} \} \right) 
= \left( T_{LM}(S_n), \{ T_{LM} (g S_n) : g \in \textbf{G} \} \right) + o_P(1). 
\end{align}
Then, we obtain that 
\begin{align}\label{eq: LR_portmanteau}
& \limsup_{n \rightarrow \infty} \mathbb{P} \left\{ LR_n > \hat{c}_{LR,n}(1-\alpha) \right\} \notag \\
&   \leq  
\limsup_{n \rightarrow \infty} \mathbb{P} \left\{ LR_n \geq \hat{c}_{LR,n}(1-\alpha); A_n = 1 \right\} \notag \\
& \leq 
\mathbb{P} \left\{ T_{LM}(S) \geq \inf \left\{ u \in \textbf{R}: \frac{1}{|\textbf{G}|} \sum_{g \in \textbf{G}} I \{ T_{LM}(gS) \leq u \} \geq 1-\alpha  \right\} \right\}, 
\end{align}
where the second inequality follows from (\ref{eq: LM_S_n}), (\ref{eq: LM_S}), (\ref{eq: P_A_n=1}), (\ref{eq: LR_n_LR^*_n_T_S_n_T_gS_n}), the continuous mapping theorem and Portmanteau's theorem. 
Finally, the upper and lower bounds for the studentized bootstrap CQLR test follows from the previous arguments for the bootstrap LM test.
$\blacksquare$

\section{Further Monte Carlo Simulation Results}\label{sec: further-simu}

\subsection{Further Simulation Results for DGP 1}\label{subsec: further-simu-DGP1}

In this section, we report further simulation results for DGP 1. We notice that the overall pattern is very similar to that observed in Section \ref{sec: simu}.

First, we report the results for DGP 1 with the number of clusters $J$ equal to $9$ or $20$. The size results with $d_z=1$ and $d_z=3$ are reported in Tables \ref{tab:dgp1_size_1_9_20} and \ref{tab:dgp1_size_2_9_20}, respectively. 
In addition, we report the further power results in Figures \ref{fig:power-dgp1-wald-K1-J920} - \ref{fig:power-dgp1-ar-K1-J920} for $d_z=1$ and $J=9$ or $20$, 
Figures \ref{fig:power-dgp1-wald-K3-J612} - \ref{fig:power-dgp1-ar-K3-J612} for $d_z=3$ and $J=6$ or $12$, 
and Figures \ref{fig:power-dgp1-wald-K3-J920} - \ref{fig:power-dgp1-ar-K3-J920} for $d_z=3$ and $J=9$ or $20$, respectively. For the power results, we vary the value of $\beta_0-\beta$ from $-3$ to $3$ with the true value of $\beta$ equal to 1 throughout the simulations. $\rho_{\eps v}$ is set equal to $0.5$.

\begin{figure}[H] 
	\vspace{-2cm}
	\makebox[\textwidth]{\includegraphics[width=1\paperwidth,height=0.5\textheight]{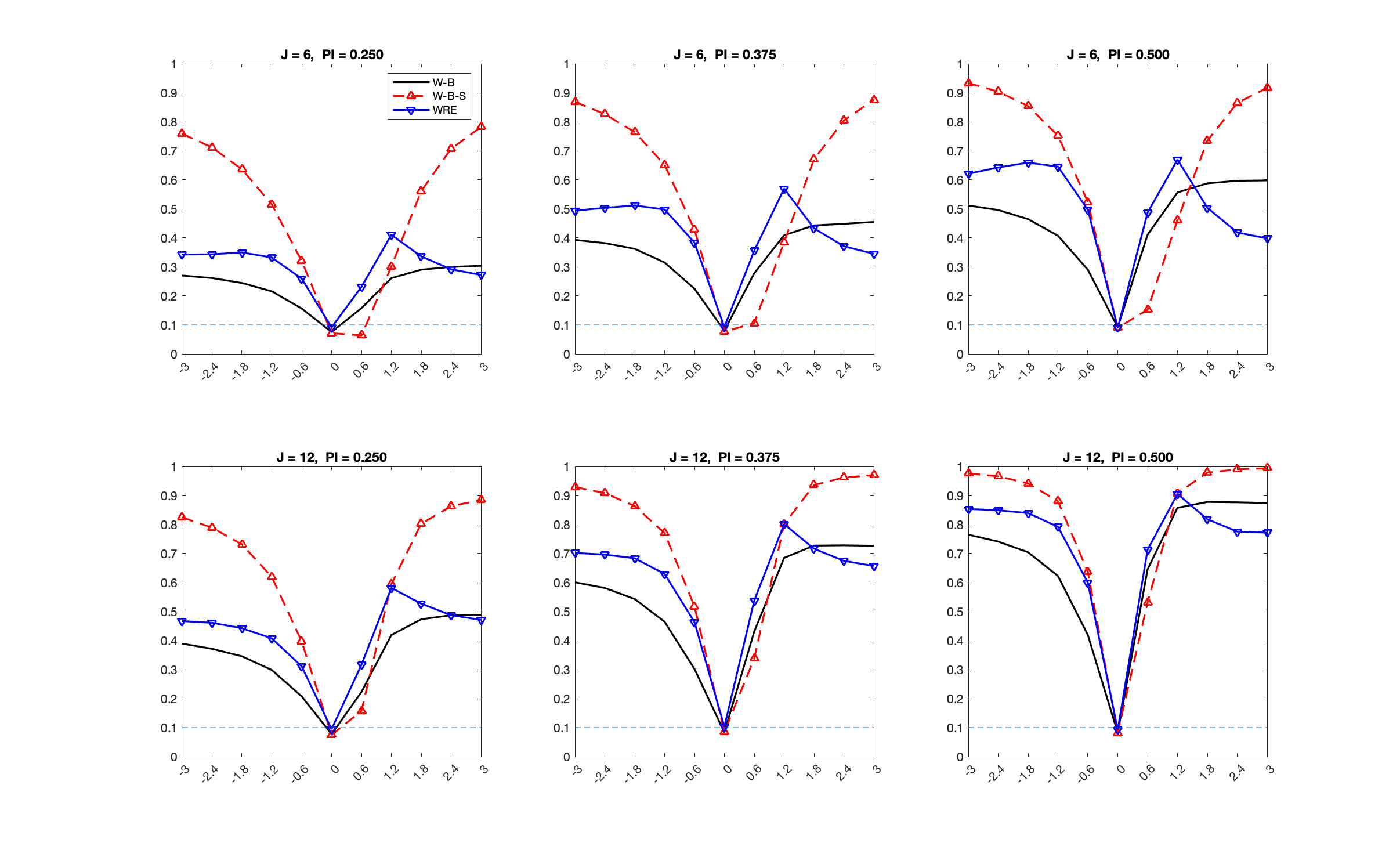}}
	\vspace*{-12mm}
	\caption{Power for Wald Tests under DGP 1 (homoskedastic $\Pi$) with $d_z=1$ and $J=6$ or $12$}
	\label{fig:power-dgp1-wald-K1-J612-homo}
	{\footnotesize{Note: W-B: dark solid line; W-B-S: red dashed line with upward-pointing triangle; WRE: blue solid line with downward-pointing triangle.}}
	
	\makebox[\textwidth]{\includegraphics[width=1\paperwidth,height=0.5\textheight]{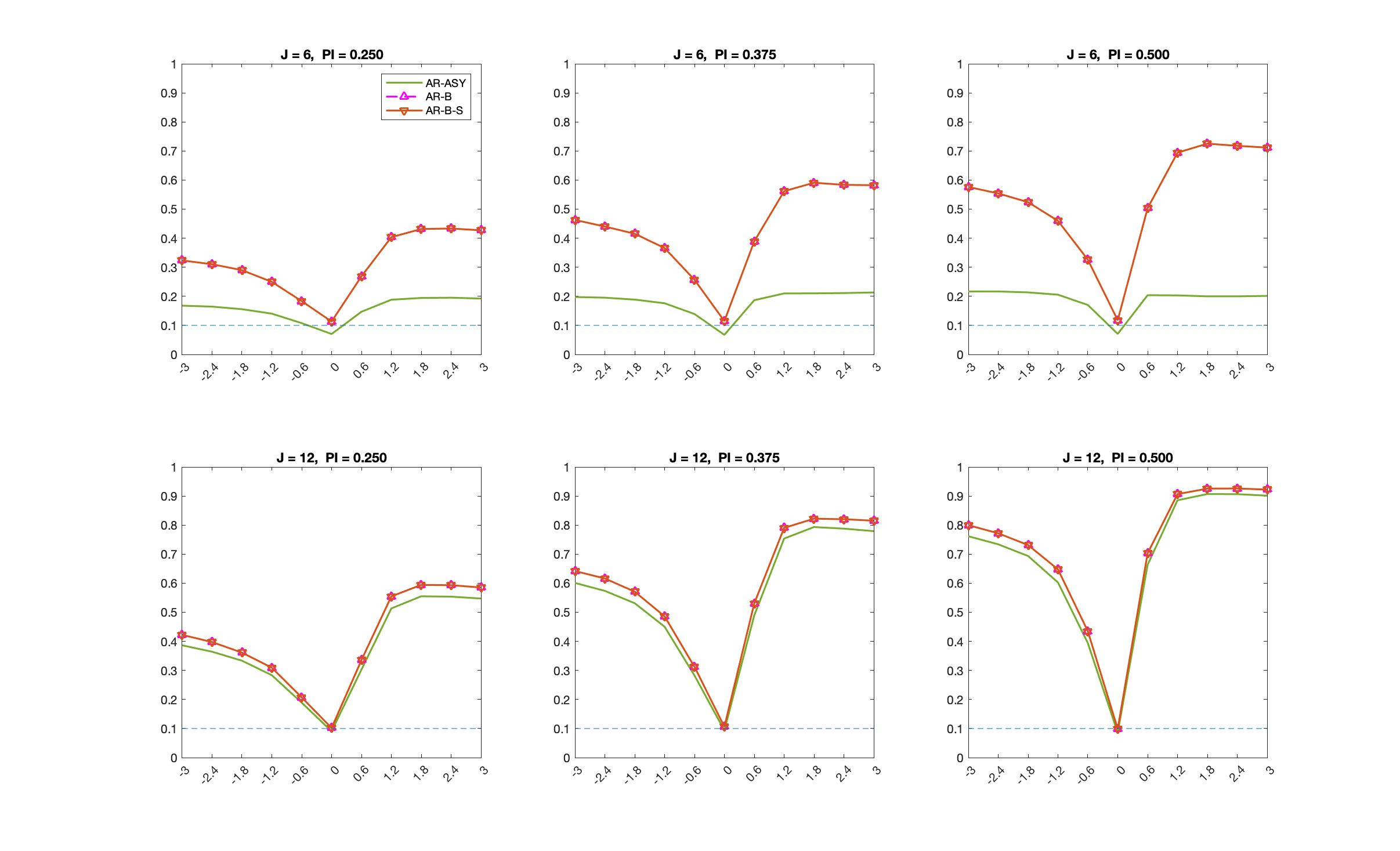}}
	\vspace*{-12mm}
	\caption{Power for AR Tests under DGP 1 (homoskedastic $\Pi$) with $d_z=1$ and $J=6$ or $12$}
	\label{fig:power-dgp1-ar-K1-J612-homo}
	{\footnotesize{Note: AR-ASY: dark green solid line; AR-B: magenta dashed line with upward-pointing triangle; AR-B-S: gold solid line with downward-pointing triangle. }}
\end{figure}

\begin{figure}[H] 
	\vspace{-2cm}
	\makebox[\textwidth]{\includegraphics[width=1\paperwidth,height=0.5\textheight]{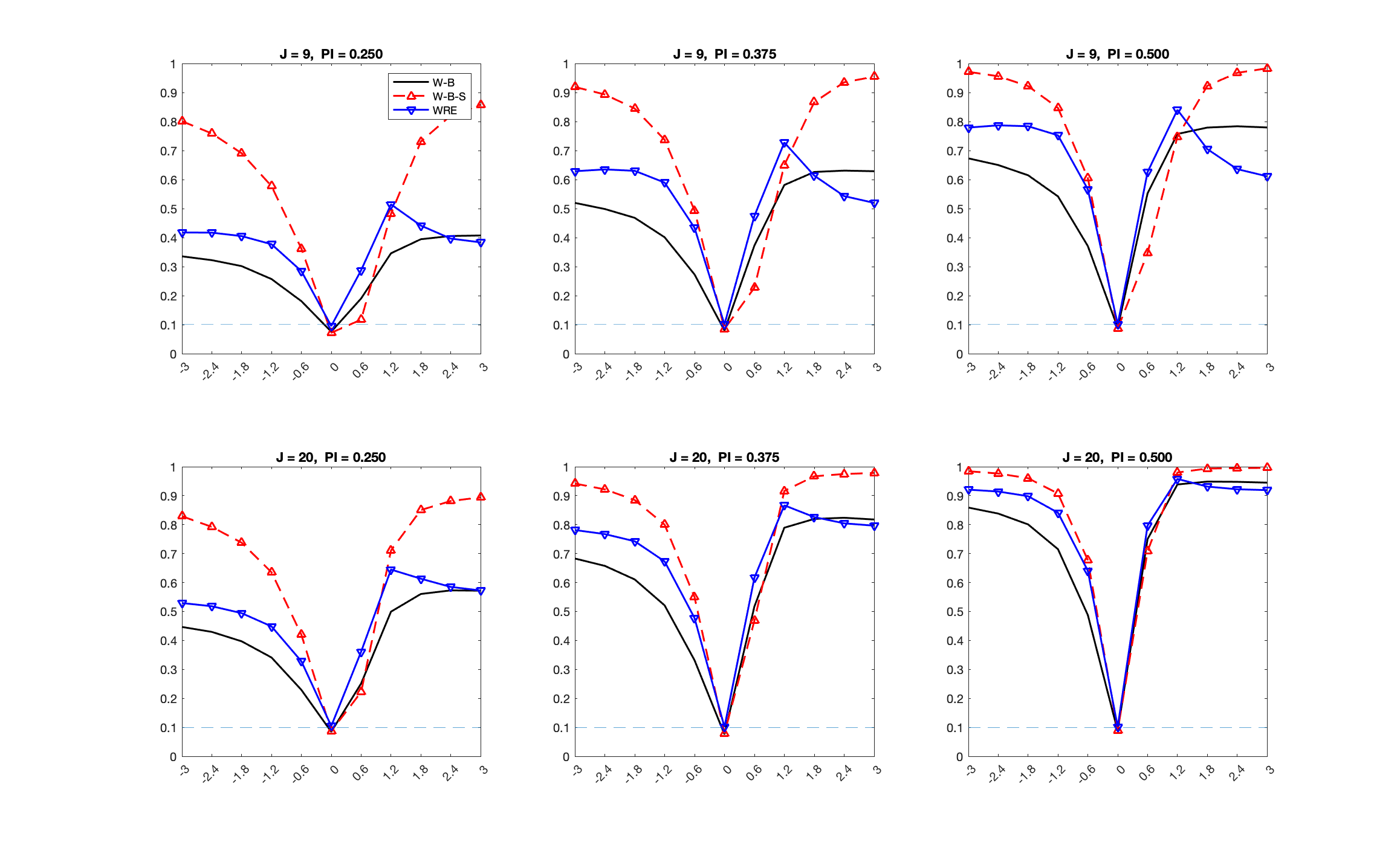}}
	\vspace*{-12mm}
	\caption{Power for Wald Tests under DGP 1 (homoskedastic $\Pi$) with $d_z=1$ and $J=9$ or $20$}
	\label{fig:power-dgp1-wald-K1-J920-homo}
	{\footnotesize{Note: W-B: dark solid line; W-B-S: red dashed line with upward-pointing triangle; WRE: blue solid line with downward-pointing triangle.}}
	
	\makebox[\textwidth]{\includegraphics[width=1\paperwidth,height=0.5\textheight]{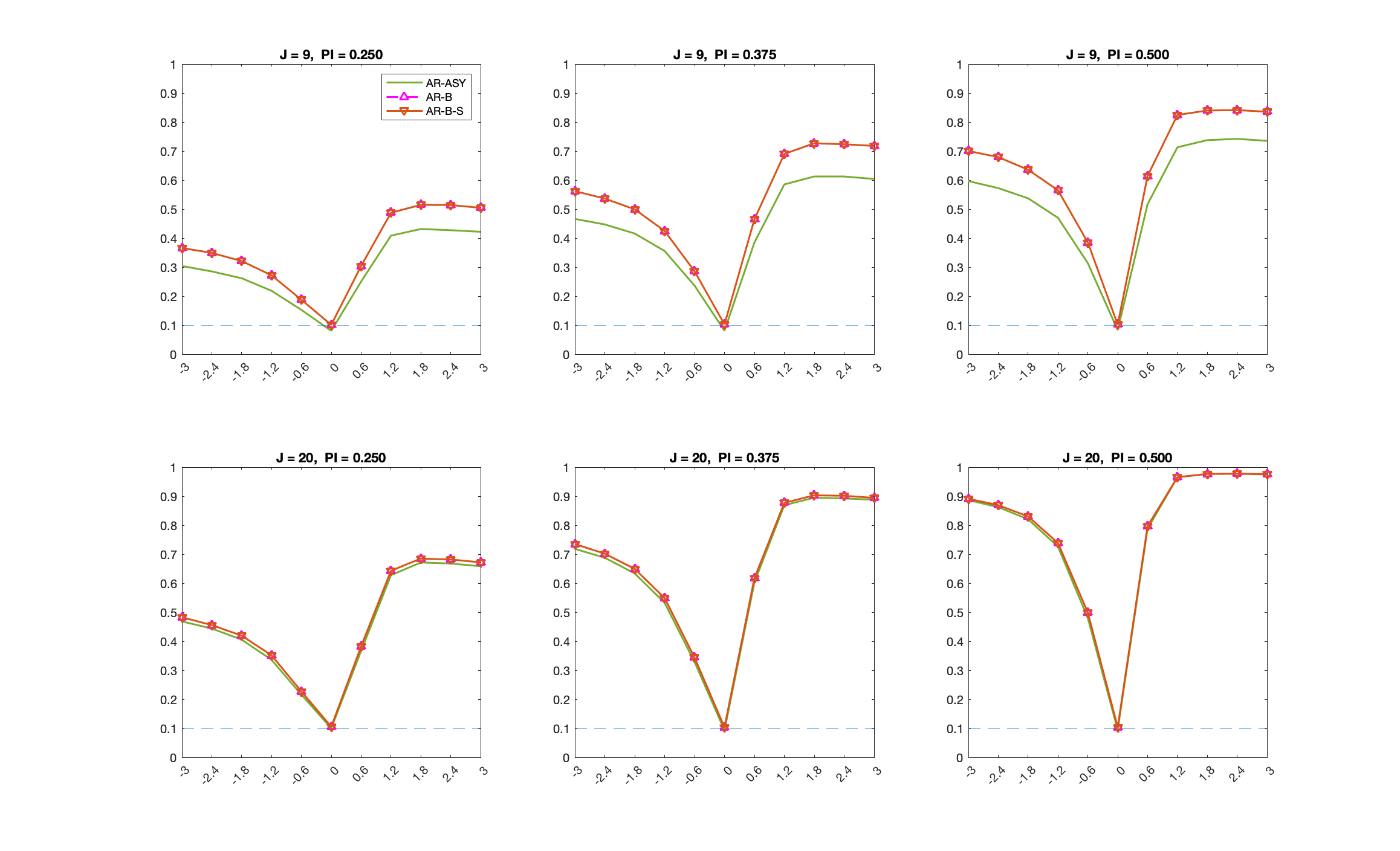}}
	\vspace*{-12mm}
	\caption{Power for AR Tests under DGP 1 (homoskedastic $\Pi$) with $d_z=1$ and $J=9$ or $20$}
	\label{fig:power-dgp1-ar-K1-J920-homo}
	{\footnotesize{Note: AR-ASY: dark green solid line; AR-B: magenta dashed line with upward-pointing triangle; AR-B-S: gold solid line with downward-pointing triangle. }}
\end{figure}

\begin{figure}[H] 
	\vspace{-2cm}
	\makebox[\textwidth]{\includegraphics[width=1\paperwidth,height=0.5\textheight]{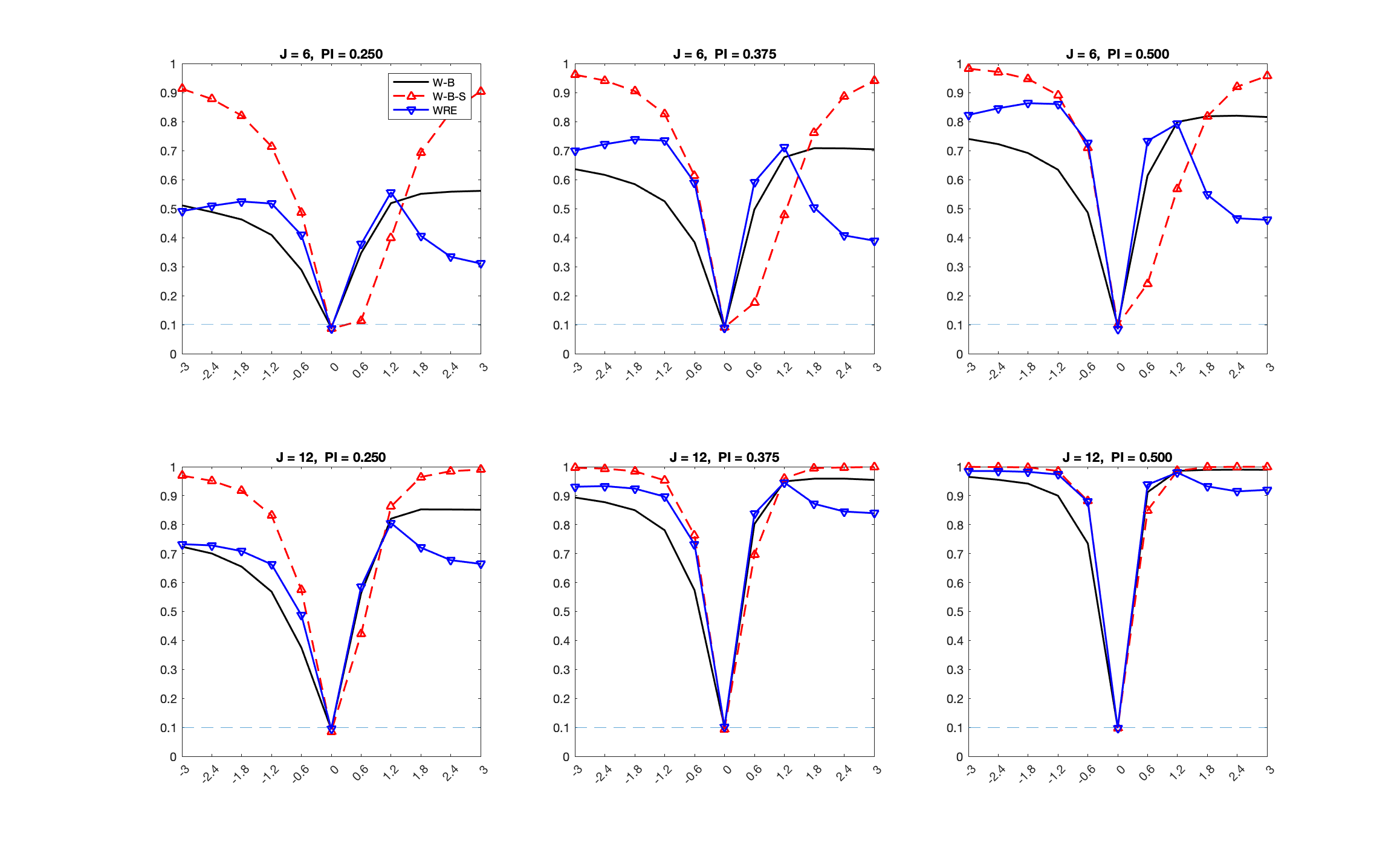}}
	\vspace*{-12mm}
	\caption{Power for Wald Tests under DGP 1 (homoskedastic $\Pi$) with $d_z=3$ and $J=6$ or $12$}
	\label{fig:power-dgp1-wald-K3-J612-homo}
	{\footnotesize{Note: W-B: dark solid line; W-B-S: red dashed line with upward-pointing triangle; WRE: blue solid line with downward-pointing triangle.}}
	
	\makebox[\textwidth]{\includegraphics[width=1\paperwidth,height=0.5\textheight]{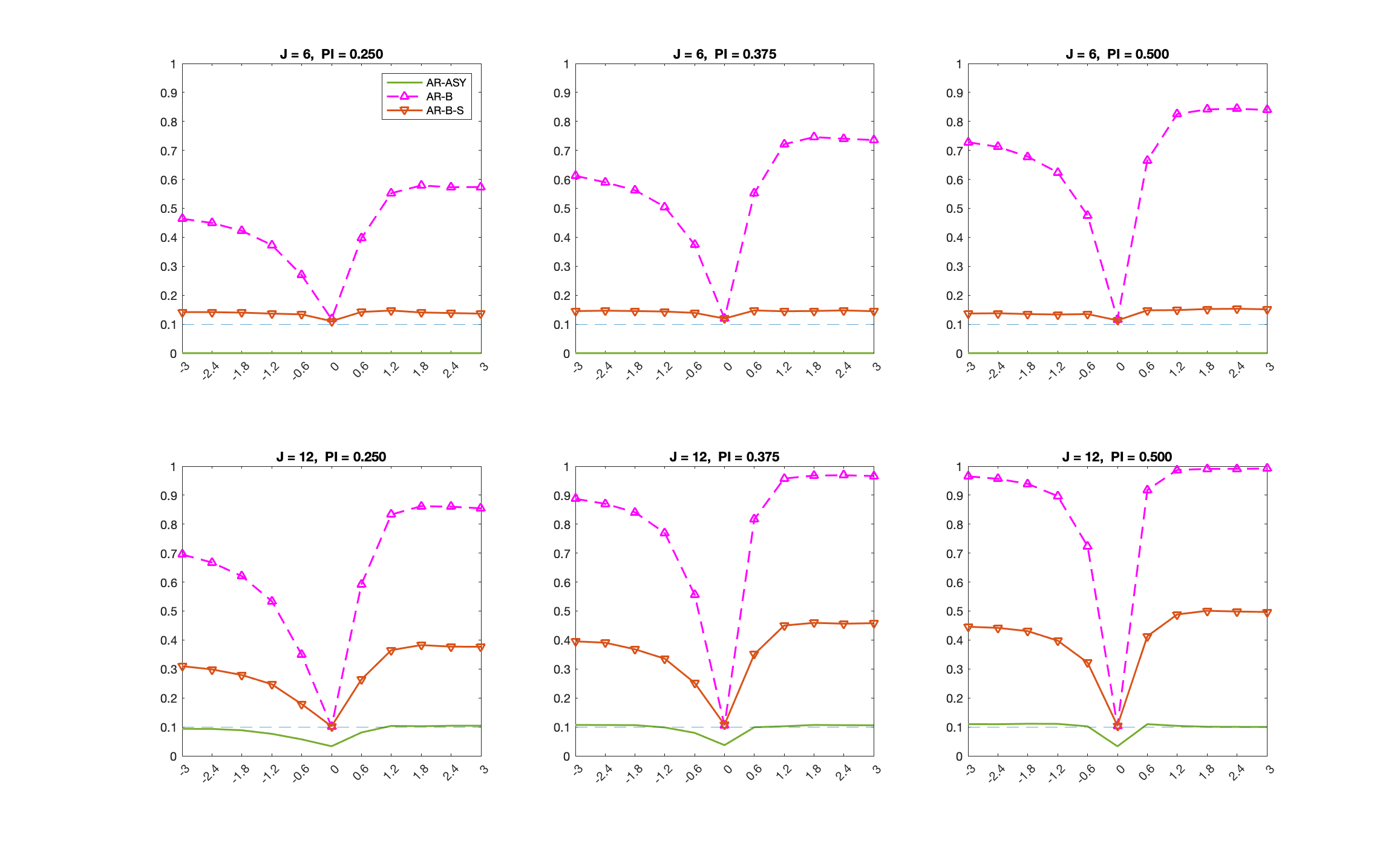}}
	\vspace*{-12mm}
	\caption{Power for AR Tests under DGP 1 (homoskedastic $\Pi$) with $d_z=3$ and $J=6$ or $12$}
	\label{fig:power-dgp1-ar-K3-J612-homo}
	{\footnotesize{Note: AR-ASY: dark green solid line; AR-B: magenta dashed line with upward-pointing triangle; AR-B-S: gold solid line with downward-pointing triangle. }}
\end{figure}

\begin{figure}[H] 
	\vspace{-2cm}
	\makebox[\textwidth]{\includegraphics[width=1\paperwidth,height=0.5\textheight]{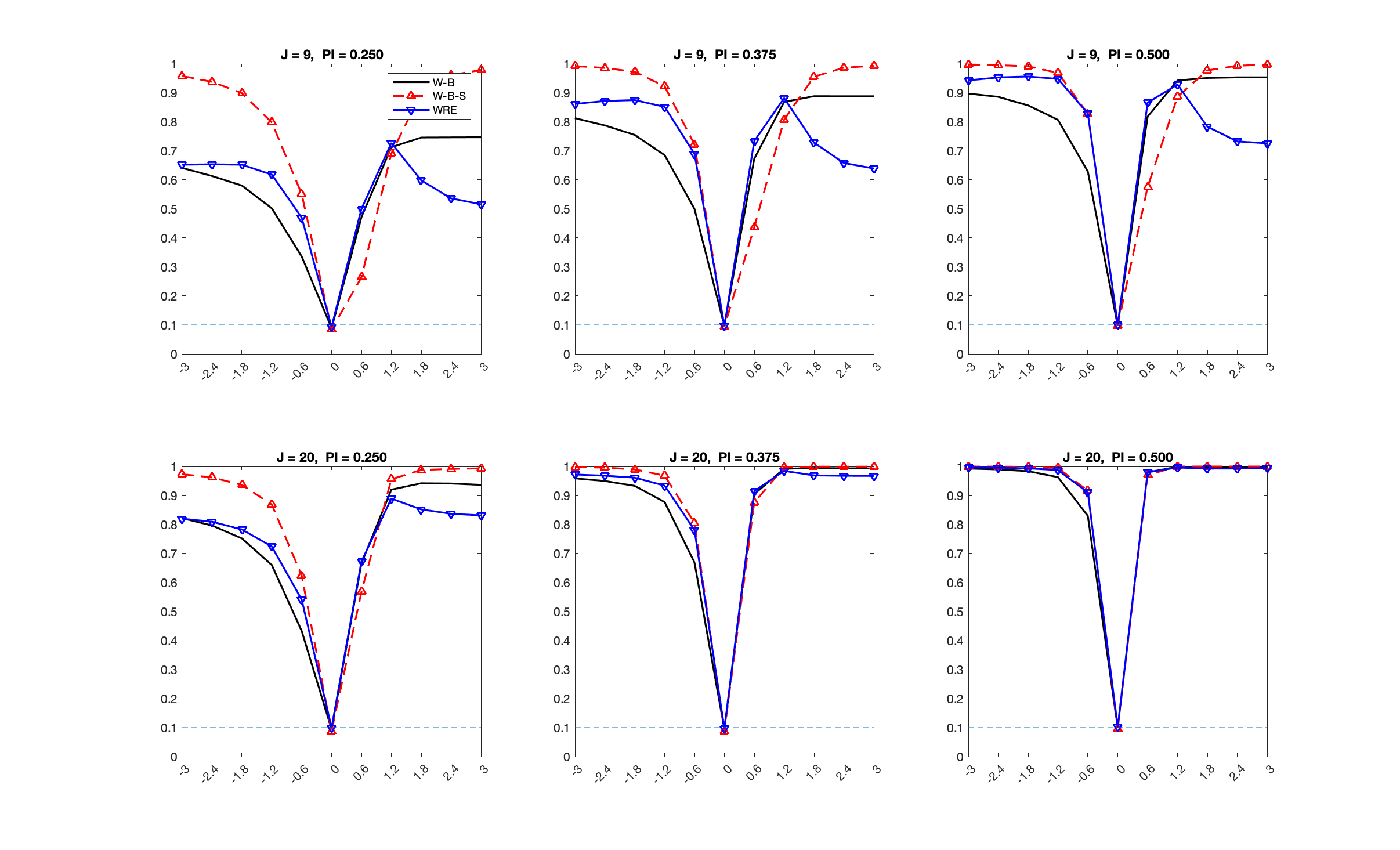}}
	\vspace*{-12mm}
	\caption{Power for Wald Tests under DGP 1 (homoskedastic $\Pi$) with $d_z=3$ and $J=9$ or $20$}
	\label{fig:power-dgp1-wald-K3-J920-homo}
	{\footnotesize{Note: W-B: dark solid line; W-B-S: red dashed line with upward-pointing triangle; WRE: blue solid line with downward-pointing triangle.}}
	
	\makebox[\textwidth]{\includegraphics[width=1\paperwidth,height=0.5\textheight]{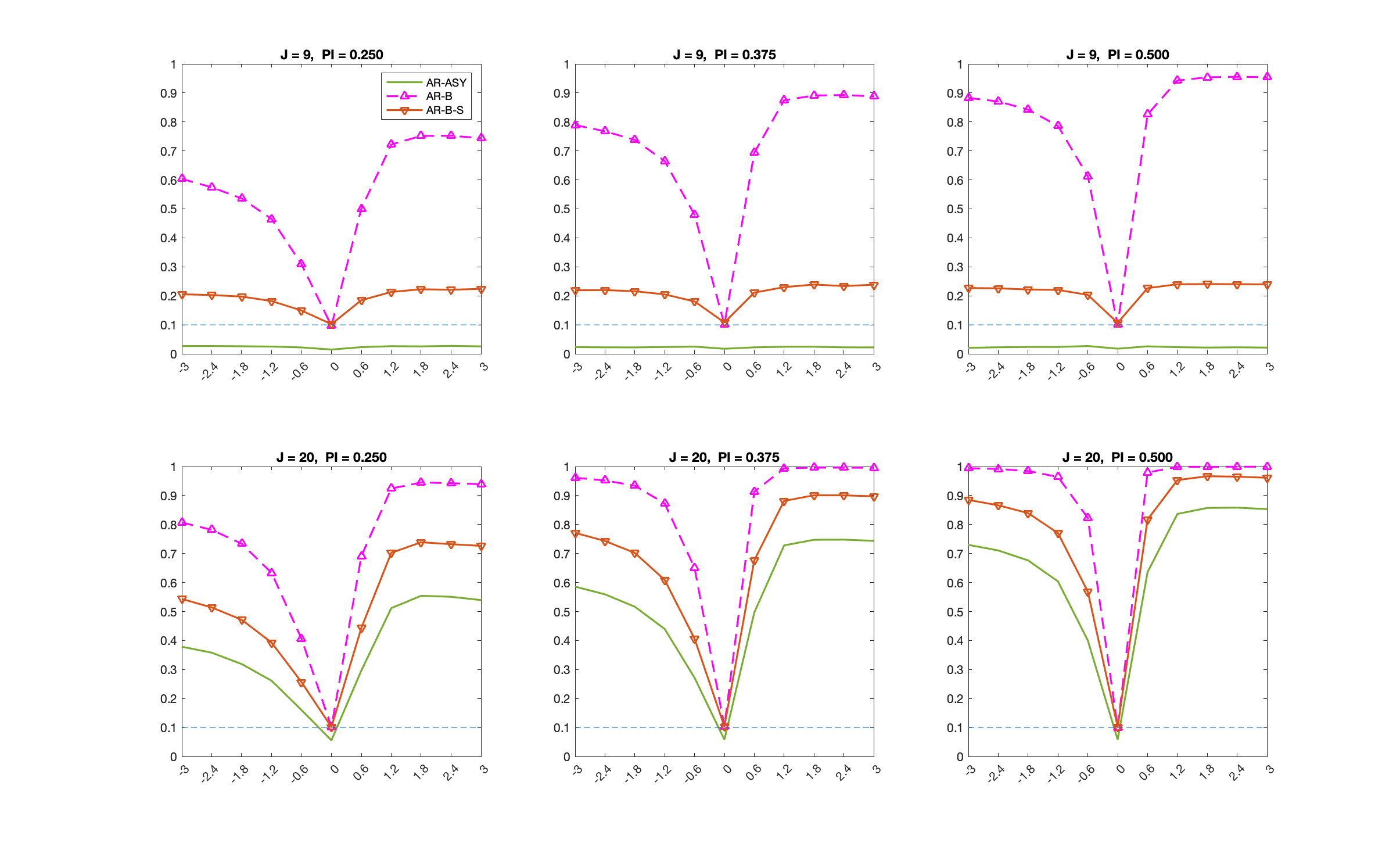}}
	\vspace*{-12mm}
	\caption{Power for AR Tests under DGP 1 (homoskedastic $\Pi$) with $d_z=3$ and $J=9$ or $20$}
	\label{fig:power-dgp1-ar-K3-J920-homo}
	{\footnotesize{Note: AR-ASY: dark green solid line; AR-B: magenta dashed line with upward-pointing triangle; AR-B-S: gold solid line with downward-pointing triangle.}}
\end{figure}

Second, we report the simulation results under DGP 1 for the case where $\Pi \in \{0.25, 0.375, 0.5\}$ remains the same across clusters (i.e., homoskedastic $\Pi$), so that the cluster-level heterogeneity in identification strength originates solely from the heterogeneity in cluster size. 
Other settings remain the same as those in DGP 1 in Section \ref{subsec:simu_design}. 
Tables \ref{tab:dgp1_size_1_homo_J612}-\ref{tab:dgp1_size_2_homo_J920} report the size properties of the ten inference methods with $d_z=1$ or $3$.
In addition, 
Figures \ref{fig:power-dgp1-wald-K1-J612-homo}-\ref{fig:power-dgp1-ar-K1-J920-homo} report the power properties of the Wald and AR tests, respectively, with $d_z=1$. 
Figures \ref{fig:power-dgp1-wald-K3-J612-homo}-\ref{fig:power-dgp1-ar-K3-J920-homo} report those for the Wald and AR tests, respectively, with $d_z=3$.

\begin{figure}[H] 
	\vspace{-2cm}
	\makebox[\textwidth]
	{\includegraphics[width=0.9\paperwidth,height=0.45\textheight]{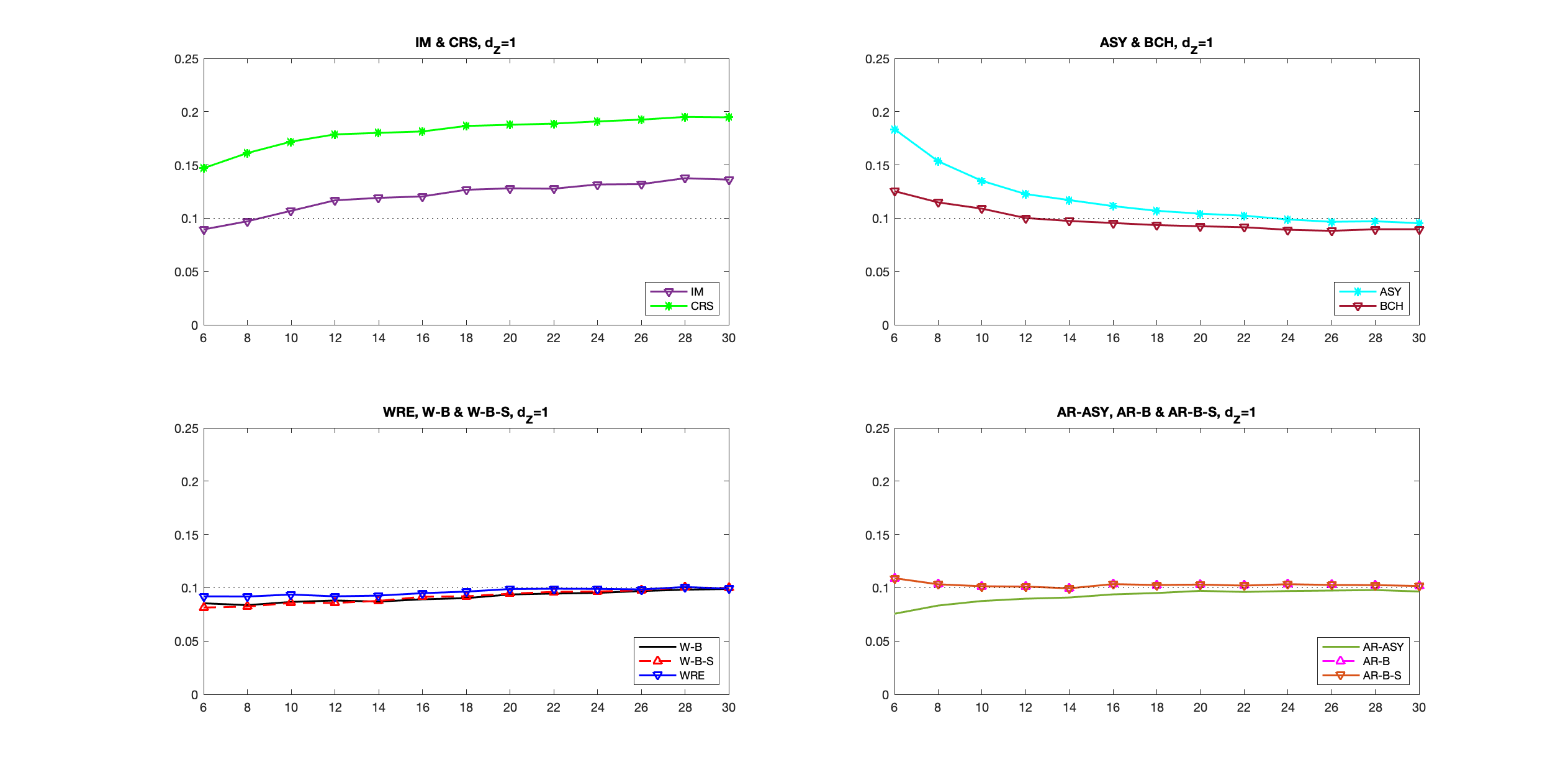}}
	\vspace*{-12mm}
	\caption{Size Comparison for DGP 1 with $d_z=1$ and $J$ varying from 6 to 30}
	\label{fig:size_DGP1_K1_G630}
	{\footnotesize{Note: IM: purple solid line with downward-pointing triangle; CRS: green solid line with asterisk; ASY: cyan solid line with asterisk; BCH: brown solid line with downward-pointing triangle; W-B: dark solid line; W-B-S: red dashed line with upward-pointing triangle; WRE: blue solid line with downward-pointing triangle; AR-ASY: dark green solid line; AR-B: magenta dashed line with upward-pointing triangle; AR-B-S: gold solid line with downward-pointing triangle.}}
\end{figure}

\begin{figure}[H] 
	\vspace{-0.3cm}
	\makebox[\textwidth]{\includegraphics[width=0.9\paperwidth,height=0.45\textheight]{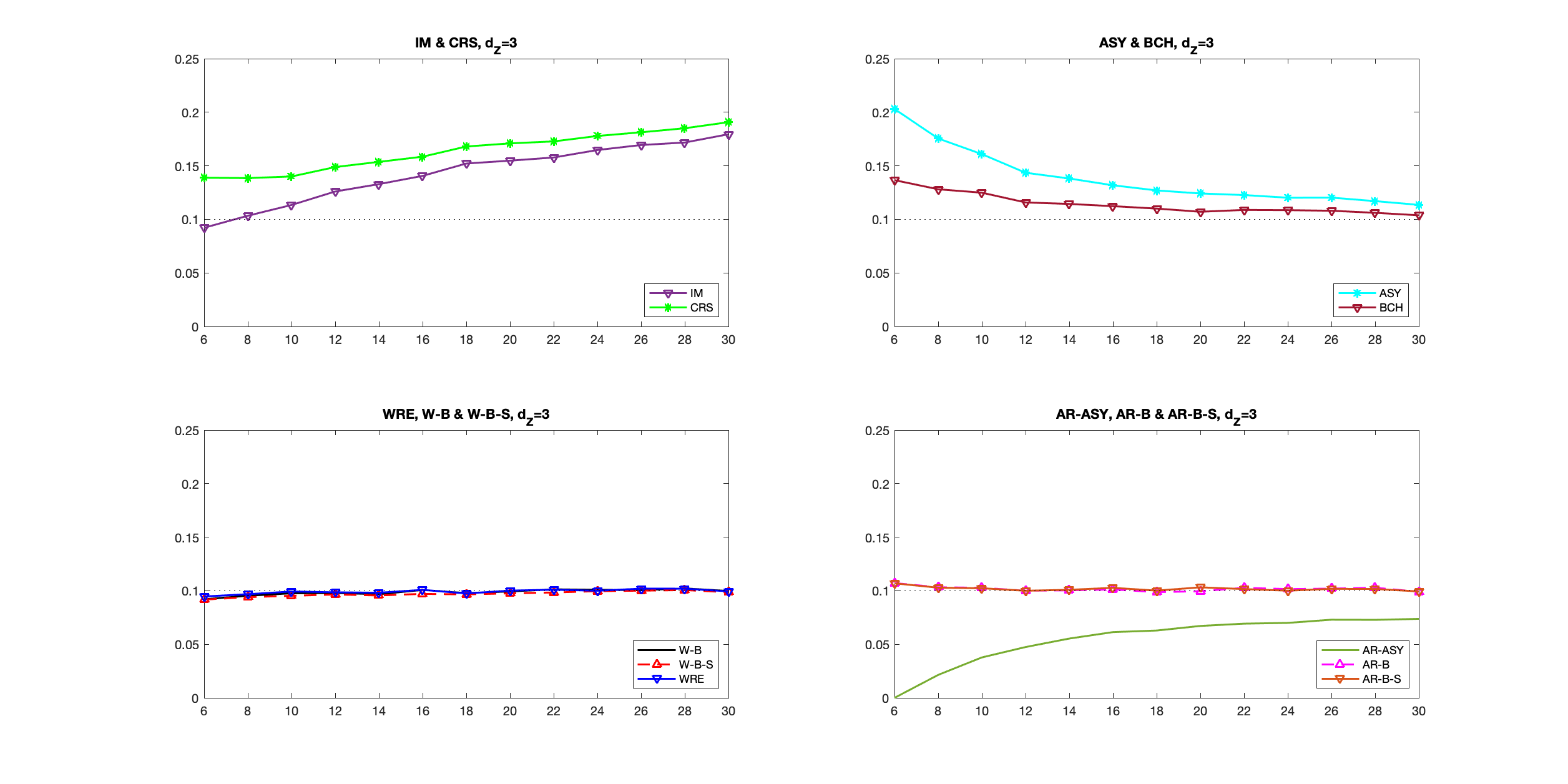}}
	\vspace*{-12mm}
	\caption{Size Comparison for DGP 1 with $d_z=3$ and $J$ varying from 6 to 30}
	\label{fig:size_DGP1_K3_G630}
	{\footnotesize{Note: IM: purple solid line with downward-pointing triangle; CRS: green solid line with asterisk; ASY: cyan solid line with asterisk; BCH: brown solid line with downward-pointing triangle; W-B: dark solid line; W-B-S: red dashed line with upward-pointing triangle; WRE: blue solid line with downward-pointing triangle; AR-ASY: dark green solid line; AR-B: magenta dashed line with upward-pointing triangle; AR-B-S: gold solid line with downward-pointing triangle.}}
\end{figure}

Furthermore, we investigate the relationship between the performance of alternative procedures and the number of clusters $J$. Specifically, we plot the size results as a function of $J$ (from $J=6$ to $30$) for $d_z=1$ in Figure \ref{fig:size_DGP1_K1_G630} and $d_z=3$ in Figure \ref{fig:size_DGP1_K3_G630}, respectively. 
In order to handle the cases with relatively large values of $J$ (e.g., $J=30$), we let the cluster heterogeneity parameter $r$ equal to 3 for $d_z=1$ and 2.4 for $d_z=3$, respectively, so that inference methods that are based on cluster-level estimators (IM and CRS) can also be implemented even with $J=30$.\footnote{As the total number of observations $n$ is equal to 500, the smallest and largest cluster sizes are 2 and 63 for $r=3$ (4 and 56 for $r=2.4$), respectively, under $J=30$.} 
We focus on DGP 1 in \ref{subsec:simu_design} of the main text 
with $\Pi=0.25$ and $\rho_{\epsilon v}=0.5$ throughout these simulations. 

We observe that similar to the findings in previous simulations, IM and CRS tests can have considerable over-rejections when the number of clusters is relatively large,  while ASY and BCH tests typically have large size distortions when the number of clusters is relatively small.
Among the seven Wald-based inference methods, only the three wild bootstrap tests (WRE, W-B, and W-B-S) have reasonable size control across different values of $J$. Although slightly lower than the nominal level when $J$ is small, their null rejection frequencies tend to become rather close to $10\%$ as the number of clusters becomes sufficiently large (e.g., when $J$ is larger than 20). 
We also notice that all the three AR-based inference methods (AR-ASY, AR-B, and AR-B-S) tend to control size. The two bootstrap AR tests over-reject slightly when $J$ is small, while the asymptotic AR test (AR-ASY) can be rather conservative in the over-identified case (Figure \ref{fig:size_DGP1_K3_G630}). 

\begin{figure}[H] 
	\makebox[\textwidth]{\includegraphics[width=0.9\paperwidth,height=0.3\textheight]{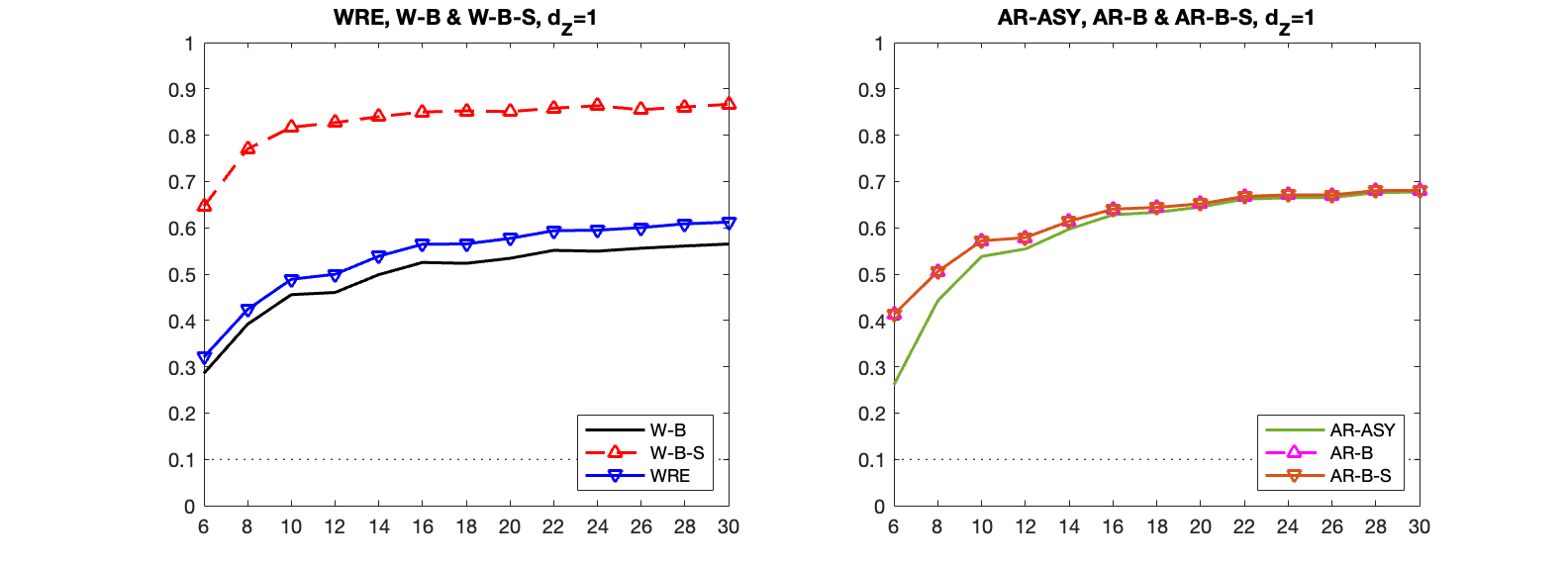}}
	\vspace*{-5mm}
	\caption{Power Comparison for DGP 1 with $d_z=1$, $J$ varying from 6 to 30, and $\beta_0 = 3$
	}
	\label{fig:power_DGP1_K1_G630_plus2}
	{\footnotesize{Note: W-B: dark solid line; W-B-S: red dashed line with upward-pointing triangle; WRE: blue solid line with downward-pointing triangle; AR-ASY: dark green solid line; AR-B: magenta dashed line with upward-pointing triangle; AR-B-S: gold solid line with downward-pointing triangle.}}
\end{figure}

\begin{figure}[H] 
	\vspace{-0.3cm}
	\makebox[\textwidth]{\includegraphics[width=0.9\paperwidth,height=0.3\textheight]{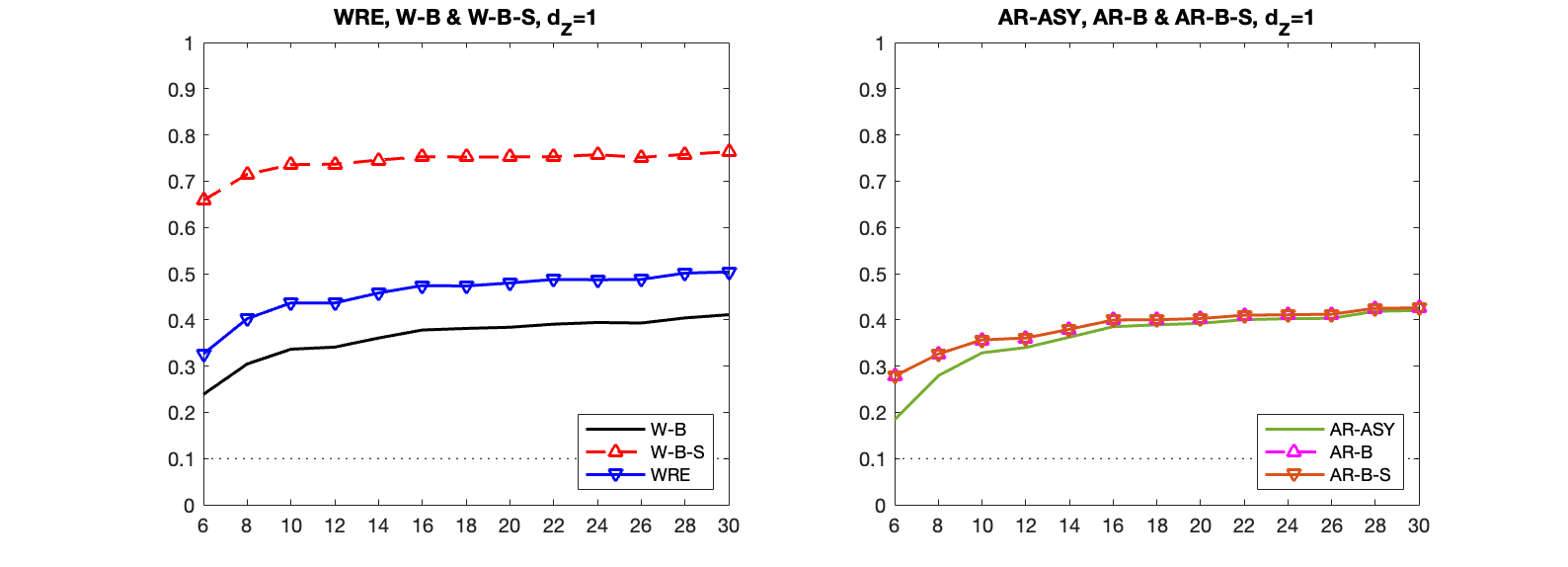}}
	\vspace*{-5mm}
	\caption{Power Comparison for DGP 1 with $d_z=1$, $J$ varying from 6 to 30, and $\beta_0 = -1$}
	\label{fig:power_DGP1_K1_G630_negative2}
	{\footnotesize{Note: W-B: dark solid line; W-B-S: red dashed line with upward-pointing triangle; WRE: blue solid line with downward-pointing triangle; AR-ASY: dark green solid line; AR-B: magenta dashed line with upward-pointing triangle; AR-B-S: gold solid line with downward-pointing triangle.}}
	
\end{figure}

\begin{figure}[H] 
	\makebox[\textwidth]{\includegraphics[width=0.9\paperwidth,height=0.3\textheight]{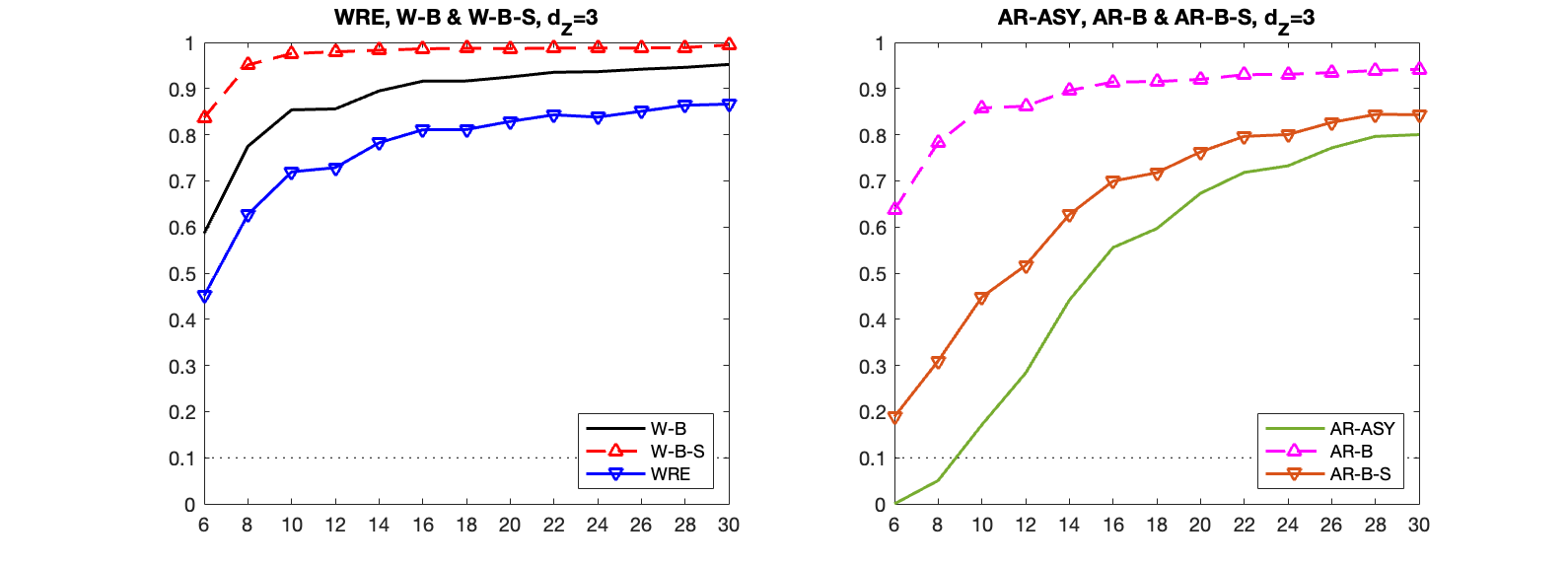}}
	\vspace*{-5mm}
	\caption{Power Comparison for DGP 1 with $d_z=3$, $J$ varying from 6 to 30, and $\beta_0=3$}
	\label{fig:power_DGP1_K3_G630_plus2}
	{\footnotesize{Note: W-B: dark solid line; W-B-S: red dashed line with upward-pointing triangle; WRE: blue solid line with downward-pointing triangle; AR-ASY: dark green solid line; AR-B: magenta dashed line with upward-pointing triangle; AR-B-S: gold solid line with downward-pointing triangle.}}
\end{figure}

\begin{figure}[H] 
	\makebox[\textwidth]{\includegraphics[width=0.9\paperwidth,height=0.3\textheight]{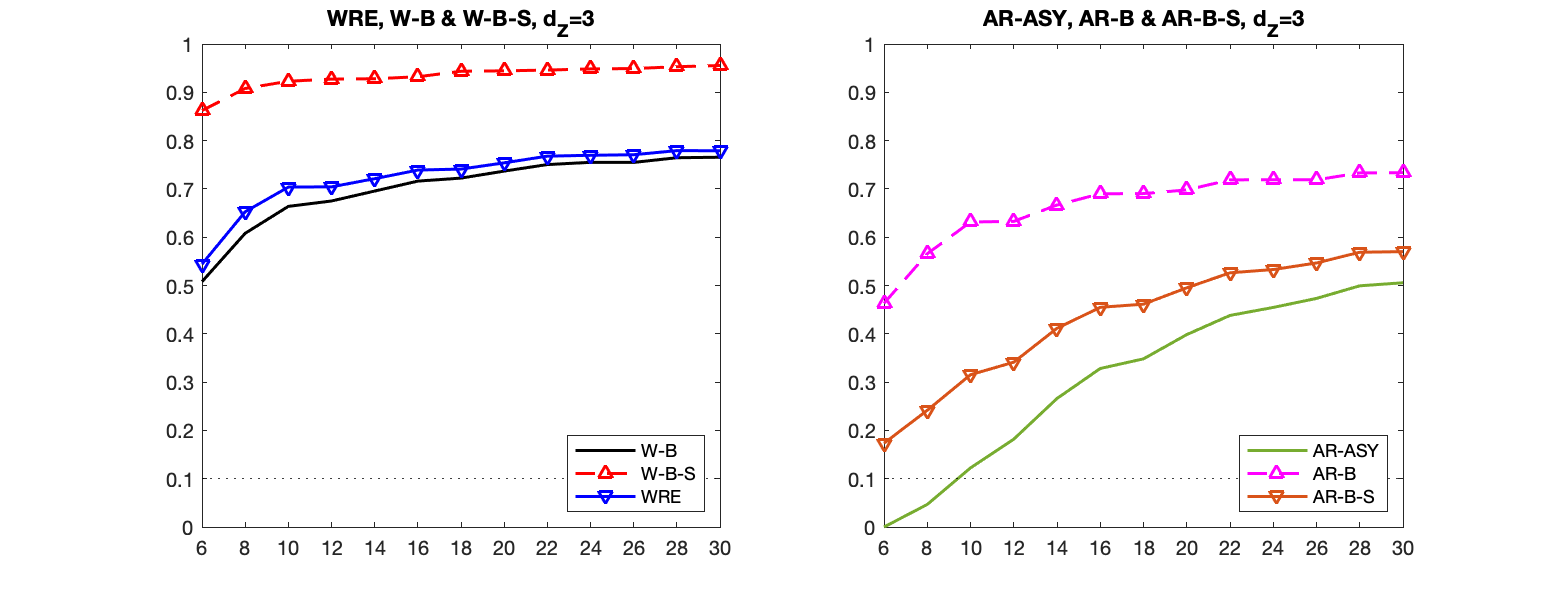}}
	\vspace*{-5mm}
	\caption{Power Comparison for DGP 1 with $d_z=3$, $J$ varying from 6 to 30, and $\beta_0=-1$}
	\label{fig:power_DGP1_K3_G630_negative2}
	{\footnotesize{Note: W-B: dark solid line; W-B-S: red dashed line with upward-pointing triangle; WRE: blue solid line with downward-pointing triangle; AR-ASY: dark green solid line; AR-B: magenta dashed line with upward-pointing triangle; AR-B-S: gold solid line with downward-pointing triangle.}}
\end{figure}

Finally, we investigate the power properties of alternative procedures as a function of $J$, from $J=6$ to $30$. Similar to previous simulations, we focus on the power performance of the inference methods that have good size control (namely, WRE, W-B, W-B-S, AR-B, AR-B-S, and AR-ASY). 
The true value of $\beta$ is set equal to 1 throughout the simulations. 
Figures \ref{fig:power_DGP1_K1_G630_plus2} and \ref{fig:power_DGP1_K1_G630_negative2} report the results for $d_z=1$ with $\beta_0 = 3$ and $-1$, respectively. In addition, Figures \ref{fig:power_DGP1_K3_G630_plus2} and \ref{fig:power_DGP1_K3_G630_negative2} report those for $d_z=3$ with $\beta_0 = 3$ and $-1$, respectively.
We highlight several observations below. 
First, the power of all the six inference methods tends to increase with the number of clusters. Second, our recommended procedure W-B-S has the highest power among the six methods across different values of $J$. Third, among the AR tests, the unstudentized bootstrap AR test (AR-B) has 
the best power performance while AR-ASY has the lowest power. The three observations are in line with our theoretical results in the main text.

\vspace{0.1in}

\begin{table}[h]
	\adjustbox{max width=\textwidth}{%
		\centering
		\begin{tabular}{ll|lll|lll|lll}
			&        &         & $\rho_{\eps v} = 0.3$  &        &         & $\rho_{\eps v} = 0.5$  &        &         & $\rho_{\eps v} = 0.7$  &        \\
			&        & $\Pi=0.25$ & $\Pi=0.375$ & $\Pi=0.5$ & $\Pi=0.25$ & $\Pi=0.375$ & $\Pi=0.5$ & $\Pi=0.25$ & $\Pi=0.375$ & $\Pi=0.5$ \\
			\hline
			$J=9$  & IM        & 0.071 & 0.069 & 0.065 & 0.103 & 0.094 & 0.084 & 0.161 & 0.136 & 0.126 \\
			& CRS       & 0.120 & 0.125 & 0.119 & 0.135 & 0.138 & 0.127 & 0.133 & 0.134 & 0.127 \\
			& ASY       & 0.157 & 0.178 & 0.201 & 0.168 & 0.188 & 0.189 & 0.177 & 0.193 & 0.192 \\
			& BCH       & 0.118 & 0.138 & 0.154 & 0.130 & 0.148 & 0.148 & 0.146 & 0.157 & 0.151 \\
			& WRE       & 0.098 & 0.094 & 0.097 & 0.097 & 0.102 & 0.092 & 0.091 & 0.099 & 0.094 \\
			& W-B & 0.080 & 0.081 & 0.089 & 0.077 & 0.080 & 0.085 & 0.076 & 0.079 & 0.080 \\
			& W-B-S   & 0.063 & 0.071 & 0.073 & 0.078 & 0.084 & 0.085 & 0.104 & 0.106 & 0.096 \\
			& AR-ASY    & 0.082 & 0.079 & 0.081 & 0.080 & 0.083 & 0.080 & 0.083 & 0.086 & 0.079 \\
			& AR-B & 0.099 & 0.099 & 0.101 & 0.100 & 0.103 & 0.098 & 0.103 & 0.106 & 0.097 \\
			& AR-B-S   & 0.099 & 0.099 & 0.101 & 0.100 & 0.103 & 0.098 & 0.103 & 0.106 & 0.097 \\
			\hline
			$J=20$ & IM        & 0.085 & 0.076 & 0.078 & 0.130 & 0.116 & 0.100 & 0.209 & 0.174 & 0.153 \\
			& CRS       & 0.131 & 0.127 & 0.124 & 0.142 & 0.134 & 0.132 & 0.141 & 0.136 & 0.135 \\
			& ASY       & 0.095 & 0.124 & 0.133 & 0.117 & 0.131 & 0.137 & 0.134 & 0.138 & 0.136 \\
			& BCH       & 0.083 & 0.107 & 0.117 & 0.107 & 0.117 & 0.122 & 0.123 & 0.123 & 0.121 \\
			& WRE       & 0.093 & 0.105 & 0.100 & 0.103 & 0.102 & 0.100 & 0.092 & 0.101 & 0.096 \\
			& W-B & 0.071 & 0.083 & 0.087 & 0.083 & 0.078 & 0.083 & 0.080 & 0.070 & 0.074 \\
			& W-B-S   & 0.065 & 0.075 & 0.078 & 0.081 & 0.086 & 0.089 & 0.105 & 0.102 & 0.094 \\
			& AR-ASY    & 0.090 & 0.096 & 0.093 & 0.097 & 0.097 & 0.091 & 0.093 & 0.093 & 0.087 \\
			& AR-B & 0.097 & 0.103 & 0.097 & 0.108 & 0.103 & 0.099 & 0.100 & 0.101 & 0.094 \\
			& AR-B-S   & 0.097 & 0.103 & 0.097 & 0.108 & 0.103 & 0.099 & 0.100 & 0.101 & 0.094
	\end{tabular}}
	\caption{Size Comparison for DGP 1 with $d_z=1$ and $J=9$ or $20$}
	\label{tab:dgp1_size_1_9_20}
	{\footnotesize{Note: IM, CRS, ASY, BCH, WRE, W-B, W-B-S, AR-ASY, AR-B, and AR-B-S denote the ten inference methods described in Section \ref{sec: inf-method}. $J$ denotes the number of clusters. The nominal level is 10\%. }}
\end{table}

\begin{table}[h]
	\adjustbox{max width=\textwidth}{%
		\centering
		\begin{tabular}{ll|lll|lll|lll}
			&        &         & $\rho_{\eps v} = 0.3$  &        &         & $\rho_{\eps v} = 0.5$  &        &         & $\rho_{\eps v} = 0.7$  &        \\
			&        & $\Pi=0.25$ & $\Pi=0.375$ & $\Pi=0.5$ & $\Pi=0.25$ & $\Pi=0.375$ & $\Pi=0.5$ & $\Pi=0.25$ & $\Pi=0.375$ & $\Pi=0.5$ \\
			\hline
			$J=9$ & IM     & 0.071 & 0.073 & 0.064 & 0.106 & 0.094 & 0.074 & 0.175 & 0.129 & 0.111 \\
			& CRS    & 0.103 & 0.105 & 0.101 & 0.141 & 0.132 & 0.114 & 0.213 & 0.169 & 0.155 \\
			& ASY    & 0.201 & 0.214 & 0.212 & 0.197 & 0.213 & 0.215 & 0.206 & 0.205 & 0.213 \\
			& BCH    & 0.157 & 0.165 & 0.167 & 0.153 & 0.166 & 0.168 & 0.163 & 0.163 & 0.168 \\
			& WRE    & 0.097 & 0.097 & 0.094 & 0.093 & 0.098 & 0.095 & 0.094 & 0.093 & 0.094 \\
			& W-B    & 0.094 & 0.096 & 0.095 & 0.093 & 0.093 & 0.094 & 0.090 & 0.092 & 0.093 \\
			& W-B-S  & 0.069 & 0.080 & 0.083 & 0.086 & 0.093 & 0.093 & 0.102 & 0.096 & 0.096 \\
			& AR-ASY & 0.016 & 0.014 & 0.017 & 0.020 & 0.017 & 0.016 & 0.016 & 0.015 & 0.019 \\
			& AR-B   & 0.104 & 0.109 & 0.098 & 0.104 & 0.100 & 0.101 & 0.102 & 0.101 & 0.101 \\
			& AR-B-S & 0.104 & 0.104 & 0.102 & 0.099 & 0.094 & 0.104 & 0.102 & 0.101 & 0.102 \\
			\hline
			$J=20$ & IM     & 0.097 & 0.090 & 0.086 & 0.156 & 0.130 & 0.110 & 0.286 & 0.213 & 0.176 \\
			& CRS    & 0.118 & 0.108 & 0.107 & 0.178 & 0.152 & 0.133 & 0.302 & 0.235 & 0.200 \\
			& ASY    & 0.142 & 0.152 & 0.150 & 0.147 & 0.150 & 0.151 & 0.144 & 0.149 & 0.157 \\
			& BCH    & 0.123 & 0.131 & 0.130 & 0.131 & 0.132 & 0.135 & 0.128 & 0.131 & 0.138 \\
			& WRE    & 0.097 & 0.100 & 0.095 & 0.104 & 0.095 & 0.100 & 0.092 & 0.095 & 0.100 \\
			& W-B    & 0.099 & 0.103 & 0.097 & 0.107 & 0.099 & 0.100 & 0.093 & 0.098 & 0.101 \\
			& W-B-S  & 0.079 & 0.091 & 0.090 & 0.097 & 0.091 & 0.098 & 0.099 & 0.098 & 0.103 \\
			& AR-ASY & 0.053 & 0.055 & 0.051 & 0.055 & 0.056 & 0.050 & 0.054 & 0.054 & 0.056 \\
			& AR-B   & 0.103 & 0.102 & 0.098 & 0.104 & 0.102 & 0.100 & 0.093 & 0.101 & 0.104 \\
			& AR-B-S & 0.100 & 0.103 & 0.097 & 0.102 & 0.098 & 0.097 & 0.097 & 0.102 & 0.101
	\end{tabular}}
	\caption{Size Comparison for DGP 1 with $d_z=3$ and $J=9$ or $20$}
	\label{tab:dgp1_size_2_9_20}
	{\footnotesize{Note: IM, CRS, ASY, BCH, WRE, W-B, W-B-S, AR-ASY, AR-B, and AR-B-S denote the ten inference methods described in Section \ref{sec: inf-method}. $J$ denotes the number of clusters. The nominal level is 10\%. }}
\end{table}

\subsection{Further Simulation Results for DGP 2}\label{subsec: further-simu-DGP2}

\begin{figure}[H] 
	\vspace{-2cm}
	\makebox[\textwidth]{\includegraphics[width=1\paperwidth,height=0.5\textheight]{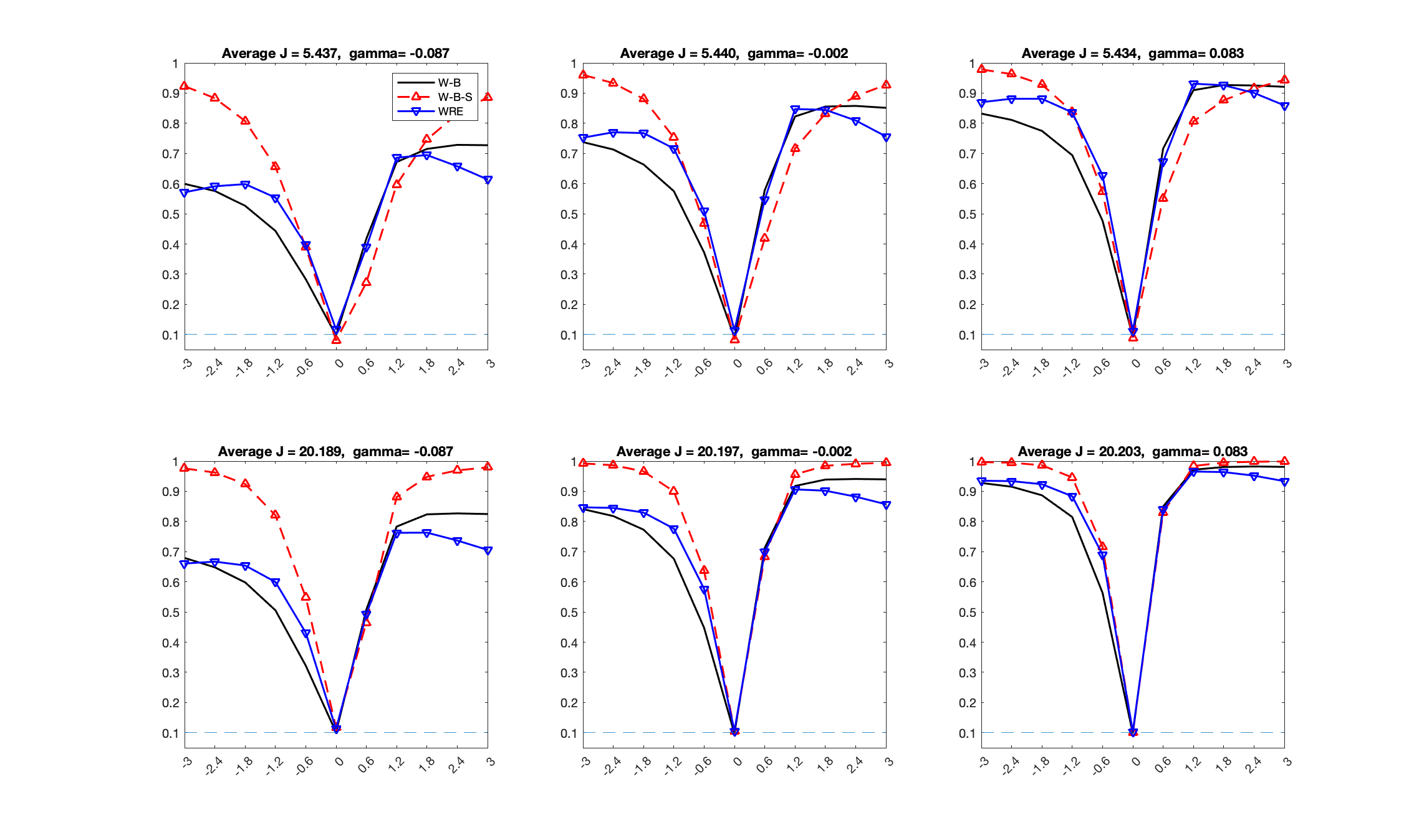}}
	\vspace*{-12mm}
	\caption{Power Comparison for Wald Tests in DGP 2 with Average $J = 5.437$ or $20.189$}
	\label{fig: power-dgp2-wald-J520}
	{\footnotesize{Note: 
			W-B: dark solid line; W-B-S: red dashed line with upward-pointing triangle; WRE: blue solid line with downward-pointing triangle. }}
	
	\makebox[\textwidth]{\includegraphics[width=1\paperwidth,height=0.5\textheight]{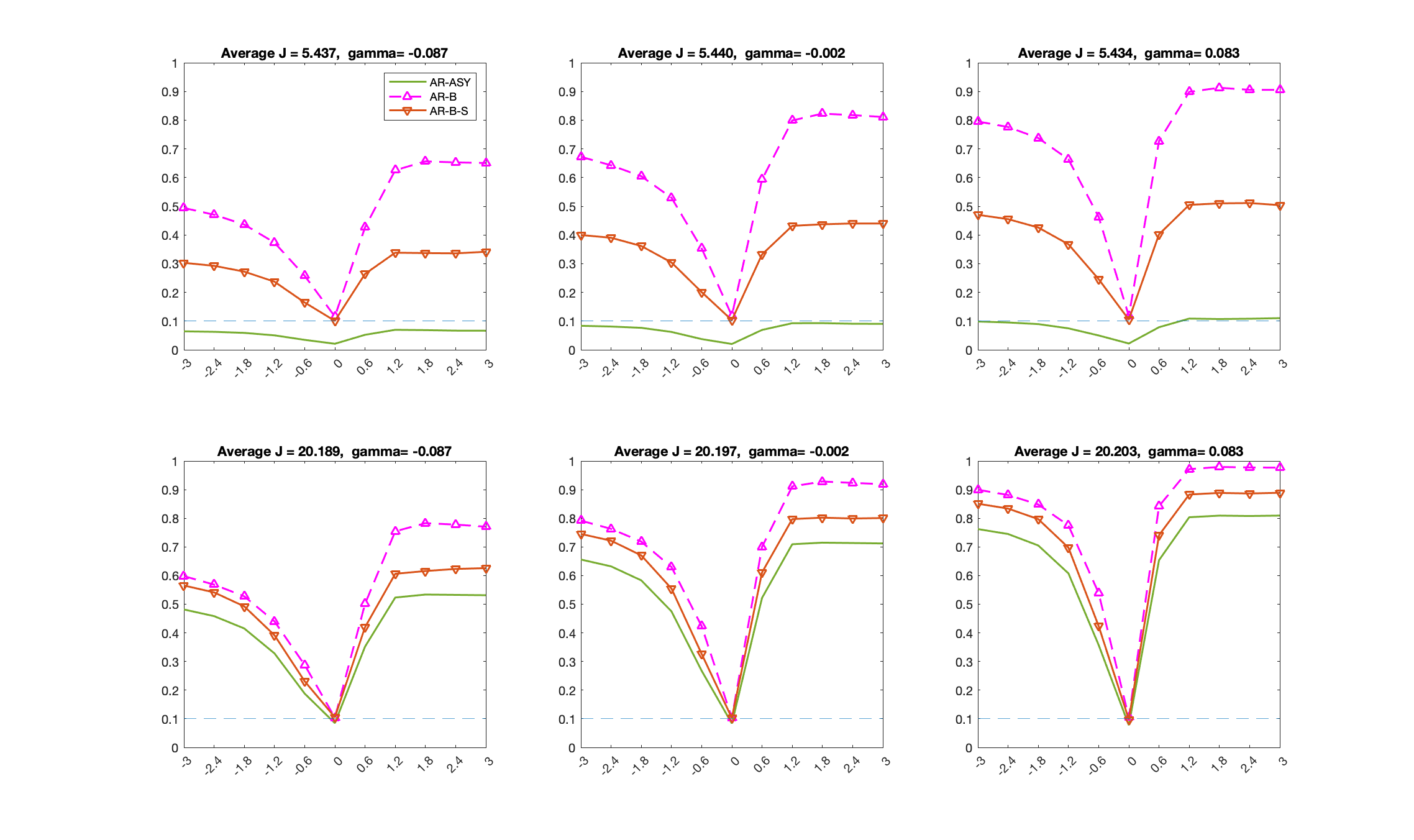}}
	\vspace*{-12mm}
	\caption{Power Comparison for AR Tests in DGP 2 with Average $J = 5.437$ or $20.189$}
	\label{fig: power-dgp2-ar-J520}
	{\footnotesize{Note: 
			AR-ASY: dark green solid line; AR-B: magenta dashed line with upward-pointing triangle; AR-B-S: gold solid line with downward-pointing triangle.}}
\end{figure}

\begin{figure}[H] 
	\vspace{-2cm}
	\makebox[\textwidth]{\includegraphics[width=1\paperwidth,height=0.5\textheight]{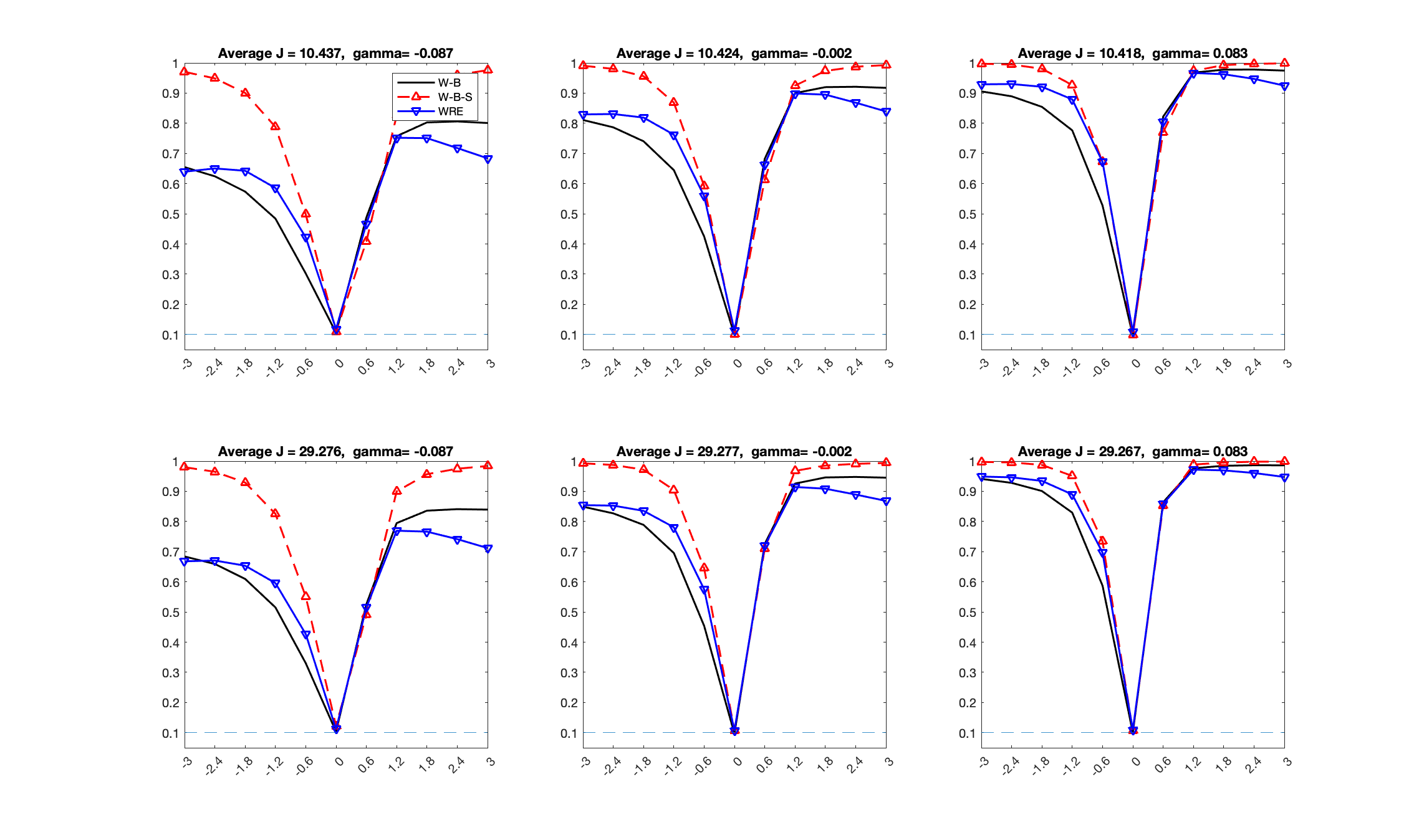}}
	\vspace*{-12mm}
	\caption{Power Comparison for Wald Tests in DGP 2 with Average $J = 10.437$ or $29.276$}
	\label{fig: power-dgp2-wald-J1030}
	{\footnotesize{Note: 
			W-B: dark solid line; W-B-S: red dashed line with upward-pointing triangle; WRE: blue solid line with downward-pointing triangle. }}
	
	\makebox[\textwidth]{\includegraphics[width=1\paperwidth,height=0.5\textheight]{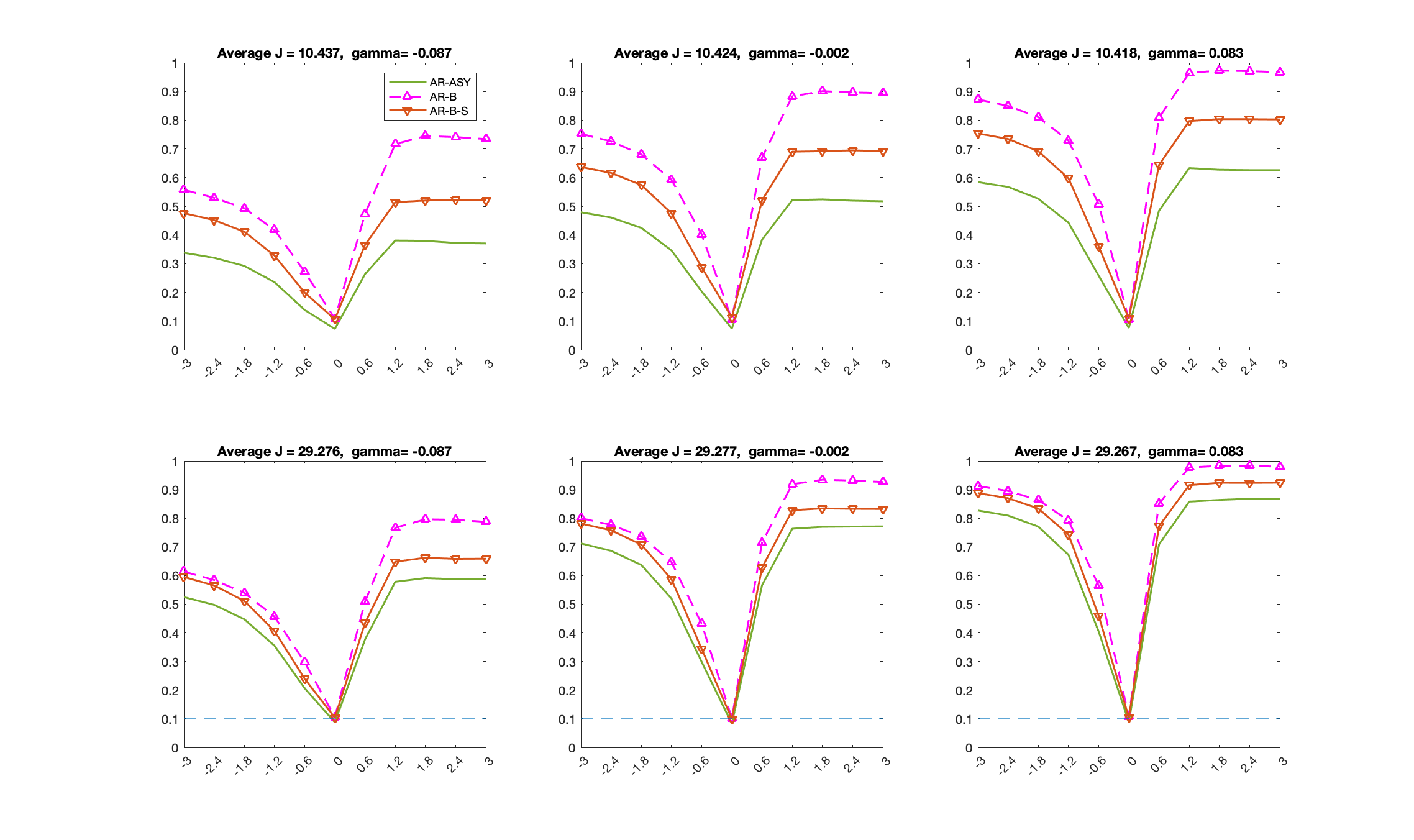}}
	\vspace*{-12mm}
	\caption{Power Comparison for AR Tests in DGP 2 with Average $J = 10.437$ or $29.276$}
	\label{fig: power-dgp2-ar-J1030}
	{\footnotesize{Note: 
			AR-ASY: dark green solid line; AR-B: magenta dashed line with upward-pointing triangle; AR-B-S: gold solid line with downward-pointing triangle.}}
\end{figure}

In this section, we present the power results for DGP 2. 
Specifically, Figures \ref{fig: power-dgp2-wald-J520} and \ref{fig: power-dgp2-ar-J520} present the results for the Wald and AR tests, respectively, with average $J$ equal to 5.437 or 20.189.
In addition, Figures \ref{fig: power-dgp2-wald-J1030} and \ref{fig: power-dgp2-ar-J1030} present those with average $J$ equal to 10.437 or 29.276.

\subsection{Simulation Results for DGPs 3 and 4}\label{subsec: further-simu-DGP34}

In this section, we introduce two more simulation designs: DGPs 3 and 4. 

\noindent \textbf{DGP 3.} We consider a time series IV model. 
\begin{align*}
y_t &= \gamma + \beta X_t + \eps_t, \; \eps_t = \rho_{_{TS}} \eps_{t-1} + u_t,\\
X_t &= \Pi \iota_{d_z}^{\top}Z_t + v_t, \; Z_t = \rho_{_{TS}} Z_{t-1} + w_t, \; w_t \sim N(0, I_{d_z}),\\
\begin{pmatrix}
u_t \\
v_t
\end{pmatrix}
& \sim N \left( \begin{pmatrix}
0 \\
0
\end{pmatrix}, 
\frac{1}{1-\rho^2_{_{TS}}} \begin{pmatrix}
1 & 0.7 \\
0.7 & 1
\end{pmatrix}
\right),
\end{align*}
where $\iota_{d_z}$ is a $d_z$-dimensional vector of ones. 
We vary the IV strength $\Pi \in \{0.25, 0.375, 0.5\}$, 
the number of IVs $d_z \in \{1,5\}$, and the AR(1) parameter $\rho_{_{TS}} \in \{0.3, 0.5, 0.7\}$. We let $\beta=1$ and $\gamma =0$.  
The number of observations is set at 160.
The simulation design is similar to the one for time series data considered by \cite{Ibragimov-Muller(2010)}, \cite{Bester-Conley-Hansen(2011)},  and \cite{Canay-Romano-Shaikh(2017)}. 
We divide the data into $J$ consecutive blocks (clusters) of equal size, where $J \in  \{8, 10, 16\}$. 
\vspace{0.1in}

\noindent \textbf{DGP 4.} We consider a spatial IV model. 
\begin{align*}
y_i &= \gamma + \beta X_i + u_i, \; X_i = \Pi \iota^{\top}_{d_z} Z_i + v_i,
\end{align*}
where $\beta = 1$ and $\gamma = 0$. 
The number of observations is 160 and the observations lie within a square of dimension $40 \times 40$. The location for each observation $s_i = (s_{i_1}, s_{i_2})$ are drawn uniformly within the square, i.e., 
we let $s_{i_1} \sim U[0, 40]$ independently of $s_{i_2} \sim U[0, 40]$. 
The distance between observations at locations $s_i$ and $s_j$ is Euclidean: 
$d_{ij} = \sqrt{(s_{i_1} - s_{j_1})^2 + (s_{i_2} - s_{j_2})^2}$.
The simulation design is similar to those considered in 
\cite{Bester-Conley-Hansen(2011)}, \cite{sun2015}, and \cite{lee2016}. The errors $(u_i, v_i)^{\top}$ are drawn from
$N\left(\begin{pmatrix}
0 \\
0
\end{pmatrix}, 
\begin{pmatrix}
1 & 0.7 \\
0.7 & 1
\end{pmatrix}\right)$. 
In addition, $u_i$ and $u_j$ have a correlation equal to $\rho_{_{SP}}^{d_{ij}}$ between them. 
Similarly, $v_i$ and $v_j$ have a correlation equal to $\rho_{_{SP}}^{d_{ij}}$. 
Furthermore, the IVs $Z_i$ are drawn from $N(0, I_{d_z})$, with a correlation $\rho_{_{SP}}^{d_{ij}}$ between each element of $Z_i$ and its corresponding element of $Z_j$.
Therefore, $\rho_{SP}$ controls the degree of spatial dependence and we let $\rho_{_{SP}} \in \{0.3, 0.5, 0.7\}$.
We vary the IV strength $\Pi \in \{0.25, 0.375, 0.5\}$ and the number of IVs $d_z \in \{1,5\}$.
To implement inference, we split the data into $J$ groups (clusters) with equal area, where $J \in \{6, 9, 12\}$.\footnote{Specifically, for $J=9$, we split the data into 9 groups formed by splitting into three equal parts both vertically and horizontally. For $J=6$, we split into two equal parts vertically and three horizontally, while for $J=12$, we split into three equal parts vertically and four horizontally.}
\vspace{0.1in}

\begin{table}[H]
	\adjustbox{max width=\textwidth}{%
		\centering
		\begin{tabular}{ll|lll|lll|lll}
			&        &         & $\rho_{_{\eps v}} = 0.3$  &        &         & $\rho_{_{\eps v}} = 0.5$  &        &         & $\rho_{_{\eps v}} = 0.7$  &        \\
			&        & $\Pi=0.25$ & $\Pi=0.375$ & $\Pi=0.5$ & $\Pi=0.25$ & $\Pi=0.375$ & $\Pi=0.5$ & $\Pi=0.25$ & $\Pi=0.375$ & $\Pi=0.5$ \\
			\hline
			$J=6$  & IM     & 0.063 & 0.054 & 0.048 & 0.096 & 0.079 & 0.068 & 0.132 & 0.112 & 0.095 \\
			& CRS    & 0.112 & 0.111 & 0.102 & 0.131 & 0.128 & 0.125 & 0.132 & 0.124 & 0.122 \\
			& ASY    & 0.206 & 0.234 & 0.252 & 0.221 & 0.247 & 0.256 & 0.228 & 0.245 & 0.257 \\
			& BCH    & 0.141 & 0.164 & 0.176 & 0.155 & 0.178 & 0.183 & 0.174 & 0.178 & 0.189 \\
			& WRE    & 0.093 & 0.088 & 0.085 & 0.095 & 0.094 & 0.091 & 0.099 & 0.094 & 0.096 \\
			& W-B    & 0.079 & 0.081 & 0.088 & 0.076 & 0.087 & 0.092 & 0.078 & 0.081 & 0.087 \\
			& W-B-S  & 0.068 & 0.064 & 0.066 & 0.072 & 0.086 & 0.090 & 0.112 & 0.103 & 0.109 \\
			& AR-ASY & 0.068 & 0.065 & 0.071 & 0.071 & 0.070 & 0.067 & 0.071 & 0.071 & 0.070 \\
			& AR-B   & 0.112 & 0.108 & 0.109 & 0.114 & 0.119 & 0.120 & 0.118 & 0.117 & 0.119 \\
			& AR-B-S & 0.112 & 0.108 & 0.109 & 0.114 & 0.119 & 0.120 & 0.118 & 0.117 & 0.119 \\
			\hline
			$J=12$ & IM     & 0.076 & 0.069 & 0.068 & 0.114 & 0.105 & 0.092 & 0.187 & 0.149 & 0.136 \\
			& CRS    & 0.125 & 0.122 & 0.119 & 0.136 & 0.132 & 0.132 & 0.132 & 0.130 & 0.125 \\
			& ASY    & 0.120 & 0.148 & 0.164 & 0.140 & 0.162 & 0.165 & 0.163 & 0.164 & 0.173 \\
			& BCH    & 0.097 & 0.119 & 0.135 & 0.116 & 0.132 & 0.132 & 0.143 & 0.141 & 0.145 \\
			& WRE    & 0.096 & 0.094 & 0.102 & 0.096 & 0.101 & 0.096 & 0.101 & 0.097 & 0.100 \\
			& W-B    & 0.072 & 0.077 & 0.092 & 0.082 & 0.082 & 0.085 & 0.080 & 0.074 & 0.082 \\
			& W-B-S  & 0.062 & 0.071 & 0.076 & 0.077 & 0.083 & 0.079 & 0.109 & 0.101 & 0.096 \\
			& AR-ASY & 0.089 & 0.088 & 0.089 & 0.088 & 0.094 & 0.085 & 0.090 & 0.090 & 0.088 \\
			& AR-B   & 0.104 & 0.098 & 0.103 & 0.102 & 0.106 & 0.099 & 0.102 & 0.103 & 0.101 \\
			& AR-B-S & 0.104 & 0.098 & 0.103 & 0.102 & 0.106 & 0.099 & 0.102 & 0.103 & 0.101 \\
	\end{tabular}}
	\caption{Size Comparison for DGP 1 (homoskedastic $\Pi$) with $d_z=1$ and $J=6$ or $12$}
	\label{tab:dgp1_size_1_homo_J612}
	{\footnotesize{Note: IM, CRS, ASY, BCH, WRE, W-B, W-B-S, AR-ASY, AR-B, and AR-B-S denote the ten inference methods described in Section \ref{sec: inf-method}. $J$ denotes the number of clusters. The nominal level is 10\%. }}
\end{table}

\begin{table}[H]
	\adjustbox{max width=\textwidth}{%
		\centering
		\begin{tabular}{ll|lll|lll|lll}
			&        &         & $\rho_{_{\eps v}} = 0.3$  &        &         & $\rho_{_{\eps v}} = 0.5$  &        &         & $\rho_{_{\eps v}} = 0.7$  &        \\
			&        & $\Pi=0.25$ & $\Pi=0.375$ & $\Pi=0.5$ & $\Pi=0.25$ & $\Pi=0.375$ & $\Pi=0.5$ & $\Pi=0.25$ & $\Pi=0.375$ & $\Pi=0.5$ \\
			\hline
			$J=9$  & IM     & 0.071 & 0.065 & 0.063 & 0.108 & 0.096 & 0.082 & 0.173 & 0.138 & 0.124 \\
			& CRS    & 0.128 & 0.121 & 0.116 & 0.138 & 0.132 & 0.126 & 0.135 & 0.126 & 0.127 \\
			& ASY    & 0.147 & 0.176 & 0.193 & 0.163 & 0.189 & 0.192 & 0.180 & 0.189 & 0.193 \\
			& BCH    & 0.111 & 0.136 & 0.150 & 0.127 & 0.150 & 0.151 & 0.145 & 0.152 & 0.154 \\
			& WRE    & 0.089 & 0.094 & 0.096 & 0.094 & 0.097 & 0.096 & 0.092 & 0.100 & 0.092 \\
			& W-B    & 0.073 & 0.079 & 0.091 & 0.078 & 0.081 & 0.088 & 0.077 & 0.074 & 0.082 \\
			& W-B-S  & 0.067 & 0.068 & 0.070 & 0.073 & 0.081 & 0.084 & 0.105 & 0.101 & 0.095 \\
			& AR-ASY & 0.079 & 0.078 & 0.083 & 0.083 & 0.085 & 0.081 & 0.088 & 0.086 & 0.085 \\
			& AR-B   & 0.101 & 0.097 & 0.102 & 0.100 & 0.106 & 0.101 & 0.103 & 0.102 & 0.100 \\
			& AR-B-S & 0.101 & 0.097 & 0.102 & 0.100 & 0.106 & 0.101 & 0.103 & 0.102 & 0.100 \\
			\hline
			$J=20$ & IM     & 0.079 & 0.075 & 0.074 & 0.132 & 0.110 & 0.100 & 0.205 & 0.176 & 0.155 \\
			& CRS    & 0.130 & 0.123 & 0.122 & 0.138 & 0.133 & 0.131 & 0.134 & 0.131 & 0.129 \\
			& ASY    & 0.098 & 0.119 & 0.138 & 0.110 & 0.121 & 0.144 & 0.145 & 0.134 & 0.139 \\
			& BCH    & 0.086 & 0.104 & 0.119 & 0.097 & 0.105 & 0.128 & 0.130 & 0.122 & 0.124 \\
			& WRE    & 0.100 & 0.099 & 0.100 & 0.094 & 0.090 & 0.105 & 0.097 & 0.096 & 0.095 \\
			& W-B    & 0.080 & 0.080 & 0.093 & 0.075 & 0.072 & 0.089 & 0.082 & 0.065 & 0.076 \\
			& W-B-S  & 0.069 & 0.069 & 0.080 & 0.077 & 0.074 & 0.093 & 0.107 & 0.098 & 0.095 \\
			& AR-ASY & 0.098 & 0.094 & 0.097 & 0.097 & 0.091 & 0.096 & 0.097 & 0.084 & 0.093 \\
			& AR-B   & 0.107 & 0.100 & 0.105 & 0.105 & 0.096 & 0.104 & 0.107 & 0.090 & 0.099 \\
			& AR-B-S & 0.107 & 0.100 & 0.105 & 0.105 & 0.096 & 0.104 & 0.107 & 0.090 & 0.099
	\end{tabular}}
	\caption{Size Comparison for DGP 1 (homoskedastic $\Pi$) with $d_z=1$ and $J=9$ or $20$}
	\label{tab:dgp1_size_1_homo_J920}
	{\footnotesize{Note: IM, CRS, ASY, BCH, WRE, W-B, W-B-S, AR-ASY, AR-B, and AR-B-S denote the ten inference methods described in Section \ref{sec: inf-method}. $J$ denotes the number of clusters. The nominal level is 10\%. }}
\end{table}

\begin{table}[H]
	\adjustbox{max width=\textwidth}{%
		\centering
		\begin{tabular}{ll|lll|lll|lll}
			&        &         & $\rho_{_{\eps v}} = 0.3$  &        &         & $\rho_{_{\eps v}} = 0.5$  &        &         & $\rho_{_{\eps v}} = 0.7$  &        \\
			&        & $\Pi=0.25$ & $\Pi=0.375$ & $\Pi=0.5$ & $\Pi=0.25$ & $\Pi=0.375$ & $\Pi=0.5$ & $\Pi=0.25$ & $\Pi=0.375$ & $\Pi=0.5$ \\
			\hline
			$J=6$  & IM     & 0.067 & 0.058 & 0.053 & 0.083 & 0.074 & 0.061 & 0.127 & 0.099 & 0.074 \\
			& CRS    & 0.113 & 0.108 & 0.106 & 0.128 & 0.127 & 0.113 & 0.184 & 0.159 & 0.137 \\
			& ASY    & 0.266 & 0.268 & 0.274 & 0.258 & 0.263 & 0.276 & 0.259 & 0.270 & 0.278 \\
			& BCH    & 0.192 & 0.192 & 0.199 & 0.184 & 0.189 & 0.197 & 0.188 & 0.194 & 0.198 \\
			& WRE    & 0.095 & 0.085 & 0.080 & 0.085 & 0.087 & 0.084 & 0.097 & 0.093 & 0.084 \\
			& W-B    & 0.088 & 0.086 & 0.090 & 0.088 & 0.089 & 0.096 & 0.094 & 0.092 & 0.097 \\
			& W-B-S  & 0.069 & 0.075 & 0.082 & 0.087 & 0.092 & 0.099 & 0.111 & 0.111 & 0.104 \\
			& AR-ASY & 0.000 & 0.000 & 0.000 & 0.000 & 0.000 & 0.000 & 0.000 & 0.000 & 0.000 \\
			& AR-B   & 0.124 & 0.113 & 0.114 & 0.115 & 0.115 & 0.117 & 0.114 & 0.117 & 0.118 \\
			& AR-B-S & 0.125 & 0.119 & 0.117 & 0.113 & 0.118 & 0.116 & 0.116 & 0.117 & 0.115 \\
			\hline
			$J=12$ & IM     & 0.083 & 0.080 & 0.080 & 0.122 & 0.103 & 0.088 & 0.215 & 0.157 & 0.125 \\
			& CRS    & 0.111 & 0.107 & 0.108 & 0.146 & 0.131 & 0.120 & 0.246 & 0.186 & 0.163 \\
			& ASY    & 0.170 & 0.188 & 0.192 & 0.177 & 0.178 & 0.184 & 0.180 & 0.185 & 0.195 \\
			& BCH    & 0.139 & 0.153 & 0.159 & 0.145 & 0.149 & 0.154 & 0.151 & 0.154 & 0.159 \\
			& WRE    & 0.097 & 0.100 & 0.099 & 0.097 & 0.100 & 0.097 & 0.095 & 0.098 & 0.102 \\
			& W-B    & 0.092 & 0.098 & 0.099 & 0.099 & 0.099 & 0.099 & 0.094 & 0.096 & 0.098 \\
			& W-B-S  & 0.076 & 0.087 & 0.093 & 0.088 & 0.094 & 0.093 & 0.104 & 0.100 & 0.104 \\
			& AR-ASY & 0.035 & 0.031 & 0.036 & 0.033 & 0.033 & 0.031 & 0.033 & 0.031 & 0.036 \\
			& AR-B   & 0.102 & 0.103 & 0.103 & 0.102 & 0.104 & 0.104 & 0.097 & 0.105 & 0.100 \\
			& AR-B-S & 0.103 & 0.093 & 0.104 & 0.106 & 0.102 & 0.102 & 0.096 & 0.099 & 0.104 \\
	\end{tabular}}
	\caption{Size Comparison for DGP 1 (homoskedastic $\Pi$) with $d_z=3$ and $J=6$ or $12$}
	\label{tab:dgp1_size_2_homo_J612}
	{\footnotesize{Note: IM, CRS, ASY, BCH, WRE, W-B, W-B-S, AR-ASY, AR-B, and AR-B-S denote the ten inference methods described in Section \ref{sec: inf-method}. $J$ denotes the number of clusters. The nominal level is 10\%. }}
\end{table}

\begin{table}[H]
	\adjustbox{max width=\textwidth}{%
		\centering
		\begin{tabular}{ll|lll|lll|lll}
			&        &         & $\rho_{_{\eps v}} = 0.3$  &        &         & $\rho_{_{\eps v}} = 0.5$  &        &         & $\rho_{_{\eps v}} = 0.7$  &        \\
			&        & $\Pi=0.25$ & $\Pi=0.375$ & $\Pi=0.5$ & $\Pi=0.25$ & $\Pi=0.375$ & $\Pi=0.5$ & $\Pi=0.25$ & $\Pi=0.375$ & $\Pi=0.5$ \\
			\hline
			$J=9$  & IM     & 0.078 & 0.070 & 0.071 & 0.100 & 0.090 & 0.077 & 0.166 & 0.134 & 0.111 \\
			& CRS    & 0.105 & 0.102 & 0.105 & 0.135 & 0.124 & 0.115 & 0.203 & 0.180 & 0.156 \\
			& ASY    & 0.204 & 0.208 & 0.217 & 0.199 & 0.200 & 0.212 & 0.202 & 0.205 & 0.217 \\
			& BCH    & 0.155 & 0.160 & 0.170 & 0.157 & 0.159 & 0.164 & 0.160 & 0.166 & 0.171 \\
			& WRE    & 0.097 & 0.093 & 0.097 & 0.096 & 0.090 & 0.100 & 0.095 & 0.097 & 0.100 \\
			& W-B    & 0.093 & 0.093 & 0.099 & 0.095 & 0.089 & 0.101 & 0.092 & 0.098 & 0.100 \\
			& W-B-S  & 0.072 & 0.076 & 0.087 & 0.085 & 0.085 & 0.097 & 0.101 & 0.099 & 0.103 \\
			& AR-ASY & 0.018 & 0.018 & 0.017 & 0.020 & 0.016 & 0.017 & 0.017 & 0.016 & 0.018 \\
			& AR-B   & 0.104 & 0.103 & 0.103 & 0.101 & 0.099 & 0.104 & 0.102 & 0.106 & 0.105 \\
			& AR-B-S & 0.104 & 0.101 & 0.099 & 0.107 & 0.102 & 0.101 & 0.104 & 0.099 & 0.103 \\
			\hline
			$J=20$ & IM     & 0.092 & 0.085 & 0.081 & 0.138 & 0.113 & 0.091 & 0.237 & 0.170 & 0.132 \\
			& CRS    & 0.109 & 0.103 & 0.104 & 0.154 & 0.134 & 0.113 & 0.258 & 0.191 & 0.156 \\
			& ASY    & 0.135 & 0.150 & 0.161 & 0.140 & 0.155 & 0.150 & 0.152 & 0.141 & 0.162 \\
			& BCH    & 0.116 & 0.129 & 0.138 & 0.124 & 0.136 & 0.132 & 0.134 & 0.124 & 0.139 \\
			& WRE    & 0.097 & 0.100 & 0.106 & 0.095 & 0.104 & 0.097 & 0.097 & 0.093 & 0.107 \\
			& W-B    & 0.088 & 0.097 & 0.099 & 0.093 & 0.098 & 0.097 & 0.097 & 0.089 & 0.102 \\
			& W-B-S  & 0.076 & 0.086 & 0.092 & 0.087 & 0.094 & 0.091 & 0.101 & 0.093 & 0.104 \\
			& AR-ASY & 0.052 & 0.057 & 0.052 & 0.056 & 0.059 & 0.052 & 0.055 & 0.053 & 0.059 \\
			& AR-B   & 0.096 & 0.103 & 0.103 & 0.100 & 0.105 & 0.101 & 0.105 & 0.098 & 0.107 \\
			& AR-B-S & 0.097 & 0.105 & 0.100 & 0.100 & 0.105 & 0.101 & 0.099 & 0.101 & 0.107
	\end{tabular}}
	\caption{Size Comparison for DGP 1 (homoskedastic $\Pi$) with $d_z=3$ and $J=9$ or $20$}
	\label{tab:dgp1_size_2_homo_J920}
	{\footnotesize{Note: IM, CRS, ASY, BCH, WRE, W-B, W-B-S, AR-ASY, AR-B, and AR-B-S denote the ten inference methods described in Section \ref{sec: inf-method}. $J$ denotes the number of clusters. The nominal level is 10\%. }}
\end{table}

\begin{figure}[H] 
	\makebox[\textwidth]{\includegraphics[width=1\paperwidth,height=0.6\textheight]{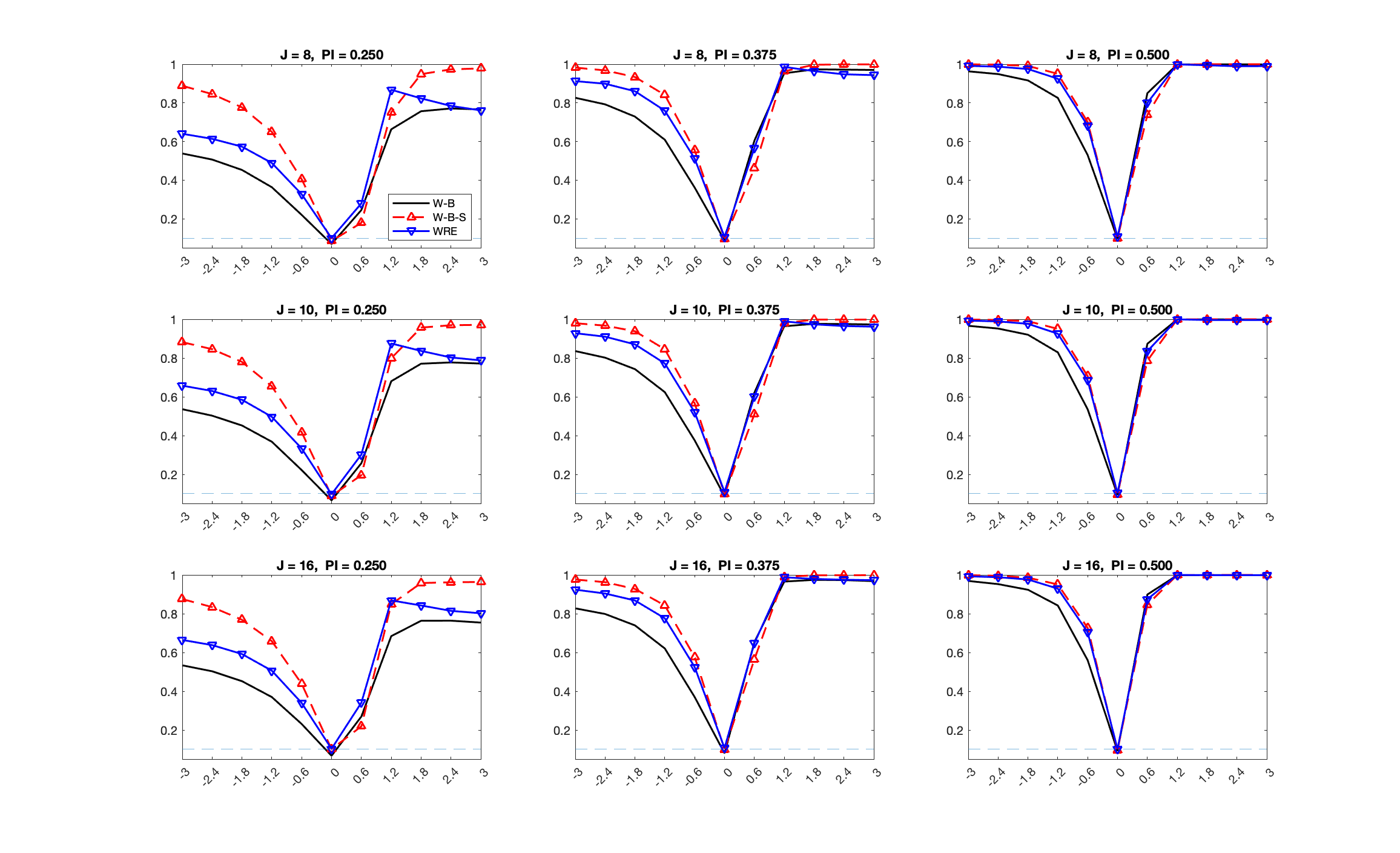}}
	\vspace*{-12mm}
	\caption{Power Comparison for Wald Tests under DGP 3 with $d_z=1$}
	\label{fig:power-dgp2-wald}
	{\footnotesize{Note: W-B: dark solid line; W-B-S: red dashed line with upward-pointing triangle; WRE: blue solid line with downward-pointing triangle.}}
\end{figure}

\begin{figure}[H] 
	\makebox[\textwidth]{\includegraphics[width=1\paperwidth,height=0.6\textheight]{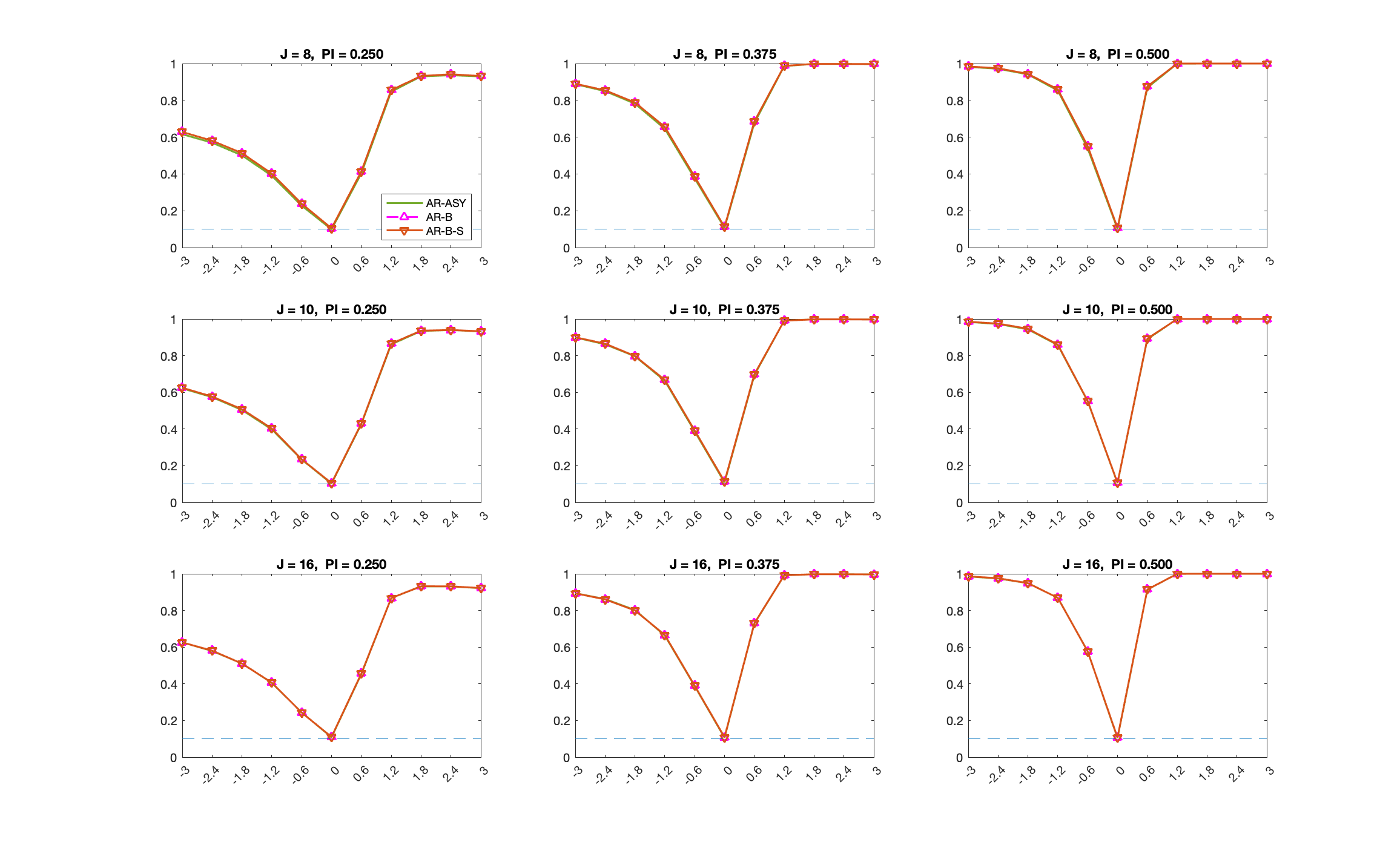}}
	\vspace*{-12mm}
	\caption{Power Comparison for AR Tests under DGP 3 with $d_z=1$}
	\label{fig:power-dgp2-ar}
	{\footnotesize{Note: AR-ASY: dark green solid line; AR-B: magenta dashed line with upward-pointing triangle; AR-B-S: gold solid line with downward-pointing triangle. }}
\end{figure}

\noindent \textbf{Results for DGP 3.}
Tables \ref{tab:dgp2_size_1}-\ref{tab:dgp2_size_2} report the size results for the time series IV model with $d_z=1$ and $d_z=5$, respectively. 
We observe that the standard Wald test (ASY) does not control size when the number of clusters is small, especially in the over-identified case.  
For example, when $d_z=5$, $J=8$, and $\Pi=0.25$, its null rejection frequencies are $0.183, 0.196$, and $0.251$ for $\rho_{TS} = 0.3, 0.5,$ and $0.7$, respectively.
BCH improves upon ASY but still has null rejection frequencies equal to $0.135$, $0.143$, and $0.186$, respectively, under such settings. 
On the other hand, IM and CRS typically have large size distortion when the number of clusters is large. 
For example, when $d_z=5$, $J=16$, and $\Pi=0.25$, the null rejection frequencies for IM are 
$0.259, 0.259,$ and $0.249$, for $\rho_{TS} = 0.3, 0.5,$ and $0.7$, respectively, while those for CRS are 
$0.267, 0.266,$ and $0.256$. 
By contrast, the wild bootstrap-based Wald inference methods (WRE, W-B, W-B-S) have good size control regardless of the number of clusters and the number of IVs. Therefore, we focus on the bootstrap methods when comparing power. Figures \ref{fig:power-dgp2-wald} and \ref{fig:power-dgp2-ar} report the power properties of the Wald and AR tests, respectively, with $d_z=1$. We find that the power curves are very similar to those in DGPs 1 and 2. Overall, our W-B-S has the best power, especially against distant alternatives. 
\vspace{0.1in}

\begin{table}[H]
	\adjustbox{max width=\textwidth}{%
		\centering
		\begin{tabular}{ll|lll|lll|lll}
			&        &         & $\rho_{_{TS}} = 0.3$  &        &         & $\rho_{_{TS}} = 0.5$  &        &         & $\rho_{_{TS}} = 0.7$  &        \\
			&        & $\Pi=0.25$ & $\Pi=0.375$ & $\Pi=0.5$ & $\Pi=0.25$ & $\Pi=0.375$ & $\Pi=0.5$ & $\Pi=0.25$ & $\Pi=0.375$ & $\Pi=0.5$ \\
			\hline
			$J = 8$ & IM & 0.084 & 0.062 & 0.070 & 0.088 & 0.065 & 0.071 & 0.085 & 0.069 & 0.070 \\
			& CRS & 0.144 & 0.104 & 0.103 & 0.150 & 0.125 & 0.125 & 0.148 & 0.122 & 0.122 \\
			& ASY & 0.143 & 0.155 & 0.161 & 0.147 & 0.159 & 0.170 & 0.151 & 0.172 & 0.180 \\
			& BCH & 0.109 & 0.112 & 0.115 & 0.110 & 0.116 & 0.122 & 0.112 & 0.125 & 0.130 \\
			& WRE & 0.100 & 0.103 & 0.101 & 0.103 & 0.102 & 0.104 & 0.106 & 0.108 & 0.107 \\
			& W-B & 0.070 & 0.085 & 0.094 & 0.071 & 0.085 & 0.095 & 0.067 & 0.085 & 0.094 \\
			& W-B-S & 0.096 & 0.095 & 0.096 & 0.087 & 0.092 & 0.095 & 0.080 & 0.081 & 0.089 \\
			& AR-ASY & 0.100 & 0.100 & 0.100 & 0.100 & 0.100 & 0.100 & 0.100 & 0.100 & 0.100 \\
			& AR-B & 0.105 & 0.104 & 0.104 & 0.108 & 0.107 & 0.106 & 0.114 & 0.113 & 0.110 \\
			& AR-B-S & 0.105 & 0.104 & 0.104 & 0.108 & 0.107 & 0.106 & 0.114 & 0.113 & 0.110 \\
			\hline
			$J = 10$ & IM & 0.100 & 0.064 & 0.065 & 0.100 & 0.068 & 0.066 & 0.095 & 0.073 & 0.069 \\
			& CRS & 0.164 & 0.111 & 0.105 & 0.165 & 0.124 & 0.125 & 0.166 & 0.122 & 0.123 \\
			& ASY & 0.133 & 0.137 & 0.147 & 0.132 & 0.146 & 0.156 & 0.138 & 0.155 & 0.165 \\
			& BCH & 0.107 & 0.105 & 0.113 & 0.106 & 0.112 & 0.121 & 0.108 & 0.120 & 0.128 \\
			& WRE & 0.100 & 0.099 & 0.101 & 0.100 & 0.103 & 0.105 & 0.107 & 0.109 & 0.109 \\
			& W-B & 0.067 & 0.083 & 0.093 & 0.070 & 0.084 & 0.096 & 0.067 & 0.082 & 0.094 \\
			& W-B-S & 0.099 & 0.093 & 0.096 & 0.091 & 0.093 & 0.099 & 0.076 & 0.083 & 0.091 \\
			& AR-ASY & 0.100 & 0.101 & 0.099 & 0.100 & 0.101 & 0.099 & 0.100 & 0.101 & 0.099 \\
			& AR-B & 0.102 & 0.102 & 0.101 & 0.108 & 0.106 & 0.107 & 0.113 & 0.112 & 0.110 \\
			& AR-B-S & 0.102 & 0.102 & 0.101 & 0.108 & 0.106 & 0.107 & 0.113 & 0.112 & 0.110 \\
			\hline
			$J = 16$ & IM & 0.149 & 0.089 & 0.064 & 0.135 & 0.093 & 0.066 & 0.121 & 0.091 & 0.073 \\
			& CRS & 0.216 & 0.139 & 0.107 & 0.208 & 0.148 & 0.121 & 0.207 & 0.148 & 0.120 \\
			& ASY & 0.115 & 0.120 & 0.127 & 0.117 & 0.125 & 0.136 & 0.115 & 0.130 & 0.145 \\
			& BCH & 0.101 & 0.103 & 0.107 & 0.103 & 0.107 & 0.114 & 0.099 & 0.110 & 0.122 \\
			& WRE & 0.098 & 0.102 & 0.102 & 0.103 & 0.106 & 0.108 & 0.110 & 0.109 & 0.113 \\
			& W-B & 0.063 & 0.080 & 0.093 & 0.067 & 0.081 & 0.095 & 0.066 & 0.079 & 0.094 \\
			& W-B-S & 0.103 & 0.099 & 0.098 & 0.101 & 0.099 & 0.102 & 0.090 & 0.091 & 0.098 \\
			& AR-ASY & 0.099 & 0.103 & 0.102 & 0.099 & 0.103 & 0.102 & 0.099 & 0.103 & 0.102 \\
			& AR-B & 0.100 & 0.104 & 0.103 & 0.108 & 0.110 & 0.110 & 0.115 & 0.117 & 0.116 \\
			& AR-B-S & 0.100 & 0.104 & 0.103 & 0.108 & 0.110 & 0.110 & 0.115 & 0.117 & 0.116 \\
	\end{tabular}}
	\caption{Size Comparison for DGP 3 with $d_z=1$}
	\label{tab:dgp2_size_1}
	{\footnotesize{Note: IM, CRS, ASY, BCH, WRE, W-B, W-B-S, AR-ASY, AR-B, and AR-B-S denote the ten inference methods described in Section \ref{sec: inf-method}. $J$ denotes the number of clusters. The nominal level is 10\%. }}
\end{table}

\begin{table}[H]
	\adjustbox{max width=\textwidth}{%
		\centering
		\begin{tabular}{ll|lll|lll|lll}
			&        &         & $\rho_{_{TS}} = 0.3$  &        &         & $\rho_{_{TS}} = 0.5$  &        &         & $\rho_{_{TS}} = 0.7$  &        \\
			&        & $\Pi=0.25$ & $\Pi=0.375$ & $\Pi=0.5$ & $\Pi=0.25$ & $\Pi=0.375$ & $\Pi=0.5$ & $\Pi=0.25$ & $\Pi=0.375$ & $\Pi=0.5$ \\
			\hline
			$J=8$  & IM     & 0.094 & 0.087 & 0.095 & 0.095 & 0.089 & 0.096 & 0.110 & 0.091 & 0.089 \\
			& CRS    & 0.112 & 0.103 & 0.107 & 0.109 & 0.106 & 0.112 & 0.122 & 0.106 & 0.107 \\
			& ASY    & 0.183 & 0.177 & 0.178 & 0.196 & 0.190 & 0.189 & 0.251 & 0.221 & 0.210 \\
			& BCH    & 0.135 & 0.128 & 0.129 & 0.143 & 0.139 & 0.138 & 0.186 & 0.163 & 0.157 \\
			& WRE    & 0.080 & 0.089 & 0.094 & 0.083 & 0.094 & 0.100 & 0.106 & 0.107 & 0.110 \\
			& W-B    & 0.100 & 0.100 & 0.103 & 0.091 & 0.100 & 0.105 & 0.074 & 0.091 & 0.102 \\
			& W-B-S  & 0.100 & 0.100 & 0.101 & 0.092 & 0.100 & 0.104 & 0.078 & 0.091 & 0.102 \\
			& AR-ASY & 0.000 & 0.000 & 0.000 & 0.000 & 0.000 & 0.000 & 0.000 & 0.000 & 0.000 \\
			& AR-B   & 0.106 & 0.108 & 0.106 & 0.113 & 0.113 & 0.115 & 0.124 & 0.124 & 0.122 \\
			& AR-B-S & 0.106 & 0.107 & 0.105 & 0.107 & 0.109 & 0.109 & 0.109 & 0.112 & 0.112 \\
			\hline
			$J=10$  & IM     & 0.113 & 0.089 & 0.091 & 0.121 & 0.089 & 0.089 & 0.141 & 0.100 & 0.093 \\
			& CRS    & 0.129 & 0.105 & 0.108 & 0.132 & 0.104 & 0.105 & 0.150 & 0.110 & 0.107 \\
			& ASY    & 0.171 & 0.167 & 0.163 & 0.184 & 0.179 & 0.173 & 0.246 & 0.213 & 0.201 \\
			& BCH    & 0.133 & 0.128 & 0.124 & 0.144 & 0.138 & 0.135 & 0.194 & 0.166 & 0.158 \\
			& WRE    & 0.079 & 0.089 & 0.095 & 0.084 & 0.095 & 0.098 & 0.113 & 0.111 & 0.112 \\
			& W-B    & 0.100 & 0.104 & 0.102 & 0.095 & 0.104 & 0.103 & 0.085 & 0.098 & 0.106 \\
			& W-B-S  & 0.098 & 0.101 & 0.101 & 0.091 & 0.100 & 0.102 & 0.081 & 0.095 & 0.104 \\
			& AR-ASY & 0.007 & 0.007 & 0.007 & 0.007 & 0.007 & 0.006 & 0.006 & 0.006 & 0.006 \\
			& AR-B   & 0.105 & 0.105 & 0.105 & 0.112 & 0.112 & 0.112 & 0.126 & 0.126 & 0.125 \\
			& AR-B-S & 0.101 & 0.103 & 0.104 & 0.108 & 0.108 & 0.109 & 0.113 & 0.112 & 0.113 \\
			\hline
			$J=16$ & IM     & 0.259 & 0.138 & 0.103 & 0.259 & 0.158 & 0.119 & 0.249 & 0.183 & 0.141 \\
			& CRS    & 0.267 & 0.150 & 0.113 & 0.266 & 0.167 & 0.127 & 0.256 & 0.190 & 0.150 \\
			& ASY    & 0.153 & 0.146 & 0.146 & 0.172 & 0.162 & 0.157 & 0.229 & 0.195 & 0.184 \\
			& BCH    & 0.130 & 0.123 & 0.123 & 0.147 & 0.138 & 0.135 & 0.199 & 0.166 & 0.158 \\
			& WRE    & 0.077 & 0.088 & 0.096 & 0.086 & 0.097 & 0.102 & 0.122 & 0.116 & 0.119 \\
			& W-B    & 0.105 & 0.103 & 0.106 & 0.107 & 0.111 & 0.111 & 0.108 & 0.109 & 0.115 \\
			& W-B-S  & 0.098 & 0.100 & 0.103 & 0.096 & 0.103 & 0.106 & 0.084 & 0.098 & 0.111 \\
			& AR-ASY & 0.055 & 0.055 & 0.056 & 0.057 & 0.055 & 0.055 & 0.055 & 0.056 & 0.058 \\
			& AR-B   & 0.106 & 0.105 & 0.106 & 0.118 & 0.118 & 0.117 & 0.125 & 0.124 & 0.128 \\
			& AR-B-S & 0.105 & 0.105 & 0.105 & 0.114 & 0.111 & 0.113 & 0.121 & 0.120 & 0.124
	\end{tabular}}
	\caption{Size Comparison for DGP 3 with $d_z=5$}
	\label{tab:dgp2_size_2}
	{\footnotesize{Note: IM, CRS, ASY, BCH, WRE, W-B, W-B-S, AR-ASY, AR-B, and AR-B-S denote the ten inference methods described in Section \ref{sec: inf-method}. $J$ denotes the number of clusters. The nominal level is 10\%. }}
\end{table}

\begin{figure}[H] 
	\makebox[\textwidth]{\includegraphics[width=1\paperwidth,height=0.6\textheight]{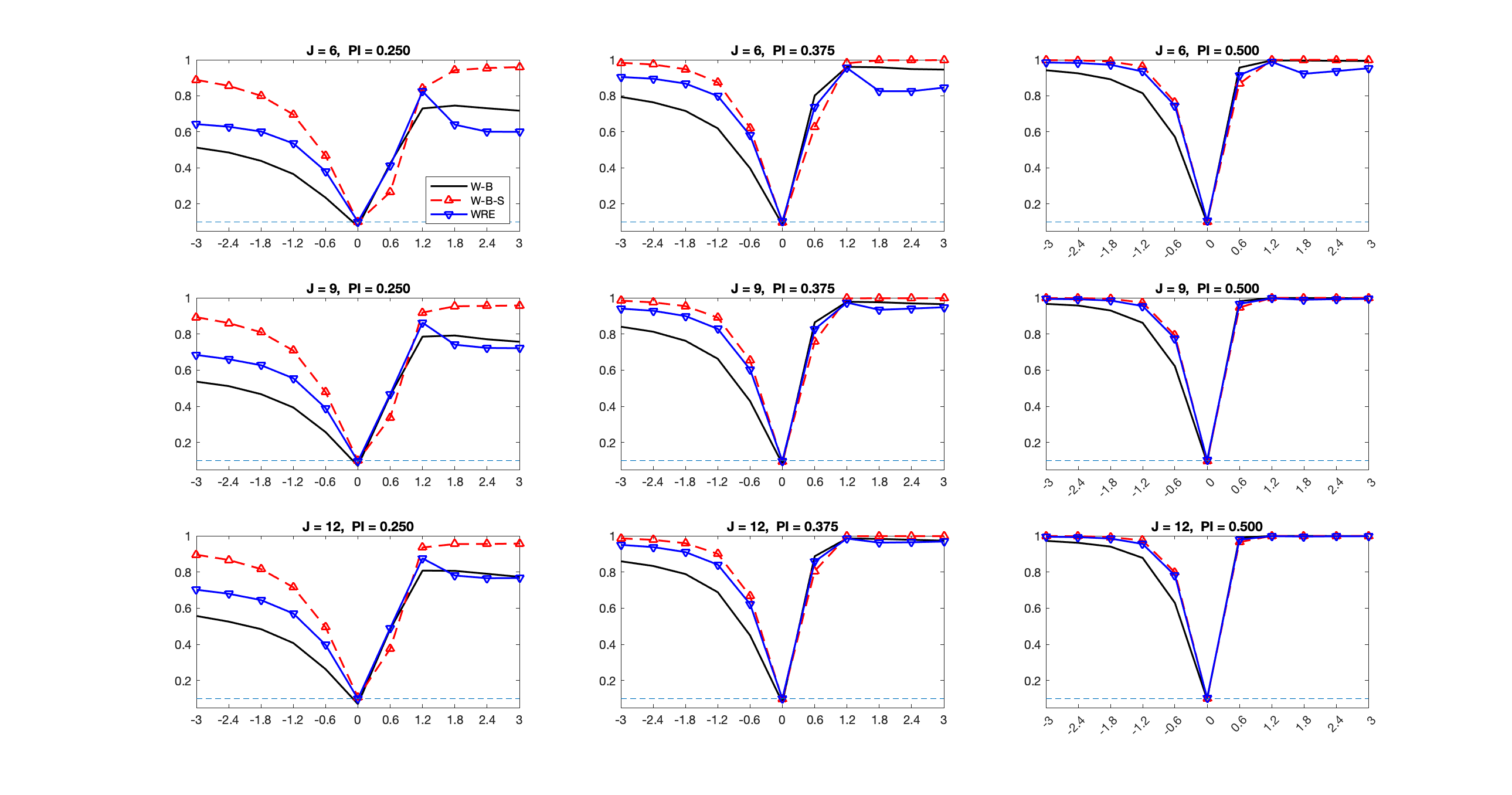}}
	\vspace*{-12mm}
	\caption{Power Comparison for Wald Tests under DGP 4 with $d_z=1$}
	\label{fig:power-dgp3-wald}
	{\footnotesize{Note: W-B: dark solid line; W-B-S: red dashed line with upward-pointing triangle; WRE: blue solid line with downward-pointing triangle.}}
\end{figure}

\begin{figure}[H] 
	\makebox[\textwidth]{\includegraphics[width=1\paperwidth,height=0.6\textheight]{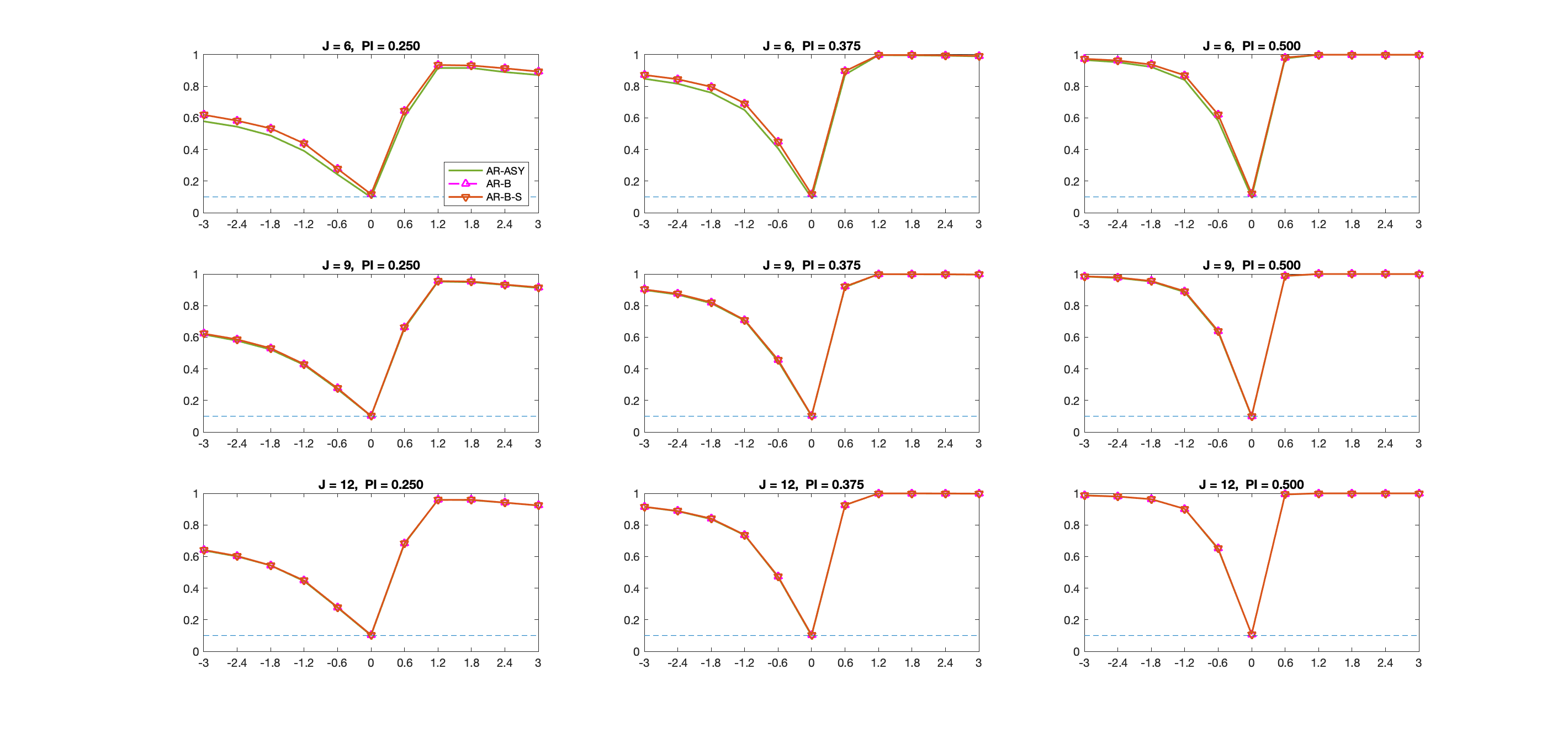}}
	\vspace*{-12mm}
	\caption{Power Comparison for AR Tests under DGP 4 with $d_z=1$}
	\label{fig:power-dgp3-ar}
	{\footnotesize{Note: AR-ASY: dark green solid line; AR-B: magenta dashed line with upward-pointing triangle; AR-B-S: gold solid line with downward-pointing triangle. }}
\end{figure}

\noindent \textbf{Results for DGP 4.}
Tables \ref{tab:dgp3_size_1}-\ref{tab:dgp3_size_2} report the size for the spatial IV model with $d_z=1$ and $d_z=5$, respectively. 
The patterns are similar to those observed in other DGPs.  
ASY and BCH tend to have relatively large size distortions 
when the number of clusters $J$ is small, 
while IM and BCH tend to have large distortions when $J$ is large.
In addition, the size distortions of these four inference methods increase when the number of IVs becomes large ($d_z=5$). 
By contrast, WRE, W-B, and W-B-S have good size control across different settings. 
Figures \ref{fig:power-dgp3-wald} and \ref{fig:power-dgp3-ar} report the power for the Wald and AR tests, respectively. Again, we see that overall W-B-S has the best power among these tests. 

\begin{table}[H]
	\adjustbox{max width=\textwidth}{%
		\centering
		\begin{tabular}{ll|lll|lll|lll}
			&        &         & $\rho_{_{SP}} = 0.3$  &        &         & $\rho_{_{SP}} = 0.5$  &        &         & $\rho_{_{SP}} = 0.7$  &        \\
			&        & $\Pi=0.25$ & $\Pi=0.375$ & $\Pi=0.5$ & $\Pi=0.25$ & $\Pi=0.375$ & $\Pi=0.5$ & $\Pi=0.25$ & $\Pi=0.375$ & $\Pi=0.5$ \\
			\hline
			$J=6$  & IM     & 0.072 & 0.066 & 0.077 & 0.079 & 0.067 & 0.075 & 0.095 & 0.071 & 0.074 \\
			& CRS    & 0.124 & 0.099 & 0.099 & 0.140 & 0.131 & 0.130 & 0.142 & 0.133 & 0.137 \\
			& ASY    & 0.165 & 0.179 & 0.191 & 0.170 & 0.180 & 0.193 & 0.179 & 0.188 & 0.201 \\
			& BCH    & 0.115 & 0.119 & 0.124 & 0.119 & 0.121 & 0.127 & 0.128 & 0.131 & 0.134 \\
			& WRE    & 0.099 & 0.101 & 0.103 & 0.103 & 0.103 & 0.103 & 0.108 & 0.109 & 0.109 \\
			& W-B    & 0.073 & 0.087 & 0.097 & 0.075 & 0.086 & 0.096 & 0.084 & 0.087 & 0.098 \\
			& W-B-S  & 0.095 & 0.095 & 0.099 & 0.098 & 0.097 & 0.098 & 0.106 & 0.103 & 0.103 \\
			& AR-ASY & 0.095 & 0.097 & 0.096 & 0.099 & 0.096 & 0.097 & 0.104 & 0.101 & 0.102 \\
			& AR-B   & 0.114 & 0.116 & 0.117 & 0.121 & 0.118 & 0.118 & 0.127 & 0.124 & 0.125 \\
			& AR-B-S & 0.114 & 0.116 & 0.117 & 0.121 & 0.118 & 0.118 & 0.127 & 0.124 & 0.125 \\
			\hline
			$J=9$  & IM     & 0.099 & 0.064 & 0.066 & 0.107 & 0.064 & 0.064 & 0.129 & 0.073 & 0.065 \\
			& CRS    & 0.163 & 0.112 & 0.104 & 0.165 & 0.128 & 0.130 & 0.163 & 0.127 & 0.132 \\
			& ASY    & 0.140 & 0.146 & 0.157 & 0.143 & 0.151 & 0.160 & 0.153 & 0.154 & 0.171 \\
			& BCH    & 0.110 & 0.111 & 0.117 & 0.114 & 0.116 & 0.122 & 0.124 & 0.120 & 0.130 \\
			& WRE    & 0.100 & 0.101 & 0.103 & 0.101 & 0.104 & 0.105 & 0.107 & 0.106 & 0.110 \\
			& W-B    & 0.067 & 0.084 & 0.095 & 0.070 & 0.082 & 0.094 & 0.079 & 0.081 & 0.098 \\
			& W-B-S  & 0.100 & 0.095 & 0.097 & 0.103 & 0.099 & 0.099 & 0.113 & 0.103 & 0.105 \\
			& AR-ASY & 0.100 & 0.102 & 0.100 & 0.101 & 0.102 & 0.102 & 0.109 & 0.106 & 0.109 \\
			& AR-B   & 0.102 & 0.105 & 0.103 & 0.105 & 0.105 & 0.105 & 0.114 & 0.111 & 0.115 \\
			& AR-B-S & 0.102 & 0.105 & 0.103 & 0.105 & 0.105 & 0.105 & 0.114 & 0.111 & 0.115 \\
			\hline
			$J=12$ & IM     & 0.125 & 0.071 & 0.059 & 0.135 & 0.075 & 0.059 & 0.154 & 0.092 & 0.065 \\
			& CRS    & 0.192 & 0.121 & 0.103 & 0.188 & 0.132 & 0.122 & 0.189 & 0.131 & 0.121 \\
			& ASY    & 0.125 & 0.134 & 0.142 & 0.131 & 0.136 & 0.141 & 0.144 & 0.146 & 0.152 \\
			& BCH    & 0.107 & 0.108 & 0.114 & 0.111 & 0.111 & 0.114 & 0.124 & 0.122 & 0.126 \\
			& WRE    & 0.097 & 0.101 & 0.104 & 0.100 & 0.103 & 0.101 & 0.109 & 0.110 & 0.110 \\
			& W-B    & 0.066 & 0.081 & 0.094 & 0.067 & 0.081 & 0.092 & 0.081 & 0.083 & 0.096 \\
			& W-B-S  & 0.102 & 0.097 & 0.099 & 0.106 & 0.100 & 0.097 & 0.119 & 0.107 & 0.106 \\
			& AR-ASY & 0.102 & 0.101 & 0.103 & 0.103 & 0.104 & 0.102 & 0.112 & 0.112 & 0.112 \\
			& AR-B   & 0.103 & 0.102 & 0.104 & 0.105 & 0.105 & 0.103 & 0.115 & 0.114 & 0.114 \\
			& AR-B-S & 0.103 & 0.102 & 0.104 & 0.105 & 0.105 & 0.103 & 0.115 & 0.114 & 0.114
		\end{tabular}
	}
	\caption{Size Comparison for DGP 4 with $d_z=1$}
	\label{tab:dgp3_size_1}
	{\footnotesize{Note: IM, CRS, ASY, BCH, WRE, W-B, W-B-S, AR-ASY, AR-B, and AR-B-S denote the ten inference methods described in Section \ref{sec: inf-method}. $J$ denotes the number of clusters. The nominal level is 10\%. }}
\end{table}

\begin{table}[H]
	\adjustbox{max width=\textwidth}{%
		\centering
		\begin{tabular}{ll|lll|lll|lll}
			&        &         & $\rho_{_{SP}} = 0.3$  &        &         & $\rho_{_{SP}} = 0.5$  &        &         & $\rho_{_{SP}} = 0.7$  &        \\
			&        & $\Pi=0.25$ & $\Pi=0.375$ & $\Pi=0.5$ & $\Pi=0.25$ & $\Pi=0.375$ & $\Pi=0.5$ & $\Pi=0.25$ & $\Pi=0.375$ & $\Pi=0.5$ \\
			\hline
			$J=6$  & IM     & 0.088 & 0.088 & 0.093 & 0.091 & 0.088 & 0.091 & 0.096 & 0.090 & 0.098 \\
			& CRS    & 0.110 & 0.104 & 0.102 & 0.117 & 0.106 & 0.106 & 0.129 & 0.111 & 0.113 \\
			& ASY    & 0.206 & 0.197 & 0.202 & 0.207 & 0.204 & 0.207 & 0.220 & 0.219 & 0.226 \\
			& BCH    & 0.138 & 0.127 & 0.135 & 0.138 & 0.135 & 0.139 & 0.149 & 0.145 & 0.154 \\
			& WRE    & 0.101 & 0.097 & 0.099 & 0.100 & 0.096 & 0.105 & 0.102 & 0.104 & 0.112 \\
			& W-B    & 0.098 & 0.099 & 0.104 & 0.093 & 0.098 & 0.106 & 0.091 & 0.101 & 0.113 \\
			& W-B-S  & 0.102 & 0.098 & 0.101 & 0.101 & 0.099 & 0.106 & 0.105 & 0.104 & 0.112 \\
			& AR-ASY & 0.000 & 0.000 & 0.000 & 0.000 & 0.000 & 0.000 & 0.000 & 0.000 & 0.000 \\
			& AR-B   & 0.118 & 0.113 & 0.117 & 0.121 & 0.119 & 0.125 & 0.124 & 0.127 & 0.132 \\
			& AR-B-S & 0.120 & 0.119 & 0.118 & 0.112 & 0.119 & 0.117 & 0.118 & 0.118 & 0.120 \\
			\hline
			$J=9$  & IM     & 0.105 & 0.082 & 0.087 & 0.113 & 0.084 & 0.089 & 0.146 & 0.097 & 0.092 \\
			& CRS    & 0.123 & 0.099 & 0.102 & 0.133 & 0.104 & 0.105 & 0.165 & 0.116 & 0.110 \\
			& ASY    & 0.179 & 0.176 & 0.172 & 0.175 & 0.182 & 0.175 & 0.199 & 0.197 & 0.190 \\
			& BCH    & 0.135 & 0.134 & 0.132 & 0.135 & 0.136 & 0.133 & 0.154 & 0.153 & 0.146 \\
			& WRE    & 0.099 & 0.105 & 0.105 & 0.099 & 0.104 & 0.103 & 0.110 & 0.114 & 0.109 \\
			& W-B    & 0.099 & 0.106 & 0.102 & 0.096 & 0.106 & 0.104 & 0.104 & 0.111 & 0.111 \\
			& W-B-S  & 0.100 & 0.104 & 0.103 & 0.098 & 0.105 & 0.101 & 0.108 & 0.110 & 0.109 \\
			& AR-ASY & 0.000 & 0.000 & 0.000 & 0.000 & 0.000 & 0.000 & 0.000 & 0.000 & 0.000 \\
			& AR-B   & 0.104 & 0.109 & 0.108 & 0.111 & 0.113 & 0.105 & 0.122 & 0.126 & 0.122 \\
			& AR-B-S & 0.104 & 0.106 & 0.110 & 0.103 & 0.102 & 0.102 & 0.112 & 0.113 & 0.109 \\
			\hline
			$J=12$ & IM     & 0.156 & 0.096 & 0.089 & 0.173 & 0.099 & 0.087 & 0.212 & 0.122 & 0.096 \\
			& CRS    & 0.169 & 0.113 & 0.106 & 0.186 & 0.115 & 0.100 & 0.226 & 0.136 & 0.112 \\
			& ASY    & 0.162 & 0.166 & 0.158 & 0.170 & 0.164 & 0.161 & 0.187 & 0.177 & 0.182 \\
			& BCH    & 0.132 & 0.135 & 0.125 & 0.138 & 0.131 & 0.130 & 0.154 & 0.142 & 0.151 \\
			& WRE    & 0.103 & 0.107 & 0.103 & 0.101 & 0.104 & 0.103 & 0.114 & 0.107 & 0.116 \\
			& W-B    & 0.103 & 0.106 & 0.102 & 0.103 & 0.103 & 0.102 & 0.112 & 0.105 & 0.113 \\
			& W-B-S  & 0.101 & 0.106 & 0.101 & 0.101 & 0.102 & 0.103 & 0.110 & 0.105 & 0.114 \\
			& AR-ASY & 0.026 & 0.027 & 0.028 & 0.029 & 0.026 & 0.030 & 0.027 & 0.027 & 0.028 \\
			& AR-B   & 0.098 & 0.102 & 0.099 & 0.110 & 0.108 & 0.107 & 0.127 & 0.129 & 0.129 \\
			& AR-B-S & 0.103 & 0.105 & 0.102 & 0.104 & 0.105 & 0.105 & 0.117 & 0.118 & 0.117
		\end{tabular}
	}
	\caption{Size Comparison for DGP 4 with $d_z=5$}
	\label{tab:dgp3_size_2}
	{\footnotesize{Note: IM, CRS, ASY, BCH, WRE, W-B, W-B-S, AR-ASY, AR-B, and AR-B-S denote the ten inference methods described in Section \ref{sec: inf-method}. $J$ denotes the number of clusters. The nominal level is 10\%. }}
\end{table}

\newpage

\newpage

\bibliographystyle{ecta}
\bibliography{Biblio_boot_few_clusters}

\end{document}